\documentclass[12pt,a4paper]{JHEP3}
\usepackage{amsmath,epsfig}
\usepackage{amssymb,amsfonts}
\usepackage{latexsym}
\usepackage[latin1]{inputenc}
\usepackage{slashed}
\usepackage{empheq}
\numberwithin{equation}{section}
\usepackage{cancel}
\usepackage{overpic}
\usepackage{subcaption}
\usepackage{subeqnarray}
\usepackage{xcolor}
\let\normalcolor\relax
\usepackage{longtable}
\usepackage{color}
\usepackage{multirow}
\usepackage{epstopdf}
\usepackage{caption,graphicx,color}
\usepackage{amsmath,epsfig}
\usepackage{amssymb,amsfonts}
\usepackage{latexsym}
\usepackage[latin1]{inputenc}
\usepackage{float}
\usepackage{overpic}
\usepackage{slashed}
\usepackage{caption}
\usepackage{subcaption}
\usepackage{xcolor}
\let\normalcolor\relax
\usepackage{graphicx}
\usepackage{longtable}

\relax
\renewcommand{\theequation}{\arabic{section}.\arabic{equation}}

\def\be{\begin{equation}}
\def\ee{\end{equation}}

\newcommand{\de}{\partial}
\newcommand{\bear}{\begin{eqnarray}}
\newcommand{\bea}{\begin{eqnarray}}
\newcommand{\eear}{\end{eqnarray}}
\newcommand{\eea}{\end{eqnarray}}

\newcommand{\hepsilon}{\hat{\epsilon}}
\def\bsq{\begin{subequations}}
\def\esq{\end{subequations}}
\def\hri#1#2{\href{http://arxiv.org/abs/#1}{[ArXiv:#1]#2}}
\def\hre#1#2{\href{http://arxiv.org/abs/#1/#2}{[ArXiv:#1/#2]}}

\def\hrj#1#2{\href{https://doi.org/#1}{#2}}

\newbox\pippobox

\def\II{\relax{\rm I\kern-.18em I}}

\def\tu{{\tilde{u}}}
\def\tell{{\tilde{\ell}}}

\def\e{\epsilon}
\def\l{\lambda}
\def\m{\mu}
\def\n{\nu}

\def\g{\gamma}

\def\pa{\partial}

\def\sp{\;\;\;,\;\;\;}

\def\p{\partial}

\def\f{\varphi}
\def\z{\zeta}
\def\a{\alpha}
\def\b{\beta}

\def\k{\kappa}

\def\D{\Delta}
\def\nn{\nonumber}

\def\mR{\mathcal{R}}
\def\mb{\mathcal{B}}
\def\mbr{\mathcal{B}(\mR)}
\def\mcr{{C}(\mR)}

 \def\RR{\mathbb{R}}

\title{On holographic confining QFTs on AdS}

\author{ Ahmad Ghodsi$^a$, Elias Kiritsis$^{b,c,d}$ and Francesco Nitti$^{b}$ \\

$^a$
 Department of Physics, Faculty of Science,	Ferdowsi University of Mashhad,  Mashhad, Iran.
~\\

$^b$ \href{http://www.apc.univ-paris7.fr}{Universit\'e Paris Cit\' e, CNRS, Astroparticule et Cosmologie}, F-75013 Paris, France.
~\\

$^c$ \href{http://hep.physics.uoc.gr}{Crete Center for Theoretical Physics}, Institute for Theoretical and Computational Physics,
Department of Physics\\
University of Crete, Heraklion, Greece
~\\

$^d$ \href{https://www.theorie.physik.uni-muenchen.de/}{Arnold Sommerfeld Center for Theoretical Physics}, Ludwig-Maximilians-Universit\"at M\"unchen, 80333
M\"unchen, Germany.
}

\preprint{CCTP-2024-12\\
ITCP-2024/12}

\abstract{
Holographic quantum field theories that confine in flat space, are considered on a fixed $AdS$ space.
The space of holographic solutions for such theories is constructed and three types of regular solutions are found. Theories with two $AdS$ boundaries provide interfaces between two confining theories.
Theories with a single $AdS$ boundary correspond to ground states of a single confining theory on $AdS$. We find solutions without a boundary, whose interpretation is not obvious.
There is also a special limiting solution that oscillates an infinite number of times around the UV fixed point.
We analyze in detail the holographic dictionary for the one-boundary solutions and compute the free energy. No (quantum) phase transitions are found when we change the curvature.
We find an infinite number of pure vev solutions, but no CFT solution without a vev.
We also compute the free energy of the interface solutions. We find that the product saddle points have always lower free energy than the connected solutions. This implies that in such interfaces,  normalized cross-correlators vanish exponentially in $N_c^2$.}

\begin{document}

\maketitle 

\section{Introduction}

We study Quantum Field Theory (QFT) most of the time on flat manifolds, like Minkowski space or tori.
However, there are contexts where we need to consider QFT on curved manifolds, like in cosmology.

When a quantum field theory is placed on a curved manifold, the effects of curvature are negligible at high energy (every smooth manifold is locally flat),  but they may cause drastic qualitative changes to the infrared of the theory. In standard parlance, curvature is relevant in the IR.
For example, conformal theories in flat space become gaped on a positive curvature manifold (as it happens for $N=4$ SYM when space is taken to be a sphere \cite{Witten:1998zw}) and similarly for string theories, \cite{KK}. On the other hand,  massless theories in flat space are expected to behave as massive theories on negative curvature space-times \cite{CW}.

In this work, we shall explore, using the gauge/gravity duality, the fate of holographic confining quantum field theories (QFTs) when placed on $AdS$ space-time. We shall focus on the case when the gravity dual description is modeled in terms of gravity plus a single scalar, whose running drives the confinement dynamics when the theory is defined on flat space.

\subsection{QFT$_d$ on $AdS_d$}

It was argued already in \cite{CW} that placing a  QFT  on $AdS_d$ significantly affects the infrared regime. This is because the Laplacian has a gap, and therefore even massless propagators have an exponential fall-off at large distances.
The authors of \cite{CW} proposed this could be used as an IR regulator for perturbation theory in gauge theories.
More generally, it is expected that  $AdS$ would quench strong IR effects.

One of the features that crucially distinguishes gauge theories on $AdS$ from their Minkowski counterpart, is that the role of boundary conditions (bcs) in $AdS$ is important.  There are essentially two types of boundary conditions for gauge theories  \cite{A2}:
\begin{itemize}
\item  Dirichlet (or electric):
 electric charges (gluons) are part  of the spectrum (as in flat space), they are gaped, and there is a global $SU(N)$ symmetry, which has only boundary currents;

\item Neumann (or magnetic): electric charges are not allowed in the bulk of $AdS$, there are ${\cal O}(1)$ degrees of freedom and there is (boundary condition-induced) confinement.
\end{itemize}

In the case of electric bcs, the expectation is that asymptotically-free gauge theories display a confinement/deconfinement (quantum) phase transition, as a function of $\Lambda L_{AdS}$, \cite{A2}.
Here  $L_{AdS}$ is the $AdS$ radius of curvature while  $\Lambda$ is  the dynamical mass scale of the gauge theory:
\begin{enumerate}
\item  $\Lambda L_{AdS}\gtrsim 1$: ``confined phase", strong interaction in the IR at energies above the $AdS$ mass gap.
\item  $\Lambda L_{AdS}\lesssim 1$: ``deconfined phase", weakly coupled above the $AdS$ mass gap.
\end{enumerate}
It was argued in \cite{A2} that this is a phase transition, but
the details of this putative phase transition are unknown.
We have put quotation marks around confined or deconfined phase, as strictly speaking, there is no known order parameter for confinement in $AdS$: the standard Wilson loop test does not give a clear answer as volume and area scale similarly in $AdS$, \cite{CW}.

With magnetic boundary conditions instead, confinement occurs at all scales, with
 a free energy of ${\cal O}(1)$,  as no electric charges are allowed in the bulk (confinement occurs due to boundary conditions). In this case, no phase transition is expected.

One should emphasize that so far, the only clear criterion for confinement
/deconfinement in $AdS$  exists only  when the rank of the gauge group $N_c\to \infty$: that is, whether
free energy is   ${\cal O}(1)$ (confined phase)  or ${\cal O}(N_c^2)$ (deconfined phase).

\subsection{Exploring confinement in $AdS$ via holography}

Gauge/gravity duality has proven to be a useful laboratory to study curvature effects on strongly coupled  $d$-dimensional quantum field theories (see e.g. \cite{Maldacena:2004rf,Buchel:2002wf,Marolf:2010tg, Blackman:2011in}).  Its stripped-down incarnation, consisting in $(d+1)$-dimensional Einstein-dilaton gravity, has been particularly fruitful in exploring holographic theories on curved $d$-dimensional space-times.  A  systematic investigation of RG-flow solutions of these models,  in which the dual QFT lives on a constant-curvature background,  has been the subject of a recent series of papers \cite{C,F,Ghosh:2020qsx,s2s2,s3,Ghosh:2021lua,Ghodsi:2022umc,Ghodsi:2023pej}.  The bulk  action is schematically given by
\be
S_{d+1}  = M_p^{d-1} \int d^{d+1}x \sqrt{-g} \left [R -{\frac12}(\de \f)^2 - V(\f) \right],
\ee
where $\f$ is a scalar field and $V(\f)$ a (negative-definite) potential admitting one or more extrema. The   $(d+1)$-dimensional ansatz describing RG-flow solutions in curved $d$-dimensional space-time is:
\be \label{int1}
ds^2  = du^2 + e^{2A(u)} \zeta_{\mu\nu}(x) dx^\mu dx^\nu \sp \f = \f(u) \sp \mu,\nu = 0,\ldots, d-1\,,
\ee
where  $\zeta_{\mu\nu}$ is the metric of a  $d$-dimensional space, which can be identified with the boundary metric seen by the dual QFT, and $u$ is the holographic coordinate.

In this work,  we shall be interested in quantum field theories on constant negative curvature space-times.
One can choose $\zeta_{\mu\nu}$ to be the metric on $AdS_d$. However, any other constant negative curvature
manifold gives rise to the same holographic solution as the equations depend only on the value of the constant curvature and not the details of the slice metric.

 In \cite{Ghodsi:2022umc},  this kind of solution was systematically explored in the case where the {\em flat}-space theory admits holographic RG-flow solutions connecting two conformal fixed points. In the gravity dual, this corresponds to solutions having two asymptotically $AdS$ regions, one in the UV (near-boundary) and one in the IR (interior). When a flat boundary metric is replaced by the $AdS_d$  metric, these models were found to admit a rich variety of  $AdS$-sliced  RG-flow solutions. These share the common feature that the scale factor $e^A$ is not monotonic,  but from the UV boundary it decreases to a minimum non-zero value, then it increases again to reach a second UV-boundary. This may correspond to the same UV fixed point (like in the {\em boomerang} RG-flows discussed in \cite{Donos:2017sba}), or to a different one located elsewhere in field space (another maximum of the potential or, in special cases, the minimum which corresponded to the IR fixed point in the flat space theory).
 In all these cases, negative curvature prevents the flow from reaching the IR of the theory.

 Constant curvature manifolds, like spheres, also generate a gap in QFTs. It is interesting to contrast how QFTs on S$^d$ or $AdS_d$ have their IR removed by the gap. In the former case, the flow stops before reaching the IR fixed point, \cite{C}. In the latter, the scale factor turns around and never vanishes.

   The resulting two-boundary solutions are wormholes when the negative constant curvature slice manifold has finite volume and no boundary. They are interpreted as describing interfaces between different (deformed)  CFTs, rather than RG flows of a single theory when the slice manifold is non-compact. The scalar field behavior, as a function of the holographic coordinate $u$ may be monotonic, or exhibit any number of $\f$-{\em bounces} \cite{multirg}, i.e. inversions along the holographic direction.

The holographic theories considered in \cite{Ghodsi:2022umc} were gapless and non-confining (in flat space). The purpose of this paper is to extend this analysis to {\em confining}  holographic theories. By this, we mean holographic theories whose $d$-dimensional QFT dual {\em in flat space} exhibit confinement and a mass gap. For $(d+1)$-dimensional Einstein-dilaton theories, this condition boils down to a simple requirement, \cite{iQCD}: the solution must reach the large-field region, in which  the scalar field potential $V(\f)$ should have an exponential behavior of the form:
\be \label{intro2}
V(\f) \simeq -V_{\infty} e^{2a \f}  \sp \sqrt{\frac{1}{2(d-1)}} < a < \sqrt{\frac{d}{2(d-1)}}\,.
\ee
If $a$ falls below this window, the IR of the theory is gapless; if it falls above, the holographic theory violates Gubser's bound and it only admits ``bad'' singular solutions \cite{multirg}.
It was shown in the second reference in \cite{thermo}, that all such confining theories have a mass gap in flat space, as well as a first-order transition to a deconfined (black-hole) phase.

\subsection{Results and discussion}

In this work,  we perform a complete analysis of the space of solutions of Einstein-dilaton theories with a potential admitting a single extremum (where the solutions can flow to an $AdS$ UV fixed point) and satisfying the requirement (\ref{intro2}), with a metric ansatz as in (\ref{int1})  where $\zeta_{\mu\nu}$ is a constant negative curvature metric. Such a holographic theory is confining, \cite{KS}, when the spacetime is flat Minkowski space.

We  classify solutions according to two features:
\begin{enumerate}
\item the number of UV boundaries (ie. asymptotically $AdS_{d+1}$ boundaries);

\item the nature of the  singularity at $\f \to +\infty$, in case this region is reached.
\end{enumerate}

 Among the solutions that display a singularity, we shall identify those of a special kind, in which the singularity is acceptable in the holographic sense (for example these solutions can be uplifted to regular solutions of higher-dimensional Einstein gravity theories \cite{GK,Gouteraux:2011qh}). Solutions with this ``special'' (also known as ``good", \cite{Gubser}) singularity admit only a one-parameter deformation (which one can map to the value of the curvature of the dual field theory metric), whereas generic solutions have two free parameters. Therefore, the solutions with an acceptable singularity, lie on special lines in the full space of (singular) solutions. We shall call them, with an abuse of language, the ``regular" solutions.

The UV data that define the dual QFTs are the constant negative curvature of the space-time, $R^{UV}$ as well as the relevant coupling constant $\varphi_-$. One can form a dimensionless ratio ${\cal R}$ as in  (\ref{ifr13a}) and this is the only dimensionless (arbitrary) parameter of the dual QFT.

\paragraph{Boundary conditions}
\addcontentsline{toc}{subsubsection}{Boundary conditions}

\begin{figure}[!ht]
\begin{center}
\includegraphics[width = 8cm]{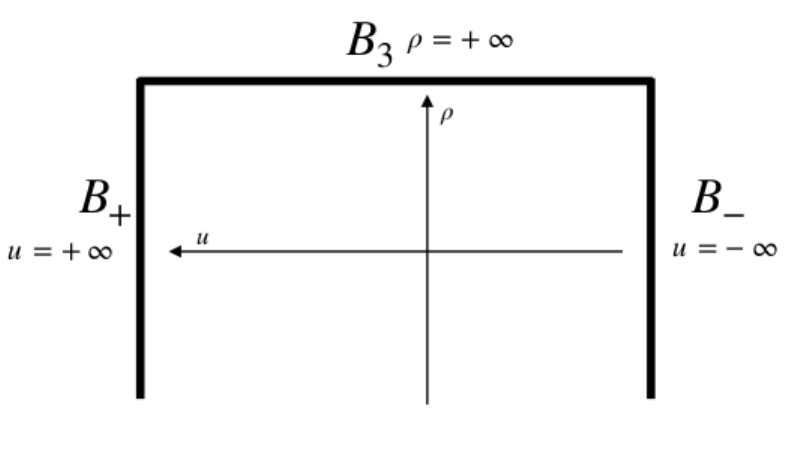}
\caption{\footnotesize{Sketch of the general boundary structure of holographic RG flow solutions with negative curvature slices. Here, $u$ indicates the ``holographic'' coordinate transverse to the slices, and $\rho$ is the radial coordinate of the slices (e.g. the radial $AdS_d$ coordinate) which runs to $+\infty$. The boundaries $B_{\pm}$ are parametrized by the same coordinates as any constant-$u$ slice; the side boundary $B_3$ is parametrized by the holographic coordinate $u$ and by $d-1$ slice coordinates.}}\label{boundaries}
\end{center}
\end{figure}

Before proceeding with the results, we would like to clarify the issue of boundary conditions.
In a space with several boundaries,  the solution of Einstein's equations will need boundary conditions on all of them.
There are at most three distinct boundary components in our solutions. A number of them (zero, one or two) are asymptotically $AdS_{d+1}$ boundaries. When they are present, we call them $B_+$ and if a second exists,  $B_-$. These are transverse to the holographic coordinate. There may be, however, a third asymptotic boundary that we call $B_3$ or side boundary. This is comprised of the constant negative curvature slice boundaries, when these slices are non-compact (see figure \ref{boundaries}). $B_3$ intersects $B_+$ and $B_-$ when they exist.
If the negative curvature slices are compact, there is no $B_3$ boundary and the solution is a wormhole.

The boundary conditions of the asymptotically $AdS_{d+1}$ boundaries, follow the dictums of the holographic correspondence.
In more detail, for the one-boundary solutions, with $B_+$ boundary, we need two boundary conditions on the single boundary $B_{+}$. One of them is interpreted as the source, while the second as the vev. The regularity of the solution typically determines the source as a function of the vev.

For two-boundary solutions, with both $B_+$ and $B_-$, the interpretation was discussed in \cite{BP}.
We need two boundary conditions on one of the boundaries, say $B_+$, and then the fixed slice geometry determines uniquely the solution that is always regular. In particular, this determines the two boundary conditions on the other boundary $B_-$. We can phrase the same in the following holographically transparent way: choosing the source on $B_+$ and the source on $B_-$ determines completely the solutions as well as the two associated vevs.

The issue is why we do not need to specify also boundary conditions on the side boundary $B_3$ if it exists.
The answer for the solutions we are looking at is that we have fixed completely the metric ansatz by demanding that the metric is a conifold metric with fixed constant curvature slices.
In the special case that these slices have $AdS_d$ metrics, it is the $O(1,d)$ symmetry that determines the $B_3$ asymptotics.
On the other hand, if one solves a fluctuation equation in the geometry, then we need also extra boundary conditions on $B_3$.

\paragraph{One-boundary solutions }
\addcontentsline{toc}{subsubsection}{One-boundary solutions}

As we shall see, in the confining case, there exist one-boundary solutions (with asymptotically $AdS_{d+1}$ boundary $B_+$), for any value of the (negative) $d$-dimensional curvature.
These solutions are expected to be dual to holographic QFTs on the negative curvature manifold.
Such a QFT is characterized by two UV couplings, the relevant coupling $\f_-$ and the constant negative curvature $R^{UV}$. They have therefore a single dimensionless coupling $\mathcal{R}$, defined in (\ref{ifr13a}), that controls all the physics.

Regular solutions having the same value of $\mathcal{R}$ are competing ground-states of the same holographic QFT.
The number of competing ground states depends very much on the value of $\mathcal{R}$. For low enough values of $\mathcal{R}$, there is a single ground-state solution. For somewhat larger values, there are three competing ground states, for even larger $\mathcal{R}$  five, and eventually after a given $\mathcal{R}$, an infinite number of ground-states in competition.
For any $\mathcal{R}$, there is always at least one solution that has a monotonic scale factor.
Many others have non-monotonic scale factors featuring many A-bounces.

\begin{figure}[!ht]
\begin{center}
\includegraphics[width = 3.5cm]{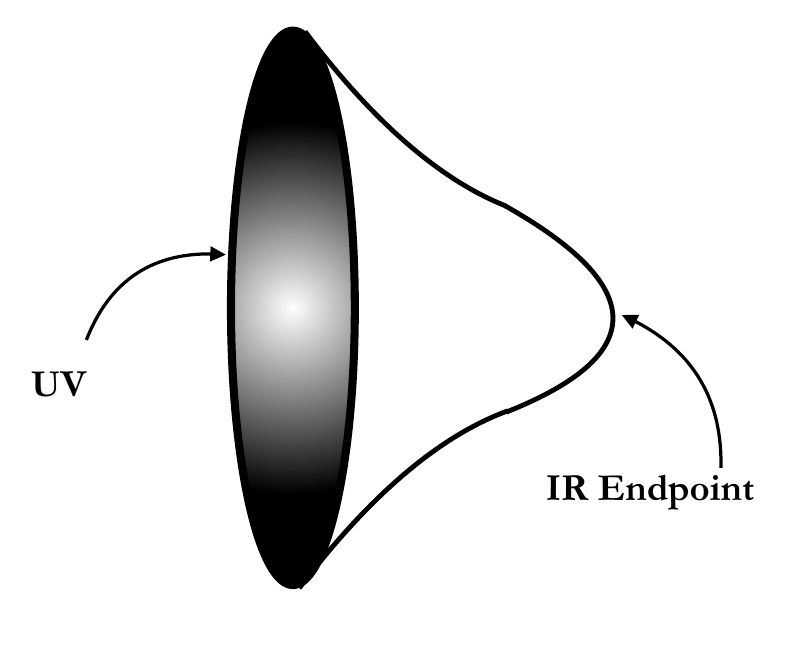} \hspace{1cm}
\includegraphics[width = 4cm]{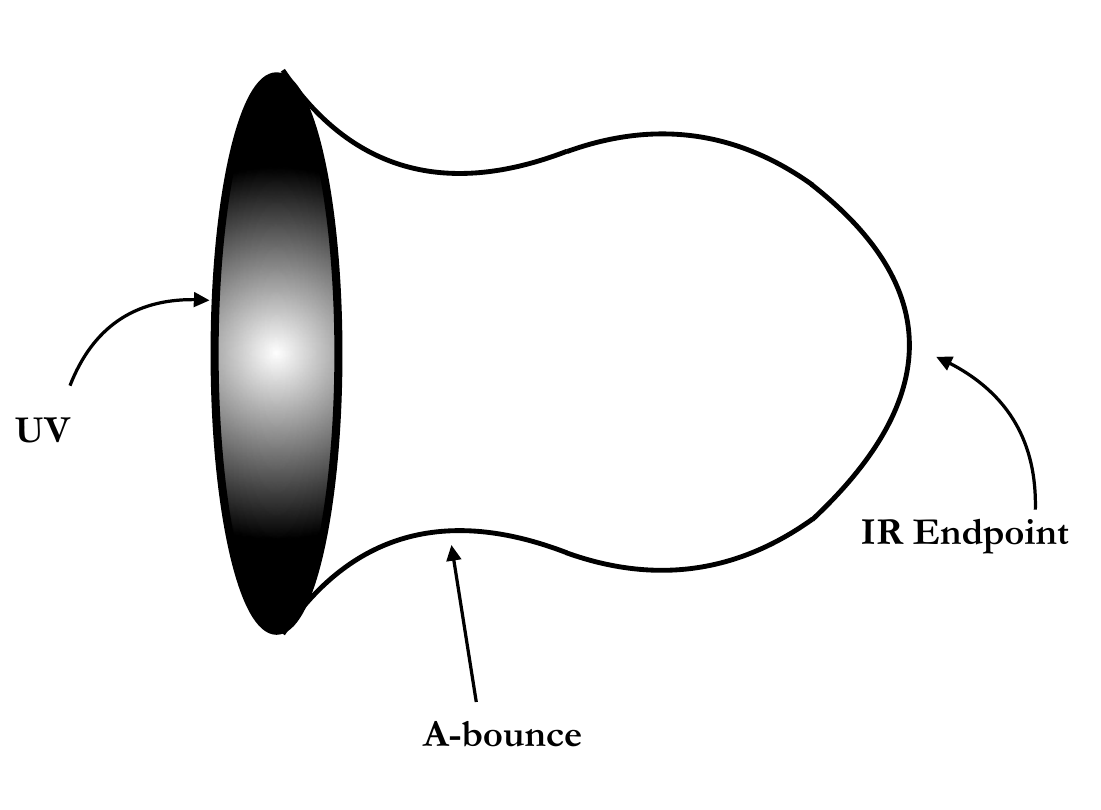} \hspace{1cm}
\includegraphics[width = 4.7cm]{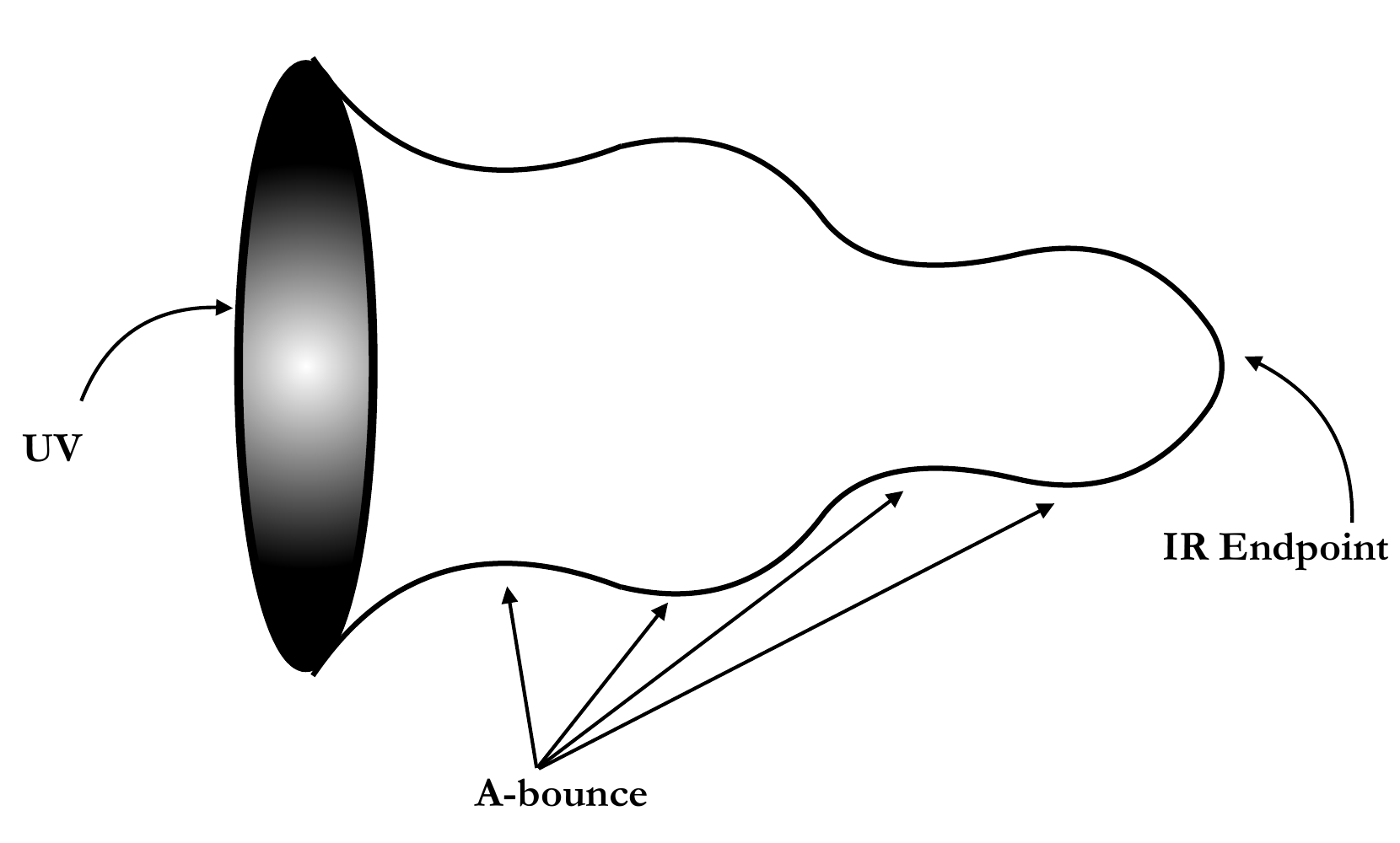}
\caption{\footnotesize{Sketch of single-boundary holographic solutions whose scale factor is monotonic (left) or features two A-bounces (middle) or four A-bounces (right).}}\label{fig1}
\end{center}
\end{figure}

All these solutions start at the $B_+$ (UV) boundary, corresponding to the maximum of the scalar potential, and after a certain number of A-bounces, the scale factor shrinks to zero at the (good) IR singularity. Along the way, these solutions can also feature several {\em $\f$-bounces}, i.e. points at which $\f(u)$ reverses its flow.
Three types of one-boundary solutions, with different numbers of $A$-bounces are schematically depicted in Figure \ref{fig1}.  We find that there are special limiting points in solution space, where the number of A-bounces can become arbitrarily large.\footnote{This phenomenon is reminiscent of $AdS_2$ condensation, \cite{Son,E1,E2}, walking behavior, \cite{JK} and Efimov scaling, \cite{E2,JK}. Apart from the existence of a discrete infinity of solutions with a distinct number of nodes, it does not seem that there is an analog of Efimov scaling in our case.}  These limiting solutions with an infinite number of oscillations, will be discussed more extensively below.

 We compare the free energy of competing solutions and find that, for any negative value of ${\mathcal R}$,  the dominant one is {\em always} the monotonic solution with no bounces, which connects to the flat-space confining background when the boundary curvature is taken to zero. The solutions featuring $A$-bounces and $\f$-bounces are always subleading saddles. These results can be seen in figure \ref{FRZ} in Section \ref{rfe}.

We therefore conclude that there is no (quantum) phase transition, as we vary the dimensionless parameter $\mathcal R$ from $0$ to $-\infty$. This seems to be in contrast to general expectations concerning confining gauge theories on $AdS$, with electric boundary conditions, \cite{A2}. There, it was argued that there should be a phase transition between the $|\mathcal R|\ll 1$ regime (``strong coupling"),  and the $|\mathcal R|\gg 1$ regime (``weak coupling") that is probably first order.
Moreover, an important order parameter in the context of electric boundary conditions is associated with the $SU(N_c)$ global symmetry and its boundary currents. There is no $SU(N_c)$ symmetry in our setup, nor associated boundary currents.

It is therefore natural to expect that the confining theory we discuss, is one with Neumann boundary conditions in $AdS$. In such a case, we do not have the $SU(N_c)$ global symmetry, nor the associated currents.
Moreover, as argued in \cite{A2}, no phase transition is expected in this case as a function of $\mathcal R$.

 It would seem though, that there is the following puzzle: the renormalized partition function of these theories scales as $N_c^2$ times a well-defined function of $\mathcal R$, that we have calculated in the present paper. However, we would expect that for a confined theory, the partition function should be ${\cal O}(1)$.
 In fact, something similar happens in flat space. A holographic calculation of the confined partition function gives a result that is ${\cal O}(N_c^2)$ times a function of the cutoff.
 A renormalization of this partition function will eventually give a constant times $N_c^2$, a result that is scheme-dependent. However, we do not worry about this, as we can always subtract it to zero by a further finite renormalization.
 We believe that a similar explanation can be applied to the theory on $AdS$.

A problem that remains unsolved is the calculation of Wilson-loop expectation values in gauge theories on $AdS$.
 In our examples, this calculation can be in principle done, and we plan to do it in future work\footnote{Circular Wilson loops in wormhole solutions with positive curvature slices have been recently computed in \cite{BP}.}.

 We should also mention that several of the non-dominant saddle points of the confining theories we discuss here have rather unusual RG properties. As mentioned, the leading saddle-point solution features a monotonic scalar field. In holography, this is expected to reflect the monotonicity of RG flow. However, we also found a large number of subleading saddles, with many $\varphi$-bounces that indicate that their RG flow is not monotonic, at least in conventional terms.
 A priori, this is not inconsistent with expectations, as the monotonicity of RG flow is expected for Minkowski ground states, but not much is known for $AdS$ ground states. However, this issue can be investigated in weakly-coupled gauge theories on $AdS$, in order to see if, at least perturbatively, a monotonicity in RG flow is expected in such cases.
The leading as well as some other subleading saddle-point solutions have a monotonic scale factor. However, we also found a large number of subleading saddles, with many $A$-bounces, where the scale factor turns around. This feature is even more puzzling than $\phi$-bouces and as far as we can tell has no interpretation in terms of field theory RG flows, at least if we insist on identifying the scale factor with the energy scale.

\paragraph{Two-boundary solutions. }
\addcontentsline{toc}{subsubsection}{Two-boundary solutions}

\begin{figure}[!ht]
\begin{center}
\includegraphics[width = 4.3cm]{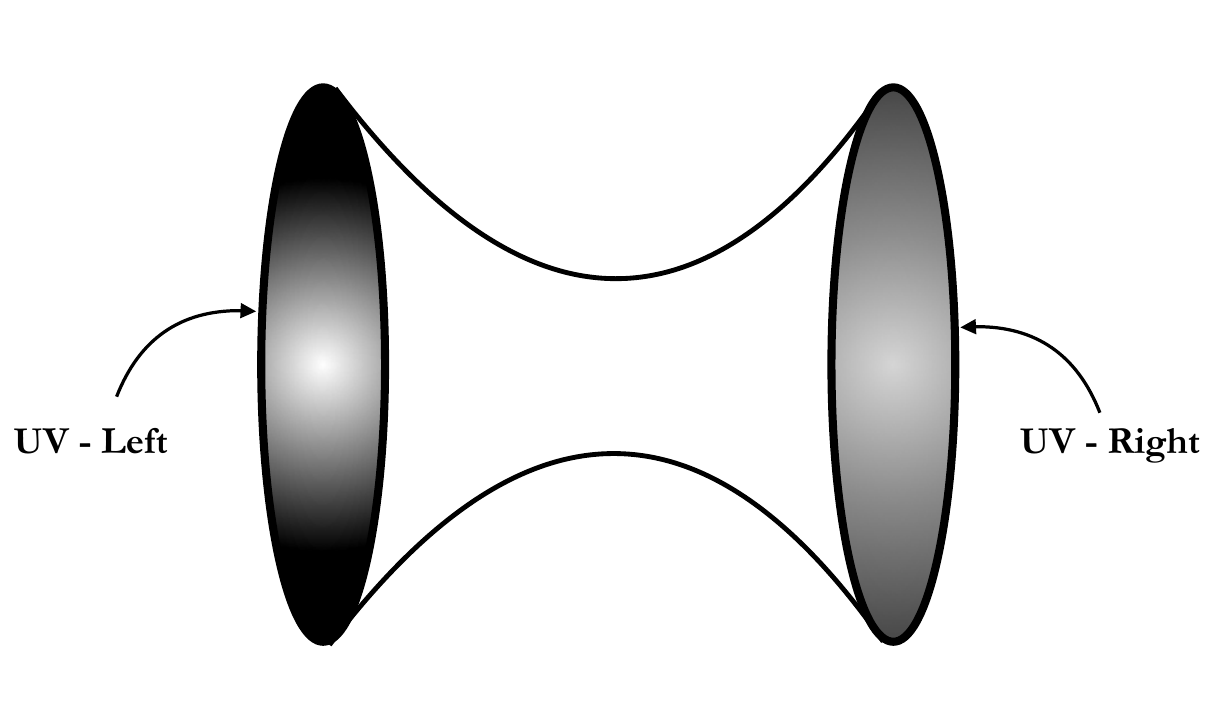} 
\includegraphics[width = 5.cm]{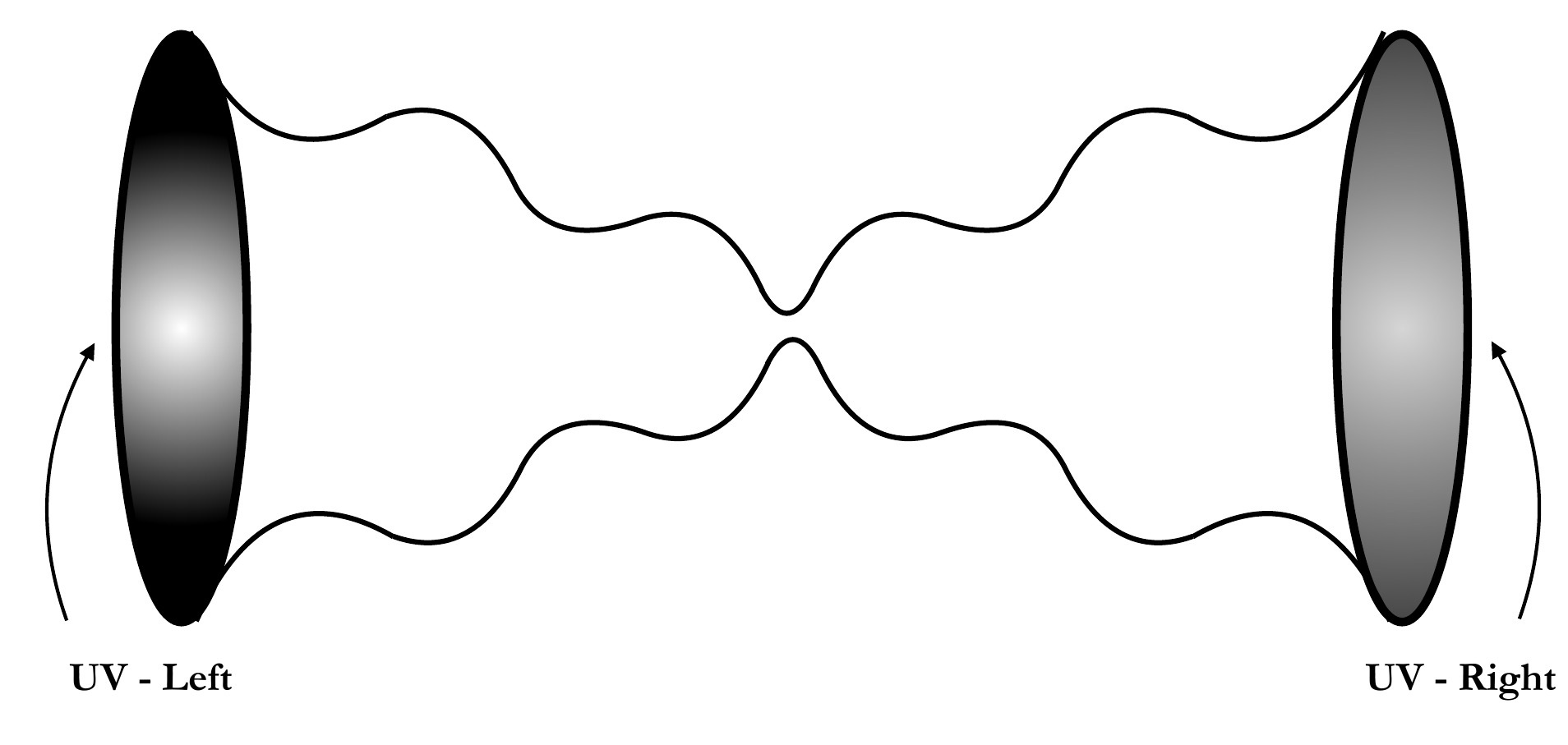}
\includegraphics[width = 5.3cm]{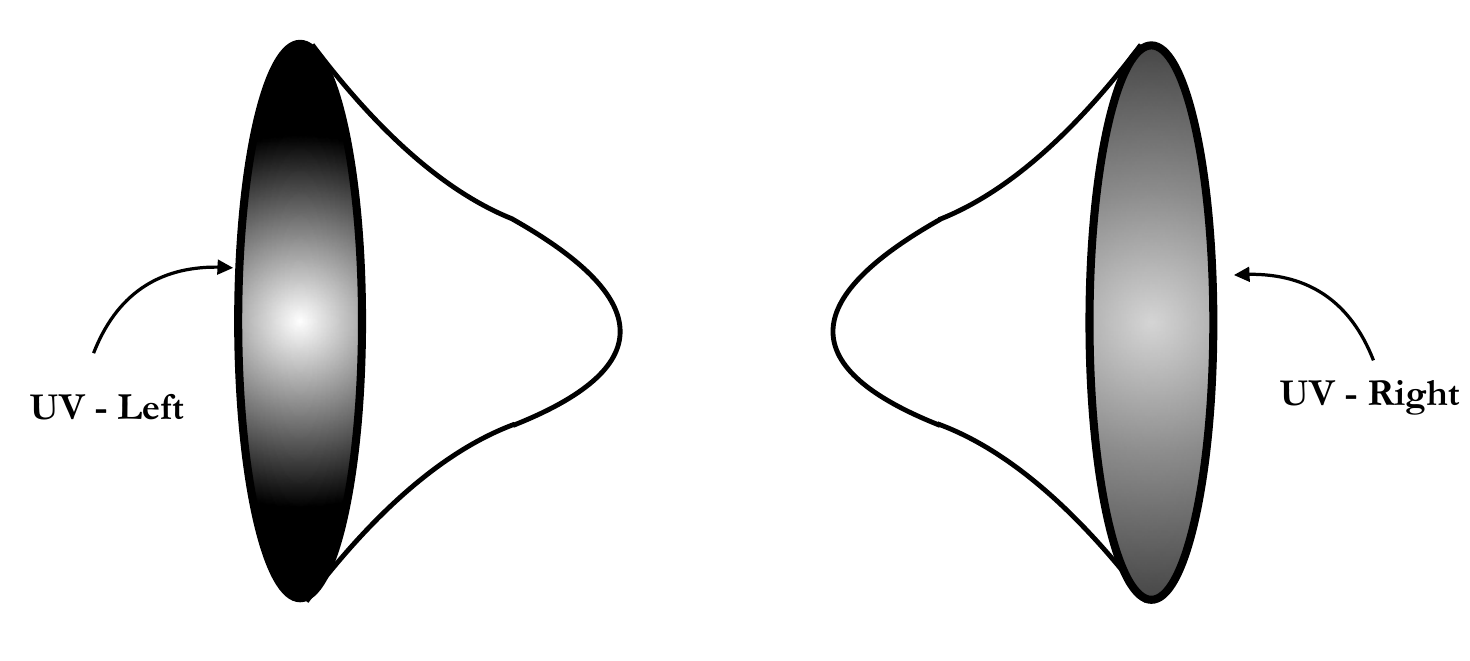}
\caption{\footnotesize{Connected (left), connected with several A-bounces (middle) and disconnected (right) two-boundary solutions.}}\label{fig2new}
\end{center}
\end{figure}

 These solutions are akin to the interface solutions found in \cite{Ghodsi:2022umc} and qualitatively similar to many other holographic interfaces in the literature, \cite{Bak}-\cite{Gut}. They connect two asymptotically $AdS_{d+1}$ UV boundaries, that we denote by $B_+$ and $B_-$, corresponding in our examples to the same UV CFT\footnote{This is because the potential in our example has a single maximum. If it had more, then solutions interpolating between different maxima would exist also, as shown in \cite{Ghodsi:2022umc}.} (but with different boundary curvature and relevant coupling). Such solutions are illustrated schematically on the left and center of figure \ref{fig2new}.

  As a function of the holographic coordinate $u$, the scalar field starts and ends at the same value (the maximum of $V(\f)$).  The scale factor reaches a non-zero minimum but never vanishes. These solutions are therefore singularity-free. As in the one-boundary case, they may feature any number of $A$-bounces and $\f$-bounces. The slice manifold can be any constant negative curvature manifold. When the slice manifold is $AdS_d$ in Poincar\'e coordinates they can be mapped by a coordinate transformation to interfaces between confining theories in flat space.

As argued above, such confining theories in flat space, do not have an $SU(N_c)$ global symmetry, and they are probably in the magnetic (confined) phase. The non-trivial interface is generated by integrating the main relevant operator of the confining theories on the interface hyperplane.
There is no issue of boundary conditions here, as everything is gauge-invariant, including the perturbing operators.

We obtain a whole family of theories that are characterized by the parameters $\mathcal R_+$ and $\mathcal R_-$ of the theories on the two sides of the interface, as well as the ratio of the two relevant couplings $\mathcal{U}=\frac{\Lambda_+}{\Lambda_-}$ in (\ref{Eratio}).

Once we fix these three dimensionless numbers, we still have a large (and in some cases infinite) number of connected two-boundary saddle-point solutions that compete in the gravitational path-integral. In general, they contain one or more $A$-bounces. Among them, there is one that has minimal free energy and it has a single $A$ bounce.
However, in this case, there are also disconnected competing solutions obtained by tensoring two one-boundary solutions, that were discussed above (illustrated on the right of figure \ref{fig2new}). Using them, we can construct a large  number of
disconnected saddle-point solutions and the one with the lowest free energy contains, on both sides, the lowest free energy single-boundary solution.

What we found is, that in all cases, the leading disconnected solution has lower free energy compared to the leading connected solution.
We also found that for sufficiently small $\mathcal{R}_+$ and/or $\mathcal{R}_{-}$, there are no connected two-boundary solutions. In that case, the only solutions are disconnected ones.

As explained in section \ref{83},  the dominance of disconnected solutions in the path-integral implies that correlation functions that contain operators on both sides of the interface, obtain no contribution from the disconnected leading solution.
This further implies that such normalized correlators are exponentially small, by factors of $e^{-cN_c^2}$ where $c$ is a constant that depends on the dimensionless curvatures.
It is not clear to us, where such effects can appear from the dual holographic gauge theory.

\paragraph{Uplift to higher dimensions}
\addcontentsline{toc}{subsubsection}{Uplift to higher dimensions}

Some regions of the solutions we discuss are related by dimensional reduction to pure $(d+1+n)$-dimensional gravity solutions where the bulk is sliced by  $AdS_{d} \times S^n$, which were recently explored in \cite{Ghodsi:2023pej}.  The dimension $n$  of the sphere is related to the exponent $a$ appearing in the large-$\f$ regime of the potential (\ref{intro2}) as in (\ref{coonf}).  The solutions in which $S^n$ shrinks to zero size in a regular manner, map to the IR region $\f \to \infty$  of the special (``regular'')  solutions of the reduced Einstein-dilaton theory.   In the present work, we find that the one-boundary solutions of the Einstein-dilaton theory, do in fact uplift to regular solutions of the $(d+1+n)$-dimensional theory. On the other hand, the two-boundary solutions do not uplift, because the higher-dimensional theory on a shrinking $S^n$ captures only the IR large-$\f$ region. More generally, the lower- and higher-dimensional theories can be related in the IR but have different UVs.

Interestingly, we find that the limit in which the solution in the lower-dimensional theory has an infinite number of bounces corresponds to a special solution of the higher-dimensional theory with factorized geometry $AdS_d \times AdS_{n+1}$, \cite{Ghodsi:2023pej}.

\paragraph{No-boundary solutions}
\addcontentsline{toc}{subsubsection}{No-boundary solutions}

We have found holographic solutions that do not have any $AdS_{d+1}$ boundary. They may however have a $B_3$ side boundary if the slice manifold is non-compact.
Such solutions may start at $\f\to -\infty$ and end at $\f\to +\infty$ being Gubser-regular on both sides.
There is an infinite set of such solutions, distinguished by the even number of $\f$-bounces.

There is also a class of similar solutions that start at $\f\to -\infty$ and end at $\f\to -\infty$
or start at $\f\to +\infty$ and end at $\f\to +\infty$. Their number is again infinite and they are distinguished by the odd number of $\f$-bounces.

As these solutions have no $AdS_{d+1}$ boundary, their holographic interpretation is unclear. We believe that they correspond to holographic interface theories when the bulk theory on both sides is ``topological" in that it has no local operators.
Moreover, there should be an associated topology, related, in the multidimensional case, to the topology of the boundary of the scalar manifold.
The interface however is dynamical, as one can in principle insert local operators on the side boundary $B_3$ and calculate non-trivial interface correlators.
In particular, one can study the fluctuation problem of any field around these solutions, and in this case, one would need boundary conditions on $B_3$. These can insert localized sources on $B_3$ and the response will define interface correlators. Although such calculations remain to be done, conceptually they appear possible.

An evaluation of the free energy of such solutions in this paper indicates that their free energy is an increasing function of their number of bounces, as shown in figure \ref{regreg4}. Therefore, the thermodynamically dominant solution is the (unique) one that interpolates between $\f=-\infty$  to $\f=+\infty$ and is shown in figure \ref{regreg4}.

\paragraph{Solutions with an infinite number of oscillations}
\addcontentsline{toc}{subsubsection}{Solutions with an infinite number of oscillations}

Such solutions appear at the border between one-boundary and two-boundary solutions.
As the parameters of the solutions approach this border in solution space, the scale factor (and the scalar) develop an ever-growing number of oscillations as can be seen in figures \ref{bound1a}, \ref{bound3a} and \ref{AA}.
These oscillations have approximately constant amplitude and take place in an area near the UV fixed point of the potential, $\f\leq |\f_{\rm max}|$, the same for all solutions. Moreover, $\dot A$ is much smaller than $\dot\f$ and the potential, and this is the starting point for developing a perturbative expansion for such solutions in appendix  \ref{loops}.
The leading solution can eventually be written in terms of elliptic functions.

Such solutions are highly unusual, as they appear as RG flow solutions where the IR endpoint features something that looks like a limit cycle (if we interpret the behavior of the scalar $\f$ as a running coupling). However, the scale factor is also oscillating, and therefore such one-boundary solutions defy a straightforward RG-flow description.
Although they look like limit-cycle solutions, that have been discussed in the past in non-unitary QFTs, \cite{limit}, they are also different. Here, the QFT is on $AdS$ and there is no no-go theorem in this case.
Similar solutions have been found recently in a different but related context, featuring a linear string theory dilaton, \cite{C2}.

We have observed that in competing solutions, the free energy increases with the number of A-bounces or $\f$-bounces. This suggests that solutions with a large number of $A$- and $\f$-bounces are strongly subleading in the path-integral.

An interesting question is whether such solutions exist in top-down effective gravity theories.
We expect the answer is yes, and we shall address this question in the near future.

\paragraph{Top-down vs bottom-up and general potentials}
\addcontentsline{toc}{subsubsection}{Top-down vs bottom-up and general potentials}

This is a relevant question: how many of the effects found in this paper, using a generic bottom-up potential, persist in top-down situations in supergravity? When the flows are in the bulk of the scalar potential, many previous works have shown, \cite{Bak}-\cite{Gut},\cite{Ghodsi:2022umc}, that there is very little qualitative difference in the behavior of relevant solutions.

In our case, however, we must consider higher-dimensional solutions that lead to confining behavior.
There are several possibilities but the simplest is the black  D$_4$ solution of Witten, \cite{Witten:1998zw}.
This theory will be investigated in the near future.

A further question is how to combine the solutions that were found here (with a potential with a single maximum) with those found in \cite{Ghodsi:2022umc}, where the flows start and end at finite maxima for the potential.
If we consider a globally negative potential for a single scalar, with an arbitrary number of finite maxima and minima, and $\f\to\pm \infty$ asymptotics that satisfy the Gubser bound, then we expect  again the following space of ``regular" solutions :

\begin{enumerate}

\item[a.] Zero-boundary solutions. These are similar to the ones found here running from $\pm\infty \to \pm \infty$.

\item[b.] One-boundary solutions. These solutions can start at {\em any} of the maxima, and end up at $\f\to \pm\infty$.

\item[c.] Two-boundary solutions. Such solutions can start at any of the finite maxima, and end at the same maximum or a different one.

\end{enumerate}
It would be interesting to verify this picture in a concrete example.

\paragraph{RG Interfaces}
\addcontentsline{toc}{subsubsection}{RG Interfaces}

A concept in interface physics is that of RG interfaces. These are interfaces between on one side a CFT driven by a relevant operator (the QFT) and on the other side the same CFT without a non-trivial coupling constant
for the same operator, \cite{RG1}-\cite{RG4},\cite{C1}.
Our solutions here, and those discussed in \cite{Ghodsi:2022umc} give many examples of such cases.
For each boundary, one of the dimensionless QFT data is the dimensionless curvature ${\cal R}$ that is defined in (\ref{ifr13a}) as the ratio of the space-time curvature scale to the relevant coupling.
This parameter ${\cal R}\to -\infty$ whenever the coupling of the QFT vanishes and this becomes a CFT.

Therefore, in all those solutions, there is a subset where  ${\cal R}_+\to -\infty$ while ${\cal R}_-$
remains finite. This is an example of an RG Interface  where the + theory is a CFT while the - one is a QFT
We also have twice-tuned solutions both here and in \cite{Ghodsi:2022umc}  where both  ${\cal R}_{\pm}\to -\infty$. In this case, both theories on the two sides are CFTs, and the only bulk breaking of conformal invariance is generated by the interface.

\paragraph{Pure vev solutions}
\addcontentsline{toc}{subsubsection}{Pure vev solutions}

There are one-boundary solutions with $\mR\to-\infty$. In such solutions, the coupling of the relevant operator vanishes, and therefore, they correspond to the CFT$_d$ defined on a constant negative curvature manifold. There is an infinite number of distinct such solutions. The scalar operator, although it does not have a source, has a non-trivial vev, which drives the RG flow of the theory. Interestingly there is no one-boundary solution where both the source and the vev are zero, i.e. with a constant scalar field. In other words, if the scalar field is trivial the solution always has two UV boundaries (the geometry is $AdS_{d+1}$ written in $AdS_d$ slices).

Although the development of a non-trivial vev is expected for scalar operators of a CFT on a constant negative curvature manifold, the non-existence of a zero vev solution indicates that there is always symmetry breaking in such cases.
For example, when the base manifold is $AdS_{d}$, the O(1,d+1) symmetry of the CFT is broken to O(1,d).
It would be interesting to find the Goldstone modes in this case.
Out of all vev solutions, we find that the dominant one is the one with the steepest scale factor and without any $A$-bounces.

For the two-boundary solutions, similar remarks apply. There is an infinite number of solutions that are vev
on both boundaries. Their interpretation is as interfaces, where one has a CFT on both sides of the interface.
There is a (in general different) vev for the scalar operator on both sides of the interface.
However, in this case, there also exists a solution without any vev: this has a trivial scalar field sitting at the conformal point. The metric is simply $AdS_{d+1}$ sliced by $AdS_d$.

The free energy of such a solution is calculated in appendix  \ref{fcft}.
We find that in the case of non-compact slices,  the leading, non-trivial, connected solution with non-zero vev on both sides has lower free-energy compared to the zero vev solution. For wormholes instead, it is the zero-vev solution that dominates. In both cases however, none of the connected geometries is the dominant solution in the path integral:  like in the case with non-zero source solutions, the dominant solution is always a disconnected solution, which necessarily has zero vev. This implies that in holographic theories on $AdS$, conformal invariance is always spontaneously broken by a vev.

\vskip 0.3cm

The structure of this paper is as follows:

In section \ref{dim2}, we explore an Einstein-dilaton model characterized by a scalar field with negative definite potential that exhibits exponential growth at high field values. We examine various limitations on this exponential growth to ensure the existence of a (de)confining dual Quantum Field Theory (QFT). Additionally, we establish the concept of a ``good'' singularity following Gubser's criteria. We present the argument that adherence to the computability limit ensures holographic computation results remain unaffected by the specifics of how the Infrared (IR) singularity is resolved. Concluding this section, we introduce the first-order formalism for the equations of motion and review the special points that emerge during the Renormalization Group (RG) solution analysis.

In section \ref{vper}, we explicitly show that at large values of the scalar field, there are three types of solutions. The generic (type 0) solution, is a solution in which the potential is considered as a perturbation. The second and third types, known as type I and type II, are primarily influenced by the potential's dominant behavior. It is shown that type I solutions with negative slice curvature, are exclusive to the deconfined phase. Conversely, type II solutions are found within the confining phase. Additionally, this section addresses the nature of the singularities at the IR end-points.

In section \ref{REGSOL}, we aim to identify both regular and singular solutions corresponding to the potential RG flows for a theory with a specific confining potential.
We find a two dimensional space of solutions, in which, each point features at least one A-bounce. This space encompasses solutions with two, one, or no $AdS$ boundaries.
Within this space, we specifically pinpoint a one-dimensional subset of solutions that are regular and extend to the Ultraviolet (UV) boundary. Furthermore, we discuss the transitions among different solutions.

In section \ref{uplsec} we consider an uplifted Einstein-dilaton gravity with a geometry containing two constant curvature spaces. In particular, we examine a sphere and an $AdS$ space.
By performing a dimensional reduction on the sphere, we uncover an action in a lower dimension that includes two scalar fields. This theory displays confinement properties within a certain range which depends on the dimensions of the two spaces.

In section \ref{comcu}, we conduct a comparative analysis of the confining solutions presented in section four with the uplifted solutions discussed in section five. We particularly demonstrate that the solution characterized by an infinite series of loops is analogous to the product space solution within the uplifted framework. Moreover, we ascertain that when the scalar field attains large values where the two theories converge, the ``regular" solutions from the confining scenario are correspondingly uplifted to regular solutions of the higher-dimensional theory.

In sections \ref{onshc} and \ref{numq}, we calculate the on-shell action for the confining solutions and determine the free energy based on the boundary QFT data. We evaluate one-boundary solutions, including those devoid of A-bounces and those with at least one A-bounce. Our findings indicate that in terms of free energy, solutions lacking an A-bounce are predominant. Additionally, we establish that among solutions with two UV boundaries, the disconnected solutions possess a lower free energy compared to the connected ones.

In section \ref{worm2} we show the results of free energy for solution with compact slice geometries (wormhole solutions).

More details of the calculations in this paper are given in Appendix \ref{UVF}-\ref{fcft}.

\section{Einstein-dilaton gravity in the confining regime\label{dim2}}

We shall study an  Einstein-dilaton theory in $(d+1)$-dimensions, governed by the action:
\be \label{eq:action}
S= \int du \, d^ dx \, \sqrt{g} \left[R - \frac{1}{2} \partial_a \f \partial^a \f - V(\f) \right]\,.
\ee
We assume that the potential $V(\f)$ is negative definite and has a maximum at $\f=0$ which supports an $AdS_{d+1}$ solution. The $AdS$ curvature length is given by:
\be\label{ell}
\ell = \sqrt{-{\frac{d(d-1)}{V(0)}}}\,.
\ee
Holographic RG flows of this theory are asymptotically $AdS$ solutions and are dual to $d$-dimensional field theories with a UV conformal fixed point corresponding to $\f=0$. Flat flows\footnote{These are flows where the metric $\zeta_{\m\n}$ in (\ref{int1}) is the Minkowski metric.} that start at $\f=0$ and end at a finite minimum of the potential, in the single scalar theory studied here, correspond to non-confining QFTs, \cite{multirg}.
When on the other hand the   RG flow, starts at $\f=0$ and ends at the boundary of the scalar space, $\f\to\infty$, \cite{multirg}, then one can obtain a confining QFT in certain cases.

We shall be interested in {\em confining} holographic theories: by this, we mean that the dual field theory in $d$-dimensional {\em flat space} (at zero temperature) exhibits confinement\footnote{For a pure glue theory without dynamical flavor, this means an area law for the Wilson loop and a gapped discrete spectrum of glueball excitations}. Our goal will then be to investigate what happens to this theory when considered on a constant negative manifold, and in particular on $AdS$.

Whether a QFT, holographically dual to (\ref{eq:action}) is confining in flat space (in the sense specified above) depends on the properties of the potential at large $\f$ \cite{iQCD,multirg}:

\vspace{0.5cm}

{\em The flat space theory is confining if (and only if)  $|V(\f)|$ grows faster than $\exp(2 a_C \f)$ as $\f \to +\infty$,} where
\be \label{ac}
a_C \equiv \sqrt{\frac{1}{2(d-1)}}\,.
\ee
In the following, motived by string-generated supergravity potentials, we  parametrize the large-$\f$ potential asymptotics by a leading exponential behavior, as
\be \label{expV}
V(\f) \simeq -V_\infty e^{2 a \f}+{\rm subleading} \sp \f \to +\infty\,,
\ee
where $V_{\infty}>0$. With this parametrization, the holographic theory is confining in flat space if $a> a_C$.  We call the value (\ref{ac})  the {\em confinement bound}.

The solutions which exhibit confinement, extend to $\f \to +\infty$, where the metric becomes singular. However, this naked singularity may be acceptable if it satisfies certain criteria, \cite{Gubser}. This singles out  two other interesting values of the exponent $a$ which mark qualitatively different properties of the  IR region:
\begin{itemize}
\item {\bf Gubser's bound}
\be \label{ag}
a_G = \sqrt{\frac{d}{2(d-1)}}\,.
\ee
For $a>a_G$ no solution exists which extends to   $\f \to +\infty$ with an acceptable singularity in the sense of Gubser \cite{Gubser}, or equivalently in the sense that it can be uplifted to a non-singular geometry by generalized dimensional uplift \cite{GK,Gouteraux:2011qh}.

For $a<a_G$, the generic solution still cannot be uplifted and has a ``bad'' singularity. However {\em special} solutions exist for specific combinations of the integration constants which have a milder singularity, which is both ``good'' in the Gubser sense and can be uplifted to a regular geometry.

\item {\bf Computability bound}
\be\label{acomp}
a_{comp} = \sqrt{\frac{d+2}{6(d-1)}}\,.
\ee
For $a>a_{comp}$, the IR singularity reached as $\f \to +\infty$  is such that the fluctuation problem around the singularity for both the scalar fluctuation and the metric fluctuation has two IR-normalizable solutions \cite{multirg}. This means that one cannot use normalizability as a criterion to single out physical perturbations, and dual field theory quantities such as e.g. holographic correlators are not computable unless we give an extra prescription on how to pick the solution in the IR. In other words, the dual geometry must be supplemented with some extra IR information (e.g. the precise way the singularity is resolved) in order to do holographic computations.

On the other hand, if $a<a_{comp}$, requiring normalizability in the IR is enough to select one of the two independent perturbations. Moreover,  radial wavefunctions of the normalizable modes are concentrated in the interior,  far from the singularity, implying that the result of holographic calculations is insensitive to the details of what (eventually) resolves the IR singularity. In this sense, the holographic model (\ref{eq:action})  is self-sufficient, at least as far as low-energy excitations are concerned.
\end{itemize}

The relevant parameters satisfy the inequality
$$a_C < a_{comp}< a_G\;.$$
 In this work we shall be mainly studying potentials with asymptotics (\ref{expV}) with $a_C < a < a_{comp}$, i.e. we shall be mostly interested in computable, confining theories driven by a single relevant operator.

It is interesting to note now (and we shall come back to this in section 5), that the confinement bound (\ref{ac}) also separates two different behaviors upon (generalized) uplift to a higher dimensional pure gravity theory: for $a<a_C$ the IR geometry uplifts to  $AdS_{d+1} \times T^n$, whereas for $a_C < a < a_G$ it uplifts to  $AdS_{d+1} \times S^n$ \cite{GK,Gouteraux:2011qh}. The value  $n$ (which can be taken to be  a real number, hence the term {\em generalized} dimensional reduction) is related to $a$ by:

\be\label{gen}
a = \left\{\begin{array}{ll} \sqrt{\frac{d+n-1}{2n(d-1)}} & \qquad a_C < a < a_G\,, \\ & \\
\sqrt{\frac{n}{2(d-1)(d+n-1)}} & \qquad 0 < a < a_C\,. \end{array} \right.
\ee

In flat space, the large-$\f$ region of a holographic RG-flow starting at $\f=0$ corresponds to the IR regime of the dual field theory. This is why confinement (which is essentially an IR feature) is determined by the large-$\f$ region of the potential.

When the same holographic field theory is defined on a curved spacetime however, things are different:  curvature is an IR-relevant deformation and may qualitatively change the IR behavior of the solution (for example it may prevent the flow from even reaching the large-$\f$ region, \cite{C,Ghodsi:2022umc}). In the rest of this paper, we systematically explore the effects of negative curvature on holographic RG-flows of confining theories.

\subsection{Ansatz and equations of motion}

We shall consider a confining holographic QFT$_d$, driven by a single relevant operator, and living on a constant negative curvature manifold. Its gravitational dual will be given by the Einstein dilaton theory in (\ref{eq:action}), where the scalar $\f$ is dual to the non-trivial relevant operator that drives the RG flow.

To find the holographic solution describing the ground state of this theory, we consider the  ansatz which is characterized by
the scale factor and dilaton as a function of the holographic coordinate $u$
\be
\label{eq:metric}ds^2 = du^2 + e^{2 A(u)} \zeta_{\mu \nu} dx^{\mu} dx^{\nu}\sp  \f = \f(u)\,.
\ee
Here, $\zeta_{\mu \nu}$ is a metric describing a space-time of constant negative curvature $M_d$, ($AdS_d$ is a special case of this). As a consequence, we have
\be
\label{eq:Rzeta}
R^{(\zeta)}_{\mu \nu} = \kappa \zeta_{\mu \nu} \sp R^{(\zeta)} = d \kappa \,,\quad  \textrm{with} \quad \kappa =
- \frac{(d-1)}{\alpha^2}\,,
\ee
where $\alpha$ is the curvature length scale of the $M_d$. In the following, we use the following shorthand notation. Derivatives with respect to $u$ will be denoted by a dot while derivatives with respect to $\f$ will be denoted by a prime, i.e.:
\be
\dot{f}(u) \equiv \frac{d f(u)}{du} \sp g'(\f) \equiv \frac{d g(\f)}{d \f} \, .
\ee
Varying the action \eqref{eq:action} with respect to the metric and the scalar $\f$ gives rise to the equations of motion:
\be
\label{eq:EOM1} 2(d-1) \ddot{A} + \dot{\f}^2 + \frac{2}{d} e^{-2A} R^{(\zeta)} =0 \, ,
\ee
\be
\label{eq:EOM2} d(d-1) \dot{A}^2 - \frac{1}{2} \dot{\f}^2 + V - e^{-2A} R^{(\zeta)} =0 \, ,
\ee
\be
\label{eq:EOM3} \ddot{\f} +d \dot{A} \dot{\f} - V' = 0 \, .
\ee
These equations have an obvious symmetry: $u\to -u$.
If we choose also a symmetric potential, $V(\f)=V(-\f)$ as we shall do later, then there is also the symmetry $\f\to -\f$.

\subsection{The first order formalism} \label{sec:1storder}
Holographic RG flows are in one-to-one correspondence with regular solutions\footnote{In confining theories without extra dimensions, regularity is interpreted more loosely, as we discussed in the previous section.} to the equations of motion
\eqref{eq:EOM1}--\eqref{eq:EOM3}.
In this paper, we are interested in solutions where the scalar field has a non-trivial dependence on the coordinate $u$. To write the equations of motion in the language of the holographic RG flows, it will be convenient to rewrite the second-order Einstein equations as a set of first-order equations. Except at special points where $\dot{\f}=0$, which we call a $\f$-bounce, it is locally always possible to invert the relation between $u$ and
$\f(u)$ and define the following functions of $\f$
\be
\label{eq:defSc} S(\f)  \equiv \dot{\f} \, , \ee
\be
\label{eq:defWc} W(\f)  \equiv -2 (d-1) \dot{A} \, , \ee
\be
\label{eq:defTc}  T(\f)  \equiv e^{-2A} R^{(\zeta)}\,.
\ee
In terms of these functions, the equations of motion \eqref{eq:EOM1}--\eqref{eq:EOM3} become
\be
\label{eq:EOM4} S^2 - SW' + \frac{2}{d} T =0 \, ,
\ee
\be
\label{eq:EOM5} \frac{d}{2(d-1)} W^2 -S^2 -2 T +2V =0 \, , \ee
\be
\label{eq:EOM6} SS' - \frac{d}{2(d-1)} SW - V' = 0 \,.
\ee
Note that  equations \eqref{eq:EOM4}--\eqref{eq:EOM6} are algebraic in $T$ and we can partially solve this system by eliminating $T$
\be
T={\frac{d}{2}}S(W'-S)=V-{\frac12}S^2+\frac{dW^2}{ 4(d-1)}\sp \frac{T'}{T}=\frac{W}{(d-1)S}\,.
\label{t}
\ee
It is important to note that $T$ never changes sign (as seen from (\ref{eq:defTc})), and if it vanishes in a generic point, it vanishes everywhere. When $T=0$ then $S=W'$ everywhere.

We are now left with the following equations
\be
\label{eq:EOM7} \frac{d}{2(d-1)} W^2 + (d-1) S^2 -d S W' + 2V =0 \, ,
\ee
\be
\label{eq:EOM8} SS' - \frac{d}{2(d-1)} SW - V' = 0 \, .
\ee
The first-order equations are invariant under two independent $Z_2$ symmetries, that reflect the associated symmetries of the second-order system in (\ref{eq:EOM1})-(\ref{eq:EOM3}):
\be
W\to -W ~~~{\rm and}~~~S\to -S\,,
\label{sym1}\ee
and
\be
S\to -S~~~{\rm and}~~~\f\to -\f~~~~{\rm iff}~~~V(\f)=V(-\f)\,.
\label{sym2}\ee
By solving the second equation \eqref{eq:EOM8} for $W$ algebraically, we obtain
\be
W=\frac{2(d-1)}{d}\big(S'-\frac{V'}{S}\big)\,.
\label{w}
\ee
Substituting (\ref{w}) into the first equation \eqref{eq:EOM7} we obtain
\be
 dS^3S''-\frac{d}{2}S^4  -S^2 (S')^2-\frac{d}{d-1}S^2V +{(d+2)}SS'V' -dS^2V''-(V')^2 =0\,.
\label{eq:EOM9}
\ee
This is a second-order equation in $S$ and its integration requires two integration constants. One
integration constant is related to the vev of the perturbing operator, and the other one is related to the UV curvature (a source in holographic parlance).

The analysis of holographic RG solutions for the metric in \eqref{eq:metric} with $M_d$ slices and a negative potential with one maximum shows three types of special points \cite{C}:

\begin{enumerate}
\item {\bf{UV fixed point} corresponding to an asymptotically $AdS_{d+1}$ boundary.}
At the maximum of the potential, there is a UV fixed point. Solutions of the first order system near this fixed point, come as a two-parameter family, parametrized  by $C$ and $\mathcal{R}$, which are dimensionless parameters related to vev of the relevant operator on the boundary
\be\label{iexpo}
\langle\mathcal{O}\rangle = \frac{C d}{4-\D}|\f_- |^{\frac{\D}{4-\D}}\,.
\ee
and the scalar curvature $R^{UV}$ of the boundary
\be \label{ifr13a}
R^{UV}=\mathcal{R} |\f_- |^{\frac{2}{4-\D}}\,,
\ee
where $\f_-$ is the source of the scalar field operator $\mathcal{O}$ in the boundary field theory
associated with $\f$ and $\D$ is the scaling dimension of the dual operator.
The near-boundary asymptotics of the solutions are given in appendix \ref{UVF}.

\item {\bf{Turning points:}}
The flows may eventually reach a turning point at a finite value of $\f_0$ where either $A(u)$ or $\f(u)$ invert their directions. We call these points as A-bounce or $\f$-bounce. At an A-bounce, the scale factor of $M_d$ space remains finite (IR throat). For more details on the expansions of the solutions near an A-bounce see appendix \ref{bounce}.

\item {\bf{IR end-points:}} At finite $u=u_*$ the scalar field can reach to $\pm \infty$. At this point, the scale factor vanishes and there is a curvature singularity. However, solutions that satisfy Gubser's criterion are good singularities, \cite{Gubser}. In the subsequent section, we analyze the behavior of the RG flows near the IR end-points.

    \item When $M_d$ has positive constant curvature, it was shown in \cite{C} that flows can end at any point on the scalar manifold. This is not the case when $M_d$ has negative constant curvature. Therefore, flows can only end at the maximum of the potential or $\f\to\pm \infty$.

\end{enumerate}

\section{The  asymptotics as $\f\to\pm\infty$}\label{vper}

In this section, we shall discuss the asymptotic solutions to our equations (\ref{eq:EOM7}), (\ref{eq:EOM8}), or equivalently (\ref{eq:EOM9}), when the scalar asymptotes to the boundaries $\f\to\pm \infty$.

We assume that as $\f\rightarrow +\infty$, the potential behaves exponentially as
\be \label{lfpot}
V\simeq -V_\infty e^{2a\f}+\cdots\,,
\ee
where $V_\infty$ is a positive constant, $a$ (in this section only) is any real number, and the ellipsis implies subleading terms.
The case $\f\to-\infty$ can be obtained by taking $a\to -a$ in (\ref{lfpot}).
We keep $a$ arbitrary here, although eventually, in the rest of this paper, we shall consider $a$ bounded as in (\ref{intro2}).

We shall solve asymptotically equation \eqref{eq:EOM9} for $S$, that we reproduce here,
\be
 dSS''-\frac{d}{2}S^2  - (S')^2=\frac{d}{d-1}V -{(d+2)}S'\frac{V'}{S} +dV''+\frac{{V'}^2}{S^2}\,.
\label{s1}
\ee
If $S$ is a solution, then always $-S$ is also a solution. The change of sign inverts the direction of the flow, by changing the direction of the holographic coordinate $u$.

There are several independent classes of solutions for (\ref{s1})  as $\f\to+\infty$.  We present these solutions in the next two subsections. The detailed analysis can be found in appendix \ref{avper}.

\subsection{The generic (type 0) solution\label{0}}

This is a leading solution near $\f\to\infty$ and contains all constants of integration.  It is a valid solution in most cases. In this solution, the potential can be treated as a perturbation.
This full solution can be written in the form,
\be
S=\bar{S}_0+\cdots\,,
\label{IR1a}
\ee
where $S_0$ is the solution to the equation \eqref{s1} without a potential (i.e. $V=0$), and the ellipsis stands for sub-leading contributions.
The leading order solution $S_0$ satisfies
\be
d\bar{S}_0\bar{S}_0''-\frac{d}{2}\bar{S}_0^2  - (\bar{S}_0')^2=0\,.
\label{s2}
\ee
The exact solution of \eqref{s2} is
\be\label{Exs0}
\bar{S}_0(\f)= C_1 \Big[\cosh\big(\sqrt{\frac{d-1}{2d}} \left(\f- C_2 \right)\big)\Big]^{\frac{d}{d-1}}\,,
\ee
where $C_1$ and $C_2$ are two constants of integration. We can keep the leading terms multiplying each constant  of integration as follows
\be \label{apps0}
\bar{S}_0(\f)=
c_1 e^{\sqrt{\frac d{2(d-1)}}\varphi}+c_2 e^{-\frac{(d-2)\varphi}{\sqrt{2d(d-1)}}}+\cdots,
\ee
where $c_{1,2}$ are functions of $C_1$ and $C_2$ in (\ref{Exs0}) and the ellipsis denotes sub-leading exponential terms.

In order to consider the potential as a perturbation in \eqref{s1}, the value of $a$ in \eqref{lfpot} should be
\be \label{algb}
\frac{V(\f)}{S^2(\f)}\Big|_{\f\rightarrow +\infty}<1 \Rightarrow 2a<2b_1 \Rightarrow a<a_G\,.
\ee
In other words, the value of $a$ should be below the Gubser bound $a_G$.
Therefore, this solution exists for all $a<a_G$, and this is what we assume henceforth.

We obtain the leading behavior of $W$ from (\ref{w})
\be
\bar{W}_0(\f)= c_1 \sqrt{\frac{2(d-1)}{d}} e^{\sqrt{\frac{d}{2(d-1)}}\f}+\cdots
- c_2 \frac{d-2}{d}\sqrt{\frac{2(d-1)}{d}} e^{-\frac{(d-2)\varphi}{\sqrt{2d(d-1)}}}+\cdots \,,
\label{s4}
\ee
and from (\ref{t})
\be
\bar{T}_0(\f)= -\frac{2c_1 c_2 (d-1)}{d} e^{\sqrt{\frac{2}{d(d-1)}}\f}+\cdots\,.
\label{s5}
\ee
Our derived solution up to this order for the equation
\eqref{s1} would then be as follows (see appendix \ref{a0} for more details)
\be\label{s13a}
S=c_1 e^{\sqrt{\frac d{2(d-1)}}\varphi}+c_2 e^{-\frac{(d-2)\varphi}{\sqrt{2d(d-1)}}}
-\frac{a V_\infty\big(4 a (d-1)\!+\!\sqrt{2d(d-1)}\!-\!d\big)}{2c_1(2 a^2 (d-1) -d)}  e^{(2a-\sqrt{\frac{d}
{2(d-1)}})\f}+\cdots.
\ee
The solution above contains three exponentials, the last one being determined by the potential. Each of these exponentials comes with a whole series of subleading exponential terms, that we do not write explicitly here, as it is not needed.
As a function of the value of $a<a_G$, there are  two possibilities:
\begin{enumerate}

\item When
\be \label{domsin}
\sqrt{\frac{1}{2d(d-1)}}<a<a_G\,,
\ee
 the third term in \eqref{s13a} is bigger than the second term.
So in this domain, the leading and sub-leading solutions for $S(\f)$ are given by
\be \label{s15}
S(\f)=c_1 e^{\sqrt{\frac d{2(d-1)}}\varphi}-\frac{a\big(4 a (d-1)+ \sqrt{2d(d-1)} -d\big)}{2c_1(2 a^2 (d-1) -d)} V_\infty e^{(2a-\sqrt{\frac{d}{2(d-1)}})\f}+\cdots.
\ee
Moreover, for $W$ and $T$ we have
\be\label{s16}
W(\f)=c_1 \sqrt{\frac{2(d-1)}{d}} e^{\sqrt{\frac{d}{2(d-1)}}\f}\!-\frac{2 a (d-1)\!+\!\sqrt{2d(d-1)}}{2c_1 (2 a^2 (d-1) - d)} V_\infty e^{(2a-\sqrt{\frac{d}{2(d-1)}})\f}+\cdots,
\ee
\be
T(\f)=-\frac{2c_1 c_2 (d-1)}{d} e^{\sqrt{\frac{2}{d(d-1)}}\f}+\cdots\,. \label{s17}
\ee

\item The other possibility is when
\be  \label{s18}
a\leq \sqrt{\frac{1}{2d(d-1)}}\,,
\ee
and the second term in \eqref{s13a} is bigger that the third term.
In this case, we  find that the leading and sub-leading terms for  $S, W$, and $T$ are given by the
\be \label{s19}
S(\f)=
c_1 e^{\sqrt{\frac d{2(d-1)}}\varphi}+c_2 e^{-\frac{(d-2)\varphi}{\sqrt{2d(d-1)}}}+\cdots\,,
\ee
\be
W(\f)= c_1 \sqrt{\frac{2(d-1)}{d}} e^{\sqrt{\frac{d}{2(d-1)}}\f}- c_2 \frac{d-2}{d}\sqrt{\frac{2(d-1)}{d}} e^{-\frac{(d-2)\varphi}{\sqrt{2d(d-1)}}}+\cdots \,,\label{s20}
\ee
\be
T(\f)= -\frac{2c_1 c_2 (d-1)}{d} e^{\sqrt{\frac{2}{d(d-1)}}\f}+\cdots\,.\label{s21}
\ee
This case also includes the generic solution for $a<0$.

\end{enumerate}

\subsection{Type I and type II solutions\label{I}}

There are special solutions whose leading exponential behavior is dictated by the leading behavior of the potential at large $\f$ in (\ref{lfpot}).
We parametrize the leading behavior of the solution  as
\be\label{s0fun}
S_0=S_\infty^{(0)} e^{b_2\f}\,.
\ee
Moreover, we have
\be \label{wt0fun}
W_0=W_\infty^{(0)} e^{b_2\f}\sp
T_0=T_\infty^{(0)} e^{2b_2\f}\,.
\ee
We obtain two possibilities by inserting \eqref{s0fun} into \eqref{s1}

\begin{itemize}

\item The type I solution:
\be \label{s1GV}
S_\infty^{(0)}=\pm\sqrt{\frac{2V_\infty}{d-1}} \sp
b_2=a \,,
\ee
with
\be\label{s1WT}
W_{\infty}^{(0)}=\pm a\sqrt{8(d-1)V_\infty}\sp
T_{\infty}^{(0)}=d(2a^2-\frac{1}{d-1})V_\infty\,.
\ee

\item The type II solution:
\be \label{s2GV}
S_\infty^{(0)}= \pm 2 a \sqrt{\frac{(d-1) V_\infty}{d-2 a^2 (d-1)}}\sp b_2=a \,,
\ee
with
\be\label{s2WT}
W_{\infty}^{(0)}=\pm  2 \sqrt{\frac{(d-1)V_\infty}{d-2 a^2 (d-1)}} \sp T_{\infty}^{(0)}=0\,.
\ee
\end{itemize}
Both solutions above are exact solutions to equation \eqref{s1}, assuming that the potential is given by its leading term in \eqref{lfpot}.
Also, note that the type II solutions exist if $|a|<a_G$.

To find whether these solutions contain any integration constants, we now consider a perturbation around the solution \eqref{s0fun} as follows
\be\label{s3GV0}
S=S_\infty^{(0)} e^{a\f}+\delta S(\f)\,.
\ee
Upon insertion \eqref{s3GV0} into \eqref{s1}, we obtain the following linearized second-order differential equation
\be
\label{s3GV1}
\delta S''-\frac{2a}{d}\Big(1+\frac{(d+2)V_\infty}{{S_\infty^{(0)}}^2}\Big) \delta S'
+\Big(a^2-1+\frac{2a^2 (d+2)V_\infty}{d{S_\infty^{(0)}}^2}+\frac{8a^2 V_\infty^2}{d{S_\infty^{(0)}}^4}\Big)\delta S\simeq 0\,.
\ee
In the equation above, we have dropped the non-linear terms in $\delta S$, as they are sub-leading and we shall check this a posteriori.

We denote by $\l_{1,2}$ the two roots of the characteristic polynomial
\be
\label{s3GV1a}
\l^2-\frac{2a}{d}\Big(1+\frac{(d+2)V_\infty}{{S_\infty^{(0)}}^2}\Big) \l
+\Big(a^2-1+\frac{2a^2 (d+2)V_\infty}{d{S_\infty^{(0)}}^2}+\frac{8a^2 V_\infty^2}{d{S_\infty^{(0)}}^4}\Big)\,.
\ee

We obtain the following cases

\be\label{s3GV}
\delta S=
\begin{cases}
S_{\infty}^{(1)} e^{\l_1\f}+S_{\infty}^{(2)} e^{\l_2\f}\,,
& \l_1 \neq \l_2\in \mathbb{R}\,,
\\ \cr
e^{\l\f}\big(S_{\infty}^{(1)}\f +S_{\infty}^{(2)}\big)\,, &
\l_1=\l_2=\l \in \mathbb{R}\,, \\  \cr
e^{\l\f}\big(S_{\infty}^{(1)}\cos(\omega\f) +S_{\infty}^{(2)}\sin(\omega\f)\big)\,, &
\l_{1,2}=\l\pm i\omega,\quad \l,\omega \in \mathbb{R}\,, \\
\end{cases}
\ee
where $S_{\infty}^{(1)}$ and $S_{\infty}^{(2)}$ are free constants of integration.
The associated expansions for $W$ and $T$ can be found in  (\ref{as4GV}) and (\ref{as5GV}).

For all of the solutions to be consistent, the subleading solutions must be smaller than the leading ones.
This requires that $Re(\l_i)<a$ for a solution to be acceptable. Otherwise, it is not. Each acceptable solution gives an extra integration constant.

\subsubsection{Type I solutions}
For the type I solution, with the values given in \eqref{s1GV}, we find the following values for $\l_1$ and $\l_2$
\begin{gather} \label{s6GV}
 \l_1= \frac{a (d+1)-\sqrt{a^2 (d-9) (d-1)+4}}{2}\,, \\
 \l_2= \frac{a (d+1)+\sqrt{a^2 (d-9) (d-1)+4}}{2}\,.\label{t1lam2}
\end{gather}
Defining
\be \label{ad}
a_d=\frac{2}{\sqrt{(9-d)(d-1)}}\,,
\ee
we classify the allowed region where the sub-leading terms are consistent, ($\l_{1,2}<a$) in tables \ref{type1d8} and \ref{type1d9}. The $\checkmark$ sign indicates an acceptable solution coming together with its integration constant.

\begin{table}[!ht]
\centering
\begin{tabular}{|c|c|c|c|c|c|}
\hline
domain  &  $a\leq \!-a_d$ & $-a_d<\!a\!<\!-a_C$ & $ |a|\!<\!a_C$ & $a_C\!<\! a\!<\!a_d$ & $a_d\!\leq a\!<\! a_G$  \\ \hline
$Re(\l_1)<a$ & \checkmark & \checkmark  & \checkmark & $\times$ &  $\times$ \\ \hline
$Re(\l_2)<a$ & \checkmark & \checkmark & $\times$ & $\times$ & $\times$ \\ \hline
Slice curvature &+ & + & $-$ & + & + \\ \hline
\end{tabular}
\caption{\footnotesize{Type I solutions. The presence of integration constants when the boundary dimension is $1<d\leq 8$. A $\checkmark$ indicates the presence of an integration constant. The third row shows the sign of slice curvature defined by $T(\f)$ in equation \eqref{s1WT}.}}\label{type1d8}
\end{table}
\begin{table}[!ht]
\centering
\begin{tabular}{|c|c|c|c|}
\hline
domain  & $ a<-a_C$ &  $|a|<a_C$ & $a_C< a<a_G$ \\ \hline
$\l_1<a$  & \checkmark  & \checkmark & $\times$ \\ \hline
$\l_2<a$  & \checkmark & $\times$  & $\times$ \\ \hline
Slice curvature  & + & $-$  & + \\ \hline
\end{tabular}
\caption{\footnotesize{Allowed domain for sub-leading terms of type I solution when the boundary dimension is  $9\leq d$.}}\label{type1d9}
\end{table}

We observe from the tables that in different ranges of $a$ we can have solutions with two integration constants, only one integration constant, or completely isolated solutions.

Knowing the expansion of $S(\f)$, we obtain the coefficients of $W(\f)$ and $T(\f)$ in \eqref{as4GV} and \eqref{as5GV} in equations (\ref{ayy1})--(\ref{ayy4}) in appendix \ref{avper}.

We should note that when both $\l_1$ and $\l_2$ are complex numbers, i.e.
\be \label{sad3}
|a|>a_d \sp 1<d\leq 8\,,
\ee
the solution for $\delta S$ is given by the third line of \eqref{s3GV}
\be  \label{sad4}
\delta S=e^{\frac{a(d+1)}{2}\f}\big(S_{\infty}^{(1)}\cos(\omega\f) +S_{\infty}^{(2)}\sin(\omega\f)\big)\sp
\omega=\sqrt{\frac{a^2}{a^2_d}-1}\,.
\ee
The above result shows that $\delta S$ is an acceptable sub-leading only if $a<-a_d$.

The boundaries of the cases in the tables above must be analyzed separately and this is done in appendix \ref{avper}.

\subsubsection{Type II solutions}
In this case,  equation \eqref{s2GV} has  real coefficients for the leading term, and  we should consider
\be \label{contii}
-a_G<a<a_G\,.
\ee
We obtain the following values for $\l_1$ and $\l_2$ in the perturbed solution \eqref{s3GV}
\be  \label{s8GV}
\l_1 = \frac{1}{a (d-1)}-a\,,
\ee
\be
\l_2 = \frac{d}{2a(d-1)}\,.  \label{t2lam2}
\ee
Here the sub-leading solutions are allowed ($\l_{1,2}<a$) according to table \ref{ttype2}.
\begin{table}[!ht]
\centering
\begin{tabular}{|c|c|c|c|c|}
\hline
domain  & $-a_G <  a  \leq  -a_C$ & $-a_C <  a  <  0$ & $0 <  a  \leq  a_C$ & $a_C <  a <  a_G$  \\ \hline
$\l_1  <  a$  & -- & \checkmark & -- &  \checkmark \\ \hline
$\l_2  <  a$  & \checkmark &  \checkmark & -- & -- \\ \hline
\end{tabular}
\caption{\footnotesize{Allowed domain for sub-leading terms of type II solution.}}\label{ttype2}
\end{table}
Using the expansion of $S(\f)$, we obtain the coefficients of $W(\f)$ and $T(\f)$ from \eqref{as4GV} and \eqref{as5GV} in (\ref{azz1})--(\ref{azz2}) in appendix \ref{avper}.

According to the table \ref{ttype2} we can have type II  solutions with two or one constant of integration or completely isolated solutions.
We should note that the sign of slice curvature is given by the sign of the next leading term $T_\infty^{(1)}$ in \eqref{azz2}, which we can always choose as a negative number.

\subsection{Properties of the confinement-deconfinement regions}
In this paper, we are interested in solutions with an acceptable singularity in the sense of Gubser and with exponentially large potential at large $\f$. Considering these, we now restrict the potential in \eqref{lfpot} to parameters
\be \label{resta}
0<a<a_G\,.
\ee
According to the above values of $a$, we have the following results:
\begin{itemize}
\item{Type I solution:}
\end{itemize}
Inside the region \eqref{resta}, according to the tables \ref{type1d8} or \ref{type1d9}, there is only one free parameter for the next to leading solution $\delta S$ if $0<a<a_C$ i.e.
\be \label{saga}
S = \pm \sqrt{\frac{2V_\infty}{d-1}} e^{a\f}+S_\infty^{(1)} e^{\l_1 \f}+\cdots\,,
\ee
where $\l_1$ is given in \eqref{s6GV}.

In the range $0<a<a_C$ we have the following non-zero coefficients for the other functions from \eqref{as4GV} and \eqref{as5GV}
\begin{gather}\label{sol1GV}
W_{\infty}^{(0)}=\pm a\sqrt{8(d-1)V_\infty}\sp
T_{\infty}^{(0)}=d(2a^2-\frac{1}{d-1})V_\infty=2d(a^2-a_C^2)V_\infty\,,
\end{gather}
and
\be\label{sol1GVX}
W_{\infty}^{(1)}= -\frac{ d-1 }{d} \big(a(d-3)+\sqrt{a^2 (d-9) (d-1)+4}
\big)S_{\infty}^{(1)}\,,
\ee
\be\label{sol1T}
T_{\infty}^{(1)}=\mp\sqrt{\frac{2V_\infty}{d-1}}\left(1+a^2(d-1)(d-3) +a(d-1)\sqrt{a^2 (d-9) (d-1)+4}\right) S_\infty^{(1)}\,.
\ee
We observe from \eqref{sol1GV} that
\be \label{s7GV}
0<a<a_C \quad\Rightarrow\quad T_{\infty}^{(0)}<0\,.
\ee
Since the sign of $T$ controls the sign of the curvature of the slice, we conclude that solution I asymptotics, have negative curvature only in the deconfined regime. For this reason, we shall not be interested in them in this paper\footnote{They are however relevant, if holographic confining theories are put on spheres or de Sitter space.}.

\begin{itemize}
\item{Type II solution:}
\end{itemize}
Table \ref{ttype2} shows that only between $a_C<a<a_G$, there is an acceptable perturbation around the type II solution. In this region we obtain
\be \label{sftii}
S=\pm 2 a \sqrt{\frac{(d-1) V_\infty}{d-2 a^2 (d-1)}} e^{a\f}+S_\infty^{(1)} e^{(\frac{1}{a(d-1)}-a)\f}+\cdots\,.
\ee
Moreover, for the first rows of the expansions in \eqref{as4GV} and \eqref{as5GV} the non-zero coefficients are
\be\label{sol2GV}
W_{\infty}^{(0)}= \pm 2 \sqrt{\frac{(d-1)V_\infty}{d-2 a^2 (d-1)}} \sp T_{\infty}^{(0)}=0\,,
\ee
\be
W_{\infty}^{(1)}=-\frac{d-2}{a d}S_\infty^{(1)} \sp
T_{\infty}^{(1)}=\mp\frac{ \left(2 a^2 (d-1)+d-2\right) }{a (d-1)}\sqrt{\frac{(d-1)V_\infty}{d-2 a^2 (d-1)}}S_{\infty}^{(1)}\,. \label{sol2GVN}
\ee
Unlike in type I, in type II the sign of the curvature can be both positive or negative, depending on the sign of the integration constant $S_{\infty}^{(1)}$. This means that at large values of $\f$ when $S\rightarrow \pm \infty$ to have a negative slice curvature, $S_\infty^{(1)}$ should be positive or negative respectively.

\begin{figure}[!ht]
\begin{center}
\includegraphics[width = 10cm]{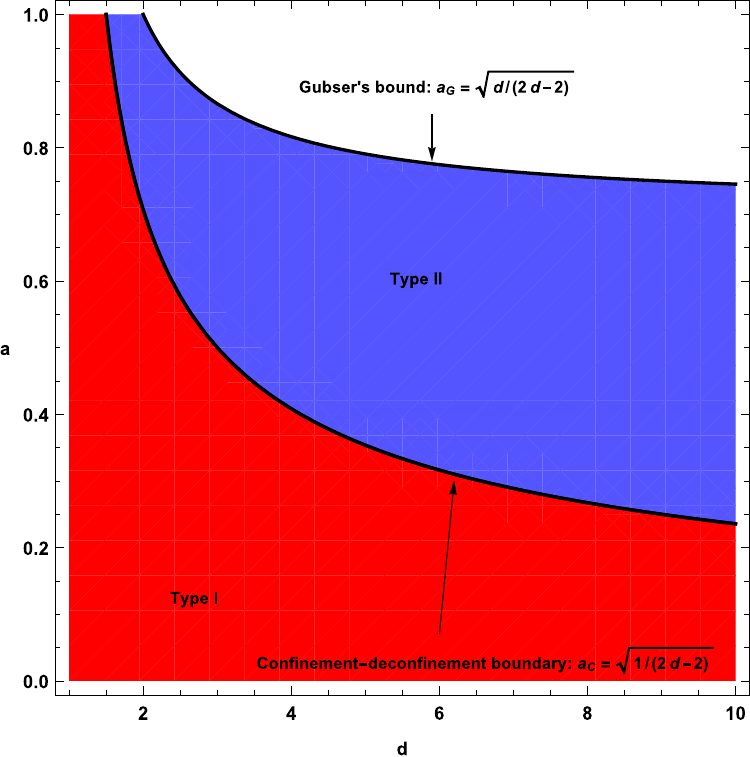}
\caption{\footnotesize{Regions of validity for type I solution (red region) $0<a<a_C$ and type II solution (blue region) $a_C<a<a_G$.
The upper bound of the type II solution is the Gubser bound defined in  \eqref{ag} and the boundary of the two regions is the confinement-deconfinement boundary \eqref{ac}.
We are assuming that the slices have negative curvature.
} }\label{CDCM}
\end{center}
\end{figure}

We illustrate the regions of validity (phases) of type I and type II solutions for $0<a<a_G$, in figure \ref{CDCM}.

The red region in figure \ref{CDCM} describes the area in the $a$-versus-$d$ diagram where the type I solution exists with the domain in $0<a<a_C$. It is limited from above by the confinement-deconfinement boundary. The space of type II solutions is the blue region and for our purposes, it is limited by $a_C<a<a_G$.
The reason is that it does not exist for $a>a_G$ and for $a<a_C$,  it cannot have negative curvature slices which is the focus of this paper.

\subsection{On the singularity of IR end-points}
The solutions we discussed so far,  have a curvature singularity at large $\f$. This can be verified by computing their scalar curvatures
\be \label{s9GV}
R = -\big(2d \ddot{A} (u)+d(d+1) \dot{A}^2 (u)\big)
+e^{-2 A(u)} R^{(\zeta)}\,,
\ee
or equivalently by using the equations \eqref{eq:defWc}--\eqref{eq:EOM6}
\be \label{s10GV}
R=\frac{1}{2}S(\f)^2+\frac{d+1}{d-1}V(\f)\,.
\ee
In leading order, for all three types (0, I, II) of solutions, the scalar curvature is
\be \label{s11GV}
R=
\left\{ \begin{array}{lll}
\displaystyle -\frac{1}{2}c_1^2e^{2 \sqrt{\frac{d}{2(d-1)}} \f}+\cdots,&\phantom{aa} &{\rm type~~ 0},\\ \cr
\displaystyle -\frac{d}{d-1} V_\infty e^{2 a \f}+\cdots ,&\phantom{aa}&{\rm type~~ I},\\ \cr
\displaystyle  -\frac{d \left(4 a^2 -\frac{d+1}{d-1}\right) }{\left(2 a^2 (d-1)-d\right)}V_\infty e^{2 a \f}+\cdots,&\phantom{aa}&{\rm type~~ II},
\end{array}\right.
\ee
which diverges as $\f\to\infty$.
Note that the leading divergence of the curvature for the type-0 solution is due to the divergence of $\dot\f$, while for the two others, it diverges as the scalar potential.

Moreover, by integrating the first order equation \eqref{eq:defSc} and \eqref{eq:defWc} one finds the scale factor vanishes as
\be
e^{A(u)}=
\left\{ \begin{array}{lll}
\displaystyle (u_0-u)^{\frac{1}{d}}+\cdots,&\phantom{aa} &{\rm type~~ 0},\\ \cr
\displaystyle (u_0-u)+\cdots,&\phantom{aa}&{\rm type~~ I},\\ \cr
\displaystyle  (u_0-u)^{\frac{1}{2a^2(d-1)}}+\cdots,&\phantom{aa}&{\rm type~~ II},
\end{array}\right.
\ee
and the dilaton profile diverges as
\be
\f(u)=
\left\{ \begin{array}{lll}
\displaystyle -\frac{2(d-1)}{d}\log\Big[\sqrt{\frac{d}{2(d-1)}} c_1 (u_0-u)\Big]+\cdots,&\phantom{aa} &{\rm type~~ 0},\\ \cr
\displaystyle -\frac{1}{a}\log\Big[\pm\sqrt{\frac{2V_\infty}{d-1}} a (u_0-u)\Big]
+\cdots,&\phantom{aa}&{\rm type~~ I},\\ \cr
\displaystyle  -\frac{1}{a}\log\Big[\pm 2 a^2 \sqrt{\frac{(d-1) V_\infty}{d-2 a^2 (d-1)}} (u_0-u)\Big]
+\cdots,&\phantom{aa}&{\rm type~~ II},
\end{array}\right.
\ee
where $u_0$ is an integration constant that indicates the location of the singularity in the bulk.

The singularity of the type 0 solution is ``bad" in the Gubser classification, \cite{Gubser}. The singularity of the II solution is of the ``good" type, \cite{iQCD}, and therefore this is a holographically acceptable solution. Soon we shall see how this singularity can be resolved by lifting to a higher-dimensional geometry.

\section{Confining holographic theories on negative curvature manifolds  and RG flows\label{REGSOL}}


In this section, we shall find the regular solutions, corresponding to potential RG flows,  for a theory with a specific confining potential. We start with the following action
\be\label{actd}
S_{d+1}=\int d^{d+1}x\sqrt{ g}\left[ R-{\frac12}(\pa{\f})^2- V({\f})\right]\,.
\ee
We assume  a metric with constant curvature slices, $M_d$ with metric $\zeta_{\m\n}$, and a non-constant scalar field
\be
ds^2=du^2+e^{2A(u)}\zeta_{\m\n}dx^\m dx^\nu\sp \f=\f(u)\,.
\label{Md}\ee
We consider the following scalar potential\footnote{We do not expect qualitative differences by varying this potential at intermediate ranges of $\f$.}
\be\label{ppp}
{V}(\f)=
-\frac{d(d-1)}{{{\ell}}^2}\left(b\f^2+\cosh^2(a\f)\right)\sp
b=\frac{\Delta(d-\Delta)}{2 d (d-1)}-a^2\,.
\ee
As $\f\rightarrow \pm \infty$, the above potential diverges as
\be \label{dpoti}
V(\f)\rightarrow - \frac{d(d-1)}{4\ell^2}e^{\pm 2a \f}\,,
\ee
where we assumed  that $a<a_G$, the Gubser's bound. The shape of this potential is shown in figure \ref{pot}.
This potential has a maximum at $\f=0$ (UV fixed point) and near this point, it can be expanded as
\be \label{dpot0}
V(\f)=-\frac{d(d-1)}{\ell^2}-\frac12 m^2\f^2+\mathcal{O}(\f^4)\sp m^2=\frac{\Delta  (d-\Delta)}{\ell^2}\,.
\ee

$\ell$ determines the length scale of asymptotically $AdS$ solutions, $\Delta$ determines  $m^2$ and is the scaling dimension of the operator dual to the scalar $\f$ near the UV fixed point.
$a$  determines the asymptotic behavior of the potential (confinement or deconfinement).
By comparing \eqref{dpoti} with \eqref{lfpot} we observe that
\be
V_\infty=\frac{d(d-1)}{4\ell^2}\,,
\ee
therefore we can use the same arguments of section \ref{vper} for IR asymptotics of the solutions.
More details on the regular solutions of the theory are given in appendix \ref{secRS}.

\subsection{The space of solutions}
Since we are interested in solutions in the confining region of figure \ref{CDCM}, and for numerical proposes, we fix the constants of the theory as follows
\be\label{numbers}
d=4\sp \Delta=\frac32\sp {\ell}=1 \sp a=\sqrt{\frac{7}{24}}\sp b=-\frac{13}{96}\,.
\ee
For the specific choice $d=4$ we have
\be \label{numbers-2}
a_C=\frac{1}{\sqrt{6}}\sp a_G=\frac{2}{\sqrt{6}}\,,
\ee
so our choice for $a$ is lying in the confining region i.e. $a_C<a<a_G$. Moreover, the computability bound \eqref{acomp} shows that $a_{comp}=\sqrt{\frac{1}{3}}$ and therefore $a<a_{comp}$ in our case.

\begin{figure}[!ht]
\begin{center}
\includegraphics[width = 10cm]{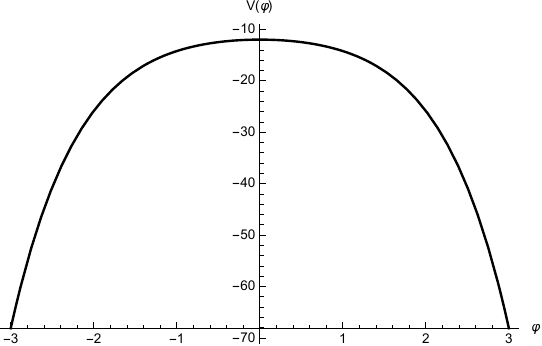}
\caption{\footnotesize{The potential \eqref{ppp} for the values of the parameters shown in \eqref{numbers}. The solutions of the equation of motion for this potential are in the confining regime. }}\label{pot}
\end{center}
\end{figure}

For the values in \eqref{numbers}, the potential \eqref{ppp} has a maximum at $\varphi =0$, see figure \ref{pot}. This corresponds to a UV fixed point, around which the operator dual to $\f$ has dimension $\Delta=\frac32$.

In general, all solutions flow between IR end-points and/or boundaries. We classify solutions according to their regularity. We use ``{\bf{Reg}}" for a regular end-point (which from the results in Section \ref{vper} can only be of type II), ``{\bf{Sing}}" for a singular end-point, and ``{\bf{UV}}" when there is a UV fixed point ($AdS$-like boundary). We have the following classes:

\begin{itemize}
\item {\bf{Regular solutions}}
\begin{enumerate}
\item {\bf{Two-boundary solutions (UV-UV):}}
These solutions describe holographic interfaces when $M_d$ in (\ref{Md}) is non-compact and wormholes when $M_d$ is compact, \cite{Ghodsi:2022umc}. They are stretched between two asymptotically $AdS_{d+1}$ boundaries (which are located both at the maximum of the potential at $\f=0$).

\item {\bf{UV-Reg solutions:}}
There exist solutions that have one regular end-point at $\f\rightarrow \pm \infty$ and an asymptotically $AdS_{d+1}$ boundary. The regular end-point corresponds to the type II asymptotics in section \ref{I}.

\item{\bf{Reg-Reg solutions:}}
There are solutions in which both end-points are regular (type II solutions) as $\f\rightarrow \pm \infty$.

\end{enumerate}

\item

{\bf{Singular solutions}}

\begin{enumerate}

\item {\bf{UV-Sing solutions:}}
These solutions  stretch between an asymptotically $AdS_{d+1}$ boundary and a singular end-point at ($\f\rightarrow \pm \infty$). These singular asymptotics are the type 0 asymptotics, discussed in section \ref{0}.

\item {\bf{Reg-Sing solution:}}
One end-point at $\f=\pm\infty$ is regular (type II) but the other end-point also at $\f=\pm\infty$ is singular (type 0).

\item {\bf{Sing-Sing solution:}}
When both end-points at $\f=\pm\infty$ are singular (type 0).

\end{enumerate}
\end{itemize}

In practice, to find all possible solutions of the equations of motion \eqref{eq:EOM7} and \eqref{eq:EOM8}, we proceed as follows. First, we start with the regular asymptotics (type II) and construct all solutions with such asymptotics.
 This procedure does not give all possible regular solutions, but only a subset of the ones that stretch all the way to $\f\to \pm\infty$.

However, this procedure obtains all solutions that do not have an A-bounce.
The reason is that by selection,  the scale factor, $e^A$, is monotonic for such solutions.
As the flow starts at an asymptotically $AdS_{d+1}$ boundary, at some point along the flow the scale factor will vanish and the flow will end.
As we have shown in section \ref{vper}, the flows can end with a shrinking scale factor only at $\f=\pm\infty$.
Therefore, flows with a monotonic scale factor always interpolate between $\f=0$ and $\f=\pm \infty$.

   In this context all such solutions are {\em uniquely} labelled by the parameter $S_{\infty}^{(1)}$ that characterizes such asymptotics, defined in (\ref{sftii})

The rest of the solutions have necessarily at least an $A$-bounce, namely a point where $\dot A=0$.
To construct them, we assume that the A-bounce is at point $\f_0\in \RR$.
We can then solve the first-order equations numerically.

A complete set of initial conditions is $\f_0$ and the value of S at $\f_0$: $S_0\equiv S(\f_0)\in \RR$.
For a given set of real numbers $(\f_0,S_0)$ we solve the equations
with initial conditions
\be \label{inis}
W(\varphi_0)=0 \sp  S(\varphi_0)=S_0>0\,.
\ee
for both $\f<\f_0$ and $\f>\f_0$.
 Note that we have $S\rightarrow -S$ and $W\rightarrow -W$ symmetry in the first-order equations.
Therefore, the parameters $\f_0$ and $S_0$ correspond to the two integration constants of the equations, (\ref{eq:EOM7}), (\ref{eq:EOM8}).

In this way, we obtain all solutions containing at least one $A$-bounce.
There is an overlap of this set of solutions with solutions found with the first method: some solutions can end at $\f=\pm\infty$ and have at least one $A$-bounce.
In the second method, when $|S_0|>S_0^c$, where $S_0^c$ a concrete positive real number that can be determined numerically, then the solutions interpolate between the maximum at $\f=0$ and $\f=\pm\infty$.

Although solutions obtained by the first method, are uniquely labeled by the parameter $S^{(1)}_{\infty}$, solutions obtained by the second method are not uniquely labeled by their initial data $(\f_0, S_0)$.
This is because the solution may have more than on $A$-bounce. For example a solution with one A-bounce at $\f_0$ and another at $\f_1$ will be found twice, once with data $(\f_0,S_0)$ and another with $(\f_1,S_1)$.
Therefore, a solution with $n$ A-bounces appears, generically, with multiplicity $n$ in our set.

\begin{figure}[!t]
\centering
\includegraphics[width = 14cm]{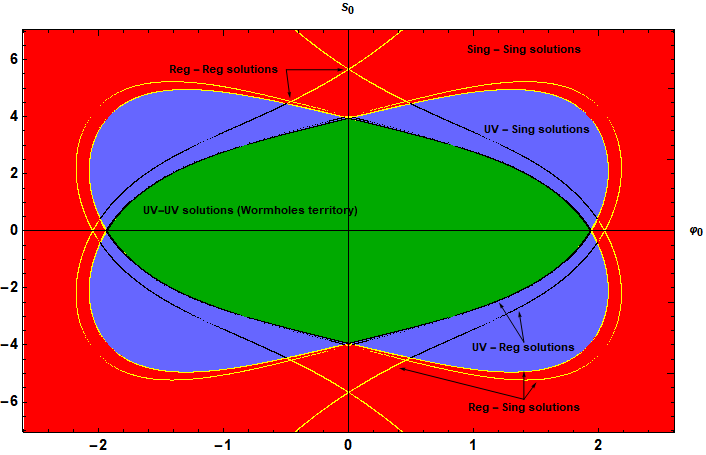}
\caption{\footnotesize{Space of solutions with at least one A-bounce. Different colors represent different classes of solutions as explained in the text.}}\label{map}
\centering
\end{figure}
\begin{figure}[!t]
\centering
\begin{subfigure}{0.49\textwidth}
\includegraphics[width=1\textwidth]{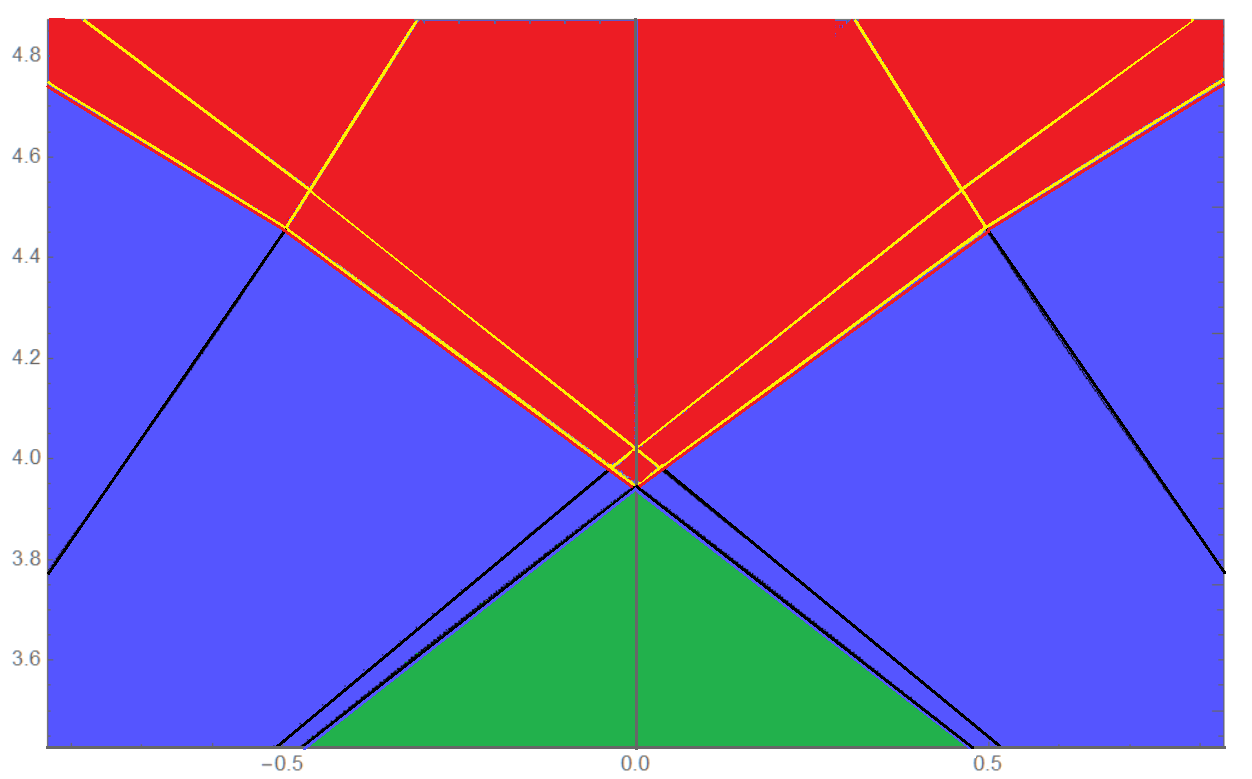}
\caption{}\label{boundarymapup}
\end{subfigure}
\centering
\begin{subfigure}{0.49\textwidth}
\includegraphics[width=1\textwidth]{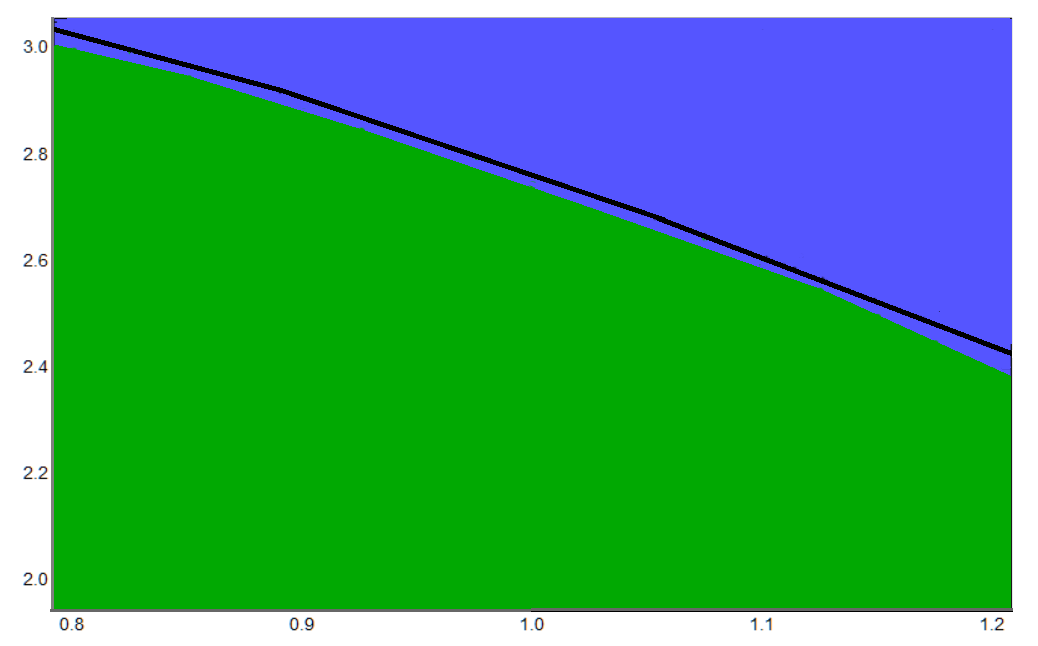}
\caption{\vspace*{0cm}}\label{boundarymap}
\end{subfigure}
\caption{\footnotesize{Some parts of the map \ref{map} with a higher resolution. (a) The intersection area of red, blue, and green regions. (b) A region very near to blue and green regions. The black curve is a small distance from the boundary between the two.}}
\end{figure}

Figure \ref{map} shows the space of solutions with at least one A-bounce. The different colored regions in this map belong to different classes of solutions:

\begin{itemize}

\item In the green region, only the two-boundary  (UV-UV) solutions exist.

\item In the blue regions, we find the generic UV-Sing solutions.

\item The red region is the space of solutions of the generic Sing-Sing type.

\item Inside this space, there are small subspaces that contain regular solutions. All the black curves represent the UV-Reg solutions. Although in figure \ref{map} two such black lines are visible, upon zooming into this map one can observe an infinite number of black lines as one approaches the green-blue boundary. This can be partly seen in the figures \ref{boundarymapup} and \ref{boundarymap} which present a (partial) zoom in this region. A more detailed discussion of these curves will appear in section \ref{rfe}.

\item The yellow curves belong to the solutions of the Reg-Sing type.

\item At the points where the yellow curves cross we have the Reg-Reg solutions.

\end{itemize}

We emphasize again that since each solution may have many A-bounces,
not all points on this plot, \ref{map}, correspond to different solutions. The distinct solutions are a subset of \ref{map}.

We present some examples of these solutions in the subsequent subsections.

\subsection{Solutions with two $AdS$ boundaries}

Such solutions start at a UV fixed point ($AdS$ boundary)  and end at the same UV fixed point\footnote{Here this is the only possibility, as our potential has a single UV fixed point. In \cite{Ghodsi:2022umc} we have shown that if more UV fixed points exist, then there are solutions connecting all of them.}. In the theory that we have considered here, there is only one extremum for the potential at $\f=0$ so the UV fixed point at both ends is at the same value of $\f$ but one of the end-points is at $u\rightarrow +\infty$, and the other one at $u\rightarrow -\infty$.

$\bullet$ In the two-boundary solutions  $\f$ varies over a compact region and never becomes asymptotically large.
They all have one or more  A-bounces as well as $\f$-bounces.

$\bullet$ By approaching the boundary of the green region from the inside, the number of A-bounces and $\f$-bounces increases and asymptotes to infinity while the values of $\f$ remain bounded. This can be observed for example in figure  \ref{bound2}  where indicative solutions are plotted by plotting their functions $W(\f)$ and $S(\f)$. Many more details about such solutions can be found in appendix \ref{loops}.

For example, in figures \ref{W12} and \ref{W13} we have sketched two different types of these two-boundary solutions. The difference between these two examples is in the number of A-bounces ($W=0$, $S\not=0$) and $\f$-bounces (the point where $S=0$ but $W\neq 0$). Since these solutions are bridges between two UV fixed points on the top and bottom ($W=\pm 6, S=0$), we should have at least one A-bounce for each solution. Therefore all such solutions are located in the green region of the map \ref{map}. Figure \ref{symsol} shows a symmetric solution of this type.

\begin{figure}[!ht]
\centering
\begin{subfigure}{0.48\textwidth}
\includegraphics[width=1\textwidth]{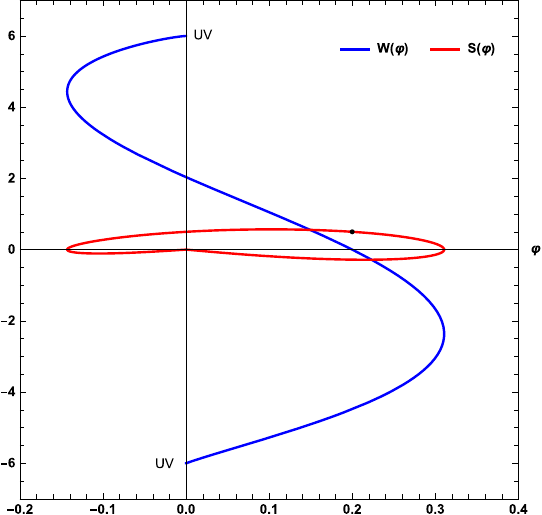}
\caption{}\label{W12}
\end{subfigure}
\centering
\begin{subfigure}{0.48\textwidth}
\includegraphics[width=1\textwidth]{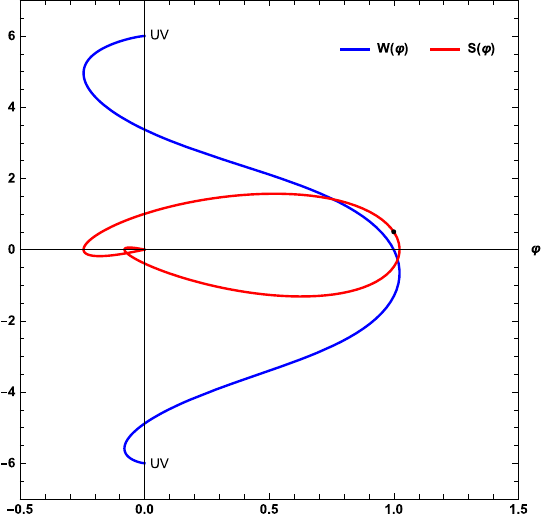}
\caption{\vspace*{0cm}}\label{W13}
\end{subfigure}
\begin{subfigure}{0.48\textwidth}
\includegraphics[width=1\textwidth]{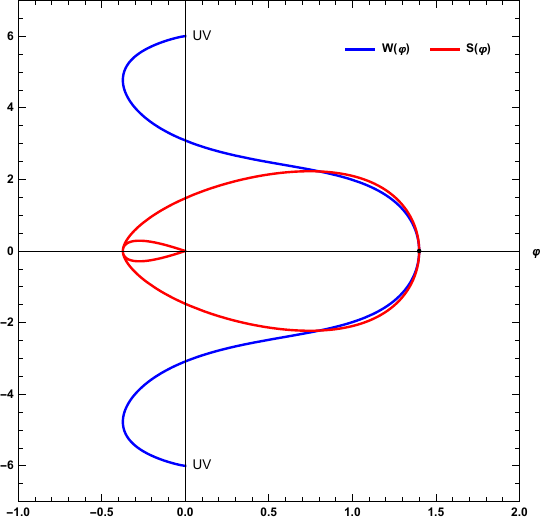}
\caption{\vspace*{0cm}}\label{symsol}
\end{subfigure}
\caption{\footnotesize{Two examples of two-boundary solutions from the green region of figure \ref{map}. A zero of $W$ corresponds to an A-bounce. A zero of $S$ corresponds to a $\f$-bounce except at the start and end of the flow. (a) A flow with one A-bounce and two $\f$-bounces. The initial data that correspond to this solution are $\f_0=0.2$ and $S_0=0.5$ (black dot in the figure). (b): A two-boundary flow with one A-bounce and three $\f$-bounces. The initial data of this solution are $\f_0=1$ and $S_0=0.5$. (c): A symmetric solution with $\f_0=1.4$ and $S_0=0$.}}
\label{W13a}
\end{figure}

\subsection{Solutions with one $AdS$ boundary}
\begin{itemize}

\item UV-Reg solutions:

The UV-Reg solutions can be found by imposing the regularity conditions (i.e. the type II asymptotics of section \ref{I}) on the equations of motion at an arbitrary point $u=u_0$ where $\f\rightarrow \pm\infty$. As we showed already (see subsection \ref{I} and in more detail appendix \ref{secRS}), we have the following expansions near the regular end-point as $\f\rightarrow +\infty$
\be\label{infexp1}
W = W^{(0)}_\infty e^{ a \f} + W^{(1)}_\infty e^{ a \l \f}+\cdots\,,
\ee
\be\label{infexp2}
S = S^{(0)}_\infty e^{ a \f}+S^{(1)}_\infty e^{ a \l \f}+\cdots\,,
\ee
\be\label{infexp3}
T= T^{(0)}_\infty e^{2 a \f}+T^{(1)}_\infty e^{ a (\l+1) \f}+\cdots\,,
\ee
where
\be\label{infs1}
S^{(0)}_\infty =\pm\frac{a (d-1)}{\ell \sqrt{2a^2 \left(\frac{1}{d}-1\right)+1}}\sp W^{(0)}_\infty = \frac{S^{(0)}_\infty}{a}\sp T^{(0)}_\infty =0\,,
\ee
\be
W^{(1)}_\infty = -\frac{(d-2)}{d a}S^{(1)}_\infty\sp
T^{(1)}_\infty =
\mp\frac{\left(2 a^2 (d-1)+d-2\right)}{2 \big( \ell a \sqrt{2a^2 \left(\frac{1}{d}-1\right)+1}\big)}S^{(1)}_\infty\,.\label{infs2}
\ee

\begin{figure}[!ht]

\centering
\begin{subfigure}{0.48\textwidth}
\includegraphics[width=1\textwidth]{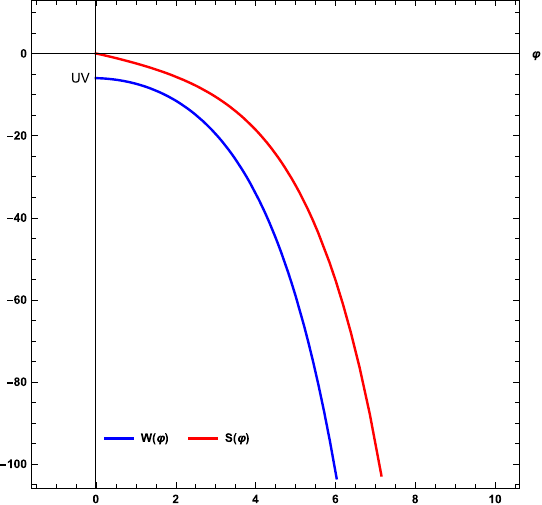}
\caption{}\label{R00}
\end{subfigure}
\centering
\begin{subfigure}{0.48\textwidth}
\includegraphics[width=1\textwidth]{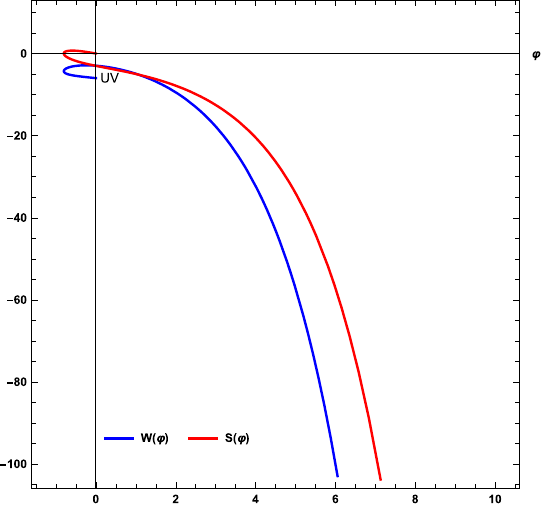}
\caption{}\label{R01}
\end{subfigure}
\centering
\begin{subfigure}{0.48\textwidth}
\includegraphics[width=1\textwidth]{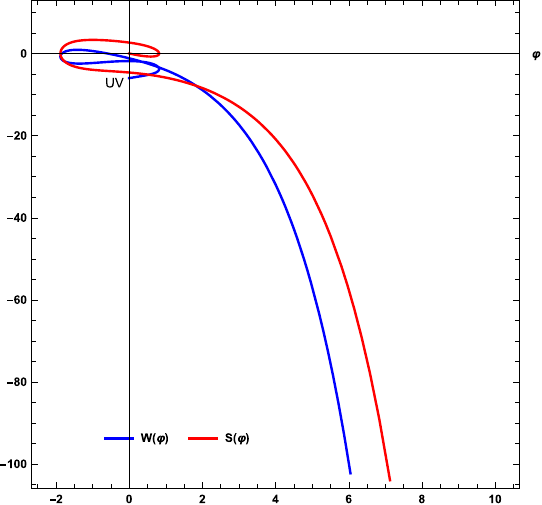}
\caption{}\label{R22}
\end{subfigure}
\caption{\footnotesize{Three examples of solutions with an $AdS$-like boundary from one side and regular on the other side as $\f\rightarrow +\infty$.  (a), (b) and (c)  correspond to  $S_\infty^{(1)}=0, -1.2, -1.22$ respectively.}}
\label{R22a}
\end{figure}
The only free parameter (integration constant)  is  $S_\infty^{(1)}$, which varies from zero to minus infinity if we choose the lower sign for $S^{(0)}_\infty$ in \eqref{infs1}. By this choice, $T^{(1)}_\infty$ is always negative and corresponds to negative curvature slices.

Figures \ref{R00} - \ref{R22} describe three different regular solutions in this class,
for three different values of $S_\infty^{(1)}=0, -1.2, -1.22$. As we decrease $S_\infty^{(1)}$ the number of  A-bounces increases. For example (6a) has no A- or $\f$-bounces, (6b) has one $\f$-bounce, while (6c) has two A-bounces and two $\f$-bounces.
At a critical value for  $S_\infty^{(1)}\approx -1.25$ the number of bounces diverges.

We can also find the location of these regular solutions with at least one A-bounce on the map \ref{map} by finding the location of A-bounces of each solution. These are indicated with the black curves in figure \ref{map}.

\item UV-Sing solutions

These are solutions in the blue region of the map \ref{map}. In figures \ref{S23} and \ref{S24} we have sketched two examples of these solutions. In the figure \ref{S23}, we have a singular solution with two A-bounces and three $\f$-bounces. In \ref{S24}, there are two A-bounces but four $\f$-bounces.  In both solutions, the A-bounce point $\f_0$ is the same, but they correspond to different values of $S_0$.
\begin{figure}[!ht]
\centering
\begin{subfigure}{0.48\textwidth}
\includegraphics[width=1\textwidth]{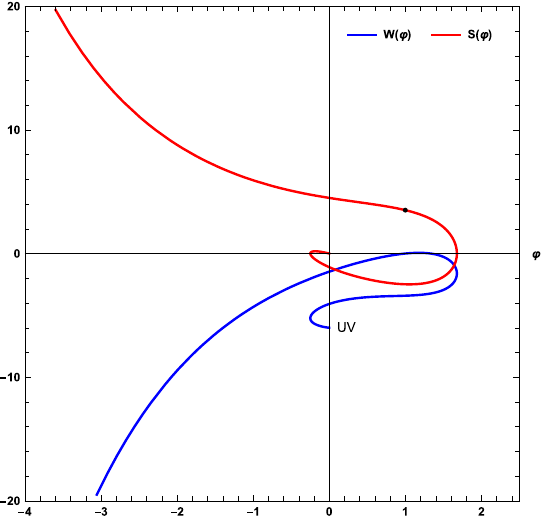}
\caption{}\label{S23}
\end{subfigure}
\begin{subfigure}{0.48\textwidth}
\includegraphics[width=1\textwidth]{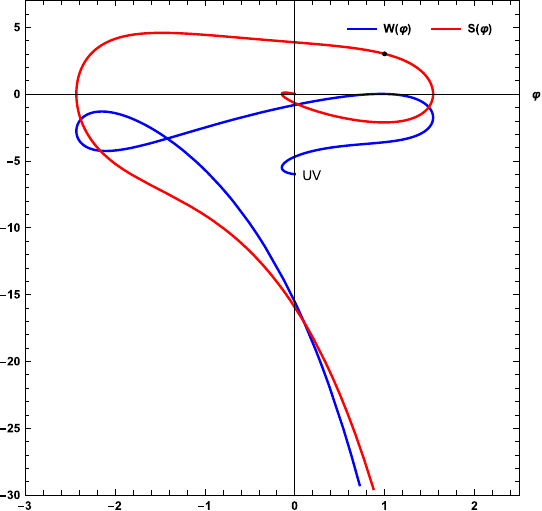}
\caption{}\label{S24}
\end{subfigure}
\centering
\caption{\footnotesize{(a), (b): Two examples of UV-Sing solutions with different numbers of $\f$-bounces.}}
\end{figure}
\end{itemize}
The IR asymptotic behavior of these solutions is given in section \ref{0}. In other words, these are solutions of type 0.

We should remember, that some solutions cannot be fitted in the map \ref{map} because they do not have any A-bounce. These solutions are either UV-Reg or UV-Sing.
\begin{figure}[!t]
\centering
\begin{subfigure}{0.48\textwidth}
\includegraphics[width=1\textwidth]{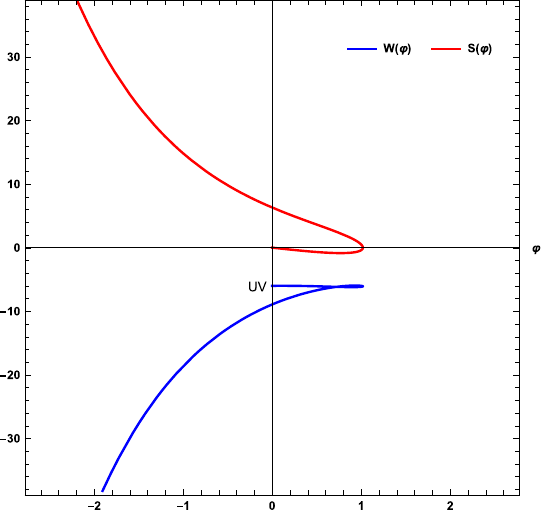}
\caption{}\label{S01}
\end{subfigure}
\centering
\begin{subfigure}{0.48\textwidth}
\includegraphics[width=1\textwidth]{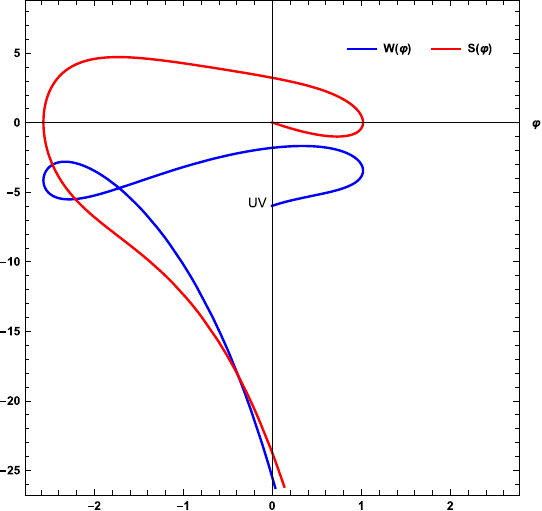}
\caption{\vspace*{0cm}}\label{S02}
\end{subfigure}
\caption{\footnotesize{Singular solutions without A-bounce.}}
\end{figure}
The UV-Reg solutions without A-bounce exist for negative values of $S_{\infty}^{(1)}$ in the region $-1.25\lesssim S_{\infty}^{(1)}<0$, for example see \ref{R00} and \ref{R01}.
In figures \ref{S01} and \ref{S02} we have sketched two singular solutions without A-bounce. The difference between these two solutions is in their number of $\f$-bounces.

\subsection{Solutions with no boundary}\label{noboso}
Another class of solutions are those which do not have any boundaries. We have three types of such solutions:
\begin{itemize}

\item Reg-Sing solutions

Solutions of Reg-Sing type correspond to the yellow curves in the map \ref{map}.
As already mentioned, for UV-Reg solutions at a critical value of  $S_\infty^{(1)}$, which we shall denote by $S_\infty^{(1)c}\simeq -1.25$,  the number of bounces asymptotes to infinity. By decreasing $S_\infty^{(1)}$ further, we still have one regular IR end-point, but the other end-point is no longer the UV fixed point but it asymptotes again to either plus or minus infinity, and it has a (generically) singular behavior. Figure \eqref{Rmn} shows an example of a Reg-Sing solution.
\begin{figure}[!ht]
\centering
\begin{subfigure}{0.48\textwidth}
\includegraphics[width=1\textwidth]{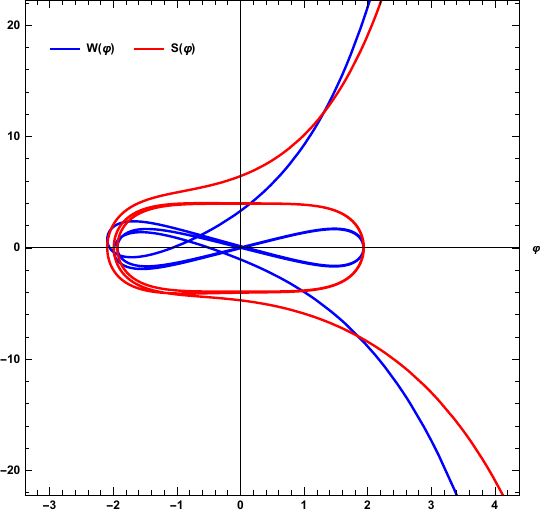}
\caption{\vspace*{0cm}}\label{Rmn}
\end{subfigure}
\begin{subfigure}{0.48\textwidth}
\includegraphics[width=1\textwidth]{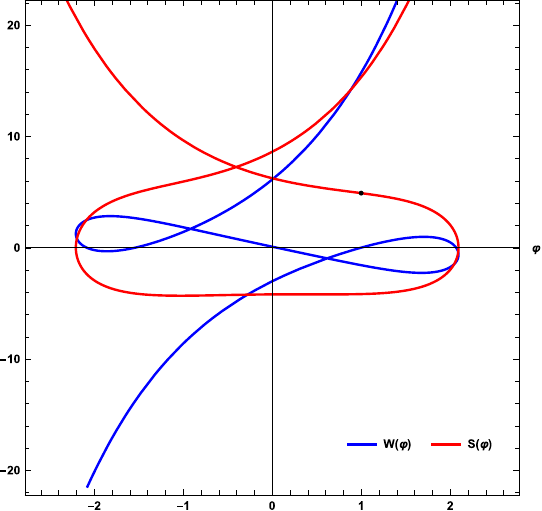}
\caption{\vspace*{0cm}}\label{S52}
\end{subfigure}
\caption{\footnotesize{(a): A Reg-Sing type of solution. The lower part with $W, S<0$ asymptotes to $+\infty$ regularly. The upper part asymptotes to $+\infty$  singularly.  (b): A Sing-Sing solution.}}
\end{figure}

\item Reg-Reg solutions

Among those solutions that are living on the yellow curves on the map \ref{map} there are special cases
for which both end-points are regular. For every number of $\f$-bounces, there is a unique such solution.
If the number of $\f$-bounces is even, then this is a solution that interpolates between $\f=-\infty$ and $\f=+\infty$. If the number of $\f$-bounces is odd, then this is a solution that interpolates between $\f=-\infty$ and $\f=-\infty$ or $\f=+\infty$ and $\f=+\infty$.

Four examples of these solutions are given in figures \ref{regreg} and \ref{regreg3}. Their difference is in their $S_\infty^{(1)}$ as $\f\rightarrow +\infty$.

\begin{figure}[!t]
\begin{subfigure}{0.48\textwidth}
\includegraphics[width=1\textwidth]{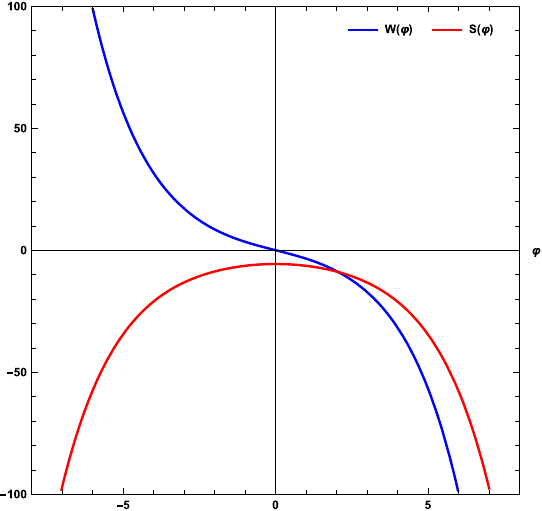}
\caption{}\label{regreg}
\end{subfigure}
\centering
\begin{subfigure}{0.48\textwidth}
\includegraphics[width=1\textwidth]{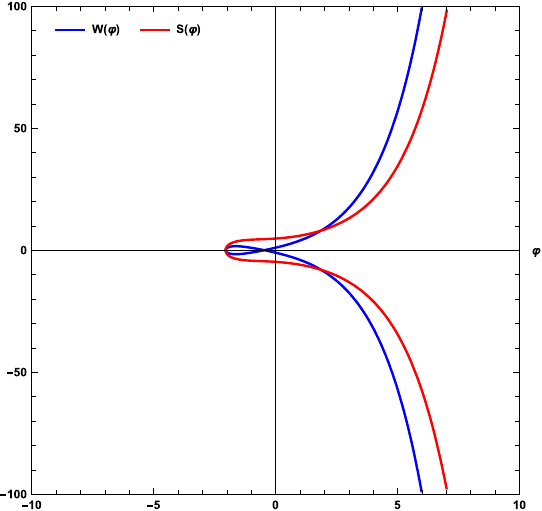}
\caption{}\label{regreg1}
\end{subfigure}
\begin{subfigure}{0.48\textwidth}
\includegraphics[width=1\textwidth]{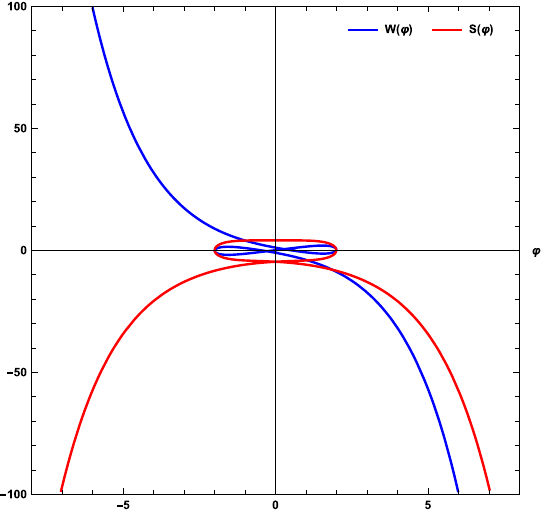}
\caption{}\label{regreg2}
\end{subfigure}
\centering
\begin{subfigure}{0.48\textwidth}
\includegraphics[width=1\textwidth]{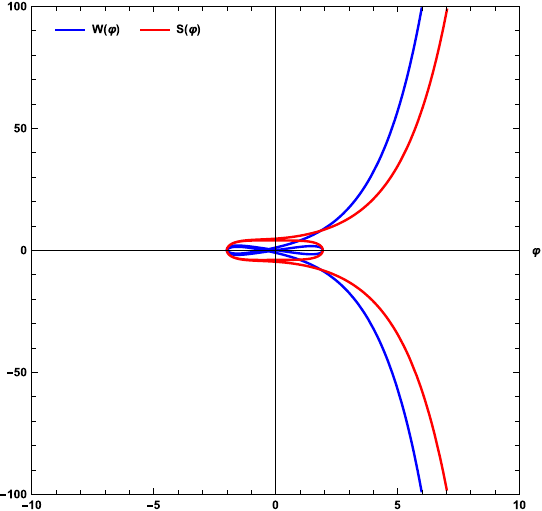}
\caption{}\label{regreg3}
\end{subfigure}
\caption{\footnotesize{Four examples of the Reg-Reg solutions. In all solutions both end-points are regular but with different values of $S_{\infty}^{(1)}$. (a) and (c) interpolate between $\f=-\infty$ and $\f=+\infty$ but with different number of $\f$-bounces. (b) and (d) interpolate between $\f=+\infty$ and $\f=+\infty$.}}
\end{figure}

\item Sing-Sing solutions

At a fixed value of $\f_0$, if we increase $S_0$,  we find the Sing-Sing type solutions where both end-points go to infinity singularly. Figure \ref{S52} shows an example of this solution. All these solutions are located on the red region of the map \ref{map}.

We should note that we have different types of the above solutions where they start at $+\infty$ and end at $+\infty$ for example see figure \ref{Rmn} or start at $+\infty$ but end at $-\infty$, see figure \ref{S52}.
\end{itemize}

\subsection{The boundaries between different solution types.}\label{bondsol}
\begin{itemize}
\item Crossing the black curves in the blue region (UV-Reg to UV-Sing transition)

Inside the blue region of figure \ref{map}, where we have the UV-Sing solutions, the black curves represent the regular, UV-Reg solutions. Figure \ref{tranz1} shows three different solutions at a fixed value of $\f_0=1$ but with different initial values of $S_0$. One solution (solid curves) is a UV-Reg solution and the others (dotted and dashed curves) are two neighboring singular solutions with a slightly smaller or bigger $S_0$ than the regular solution.
By an infinitesimal move away from any of the black curves in the map \ref{map}, the asymptotic ($\f\rightarrow +\infty$) behavior of the $W$ and $S$ curve changes. Only the solid curve has a regular behavior similar to what we found in section \ref{REGSOL}.
\begin{figure}[!ht]
\centering
\includegraphics[width=0.48\textwidth]{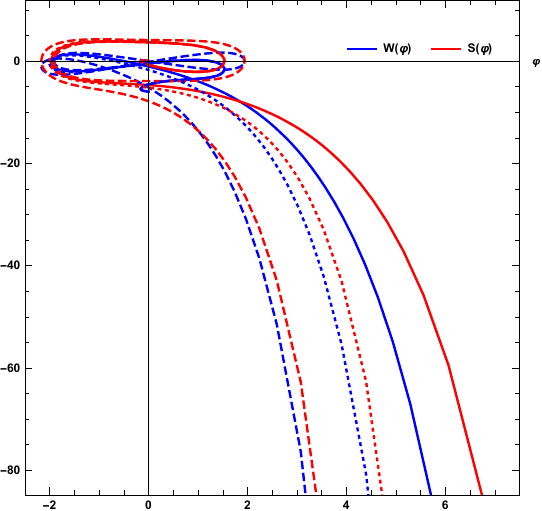}
\caption{\footnotesize{Crossing a black curve in the blue region of map \ref{map}. The solid curve is a UV-Reg solution exactly on the black curves of the map \ref{map}. The dashed and dotted solutions belong to two UV-Sing neighborhood solutions in the blue region, the dashed with a higher value of $S_0$ and the
dotted with a lower value of $S_0$. For all three solutions $\f_0=1$. The regular solution diverges slower than any of the singular solutions, as $\f\to +\infty$.}}\label{tranz1}
\end{figure}
\item Crossing the green-blue border (UV-UV to UV-Sing transition)

At a fixed value of $\f_0$ in the map \ref{map} and by increasing the value of $S_0$ we can move
from the green region with two-boundary solutions, into the blue region with one-boundary, UV-Sing solutions.
As we move towards the joint border of these two regions from both sides, the number of A-bounces (loops around the $(\f_0, S_0)=(0,0)$) increases. This is illustrated in figures \ref{bound1} and \ref{bound2}. In figure \ref{bound1} all solutions are in the blue region and from up-left to down-right gradually the value of $S_0$ decreases at a fixed value of $\f_0=1$. Inside the blue region, as we move towards the green-blue boundary in the map \ref{map}, the number of loops increases and finally tends to infinity. The same behavior exists in the green region as we see from figure \ref{bound2}. In the neighborhood of the border, we have wormhole solutions with many loops. By decreasing $S_0$ we move away from the boundary and the number of loops is decreasing. For an analytic study of solutions with many loops see appendix \ref{loops}.

\definecolor{darkgreen}{rgb}{0.0, 0.5, 0.0}
\begin{figure}[!ht]
\centering
\begin{subfigure}{0.9\textwidth}
\includegraphics[width=1\textwidth]{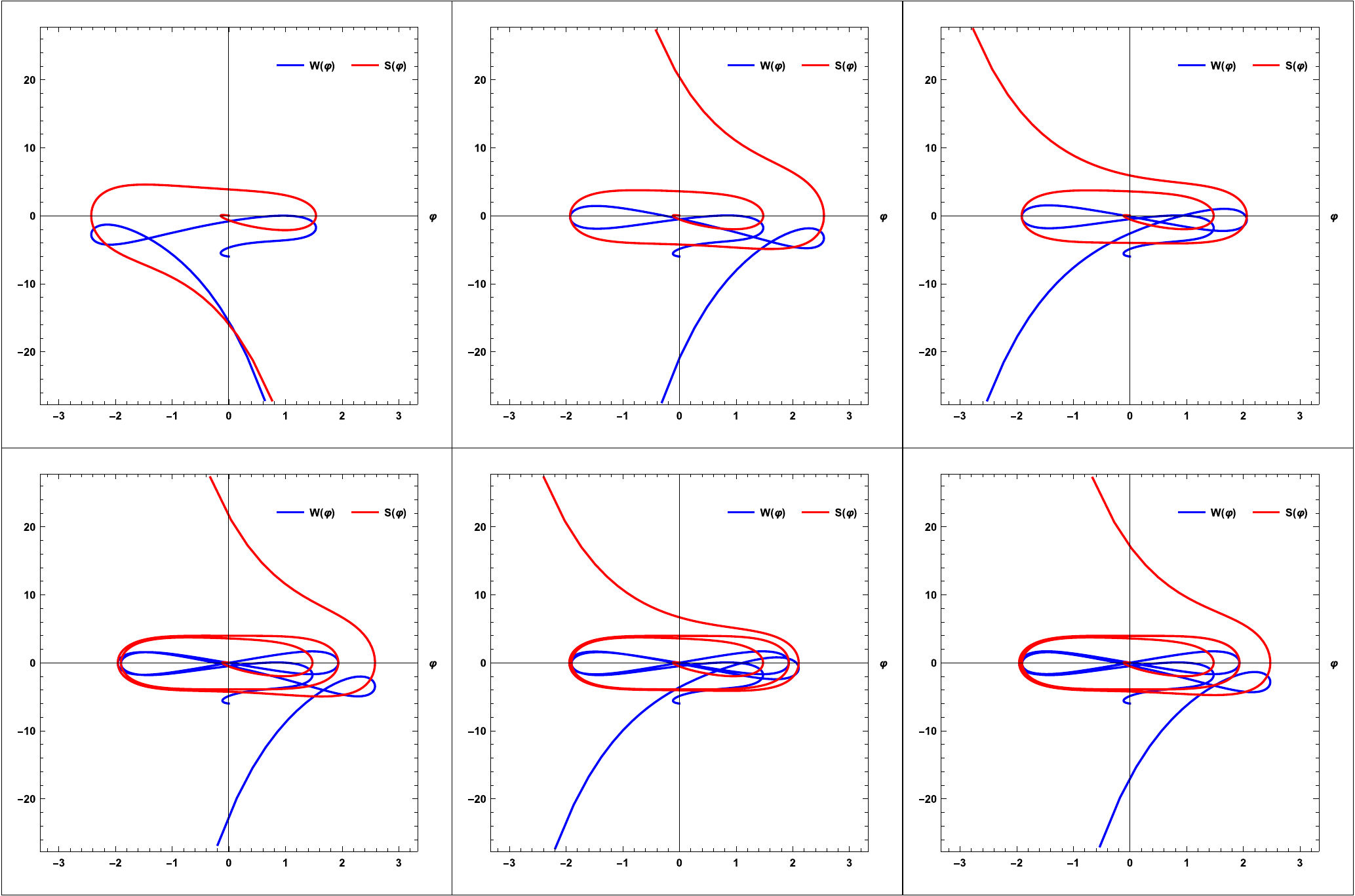}
\caption{\footnotesize{From up-left to down-right when $S_0$ decreases at fixed $\f_0=1$ in the {\color{blue}{blue}} region of the map \ref{map}, the number of bounces (loops) increases.
}}\label{bound1}
\end{subfigure}
\centering
\begin{subfigure}{0.9\textwidth}
\includegraphics[width=1\textwidth]{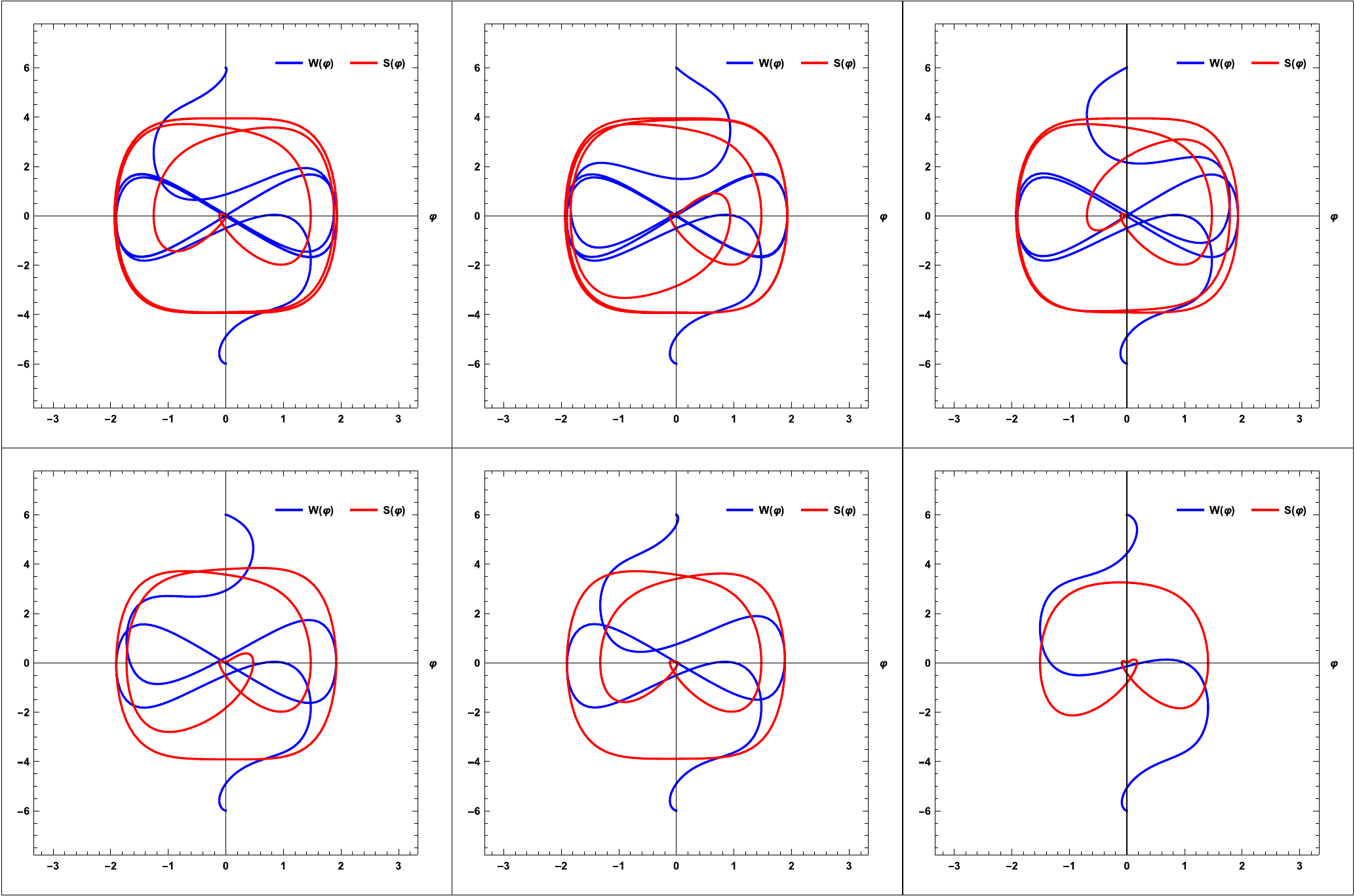}
\caption{\footnotesize{In the {\color{darkgreen}{green}} region of the map \ref{map} from up-left to down-right when $S_0$ decreases at fixed $\f_0=1$ the number of loops decreases.}}\label{bound2}
\end{subfigure}
\caption{\footnotesize{A transition from UV-Sing solutions in the blue region to wormholes in the green region. As we move towards the boundary of the blue-green region the number of loops in both sides goes to infinity.}}
\label{bound1a}
\end{figure}
\item Crossing the blue-red border (one-boundary solutions transiting to the no-boundary solutions or UV-Sing to Sing-Sing transition)

At a fixed value of $\f_0$ in the map \ref{map} and by increasing the value of $S_0$ we can move
from the blue region with one-boundary solutions, into the red region with no-boundary solutions.
As we move towards the joint border of these two regions, from both sides, the number of A-bounces increases without bound again. This is illustrated in figures \ref{bound3} and \ref{bound4}. In figure \ref{bound3} from up-left to down-right gradually the value of $S_0$ decreases at a fixed value of $\f_0=1$. As we move towards the blue-red boundary in the map \ref{map} the number of loops increases and finally tends to infinity. The same behavior exists in the blue region as we see from figure \ref{bound4}.
\end{itemize}
\begin{figure}[!ht]
\centering
\begin{subfigure}{0.9\textwidth}
\includegraphics[width=1\textwidth]{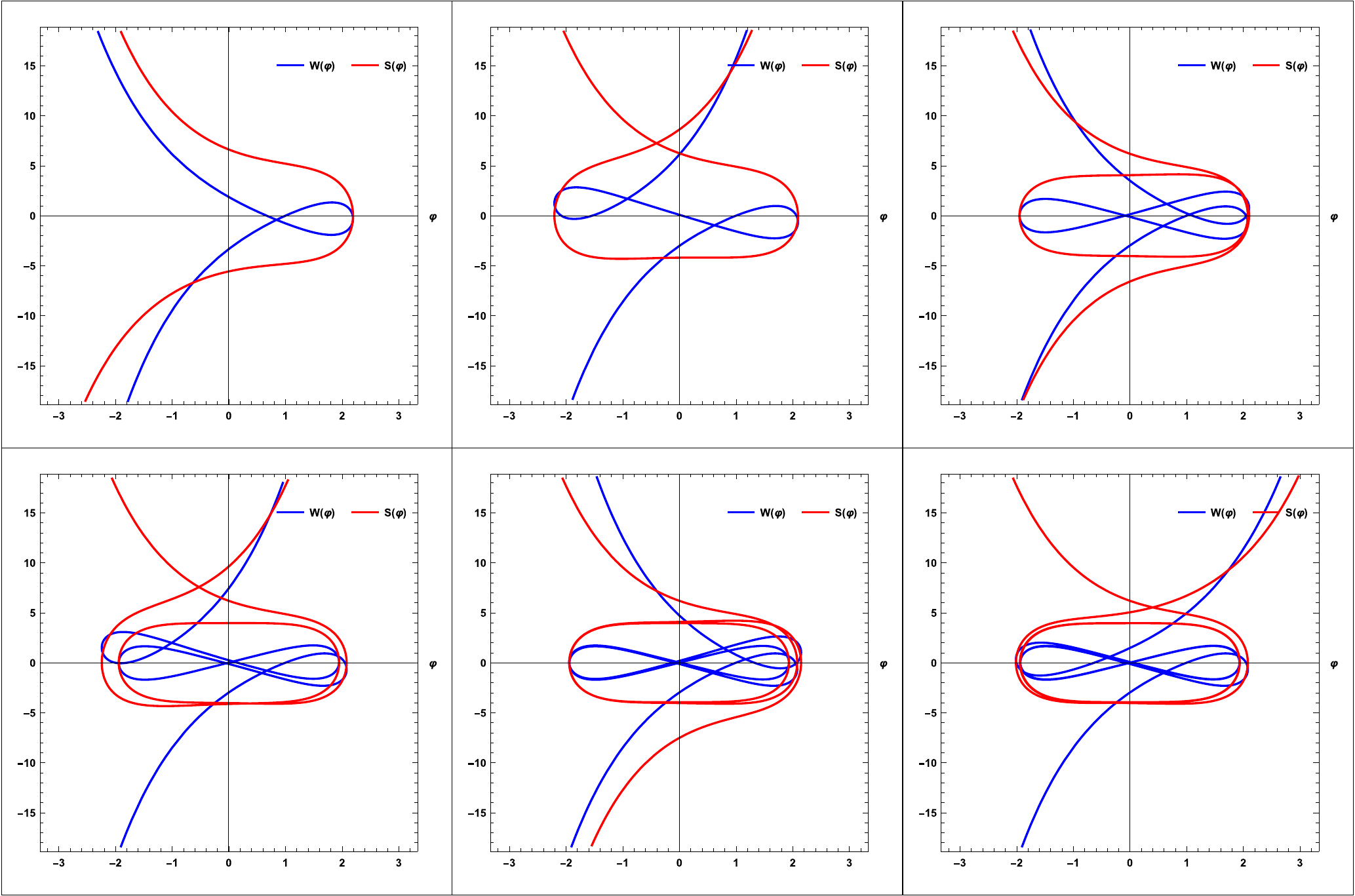}
\caption{\footnotesize{From up-left to down-right when $S_0$ decreases at fixed $\f_0=1$ in {\color{red}{red}} region of the map \ref{map}.}}\label{bound3}
\end{subfigure}
\centering
\begin{subfigure}{0.9\textwidth}
\includegraphics[width=1\textwidth]{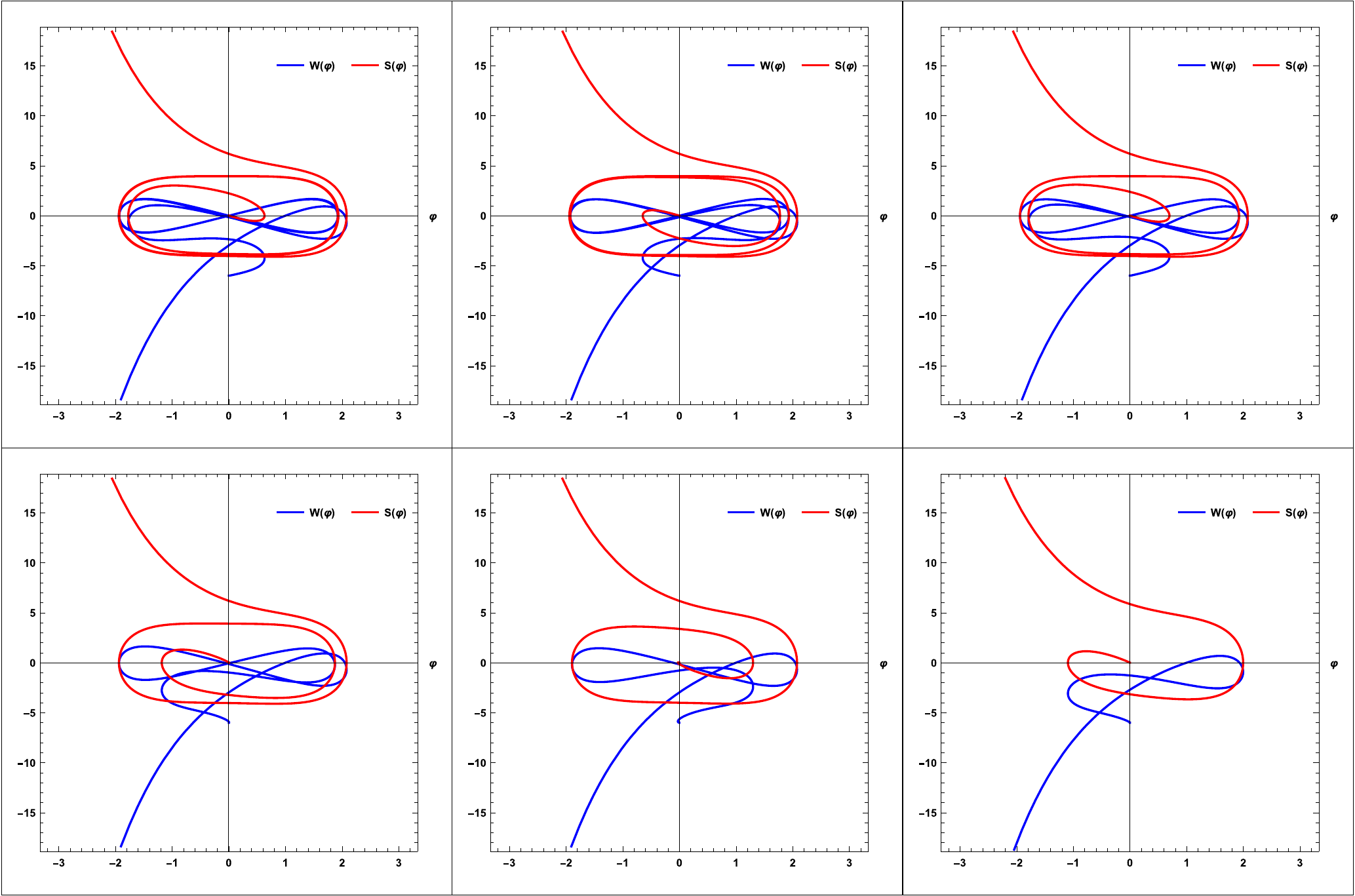}
\caption{\footnotesize{From up-left to down-right when $S_0$ decreases at fixed $\f_0=1$ in the {\color{blue}{blue}} region of the map \ref{map}.}}\label{bound4}
\end{subfigure}
\caption{\footnotesize{A transition from Sing-Sing solutions in the red region to UV-Sing solution in the blue region. As we move towards the boundary of the red-blue region the number of loops in both sides goes to infinity.}}
\label{bound3a}
\end{figure}
\begin{itemize}

\item Crossing from the black curve to the yellow curve in figure \ref{map}, (UV-Reg to Reg-Reg to Reg-Sing transition)

Consider solutions on the black curve i.e. UV-Reg solutions.
If we decrease the value of $S_{\infty}^{(1)}$ gradually the number of A-bounces (loops) increases and we eventually reach a point where this number goes to infinity. Beyond that point, we find Reg-Sing solutions which are located on the yellow curves in the map \ref{map}. As we decrease $S_{\infty}^{(1)}$ further the number of A-bounce decreases.

This transition is shown in figures \ref{reg3}, \ref{prod1} and \ref{reg4}. The first solution is a UV-Reg solution with a finite number of A-bounces. The second one has many loops and the last one is a Reg-Sing solution.

\item As we already discussed, at the intersection of two yellow curves in the map \ref{map} we have the Reg-Reg solutions.

\end{itemize}

\section{A related  uplifted theory and singularity resolution}\label{uplsec}
Consider a $(d+n+1)-$dimensional Einstein-dilaton gravity with the following Lagrangian
\be
\mathcal{L}=\sqrt{-\tilde g} \left[\tilde R - \frac12(\pa{\chi})^2
- \tilde V({\chi})\right]\,.\label{la0}
\ee
We consider a holographic solution in this theory with a metric ansatz
\be
ds^2=d{\tu}^2+\sum_{n=1}^2 e^{2A_n({\tu})}\zeta^{(n)}_{ij}d\tilde{x}^{i}d\tilde{x}^{j}\,,
\label{zz41}
\ee
with  holographic coordinate $\tilde{u}$.
$A_1$ and $A_2$ are the scale factors for two Einstein (constant curvature) spaces with $\zeta^{(1)}_{ij}$ and $\zeta^{(2)}_{ij}$ metrics and with $d$ and $n$ dimensions respectively.

We shall now consider the dimensional reduction of the equations of motion of the $(d+n+1)-$dimensional theory to $d+1$ dimensions.

To reduce this metric to a $d+1$ dimensional metric we write \eqref{zz41} as
\be \label{chds}
ds^2=e^{-{\frac{2n}{d-1}}A_2(\tu)}ds_{d+1}^2+e^{2A_2(\tu)} \zeta^{(2)}_{\m\n} d\tilde{x}^{\m}d\tilde{x}^{\n}\,,
\ee
where we have defined
\be\label{upmetr}
ds_{d+1}^2=e^{\frac{2n}{d-1}A_2(\tu)} d\tu^2+e^{\frac{2n}{d-1}A_2(\tu)+2A_1(\tu)} \zeta^{(1)}_{\a\b}d\tilde{x}^{\a}d\tilde{x}^{\b}\,.
\ee
By a change of variable
\be \label{chvab}
\frac{d\tu}{d u}= e^{-{\frac{n}{d-1}}A_2(\tu)}\,,
\ee
we write the $d+1$-dimensional metric of \eqref{upmetr} as
\be
ds_{d+1}^2 ={g}_{ab}d{x}^a d{x}^b=d u^2+e^{2 A({u})}\zeta^{(1)}_{\a\b}dx^{\a}dx^{\b} \sp  A\equiv A_1+\frac{n}{d-1}A_2\,.
\label{zz1a}
\ee
By the above definitions, the reduction of the Lagrangian\footnote{This Lagrangian was constructed to
give the same equations as the one we obtain from the higher-dimensional theory.}  \eqref{la0} gives rise to
\begin{gather}\label{lred1}
\mathcal{L}_{red} =\sqrt{-{g}}\Big[{R} +
e^{-\frac{2  (d+n-1)}{d-1}A_2}R_2 -
\frac{n \left(d+n-1\right)}{d-1}\pa_{u}A_2 \pa^{u}\!{A_2} \nn \\
-\frac12 \pa_{{u}} {\chi} \pa^{{u}} {\chi}- e^{-\frac{2n }{d-1}A_2} V({\chi})
\Big]\,,
\end{gather}
where $R_2$ is the Ricci scalar curvature of the $\zeta^{(2)}_{\m\n}$ metric.
Then this theory with metric ${g}_{ab}$ in \eqref{zz1a} contains one more canonically normalized scalar field $\f$
\be
\f \equiv \pm \sqrt{\frac{2n(d+n-1)}{d-1}}A_2\,,
\label{zz2a}
\ee
where the Lagrangian would be
\be\label{lred2}
\mathcal{L}_{red} =\sqrt{-{g}}\Big[{R} +
e^{\mp\sqrt{\frac{2(d+n-1)}{n(d-1)}}\f}R_2-
\frac12\pa_{{u}}\f \pa^{{u}}\f
-\frac12 \pa_{{u}} {\chi} \pa^{{u}} {\chi}- e^{\mp\sqrt{\frac{2n}{(d-1)(d+n-1)}}\f} V({\chi})
\Big]\,.
\ee
So the potential can be read as
\be
 V(\f,\chi) \equiv e^{\mp\sqrt{\frac{2n}{(d-1)(d+n-1)}}\f} V({\chi})-e^{\mp\sqrt{\frac{2(d+n-1)}{n(d-1)}}\f}R_2\,,\label{zz3}
\ee
and the full action in $(d+1)$-dimensions is
\be\label{Sred}
S_{d+1}=\int d^{d+1}x\sqrt{ g}\left[ R-{\frac12}(\pa\f)^2-{\frac12}(\pa\chi)^2- V(\f,\chi)\right]\,.
\ee
We observe that the leading behavior of the $(d+1)$-dimensional potential as a function of $\f$ is controlled by the internal curvature  $R_2$.
In the particular case that $V(\chi)$ is a (negative) constant, we find two different types of theories in  $d+1$ dimensions, \cite{GK}:

\begin{itemize}

\item A confining theory: The reduction is performed on a $S^n$ sphere, $R_2>0$ and
\be
a_{C}\leq \sqrt{\frac{(d+n-1)}{2n(d-1)}}\leq a_{G}\,.
\label{coonf}\ee
The upper limit is attained when $n=2$ ($n=1$ has no non-trivial curvature).
The lower limit is attained when $n\to\infty$.

\item A deconfined theory: The reduction is performed on a torus $T^n$, so $R_2=0$.
In that case,  only the first term in  (\ref{zz3}) is relevant and
\be
0\leq \sqrt{\frac{n}{2(d-1)(d+n-1)}}\leq a_C\,.
\ee
The upper bound is reached when $n\to\infty$.

\end{itemize}
In the following, we shall analyze the confining theory in detail. Reduction on a torus is analyzed in appendix \ref{dcth}.
\subsection{Reduction to a confining theory}
Consider the $d+n+1$-dimensional uplifted theory  \eqref{la0} with a negative constant potential and a constant scalar field
\be \label{confscal}
V({\chi})=-\frac{(d+n)(d+n-1)}{\tilde{\ell}^2}\sp \chi=const\,.
\ee
Moreover, consider at constant slices of $\tilde{u}$ the metric \eqref{zz41} is described by $\zeta^{(1)}_{ij}$ for an $AdS_d$ space and $\zeta^{(2)}_{ij}$ for a $S^{n}$ sphere. The properties and solutions of this theory have already been analyzed in \cite{Ghodsi:2023pej}.

Now consider a $d+1$-dimensional theory \eqref{Sred} which is given by the following potential
\be \label{vt1}
{V}=
-\frac{d(d-1)}{{{\ell}}^2}\left(b\f^2+\cosh^2(a\f)\right)\sp \f =\f(u)\,,
\ee
where we have defined the constants $a$ and $b$ as follows
\be
2a =\sqrt{\frac{2(d+n-1)}{ n(d-1)}}\sp
b=\frac{\Delta(d-\Delta)}{2 d (d-1)}-a^2\,.\label{abd}
\ee
This potential is chosen in such a way that it will reproduce asymptotically the potential \eqref{zz3} as $\f\rightarrow \pm\infty$. To see this, we note that as $A_2\rightarrow -\infty$ which means that the scale factor of $S^n$ shrinks to a zero size in the uplifted theory, according to \eqref{zz2a}  $\f\rightarrow \pm\infty$ and therefore in this limit, the second term in \eqref{zz3} is dominant.

Moreover, the value of $a$ in \eqref{abd} for all values of $n,d>1$ lies in the blue region of figure \ref{CDCM} which means that the theory in $d+1$ dimensions is a confining theory.

By comparing the two potentials at \eqref{zz3} and \eqref{vt1} we find the following relation between two parameters of these theories
\be \label{barr}
R_2=\frac{d(d-1)}{4{\ell}^2}\,.
\ee

We can relate the holographic coordinates of the confining theory to the uplifted theory by a change of variable as in \eqref{chvab}
\be\label{ud1}
\int_{{u}_0}^{{u}} d {u} = \int_{\tu_0}^\tu e^{\frac{n}{d-1}A_2(\tu)} d\tu\,.
\ee
In appendix \ref{reduction} we present the careful comparison of regular and singular asymptotics between
the higher- and lower-dimensional solutions.
In particular, for the solutions in $d+1$ dimensions where $\f\to\infty$, if the leading constant vanishes, then they become regular in $d+n+1$ dimensions. This justifies their naming ``regular" solutions.

We conclude this part by saying that the $(d+n+1)$-dimensional theory in (\ref{la0}) in ansatz (\ref{zz41}) and the $(d+1)$-dimensional solutions in (\ref{vt1}), (\ref{abd}) are different theories near the $AdS$ boundaries but they coincide near $\f\to+\infty$.
One can therefore resolve the singularity of the
$(d+1)$-dimensional solution by lifting it to the $(d+n+1)$-dimensional theory.

\subsection{Comparison between confining and uplifted solutions}\label{comcu}

To compare the confining and uplifted theories we should remember that under reduction on $S^n$, we have a confining theory if the asymptotic behavior of the  potential  is given by \eqref{abd} or
\be \label{aconf}
a=\sqrt{\frac{d+n-1}{2n(d-1)}}\,,
\ee
and we have a deconfined theory if we have (see appendix \ref{dcth})
\be  \label{adconf}
a=\sqrt{\frac{n}{2(d-1)(d+n-1)}}\,.
\ee
\begin{figure}[!ht]
\begin{center}
\includegraphics[width = 10cm]{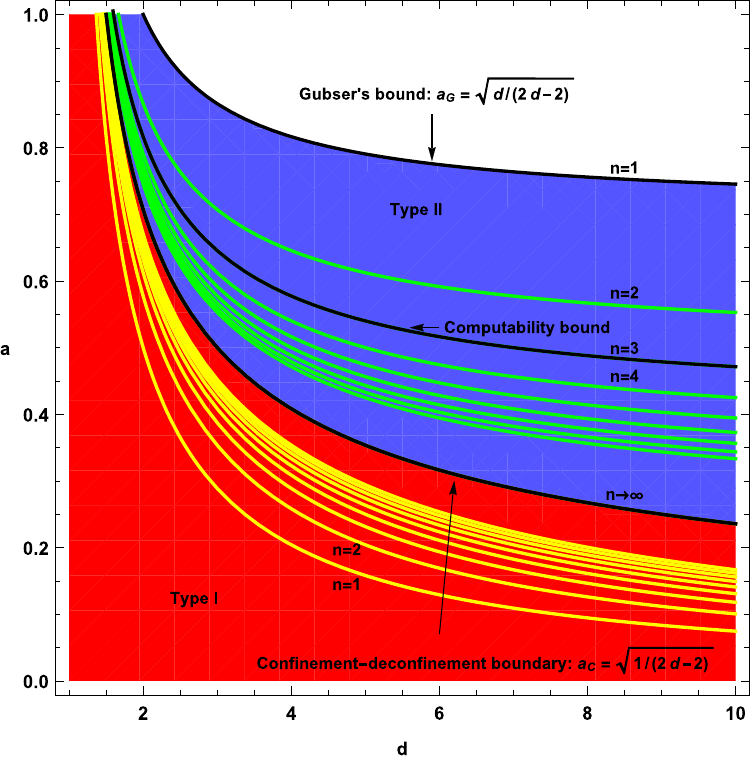}
\caption{\footnotesize{Confining and deconfining models from an uplifted theory.
The green curves are given by \eqref{aconf}. At $n=1$ (Gubser's bound) we arrive at the upper bound of the blue region. As $n\rightarrow\infty$ (confinement-deconfinement boundary) the green curves approach the lower bound of the blue region. The yellow curves are given by \eqref{adconf}. Again as $n\rightarrow +\infty$ one moves towards the confinement-deconfinement boundary. Note that the blue and red regions extend to $a\rightarrow\infty$ as $d\rightarrow 1$. The black curve in the blue region for $n=3$ shows the compatibility bound, which restricts solutions to the region defined by \eqref{ndres}.
} }\label{sol12}
\end{center}
\end{figure}
We can now observe how these two classes appear among the confining and deconfining theories in figure \ref{CDCM} for various values of $d$ and $n$. This is done in figure \ref{sol12}. In this figure, the green curves are \eqref{aconf} and the yellow ones are \eqref{adconf}. The Gubser's bound is the $n\rightarrow 1$ of \eqref{aconf} and the confinement-deconfinement boundary is the $n\rightarrow +\infty$ of the \eqref{aconf} or \eqref{adconf}.

In addition to the Gubser's bound, there is a stronger bound called the computability bound that was discussed below equation (\ref{acomp}).
This bound then restricts the dimensions $d$ and $n$ of the uplifted theory to
\be \label{ndres}
(n-3)(d-1)> 0\,.
\ee
To respect this bound we have considered our numerical results for an uplifted solution in appendix \ref{rev} with $d=n=4$.

To compare the uplifted and confining solutions we have drawn table \ref{tab1}. This table shows whether, for any solution of the confining theory, there is a similar uplifted solution or not. Note that these two theories are related only asymptotically,  at $\f\rightarrow +\infty$, by dimensional reduction.
\begin{table}[!ht]
\centering
\begin{tabular}{|ccc|ccc|}
\hline
\multicolumn{3}{|c|}{Uplifted solutions}                            & \multicolumn{3}{c|}{Confining solutions}                            \\ \hline
\multicolumn{1}{|c|}{-} & \multicolumn{1}{c|}{-} &\multicolumn{1}{c|}{-}  & \multicolumn{1}{c|}{UV-UV} & \multicolumn{1}{c|}{figures \ref{W12},  \ref{W13}} & Regular \\ \hline
\multicolumn{1}{|c|}{Regular} &\multicolumn{1}{c|}{(R, B)} & \multicolumn{1}{c|}{figure \ref{shrink1}} & \multicolumn{1}{c|}{UV-Reg} & \multicolumn{1}{c|}{figures \ref{R00} - \ref{R22}} & Regular \\ \hline
\multicolumn{1}{|c|}{Singular} &\multicolumn{1}{c|}{(S, B)} & \multicolumn{1}{c|}{figure \ref{AB1}}
& \multicolumn{1}{c|}{UV-Sing} & \multicolumn{1}{c|}{figures \ref{S24}, \ref{S01}, \ref{S02}}  & Singular  \\ \hline
\multicolumn{1}{|c|}{Singular} & \multicolumn{1}{c|}{(R, A)} &  figure \ref{shrink2} & \multicolumn{1}{c|}{Reg-Sing} & \multicolumn{1}{c|}{figure \ref{Rmn}} & Singular \\ \hline
\multicolumn{1}{|c|}{Singular} & \multicolumn{1}{c|}{(A, A)} &  figure \ref{AB2} & \multicolumn{1}{c|}{Sing-Sing} & \multicolumn{1}{c|}{ figure \ref{S52}} & Singular \\ \hline
\multicolumn{1}{|c|}{-} & \multicolumn{1}{c|}{-} & - & \multicolumn{1}{c|}{Reg-Reg} & \multicolumn{1}{c|}{figure \ref{regreg}} & Regular \\ \hline
\end{tabular}
\caption{Examples of solutions on the uplifted and confining theories.}\label{tab1}
\end{table}

\subsubsection{The analogs of the exact higher-dimensional solutions in the confining theory }

In the uplifted theory, \cite{Ghodsi:2023pej}, we found two exact solutions, the product space solution $AdS_d\times AdS_{n+1}$ and the global $AdS$ solution $AdS_{n+d+1}$, both described in section 4 of \cite{Ghodsi:2023pej} (see also appendix \ref{ESS}).  They have higher symmetries than the generic $O(d,1)\times O(n+1)$ symmetry of the ansatz. The first has $O(d,1)\times O(n+1,1)$ symmetry while the second $O(d+n+1,1)$ symmetry.

Here we describe  the  descendants  of these solutions in the confining theory:
\begin{itemize}
\item{The product space solution $AdS_d\times AdS_{n+1}$:}

As we already discussed in section \ref{bondsol}, in crossing from the black curves to yellow curves in map \ref{map}, at a specific value of $S_{\infty}^{(1)c}\approx -1.25$  we find a UV-Reg solution with an infinite number of loops, figure \ref{Nprod1}. Figure \ref{Paf} shows $A(u)$ and $\f(u)$ for this solution. For values greater than $S_{\infty}^{(1)c}$  all solutions are regular and for smaller values all solutions become singular. Figures \ref{reg3},  \ref{prod1} and \ref{reg4} show how this transition from regular solutions to singular solutions in the confining theory is happening.

\begin{figure}[!ht]
\centering
\begin{subfigure}{0.45\textwidth}
\includegraphics[width=1\textwidth]{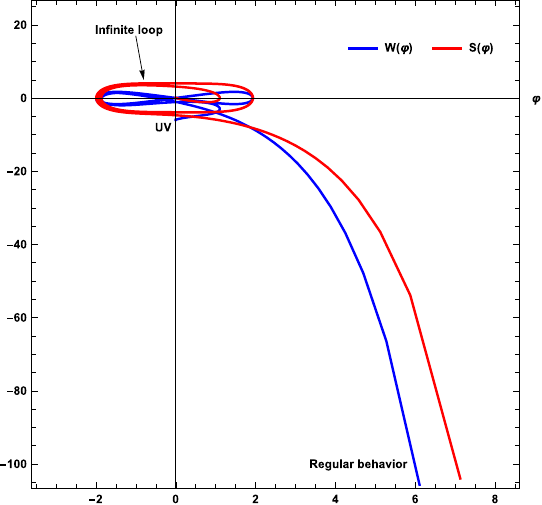}
\caption{}\label{Nprod1}
\end{subfigure}
\centering
\begin{subfigure}{0.425\textwidth}
\includegraphics[width=1\textwidth]{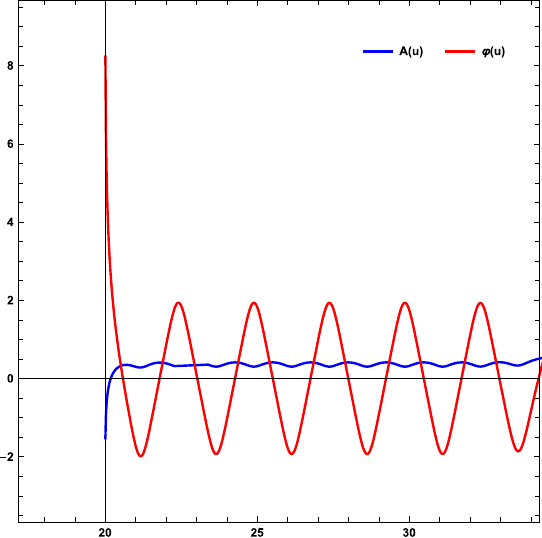}
\caption{\footnotesize{}}\label{Paf}
\end{subfigure}
\caption{\footnotesize{The critical solution with an infinite number of bounces.}}
\end{figure}

There is a similar transition in the uplifted theory, figures \ref{Nsh1}, \ref{Nsh0} and \ref{Nsh2}.
In the uplifted theory, we have a transition from regular solutions to singular solutions, and in between we have a product solution \eqref{exsol2} that has a constant scale factor for $AdS$. By comparing these two transitions, we observe that the uplifted product space solution corresponds to the value  $S^{(1)c}_{\infty}$, and this corresponds to the infinite loop UV-Reg solution in the confining theory.

\begin{figure}[!ht]
\centering
\begin{subfigure}{0.42\textwidth}
\includegraphics[width=1\textwidth]{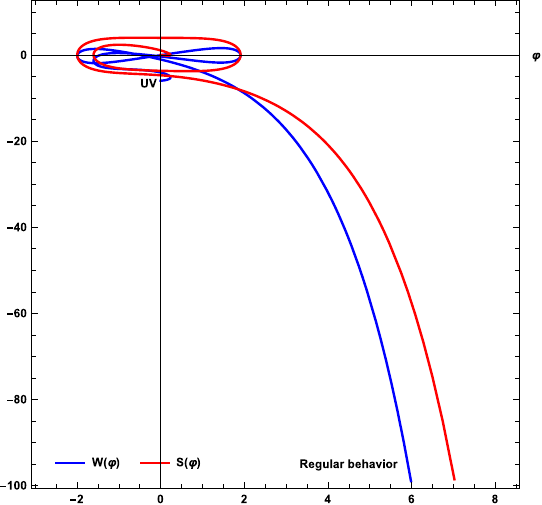}
\caption{\footnotesize{UV-Reg  solution}}\label{reg3}
\end{subfigure}\hspace{1cm}
\centering
\begin{subfigure}{0.42\textwidth}
\includegraphics[width=1\textwidth]{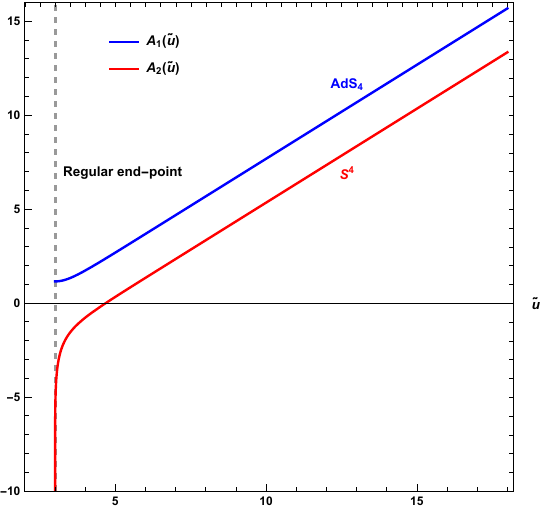}
\caption{\footnotesize{(R, B) solution}}\label{Nsh1}
\end{subfigure}
\centering
\begin{subfigure}{0.42\textwidth}
\includegraphics[width=1\textwidth]{figures/prod1}
\caption{\footnotesize{Infinite loop UV-Reg solution}}\label{prod1}
\end{subfigure}\hspace{1cm}
\centering
\begin{subfigure}{0.42\textwidth}
\includegraphics[width=1\textwidth]{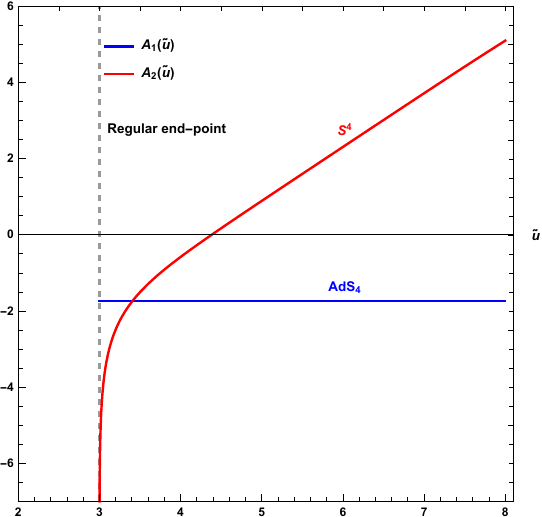}
\caption{\footnotesize{Product space solution}}\label{Nsh0}
\end{subfigure}
\centering
\begin{subfigure}{0.42\textwidth}
\includegraphics[width=1\textwidth]{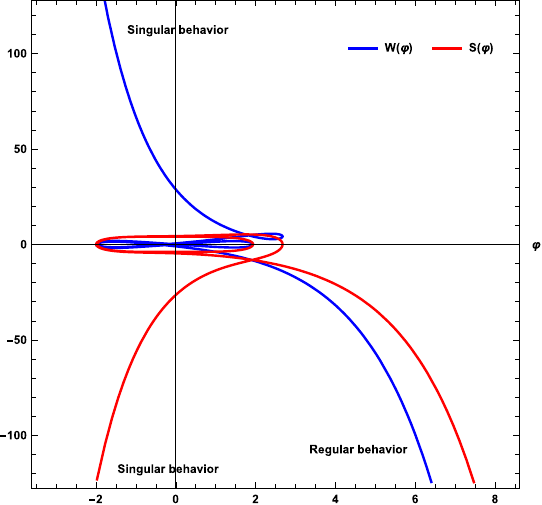}
\caption{\footnotesize{Reg-Sing solution}}\label{reg4}
\end{subfigure}\hspace{1cm}
\centering
\begin{subfigure}{0.42\textwidth}
\includegraphics[width=1\textwidth]{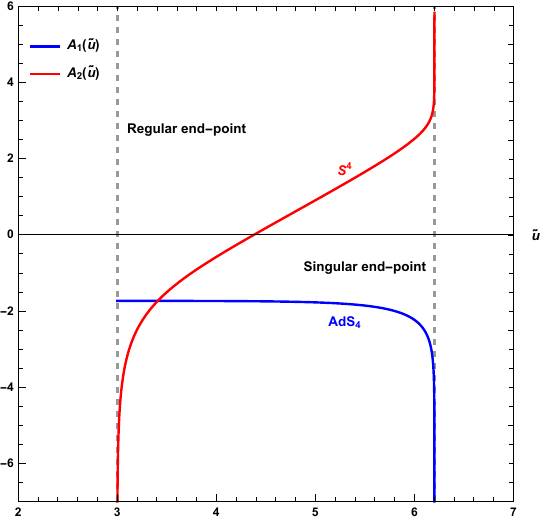}
\caption{\footnotesize{(R, A) solution}}\label{Nsh2}
\end{subfigure}
\centering
\caption{\footnotesize{A transition from a regular solution to a singular solution. The left figures are in the confining theory and the right figures are in the uplifted theory.}}
\end{figure}

\item{Global $AdS_{d+n+1}$ solution:}

The global $AdS_{d+n+1}$ is a regular solution in the uplifted theory and its metric is given in \eqref{glob0}.
To find the analogous solution to the global $AdS$ in confining theory, we need to know the relation between the $S_{\infty}^{(1)}$ in the confining theory with $a_0$ in the uplifted theory (see appendix \ref{reduction} for more details). This is given in equation \eqref{ud9}. In the uplifted theory, the product space solution corresponds to $a_0=a_0^c$ which is defined in \eqref{a0cr}. Therefore, the corresponding critical value for $S_{\infty}^{(1)}$ can be found by inserting \eqref{a0cr} into the \eqref{ud9} which is given in \eqref{sprod1} or
\be \label{tsprod1}
S^{(1)c}_\infty=\frac{2 a  \left(1-2 a^2 (d-1) d\right){\ell} }{\left(1-2 a^2 (d-1)\right)^2\tell^2}\sqrt{2a^2 (\frac{1}{d}-1)+1}\,.
\ee
We can do the same for global $AdS_{d+n+1}$ in the uplifted solution which gives rise to the relation \eqref{sglob1} or
\be \label{tsglob1}
S^{(1)glob}_\infty = -\frac{2 a (d-1){\ell}}{ \left(1-2 a^2 (d-1) \right)^2\tell^2}\sqrt{2 a^2 (\frac{1}{d}-1)+1}\,.
\ee
Dividing the results of \eqref{tsprod1} and \eqref{tsglob1} we see that
\be
\frac{S_\infty^{(1)glob}}{S_\infty^{(1)c}}=\frac{d-1}{ 2 a^2 d(d-1)-1}=\frac{n}{d+n}\,.
\ee
Therefore by finding the critical value corresponding to the product solution of the uplifted theory numerically, $S_\infty^{(1)c}\approx -1.25$, we can find $S_\infty^{(1)}$ for global solution i.e. $S^{(1)glob}_\infty\approx -0.62$.
The corresponding solution in the confining theory is sketched in figure \ref{Glob2}.
As we observe, this is a regular solution without any A-bounce.

\begin{figure}[!ht]
\begin{center}
\includegraphics[width = 7cm]{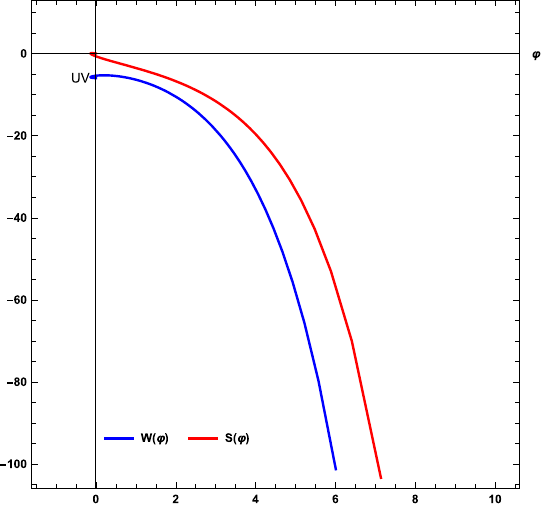}
\caption{\footnotesize{The analogous solution to the global $AdS$ solution in the uplifted theory.}}\label{Glob2}
\end{center}
\end{figure}

\end{itemize}

\subsubsection{The solutions with an infinite number of $A$-bounces}

As described in section \ref{bondsol} and appendix \ref{loops} the solutions at the boundary of the green and blue region are the direct sum of two limiting solutions. One of them starts at the maximum of the potential and oscillates forever in a fixed neighborhood of the maximum. Both the scale factor as well as the dilaton have an infinite number of oscillations.

The other starts at $\f=+\infty$ and then approaches the maximum of the potential making an infinite number of oscillations around its neighborhood.
This second solution near $\f=+\infty$, matches the factorized solution $AdS_{d}\times AdS_{n+1}$ in the upstairs theory, as shown in the previous subsection.

These solutions contain an infinite number of $\f$ bounces and as such, they appear as what in quantum field theory we would call a limit cycle. The coupling constant $\f$ oscillates indefinitely between some fixed maximal and minimal values. However, these are not accompanied by discrete scale transformations as in non-unitary models of limit cycles in the literature, \cite{limit}.

As our theories here are defined on $AdS$, there are no a priori objections to such a behavior.
However, it should be remembered that these special saddle points have really the lowest free energies as we explain in the subsequent sections.

\section{The on-shell action and free energy}\label{onshc}
As we have seen, for a given choice of UV data (spacetime curvature and relevant coupling)  there exist multiple classical saddle points. We are naturally led to address the question of which of these saddle points is the dominant one. This is answered, semiclassically,  by comparing the (Euclidean) free energy of the solutions with the same UV data.

In holography, the free energy of a classical saddle point is given by the (negative of) the Euclidean  on-shell action on the gravity side:
\be\label{fr1}
F = S_E =-M_P^{d-1}\Big(\int du d^{d}x\sqrt{ g}\left[ R-{\frac12}(\pa{\f})^2- V({\f})\right]
+2 \sum_{boundaries} \int d^d x \sqrt{\gamma} K \Big)\,.
\ee
The second term is the  Gibbons-Hawking-York terms that are required on each boundary of the solution, in which
$\gamma$ are the respective induced metrics on  boundary and $K$ its extrinsic curvature.  

As we have seen in the previous sections, there are up to three possible boundaries:
two of them are asymptotically $AdS_{d+1}$ boundaries, and we denote them by $B_+$ and $B_-$. These have satisfied the standard asymptotics of holographic theories near the UV, and we will call them ``UV boundaries''.
The third boundary, which we denote by $B_3$ arises from the   $AdS_d$ boundary of the negative curvature slices,  in the case when these are non-compact\footnote{The precise statement is that this is a d-dimensional manifold that is the $AdS_{d}$ boundary times the holographic direction.}.  We shall call the $B_3$ boundary, the {\em side boundary,} to distinguish it from the asymptotically $AdS_{d+1}$ {\em UV boundaries,} $B_{\pm}$.

We can then make the expression for the on-shell action (\ref{fr1})  more explicit
\begin{gather}
S_E =M_P^{d-1}\Big(\int_{-\infty}^{+\infty} du \int d^{d}x\sqrt{ g}\left[ R-{\frac12}(\pa{\f})^2- V({\f})\right]\nn \\
+2 \int_{{B}_+\cup{B}_-} d^d x \sqrt{\gamma} K+2 \int_{B_3} du  d^{d-1} \tilde{x} \sqrt{\tilde{\gamma}} \tilde{K}\Big)\,.\label{fr1-b}
\end{gather}
The $B_{\pm}$ boundaries are at $u=\pm\infty$ and are both included in two-sided (janus-like) solutions. For one-sided solutions (say with $B_+$),  the $B_-$ Gibbons-Hawking term is absent from (\ref{fr1-b}).
Finally, if the constant curvature manifold has compact slices, then there is no $B_3$ boundary, and the last boundary term in (\ref{fr1-b}) is absent.

In this and the next section we consider solutions with non-compact slices, so the $B_3$ boundary will be always understood to be present. We will refer to the solutions with both $B_{\pm}$ as {\em two-boundary} solutions (i.e. two holographic UV boundaries) and to the solutions with only one of either $B_{\pm}$ as  {\em one-boundary} solutions. The case with no $B_3$ will be discussed in section \ref{worm2}.

For the computation of the free energy, we classified different types of solutions according to the number of $AdS_{d+1}$ boundaries:

\vskip 0.5cm

$\bullet$ {\bf One-boundary (UV-Reg) solutions:} In this case, we have just one boundary ${B}_+$ at $u=+\infty$. There is no $B_-$ Gibbons-Hawking term.

\vskip 0.5cm

$\bullet$ {\bf Two-boundary (UV-UV) solutions:}  Here we have two boundaries. $B_-$ is located at $u=-\infty$ and ${B}_+$ at $u=+\infty$.

\vskip 0.5cm

$\bullet$ {\bf No-boundary (Reg-Reg) solutions} There is no boundary at any fixed value of $u$, however we still have the side boundary $B_3$.

\vskip 0.5cm

\subsection{On boundary conditions\label{bc}}

In a space with several boundaries, we expect that the solution of Einstein's equations will need boundary conditions on all of them.
However, in our case, we have looked for a special class of solutions, i.e. solutions with $O(1,d)$ symmetry. This symmetry was implemented in our metric ansatz in \eqref{eq:metric} by the geometry of the (Euclidean)  $AdS_d$ slices.
In such cases, the boundary conditions on the $B_3$ boundary are determined by the symmetry assumption, in this case by the slice geometry. In particular, both the scalar field and the scale factor are independent of the radial direction of the $AdS_{d}$ slice and therefore they behave as dimension $d-1$ fields under $O(1,d)$.

On the other hand, if one solves a fluctuation equation in the geometry, then we need also boundary conditions on $B_3$.

We conclude that the role of boundary conditions on $B_3$ is played by the assumption of $O(1,d)$ symmetry of our solutions.

The boundary conditions of the asymptotically $AdS_{d+1}$ boundaries follow the dictums of the holographic correspondence.

In more detail, for the one-boundary solutions, we need two boundary conditions on the single boundary $B_{+}$. One of them is interpreted as the source, while the second as vev. The regularity of the solution determines the source as a function of the vev.

For two-boundary solutions, the interpretation was discussed in \cite{BP}.
We need two boundary conditions on one of the boundaries, say $B_+$, and then $O(1,d)$ invariance determined uniquely the solution that is always regular. In particular, this determines the two boundary conditions on the other boundary $B_-$. We can phrase the same in the following holographically transparent way: choosing the source on $B_+$ and the source on $B_-$ determines completely the solutions as well as the two vevs.

\subsection{The free energy for one-boundary solutions}
The on-shell action  (\ref{fr1-b}) needs regularization because the solutions have infinite volume in the holographic directions as $u\to \pm \infty$. This is the standard holographic renormalization. Moreover, we need to regulate the infinite volume of the constant-$u$ slices.

For one-boundary solutions, the solution extends from a UV-cutoff $u_+$ (which we will eventually send to $+\infty$  after renormalization)  down to the endpoint $u=u_0$ where $e^{A(u_0)}=0$.

The regularized free energy for one-boundary solutions is computed in appendix \ref{osa} and we reproduce the result here,
\be \label{fr4}
F=-\frac{2M_P^{d-1}}{d} V_{\hepsilon} \Big(-d R^{(\zeta)}Ue^{(d-2)A}+d(d-1)e^{dA}\dot{A}\Big)\Big|^{u_+}_{u_0}\,.
\ee
Here,  $V_{\hepsilon}$ is the regularized volume of the slices. If we use  Poincar\'e coordinates defined in (\ref{H14}), it is given by\footnote{Although here we are using Poincar\'e coordinates for the metric $\zeta_{\m\n}$ of the slice for concreteness, the renormalization procedure of our solutions, does not depend on this.}:
\be \label{regvold}
V_{\hepsilon}=\int d^d x\sqrt{\zeta}=\int_{\hepsilon}^{+\infty} d\xi \int d^{d-1}x \sqrt{\zeta}=\int d^{d-1}x \frac{\a^d}{(d-1)\hepsilon^{d-1}}\,.
\ee
The  function $U(u)$ in \eqref{fr4} is defined as a solution to the first-order ODE,
\be \label{fr3}
(d-2)\dot{A}U+\dot{U}=-1\,.
\ee
Explicitly,
\be \label{fr4U}
U(u)=  e^{-(d-2)A(u)}\left(\mathfrak{b}-\int_{u_0}^u  e^{(d-2)A(u')} du'\right)\,,
\ee
where  $\mathfrak{b}$ is an integration constant. Notice that, as can be seen from \eqref{fr4}, the free energy is independent of this constant, since the contributions from $u_+$ and $u_0$ cancel out. Therefore any choice of initial condition to (\ref{fr3}) that fixes $\mathfrak{b}$ does not affect the free energy. It will be convenient to fix the solution such that $\mathfrak{b}$, which corresponds to the initial condition $U(u_0)=0$.

One important observation is that the boundary $B_3$ contributes non-trivially to (\ref{fr4}): it is responsible for the coefficient $d$  of the first term in the parenthesis, which without this contribution would be equal to one.

\subsubsection{IR contribution to the free energy}

As we have seen in section 3, regularity of the  $u=u_0$ endpoint (where $\f\rightarrow +\infty$) requires the solution to follow the type II asymptotics,
\be \label{fr5}
W=W_{\infty}^{(0)} e^{a\f}+\mathcal{O}(e^{a\l\f})\sp S=S_\infty^{(0)} e^{a\f}+\mathcal{O}(e^{a\l\f})\,,
\ee
where $W_{\infty}^{(0)}$ and $S_{\infty}^{(0)}$ are given in \eqref{wsinf0}.
Using the definitions in \eqref{eq:defSc} and \eqref{eq:defWc}, we obtain
\be \label{fr7}
e^{A(\f)}=
a_0 e^{- \frac{1}{2 a (d-1)}\f}+\cdots\,,
\ee
where $a_0$ is a constant of integration.
This shows that
\be \label{fr8a}
e^{d A} \dot{A}\sim e^{(a-\frac{d}{2a(d-1)})\f}\Big|_{\f\rightarrow +\infty}\rightarrow 0\,.
\ee
As a result of the last equation,  we can ignore the contribution from the lower limit $u_0$ of the second term in  \eqref{fr4}.

On the other hand,  solving \eqref{fr3} near $\f\rightarrow +\infty$ we obtain,
\be \label{fr8}
U(\f)= \tilde{\mathfrak{b}} e^{\frac{d-2}{2a(d-1)}\f}+\frac{2(d-1)}{(2a^2(d-1)+d-2)W_{\infty}^{(0)}} e^{-a\f}+\cdots,
\ee
where $\tilde{\mathfrak{b}}$ is the constant that is related to ${\mathfrak{b}}$ in \eqref{fr4U} as $\tilde{\mathfrak{b}}={\mathfrak{b}} a_0^{-(d-2)}$ using \eqref{fr4U} and \eqref{fr7}.
Now we can read the contribution to the free energy from the IR endpoint of the solution as
\begin{gather}\label{fr9}
{F}^{IR}\equiv \frac{2M_P^{d-1}}{d}\int d^d x\sqrt{\zeta}\Big(
- d ~R^{(\zeta)} U e^{(d-2)A}+d(d-1)e^{dA}\dot{A}\Big)\Big|_{u\to u_0}
\\
=\frac{2M_P^{d-1}}{d}\int d^d x\sqrt{\zeta}\Big(4R^{(\zeta)}
U(\f) e^{-\frac{d-2}{2(d-1)a}\f}\Big)_{\f\rightarrow +\infty}
= \frac{8M_P^{d-1} \mathfrak{b}}{d}\int d^d x\sqrt{\zeta}R^{(\zeta)} \,.\nn
\end{gather}
Here $R^{(\zeta)}$ is the curvature of $AdS_d$ slice.
\subsubsection{UV contribution to the free energy}

We consider now an asymptotically $AdS_{d+1}$ boundary $B_+$  located at $u\rightarrow +\infty$. To simplify the calculations, we consider $d=4$ and\footnote{In this notation, using standard quantization where $\varphi_-$ is the source of the relevant deformation, the dimension of the perturbing operator in the dual QFT is $d-\Delta$.  If $\D>2$ then one must Legendre transform the on-shell action.} 
\be 1<\D_-=\D<2\,.
\ee
The expansion of the scale factor and scalar field near such boundaries is given by
\begin{gather}
A(u)={A}_- +\frac{u}{\ell}-\frac{\mathcal{R}|\f_-|^{2/\D} \, \ell^2}{48} e^{- 2u/\ell} - \frac{\f_-^2 \, \ell^{2\D}}{24} e^{- 2\D u / \ell}
\nn \\
 - \frac{  (4-\D)\mcr |\f_-|^{4/\D} \, \ell^4}{12(4-2\D)}e^{- 4u/\ell} +\cdots \,,  \quad u\to +\infty, \label{fr10A}
\end{gather}
\be
\f(u)=\f_- \ell^{\D} e^{- \D u/\ell}+\frac{4\mcr\ell^{4-\D}|\f_-|^{(4-\D)/\D}}{\D(4-2\D)}  e^{- (4-\D)u/\ell} +\cdots\,,  \quad u\to +\infty, \label{fr10f}
\ee
Here $\f_-$ is the source of the scalar field operator $\mathcal{O}$ in the boundary field theory associated with $\f$. The vacuum expectation value of  $\mathcal{O}$ is given by
\be\label{expo}
\langle\mathcal{O}\rangle = \frac{d\mcr}{\D}|\f_- |^{\frac{4-\D}{\D}}\,.
\ee
We may obtain a relation between the dimensionless parameter $\mathcal{R}$ and curvature of $AdS_d$ space as follows
\be \label{fr13a}
 e^{-2A_-} R^{(\zeta)}=\mathcal{R} |\f_- |^{\frac{2}{\D}}\,.
\ee
Moreover, from \eqref{fr3} we find
\be \label{fr11}
U(u)=-\frac{\ell}{2}+ \mbr \ell^3 |\f_-|^{2/\D} e^{- 2u/\ell}+\frac{\ell^3\mathcal{R}|\f_-|^{2/\D}}{24} \frac{u}{\ell} e^{- 2u/\ell}+\cdots\,,
\ee
where $\mbr$ is a dimensionless constant of integration. This constant is not independent and is related to $\mathfrak{b}$ in \eqref{fr4U}.

The UV boundary contribution to the free energy in \eqref{fr4}, for  $d=4$, reads
\begin{gather}
{F}^{UV}
=-\frac{M_P^{3}}{2}\int d^4 x\sqrt{\zeta}\Big(
-4 R^{(\zeta)} U e^{2A}+12 e^{4 A}\dot{A}\Big)^{u= -\ell\log\epsilon}\,, \label{fr12}
\end{gather}
where $u_+=- \ell\log\epsilon$ is the regulated boundary.
Using \eqref{fr10A} and \eqref{fr11} we find
\begin{gather}
{F}^{UV} =M_P^3\int d^4 x \sqrt{\zeta}
\Big(-\frac{6 e^{4 A_-}}{\ell \epsilon^4}
-\frac{3 e^{4 A_-}\ell \mathcal{R}\f_-^{2/\D}}{4 \epsilon^2}
-\frac{e^{4 A_-} \ell^{2\D-1}(\D_- -2) \f_-^2}{2 \epsilon^{4-2\D}}\nn \\
+
\frac{1}{24} e^{4 A_-}\ell^3 \mathcal{R} |\f_-|^{4/\D} \big(48\mbr -  \mathcal{R}  (2\log\epsilon-1)\big)
+\mathcal{O}(\epsilon^{2\D-2})\Big)\,.\label{fr13}
\end{gather}
To derive the above relation, we have used \eqref{fr13a}.

To obtain a finite renormalized free energy we must add counter-terms. The following terms on the asymptotically $AdS_5$  boundary $B_+$  are enough to cancel the divergent terms in \eqref{fr13}
\be\label{fr14}
F^{UV}_{ct}= M_P^3\int d^4 x \sqrt{\g}\Big(
\frac{6}{\ell}+\frac{\D}{2\ell}\f^2+\frac{5\ell}{4} R^{(\gamma)}+\frac{\ell^3}{12}(R^{(\gamma)})^2 \log\omega\epsilon
\Big)^{u=-\ell\log\epsilon}\,,
\ee
where the parameter $\omega$ determines the scheme of the renormalized free energy\footnote{When there is no side boundary, $B_3$,  the coefficients of the first two terms in \eqref{fr14} do not change, however, the coefficients of the third and fourth terms are $\frac12$ and $\frac{1}{48}$ respectively do. Here, considering the counter-terms on the UV boundaries is sufficient to cancel the divergences corresponding to the side boundary.}.
Finally, the renormalized free energy is
\begin{gather}
{F}^{UV}_{ren}\equiv \lim_{\e\to 0}(F^{UV}+F^{UV}_{ct}) \nn \\
=\frac{M_P^3\ell^3}{96} \int d^4 x \sqrt{\zeta}
 e^{4 A_-} |\f_-|^{4/\D} \left(\mathcal{R}^2-96 \mcr+8\mathcal{R}^2\log\omega + 192 \mbr \mathcal{R}\right)\,.\label{fr15}
\end{gather}
In the most general case three scheme-dependent constants enter the free energy, as shown in \cite{Ghosh:2020qsx}. We have simplified them here as they do not play an essential role in what follows.

In the formula above we must use the regularized slice volume,  defined in (\ref{H14}).
It is also important to separate it from its curvature dependence so that we combine this with $\f_-$ above.
The end result is
\be \label{fr17}
{F}^{UV}_{ren}=\frac{3M_P^3\ell^3 \bar V_{\hepsilon}(1)}{2}
 \left(1-96 \frac{\mcr}{\mathcal{R}^2}+ 192 \frac{\mbr}{ \mathcal{R}}\right)\,,
\ee
where $\bar V_{\hepsilon}(1)$ was defined in (\ref{H14}) and we set   $\omega=1$.

The free energy \eqref{fr4} for solutions with just one boundary can now be computed.
In this case, the free energy is the subtraction of the UV part from the IR part. We choose, without loss of generality,  $\mathfrak{b}=0$ in \eqref{fr4U} as an initial condition in the IR limit of $U$ which automatically fixes the value of $\mbr$. The free energy is independent of this choice because it is the difference between UV and IR parts. In this way, the contribution from the IR side vanishes in \eqref{fr9}, and finally, the renormalized free energy is
\be \label{fr18}
{{ F}}_{ren}={ F}^{UV}_{ren}-{ F}^{IR}=
\frac{3M_P^3\ell^3 \bar V_{\hepsilon}(1)}{2}
 \left(1-96 \frac{\mcr}{\mathcal{R}^2}+ 192 \frac{\mbr}{ \mathcal{R}}\right)\,.
\ee
For numerical purposes, we define a dimensionless free energy density as follows,
\be \label{fr19}
\mathcal{F}=\frac{F_{ren}}{M_P^3\ell^3 \bar V_{\hepsilon}(1)}=
\frac{3}{2}
 \left(1-96 \frac{\mcr}{\mathcal{R}^2}+ 192 \frac{\mbr}{ \mathcal{R}}\right)\,.
\ee
We should mention here that as was first shown in \cite{F} for the three-dimensional case ($d=3$), the functions $\mcr$ and $\mbr$ satisfy a differential relation that has its origin in thermodynamics when $\mR>0$.
In  \cite{Ghosh:2020qsx} it was shown that in $d=4$, the analogous relation is\footnote{There is a different convention used here for $\mbr$ which is -two times the one in \cite{Ghosh:2020qsx}.}
\be
Y(\mR)\equiv 2\mcr'+\mbr-\mR~\mbr'-\frac{\mR}{48}=0\,,
\ee
where primes are derivatives with respect to $\mR$, and with the last term originating from the conformal anomaly.

We have analyzed the validity of this equation in our solutions.
We find that in the first branch of solutions in figure \ref{BCR}, $Y(\mR)$ is a constant approximately equal to -0.379. In the second branch, it is also a constant approximately equal to 2.903, and so on.
We do not understand this behavior.

\vskip 0.5cm
\subsection{The free energy for two-boundary solutions}
\vskip 0.5cm

In the case of a solution with two asymptotically $AdS_{d+1}$ boundaries $B_{\pm}$, the on-shell action on the gravity side is a sum of the bulk term and  all GHY boundary terms
\begin{gather}
S=M_P^{d-1}\int d^{d+1}x\sqrt{ g}\left[ R-{\frac12}(\pa{\f})^2- V({\f})\right]\nn \\
+2M_P^{d-1} \int_{B_+\cup B_-} d^d x \sqrt{\gamma} K+2M_P^{d-1} \int_{B_3} d^d \tilde{x} \sqrt{\tilde{\gamma}} \tilde{K}\,,\label{Afr1}
\end{gather}
where $B_{\pm}$ are located at $u=u_{\pm}$  and the side boundary $B_3$ at $\xi=\hepsilon$.

We do the same steps as in the appendix \ref{osa} with two changes. First, $u_0\rightarrow u_-$ and second, we should add the GHY boundary term at $u\rightarrow u_-$. Doing this,
 we find the regularized free energy as
\be \label{frw2}
{F}=-\frac{2M_P^{d-1}}{d}\int d^d x\sqrt{\zeta}\Big(
- d R^{(\zeta)} U e^{(d-2)A}+d(d-1)e^{dA}\dot{A}\Big)\Big|^{u_+}_{u_-}\,.
\ee
Using the expansions of the fields near the $B_{1,2}$ boundaries at $u=u_{\pm}\to\pm\infty$
\begin{gather}
A^{\pm}(u)={A}^{\pm}_- \pm \frac{u}{\ell}-\frac{\mathcal{R_{\pm}}|\f_-^{\pm}|^{2/\D} \, \ell^2}{48} e^{\mp 2u/\ell} - \frac{({\f_{-}^{\pm}})^2 \, \ell^{2\D}}{24} e^{\mp 2\D u / \ell} \nn \\   - \frac{ C^{\pm}(\mR_+,\mR_-) (4-\D) |\f_-^{\pm}|^{4/\D} \, \ell^4}{12(4-2\D)}e^{\mp 4u/\ell} +\cdots \,,\label{frw3}
\end{gather}
\be
\f^{\pm}(u)=\f_{-}^{\pm} \ell^{\D} e^{\mp \D u/\ell}+\frac{4C^{\pm}(\mR_+,\mR_-)\ell^{4-\D}|\f_{-}^{\pm}|^{(4-\D)/\D}}{\D(4-2\D)}  e^{\mp(4-\D)u/\ell} +\cdots\,, \label{frw4}
\ee
and
\be \label{frw5}
U^{\pm}(u)=\mp\frac{\ell}{2}+ \mathcal{B}_{\pm}(\mR_+,\mR_-) \ell^3 |\f_{-}^{\pm}|^{2/\D} e^{\mp 2u/\ell}+\frac{\ell^3\mathcal{R}_{\pm}|\f_{-}^{\pm}|^{2/\D}}{24} \frac{u}{\ell} e^{\mp 2u/\ell}+\cdots\,,
\ee
we finally obtain the renormalized free energy density of the two-boundary solutions if we subtract the following set of counterterms
\be\label{Afr14}
F^{\pm}_{ct}= \pm M_P^3\int d^4 x \sqrt{\gamma^{\pm}}\Big(
\frac{6}{\ell}+\frac{\D}{2\ell}\f^2+\frac{5\ell}{4} R^{(\gamma^\pm)}+\frac{\ell^3}{12}(R^{(\gamma^\pm)})^2 \log\epsilon
\Big)^{u_{\pm}=\mp\ell\log\epsilon}\,.
\ee
The renormalized free energy density, after taking $\e\to 0$  becomes now
\be \label{frw6}
\mathcal{F}=\frac{1}{M_P^3\ell^3 \bar{V}_{\hepsilon}(1)}\big({F}^{+}_{ren}-{F}^{-}_{ren}\big)=
3-144\Big(\frac{C^+(\mR_+,\mR_-)}{(\mathcal{R}_+)^2}+\frac{C^-(\mR_+,\mR_-)}{(\mathcal{R}_-)^2}\Big)+
\ee
$$+
288\Big(\frac{\mathcal{B}_+(\mR_+,\mR_-)}{ \mathcal{R}_+}-\frac{\mathcal{B}_-(\mR_+,\mR_-)}{ \mathcal{R}_-}\Big)\,.
$$

The strategy to calculate numerically the free energy from \eqref{frw6} is as follows: Since the solutions have two boundaries,  there is always at least one A-bounce for each solution. We choose an arbitrary point in the green region of the map \ref{map} which represents a solution with an  A-bounce and then in addition to solving the differential equations for $A(u)$ and $\f(u)$ we also solve the differential equation of $U(u)$ in \eqref{fr3}.

Knowing the solutions for $A, \f$, and $U$, we can read the parameters in \eqref{frw6} on both boundaries by using the expansions of \eqref{frw3}--\eqref{frw5}. For doing this, we need an extra initial condition for $U(u)$ at the bounce, however, the value of free energy is independent of this choice. This is because at an arbitrary A-bounce, say at $u=u_b$, we have $\dot{A}(u_b)=0$ so from \eqref{fr3} we see that for every finite value of $U(u_b)$, we should have $\dot{U}(u_b)=-1$. In practice, for the numerics, we choose $U$ at the A-bounce to vanish.

\subsection{Free energy for  solutions without $AdS_{d+1}$ boundaries}\label{FEnobo}

As we already mentioned, the Reg-Reg solutions do not have any asymptotic $AdS_{d+1}$ boundary. Therefore the on-shell action has only a bulk contribution,  plus the GHY term for the side boundary, $B_3$. For a solution in this class, we have two regular end-points at $u_1$ and $u_2$.  The bulk on-shell action is
\begin{gather}
S_1=\frac{2M_P^{d-1}}{d}\int_{\hepsilon}^{+\infty}d\xi\int d^d x\sqrt{\zeta}\Big(
 R^{(\zeta)}\int_{u_1}^{u_2} du e^{(d-2)A}- e^{dA}\dot{A}\Big|_{u_1}^{u_2}\Big)\nn\\
 = \frac{2M_P^{d-1}}{d} V_{\hepsilon}(\a)R^{(\zeta)}\int_{u_1}^{u_2} du e^{(d-2)A}\,,\label{frw7}
\end{gather}
where in the first line the contribution of the second term at both end-points vanishes,  using the same argument as in \eqref{fr8a}.

The contribution of the $B_3$ GHY term has been incorporated in the first term above, as before
\be
S_3=\frac{2(d-1) M_P^{d-1}}{d} V_{\hepsilon}(\a)R^{(\zeta)}\int_{u_1}^{u_2} du e^{(d-2)A}\,.\label{frw8}
\ee
Adding \eqref{frw7} and \eqref{frw8} we find the regularized free energy density of the solution without the $AdS_{d+1}$ boundaries
\be
\mathcal{F}=\frac{F}{M_P^{d-1} \bar{V}_{\hepsilon}(1)}=-\a^d R^{(\zeta)}\int_{u_1}^{u_2} du e^{(d-2)A}\,.\label{frw9}
\ee
We should note that the regularized free energy per unit volume above is finite and it does not need to add any counter-term.

\section{QFT data and free energy (numerical analysis)}\label{numq}

We will now focus on a specific model and the phase diagram numerically. This involves 1) extracting the UV parameter ${\mathcal R}$ of the solutions from their near-boundary asymptotics; 2) computing the free energies and comparing them for solutions with the same number of boundaries and the same values of the UV parameters.
We will focus on regular solutions with two, one, and zero UV boundaries, in the theory with the same dilaton potential we used in section \ref{REGSOL}, given by equation (\ref{ppp}) with parameters (\ref{numbers}).
We expect the results will not change qualitatively for different values of the parameters, as long as they do not cross the critical values (\ref{numbers-2}).

\subsection{Comparing the free energy of the one-boundary solutions}

In the confining theory, we found a set of regular solutions that, at one end reach a UV fixed point (associated to the boundary $B_+$), and on the other end asymptote to infinity in field space, but in a regular way. These solutions were parametrized by $S_\infty^{(1)}$ with $S_\infty^{(1)c}<S_\infty^{(1)}<0$, as we discussed in appendix \ref{secRS}.

\subsubsection{UV parameters}
We can read the information on the (unique) UV boundary for this type of solution as follows:  At the IR end-point $u=u_0$ where $\f\rightarrow +\infty$, solutions are parametrized by $S^{(1)}_\infty$ as seen in (\ref{infexp2a})  in appendix \ref{secRS}. In particular, for confined solutions we find
\be \label{UVfu}
\f(u)=-\frac{1}{a}\log \big[\k (u_0-u)\big]
-\frac{S_{\infty}^{(1)}}{\l-2}\k^{-\l}(u_0-u)^{1-\l}+\cdots\,,
\ee
\begin{gather} \label{UVAu}
A(u)=\frac{1}{2a^2(d-1)}\log \big[2(d-1)\k(u_0-u)\big]\nn \\
-\frac{S_{\infty}^{(1)} (d (\lambda -3)-2 \lambda +4)}{2 a (d-1) d (\lambda -2) (\lambda -1)}\k^{-\l}(u_0-u)^{1-\l}
+\cdots\,,
\end{gather}
where
\be\label{kaplam}
\k=\frac{a^2(d-1)}{\ell\sqrt{2a^2(\frac{1}{d}-1)+1}} \sp \l=\frac{1}{a^2(d-1)}-1\,,
\ee
where the parameter $a$ was defined in (\ref{lfpot}).

On the other hand, near the asymptotic UV boundary at $\f=0$,
 we have the following expansions as $u\rightarrow +\infty$ (see appendix \ref{UVF})
\be
 \f(u) = \f_- \ell^{\Delta}e^{-\Delta u / \ell} + \frac{\mcr d \, |\f_-|^{(d-\Delta)/\Delta}}{\Delta(d-2 \Delta)} \, \ell^{d-\Delta} e^{-(d-\Delta) u /\ell} + \ldots \,, \label{tphiLU}
\ee
\begin{gather}
 A(u) = {A}_- +\frac{u}{\ell} - \frac{\f_-^2 \, \ell^{2 \Delta}}{8(d-1)} e^{-2\Delta u / \ell}  -\frac{\mathcal{R}|\f_-|^{2/\Delta} \, \ell^2}{4d(d-1)} e^{-2u/\ell} \nn \\
  - \frac{(d-\Delta) \mcr |\f_-|^{d/\Delta} \, \ell^d}{d(d-1)(d-2 \Delta)}e^{-du/\ell} +\ldots \,,\label{tALU}
\end{gather}
where $\f_-$ and $A_-$ are constants of integration. Here $\f_-$ is the source or the coupling of the dual operator $\mathcal{O}$, $\mcr$ is proportional to the vev of the stress-energy tensor of the boundary QFT and the dimensionless curvature $\mathcal{R}$ is defined in \eqref{fr13a}.

By solving the equations of motion, and using the asymptotic expansions above, we can read the relation between the parameters $\mathcal{R}$ and $\mcr$ and $\f_-$\footnote{$\f_-$ is dimensionful and is plotted in units of the $AdS$ length $\ell$.} in terms of the IR parameter $S^{(1)}_\infty$. We have plotted these parameters in figures \ref{UVD6}--\ref{UVD8}. We observe the following properties:

\begin{figure}[!ht]
\centering
\begin{subfigure}{0.49\textwidth}
\includegraphics[width=1\textwidth]{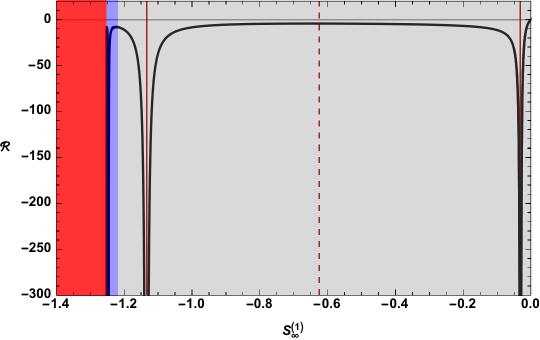}
\caption{}\label{UVD6}
\end{subfigure}
\centering
\begin{subfigure}{0.49\textwidth}
\includegraphics[width=1\textwidth]{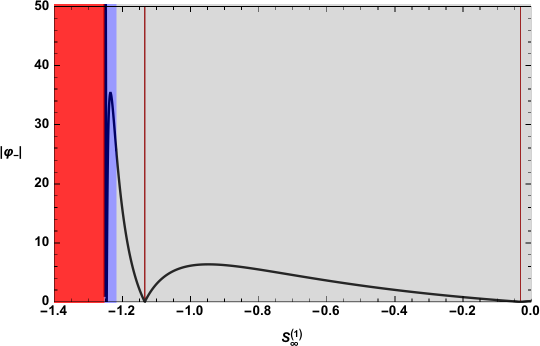}
\caption{}\label{UVD7}
\end{subfigure}
\centering
\begin{subfigure}{0.49\textwidth}
\includegraphics[width=1\textwidth]{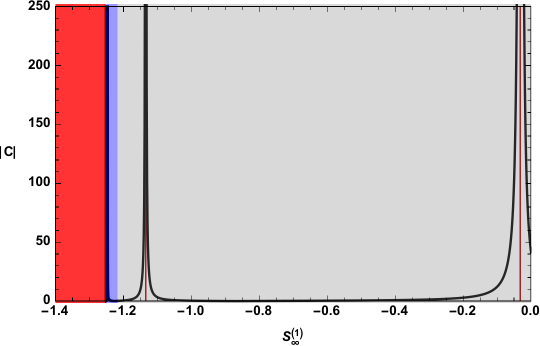}
\caption{}\label{UVD8}
\end{subfigure}
\begin{subfigure}{0.49\textwidth}
\includegraphics[width=1\textwidth]{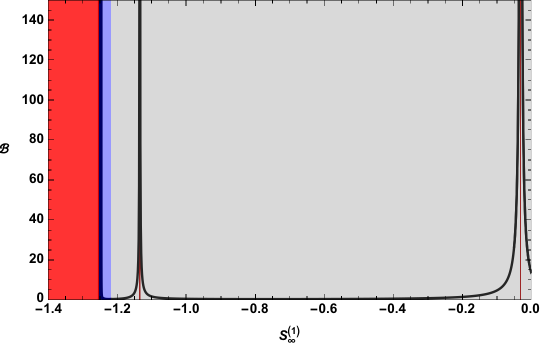}
\caption{}\label{UVD9}
\end{subfigure}
\caption{\footnotesize{(a)-(c): $\mathcal{R}$ the dimensionless curvature, $\f_-$ the coupling of operator $\mathcal{O}$ at the UV boundary and $C$ parameter of the UV boundary for UV-Reg solutions. All figures are plotted as a function of the free IR parameter  $S^{(1)}_\infty$. In each graph, the gray region belongs to the regular solutions without A-bounce and the blue region to solutions with at least one A-bounce. In the red region, we have no solutions with boundary. The vertical dashed line in figure (a) corresponds to the global $AdS$ solution in the uplifted theory and the product solution is the solution right before the blue-red boundary. Figure (d) gives $\mb$ which we need to compute the free energy of the solutions.}}
\label{UVD9a}
\end{figure}

\begin{figure}[!ht]
\centering
\begin{subfigure}{0.48\textwidth}
\includegraphics[width=1\textwidth]{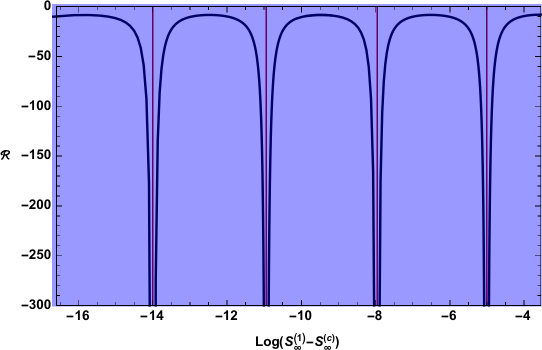}
\caption{}\label{ZD6}
\end{subfigure}
\centering
\begin{subfigure}{0.50\textwidth}
\includegraphics[width=1\textwidth]{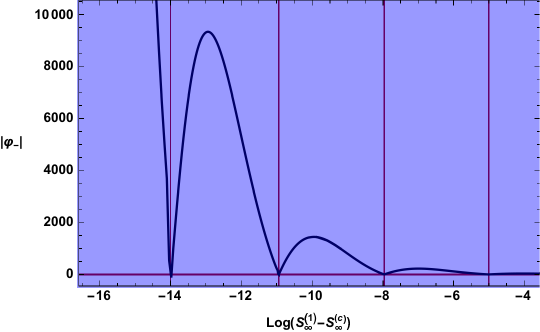}
\caption{}\label{ZD7}
\end{subfigure}
\centering
\begin{subfigure}{0.49\textwidth}
\includegraphics[width=1\textwidth]{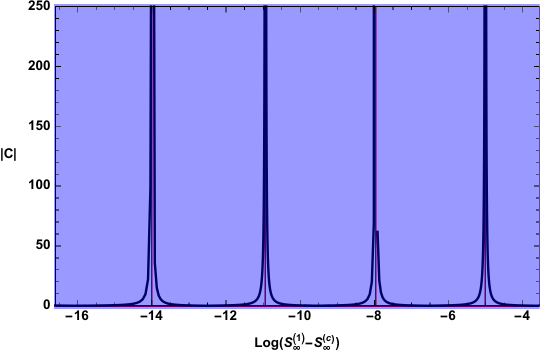}
\caption{}\label{ZD8}
\end{subfigure}
\begin{subfigure}{0.49\textwidth}
\includegraphics[width=1\textwidth]{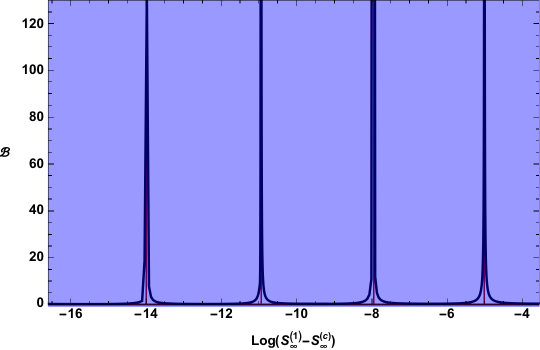}
\caption{}\label{ZD9}
\end{subfigure}
\caption{\footnotesize{The blue region in figures \ref{UVD6}--\ref{UVD9}. The horizontal axis is $\log(S_{\infty}^{(1)}-S_{\infty}^{(c)})$, where $S_{\infty}^{(c)}\approx -1.25$ is the critical value for which we have the UV-Reg solution with infinite numbers of the loops. }} \label{ZD9a}
\end{figure}

\begin{figure}[!ht]
\centering
\includegraphics[width=.6\textwidth]{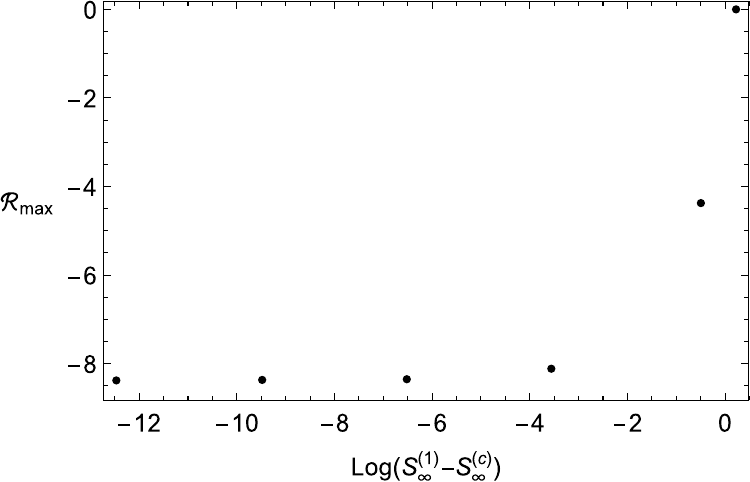}
\caption{\footnotesize{Plot of the maximum $\mathcal{R}$ values of the curves in figures  \ref{UVD6} and \ref{ZD6}. The curves there have $\cap$-shaped pieces. $\mathcal{R}_{max}$ is the maximum value of $\mathcal{R}$ in each $\cap$-shaped piece.  The horizontal axis is $\log(S_{\infty}^{(1)}-S_{\infty}^{(c)})$, where $S_{\infty}^{(c)}\approx -1.25$ is the critical value for which we have the UV-Reg solution with infinite numbers of the loops. }}
\label{Num1}
\end{figure}

\begin{enumerate}
\item
In the plots \ref{UVD9a} and \ref{Zfreea}, the gray region describes the solutions that have no A-bounce.

    \item In figure \ref{UVD6}, the vertical dashed line shows the point where in the uplifted theory we have the global $AdS$ solution.

\item A {\it vev-driven solution} is a solution where the source $\f_-$ vanishes and the flow is driven by a scalar vev. At the vev-driven solutions, $\mathcal{R}\rightarrow -\infty$ as  $\f_-\rightarrow 0$ because
\be
\mathcal{R}|\f_-|^\frac{2}{\D_-}=R^{UV}\,,
\ee
where $R^{UV}$ is the induced (source) curvature on the boundary.
Such vev-driven solutions are denoted by red vertical lines in \ref{UVD9a}.

\item Vev-driven solutions describe saddle points of the UV CFT (associated to the maximum at $\f=0$) on a constant negative curvature manifold with curvature $R^{UV}$. In such solutions, the operator dual to the scalar $\f$ has a vev and it is this vev that drives the flow.

\item  In the gray region of solutions with monotonic scale factor, we observe just two vev-driven solutions.

\item
The narrow blue region describes the regime where all regular solutions have at least one A-bounce.
In this region, we expect that as we move towards the infinite loop solution in figure \ref{Nprod1}, since the number of $\f$ and A bounces is increasing, we observe more and more points where $\f_-$ vanishes,  or $\mathcal{R}\rightarrow -\infty$. To see this clearly, we have zoomed in the blue region in figures \ref{ZD6}--\ref{ZD9}. The horizontal axis in all plots is $\log(S_{\infty}^{(1)}-S_{\infty}^{(c)})$ where $S_{\infty}^{(c)}\approx -1.25$ is the critical value for which we have the UV-Reg solution with infinite numbers of loops. Going more closer to the $S_{\infty}^{(c)}$ one moves towards $-\infty$ in the diagrams.

\item
From these plots, it is clear that there is an infinite number of saddle points, driven by non-trivial scalar vevs that compete with the CFT saddle point\footnote{Such vev-driven saddle points of a CFT, were also found for holographic CFTs on positive constant curvature
    manifolds, like spheres or dS space, \cite{C,F}. However, in all such cases the number of saddle points was finite, unlike what we find here.}.

\item As is clear from figures \ref{UVD6} and \ref{ZD6}, the curve $\mathcal{R}$ vs the IR parameter $S^{(1)}_{\infty}$, is composed of an infinite number of $\cap$-shaped pieces. The maximum value of $\mathcal{R}$ in inside each  $\cap$-shaped piece, $\mathcal{R}_{max}$, is plotted in figure \ref{Num1}, against the IR parameter $\log(S_{\infty}^{(1)}-S_{\infty}^{(c)})$,  normalized so that it asymptotes to $-\infty$  at the end-point of existence of the one-boundary solutions.
    It is clear that the discrete sequence of $\mathcal{R}_{max}$ is decreasing to the left, and seems to have a finite limit that shall call $\mathcal{R}_*$, as $S_{\infty}^{(1)}\to S_{\infty}^{(c)}$.

    We can read from this figure that when $\mathcal{R}$ is between 0 and around -4, there is a single saddle point for any given value of $\mathcal{R}$. For the next interval, between around -4 and around -8, there are three solutions per value of $\mathcal{R}$. This continues with an ever-increasing number of solutions. For $\mathcal{R}<\mathcal{R}_0$ there is an infinite number of solutions per value of $\mathcal{R}$.

    Of course, as our numerical results reach only a finite distance away from this limit, we cannot exclude the possibility that $\mathcal{R}_0=-\infty$. If this is the case, then every possible value of $\mathcal{R}$ has a finite but ever-increasing number of competing solutions. Only the CFT, has an infinite number of vev-driven competing solutions in this case.

\end{enumerate}

\subsubsection{Renormalized free energy\label{rfe}}

\begin{figure}[!ht]
\centering
\begin{subfigure}{0.45\textwidth}
\includegraphics[width=1\textwidth]{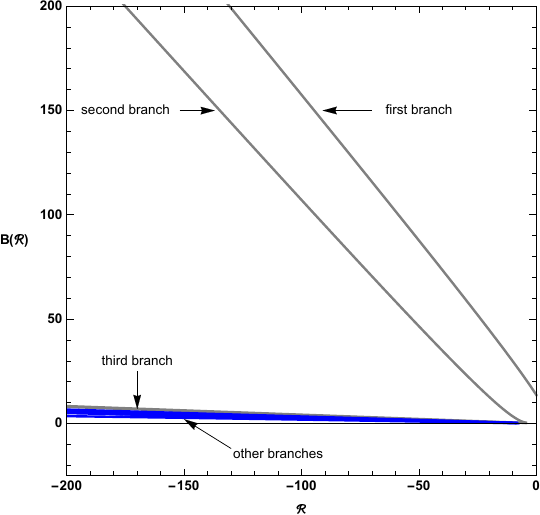}
\caption{\footnotesize{}}\label{BR}
\end{subfigure}
\begin{subfigure}{0.45\textwidth}
\includegraphics[width=1\textwidth]{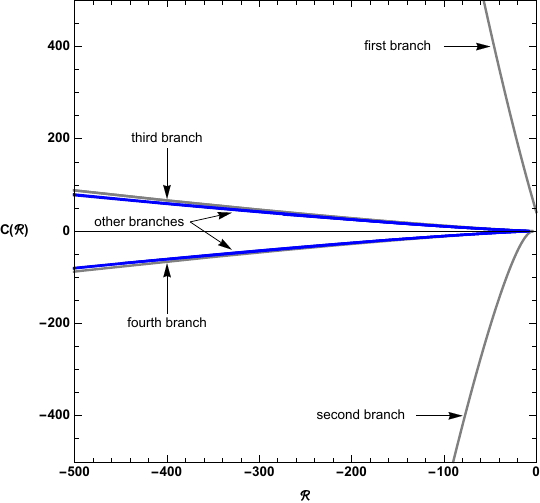}
\caption{\footnotesize{}}\label{CR}
\end{subfigure}
\caption{\footnotesize{$\mbr$ and $\mcr$ for the various branches in figure \ref{UVD6}. The gray (blue) curves correspond to branches in gray (blue) regions of the figure \ref{UVD6}.}}
\label{BCR}
\end{figure}

To compute the free energy for solutions with one boundary, we should use equation \eqref{fr19}.
Here in addition to the values of $\mcr$ and $\mathcal{R}$ which we found in the previous section, we need to know the value of $\mbr$. To read the value of $\mbr$, we should solve the differential equation for $U(u)$ in \eqref{fr3} with the initial condition that we discussed for IR end-points, i.e. $\mbr=0$ in \eqref{fr8} or equivalently at $u=u_0$ we should consider $U(u_0)=0$. Then, we can find $\mbr$ by using the expansion of $U(u)$ in \eqref{fr11} as a function $S_{\infty}^{(1)}$. The results are shown in figure \ref{UVD9} (see also figure \ref{ZD9}).

In figure \ref{BR} and \ref{CR} we plot the functions $\mbr,\mcr$ for the various branches of the one-boundary solutions in \ref{UVD6}. The various branches reflect the distinct branches in \ref{UVD6}, which are numbered starting
from the right. The functions in the range plotted are nearly linear. Nonlinearities for large $\mR$, however, exist as indicated by the behavior shown in figure \ref{Fvevall}.

\begin{figure}[!ht]
\centering
\begin{subfigure}{0.55\textwidth}
\includegraphics[width=1\textwidth]{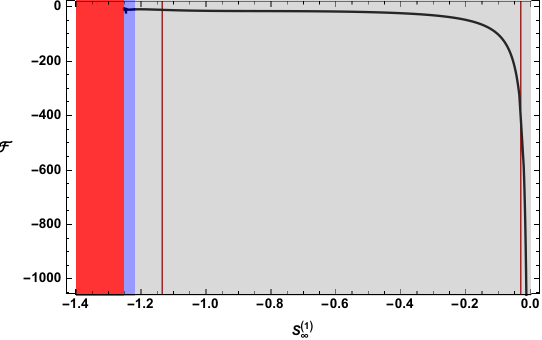}
\caption{\footnotesize{}}\label{freeUVR}
\end{subfigure}
\begin{subfigure}{0.45\textwidth}
\includegraphics[width=1\textwidth]{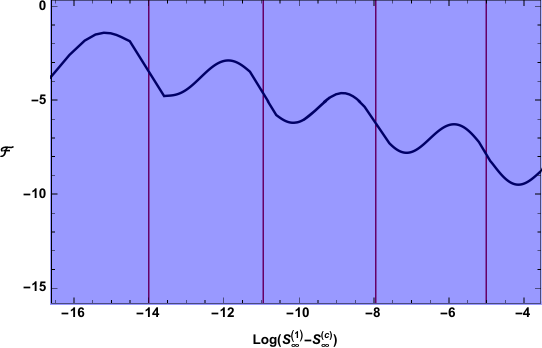}
\caption{\footnotesize{}}\label{Zfree}
\end{subfigure}
\begin{subfigure}{0.49\textwidth}
\includegraphics[width=1\textwidth]{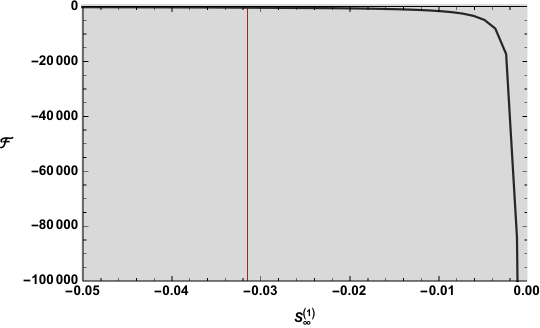}
\caption{\footnotesize{}}\label{ZfreeG}
\end{subfigure}
\caption{\footnotesize{(a): The free energy density for UV-Reg solutions living on the black curves of the map \ref{map}. (b) The blue region is zoomed in. The horizontal line is now  $\log(S_{\infty}^{(1)}-S_{\infty}^{(c)})$, where $S_{\infty}^{(c)}\approx -1.25$ is the critical value for which we have the UV-Reg solution with infinite numbers of loops. (c): The region near $S_{\infty}^{(1)}=0$ is zoomed.  In all diagrams, the vertical red lines show the locations where $\f_-\rightarrow 0$.}}
\label{Zfreea}
\end{figure}

Using the data for $\mcr, \mathcal{R}$ and $\mbr$  we can plot the free energy of the regular solutions as a function of the IR parameter $S_\infty^{(1)}$, see figures \ref{freeUVR}--\ref{ZfreeG}.

The numerical results of the renormalized free energy density \eqref{fr19} in terms of the IR parameter $S_{\infty}^{(1)}$ is sketched in figure \ref{freeUVR}. In the same figure, the gray region contains solutions without an  A-bounce and the tiny blue region is the space of solutions with at least one A-bounce. As one approaches the red region, the number of A-bounces increases. Figure \ref{Zfree} shows more detail of the blue region when we consider solutions with a larger number of A-bounces.  In this region, we have a periodic pattern in the free energy density. In the same  figure, the horizontal line is $\log(S_{\infty}^{(1)}-S_{\infty}^{(c)})$ where $S_{\infty}^{(c)}$ is the critical value of $S_{\infty}^{(1)}$ as the number of A-bounces asymptotes to  infinity. Figure \ref{ZfreeG} shows the gray region very near the $S_{\infty}^{(1)}=0$,  where at this point,  as we discuss below, the free energy asymptotes to $\mathcal{F}\rightarrow -\infty$. This is a solution with a flat boundary and it has the minimum free energy among the UV-Reg solutions.

\begin{figure}[!t]
\centering
\begin{subfigure}{0.49\textwidth}
\includegraphics[width=1\textwidth]{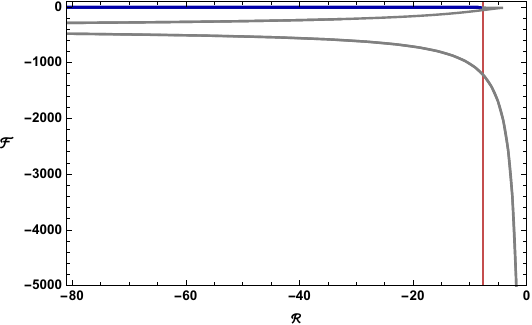}
\caption{\footnotesize{}}\label{FRT}
\end{subfigure}
\begin{subfigure}{0.475\textwidth}
\includegraphics[width=1\textwidth]{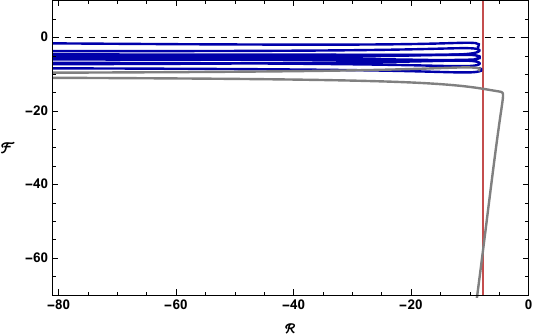}
\caption{\footnotesize{}}\label{FRZ}
\end{subfigure}
\caption{\footnotesize{(a): Free energy density in terms of dimensionless curvature. The gray/blue curves correspond to the gray/blue region in figure \ref{FRT}. Figure (b) is the zoomed region near $\mathcal{F}=0$. The vertical red line shows for $\mathcal{R}\gtrsim -7.7$ only solutions without A-bounce exist.}}
\end{figure}

\begin{figure}[!ht]
\begin{center}
\begin{subfigure}{0.47\textwidth}
\includegraphics[width=1\textwidth]{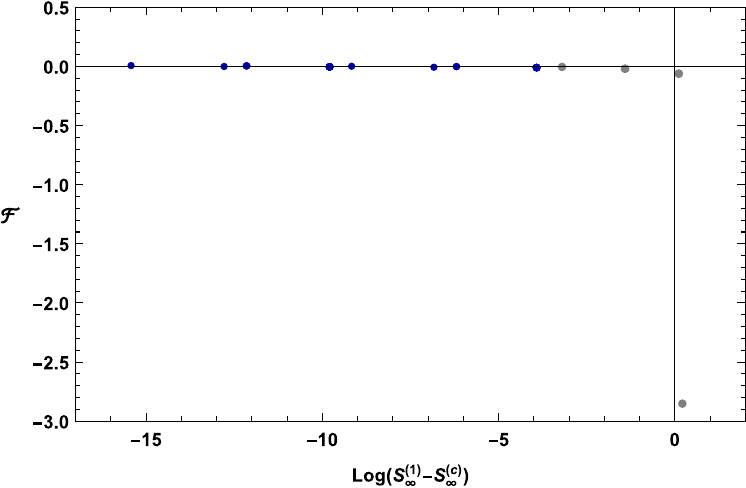}
\caption{\footnotesize{}}\label{FFR}
\end{subfigure}
\begin{subfigure}{0.46\textwidth}
\includegraphics[width=1\textwidth]{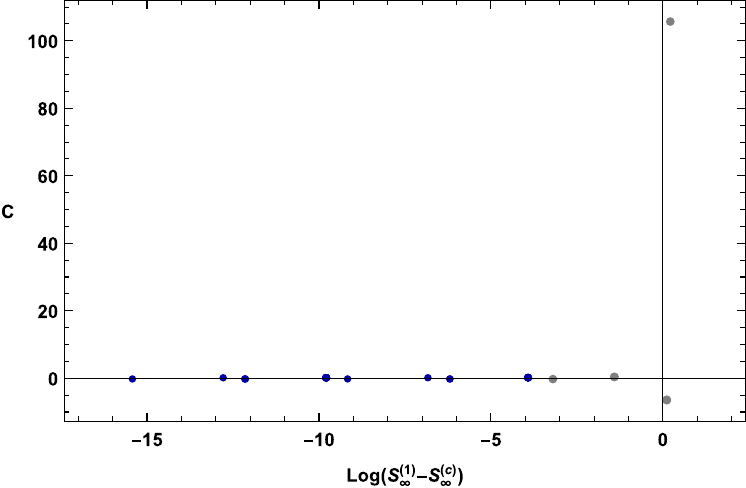}
\caption{\footnotesize{}}\label{CFFR}
\end{subfigure}
\begin{subfigure}{0.48\textwidth}
\includegraphics[width=1\textwidth]{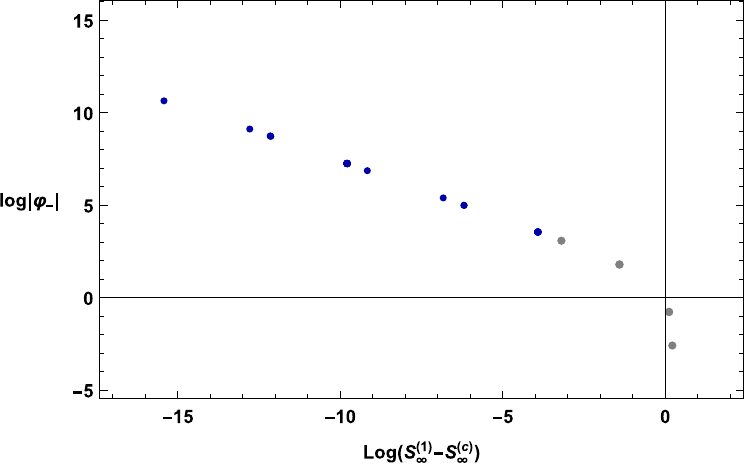}
\caption{\footnotesize{}}\label{fFFR}
\end{subfigure}
\begin{subfigure}{0.50\textwidth}
\includegraphics[width=1\textwidth]{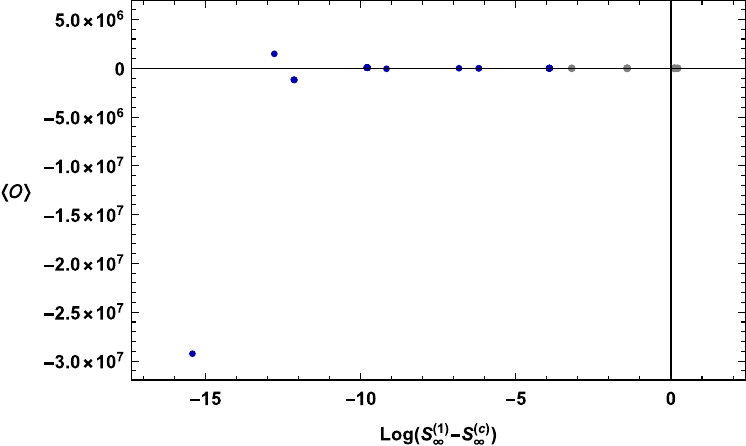}
\caption{\footnotesize{}}\label{VFFR}
\end{subfigure}
\end{center}
\caption{\footnotesize{At fixed $\mathcal{R}=-9$ (a)-(d) are the free energy, constant $C$, logarithm of the source and vacuum expectation value $\langle O\rangle =\frac{C d}{\Delta_-}|\f_-|^{\Delta_+ / \Delta_-}$. }}
\end{figure}

To find which solution is dominant from the free energy point of view, we should compare solutions corresponding to the same  QFTs (the same dimensionless curvature, $\mathcal R$, at the boundary). Figure \ref{FRT} shows the free energy in terms of the boundary curvature $\mathcal{R}$. Again, the gray and blue curves belong to the solution without(with) A-bounces. Figure \ref{FRZ} shows the region near $\mathcal{F}=0$, where we have a periodic behavior for free energy. As we move closer and closer to the solution with an infinite number of A-bounces, we have more and more oscillations for free energy.

As a specific example, we have plotted the free energy of the 12 solutions with the same value $\mathcal{R}=-9$ in figure \ref{FFR}. As we observe,  the solution with the lowest free energy is the one with the largest  $S_{\infty}^{(1)}$. 
This is the solution that does not contain  $A$-bounces nor $\varphi$-bounces.

Next, we discuss two limits of the solutions found.

\subsubsection{ The flat limit $\mathcal{R}\rightarrow 0$}

As we observed from numerical results in figures \ref{UVD6}--\ref{UVD9} in the flat limit $\mathcal{R}\rightarrow 0$ which can be reached as $S_{\infty}^{(1)}\rightarrow 0$
\be  \label{fr24}
|\f_-|\rightarrow\text{const}\sp C(\mathcal{R})\rightarrow \text{const}>0 \sp \mb(\mathcal{R})\rightarrow \text{const}>0\,,
\ee
which means that from \eqref{fr19}
\be  \label{fr25}
\mathcal{F}\rightarrow -\infty\,.
\ee
This is what we see actually in the right-hand side of figure \eqref{ZfreeG}.
The limiting theory is the confining QFT on Minkowski space-time driven by the relevant scalar operator dual to the bulk scalar $\f$.

The divergence in the free energy is a result of the fact that we normalized the free energy by the {\em unit curvature} $AdS$ volume, so what we see here is an infinite-volume limit, as in the case of the sphere \cite{F}.
The divergence is due to the infinite volume of the space in that limit. On the other hand, the free energy divided by the physical volume is finite and equal to the flat space free energy density of the confining theory.

\subsubsection{ Vev-driven solutions: $\f_-\rightarrow 0$\label{cftvev}}

When the source is zero or  $\f_-\rightarrow 0$, we have a vev-driven solution. This solution can be reached by the following scaling \cite{C}
\be \label{fr20}
{\f_-}\rightarrow 0\sp \mR\to -\infty\,,
\ee
with
\be
  C(\mathcal{R}) \sim |\mR|^{{\Delta_+}/{2}} \sp
C(\mathcal{R}) |{\f_-}|^{{\D_+} / {\D_-}}=\text{const} \sp \mathcal{R} |{\f_-}|^{2/{\D_-}}=\text{const}\,.
\ee
This limit corresponds to the vertical asymptotes in figures \ref{UVD9a}  and \ref{ZD9a}, where the relevant coupling goes to zero and the dimensionless curvature diverges at specific values of the parameter $S^{(1)}_{\infty}$. Those figures show that there are many vev-driven solutions (in fact, an infinite number in the blue region with A-bounces).

In the limit (\ref{fr20}) , near the UV boundary, from the expansion \eqref{fr10A} for $A(u)$, we observe that $\dot{A}\rightarrow \frac{1}{\ell}$. On the other hand, the expansion of \eqref{fr11} shows that $$U\rightarrow -\frac{\ell}{2}\sp \dot{U}\sim \mb(\mathcal{R}) |\f_-|^{2/\D} e^{-2u/\ell}\;.$$

By inserting these values into the equation for $U(u)$ in \eqref{fr3}, we find that when $\dot{U}\rightarrow 0$ we should have
\be \label{fr21}
{\f_-}\rightarrow 0: \qquad \mb(\mathcal{R})|\f_-|^{2/\D_-}=\text{const}\,.
\ee
The above behaviors show that the second and third terms in the free energy \eqref{fr19} are finite at $d=4$
\be \label{fr22}
{\f_-}\rightarrow 0:\quad   \frac{\mcr}{\mathcal{R}^2}\rightarrow \text{const}\times |\f_-|^{\frac{4-d+\D_-}{\D_-}}=\text{const}\times |\f_-|\rightarrow 0\sp \frac{\mbr}{\mathcal{R}}\rightarrow\text{const}\,.
\ee
Therefore the free energy density \eqref{fr19} tends to a finite constant value as one approaches the vev-driven solutions
\be  \label{fr23}
{\f_-}\rightarrow 0 :\qquad  \mathcal{F}\rightarrow\text{const}\,.
\ee
This shows that the free energy is continuous as a function of the IR parameter $S^{(1)}_{\infty}$ and remains finite as $\mR\to -\infty$, as one can see in figure \ref{Zfreea}.

For the vev-driven solution, we have the following expansions \cite{C}
\be
\varphi(u) =\varphi_{+}\ell^{\Delta_{+}}e^{-\Delta_{+}u/\ell}\left(1+\mathcal{O}\left(\bar{\mathcal{R}}|\varphi_{+}|^{2/\Delta_{+}}e^{-2u/\ell}\right)+\ldots\right)+\cdots\,,
\ee
\be
A(u) ={A}_{+}+\frac{u}{\ell}-\frac{\varphi_{+}^{2}\ell^{2\Delta_{+}}}{8(d-1)}e^{-2\Delta_{+}u/\ell}-\frac{\bar{\mathcal{R}}|\varphi_{+}|^{2/\Delta_{+}}\ell^{2}}{4d(d-1)}e^{-2u/\ell}+\cdots\,,
\ee
where ${A}_{+}$ and $\varphi_{+}$ are new constants of integration for the vev-driven solution. Similarly we find that, as $u\to +\infty$,
\be \label{Afrw5}
U(u)=-\frac{\ell}{2}+ \mb^{vev}({\bar \mR}) \ell^3 |\f_{+}|^{2/\D_+} e^{-2u/\ell}-\frac{\ell^3\bar{\mathcal{R}}|\f_{+}|^{2/\D_+}}{24} \frac{u}{\ell} e^{- 2u/\ell}+\cdots\,,
\ee
where $\mathcal{B}^{vev}({\bar \mR})$ is the constant of integration obtained from solving equation \eqref{fr3} with the initial condition $U(u_0)=0$. We should note that the new dimensionless parameter $\bar{\mathcal{R}}$,  is the proxy for the curvature in the vev case and is defined as follows
\be
\bar{\mathcal{R}}|\f_+|^\frac{2}{\D_+}=\Big|\frac{\langle O\rangle}{(2\Delta_+-d)}\Big|^\frac{2}{\D_+}      R^{UV}=e^{-2A_+}R^{(\z)}\,,
\ee
where in the second step above we used the relation between $\f_+$ and the vev of the dual scalar operator.
It is a discrete parameter, and has a concrete value, for each vev-driven solution.

Repeating  the procedure  for computing  the free energy density, we obtain:
\be \label{Afr19}
\mathcal{F}^{vev}=\frac{F^{vev}_{ren}}{M_P^3\ell^3 \bar{V}_{\hepsilon}(1)}=
\frac{3}{2}
 \left(1+ 192 \frac{\mathcal{B}^{vev}({\bar \mR})}{ \bar{\mathcal{R}}}\right)\,.
\ee

\begin{figure}[!ht]
\centering
\begin{subfigure}{0.49\textwidth}
\includegraphics[width=1\textwidth]{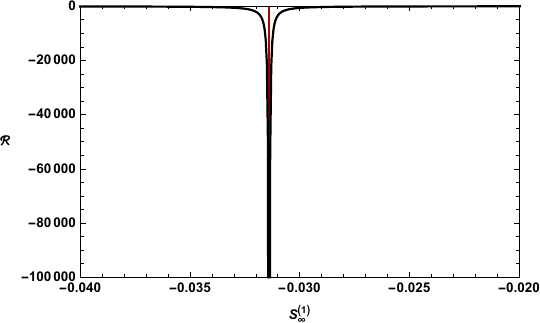}
\caption{\footnotesize{}}\label{fvev1}
\end{subfigure}
\begin{subfigure}{0.49\textwidth}
\includegraphics[width=1\textwidth]{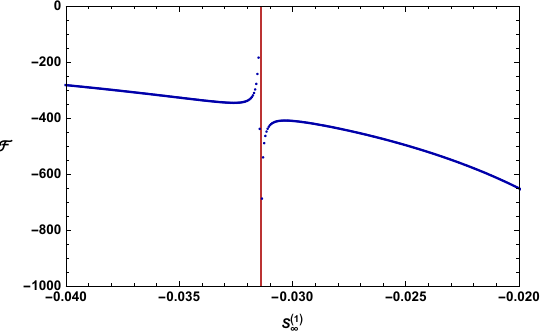}
\caption{\footnotesize{}}\label{fvev2}
\end{subfigure}
\begin{subfigure}{0.49\textwidth}
\includegraphics[width=1\textwidth]{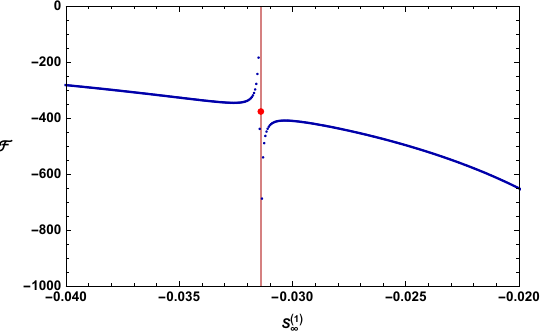}
\caption{\footnotesize{}}\label{fvev3}
\end{subfigure}
\begin{subfigure}{0.49\textwidth}
\includegraphics[width=1\textwidth]{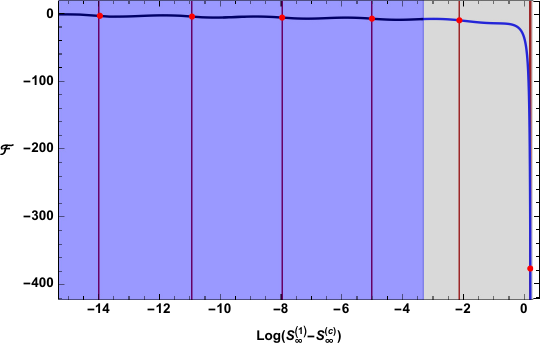}
\caption{\footnotesize{}}\label{Fvev}
\end{subfigure}
\caption{\footnotesize{(a) A zoom on the first vev driven solution: the parameter $\mR$ as a function of the IR parameter $S^{(1)}_{\infty}$. (b) The free energy $\mathcal{F}$ as function of the IR Parameter $S^{(1)}_{\infty}$ near the vev solution. It seems that the free energy diverges on both sides of the vev solution. However, this is due to the breakdown of the numerics near the vev solution. As shown in the text the curve continues at the position of the vev solution (c) The same plot as in (b), with the free energy of the vev solution, computed from (\ref{Afr19}) as red dot inserted in the plot.  (d)The free energy as a function of $\mathcal{R}$ for a much larger range of the IR parameter, that includes many more vev solutions. Here the spurious behavior near the vev solutions seen in (b) and (c) is not visible.  The solid line is the free energy computed from the general formula in (\ref{fr19}). The red vertical lines indicate the position of the vev-only solutions. The red dots indicate the free energy of the vev solutions calculated from (\ref{Afr19}). }}
\label{Fvevall}
\end{figure}

We have used this formula to compute independently the free energy of the vev solutions. The results are shown in figure \ref{Fvevall}.
In subfigure \ref{fvev1} we zoom near the first (from the right) vev-driven solution by plotting the dimensionless curvature $\mR$ as a function of the IR parameter $S^{(1)}_{\infty}$.

In subfigure \ref{fvev2} we plot in the same region the free energy $\mathcal{F}$ as a function of the IR Parameter $S^{(1)}_{\infty}$ near the vev solution. It would seem from this plot that the free energy diverges on both sides of the vev solution. However, we have shown analytically above, that this limit is finite.
The divergence is due to the breakdown of the numerics near the vev solution as the solution is losing accuracy.
This is confirmed by increasing the accuracy, by using more terms in the IR expansion to match the solution.
In subfigure \ref{fvev3} we have the same plot as in \ref{fvev2} and have added a red dot, which is the free energy of the vev solution, computed from (\ref{Afr19}). This confirms the apparent continuity of the curve.

In subfigure \ref{Fvev}  we zoom out of the first vev solution and we plot again the the free energy as a function of $\mathcal{R}$ for a much larger range of the IR parameter, that includes many more vev solutions. Here the spurious behavior near the vev solutions seen in (b) and (c) is not visible.  The solid line is the free energy computed from the general formula in (\ref{fr19}). The red vertical lines indicate the position of the vev-only solutions. The red dots indicate the free energy of the vev solutions calculated from (\ref{Afr19}).

Note that there is no one-boundary solution with $\f_-=0$ and zero vev. This would correspond to the theory on $AdS_d$ with unbroken ``conformal'' symmetry\footnote{This is short for ``the maximal number of unbroken conformal killing isomtries on $AdS_d$.''}.  This suggests that any holographic CFT$_d$ if put on $AdS_d$ and more generally on a constant negative curvature manifold with an asymptotically $AdS$ boundary, have necessarily vevs for its scalar operators. The reason is the coupling of scalar operators to the background curvature. As a simple example, we consider the (classically) scale-invariant $\phi^{4}$ theory in 4d, with the curvature-depended potential
\be
V(\phi)=\frac{\l}{12}\phi^4-\frac{1}{6\a^2}\phi^2\,,
\ee
where $\a$ is the radius of $AdS_4$.
In flat space, there is a single vacuum with $\langle\phi\rangle=0$, however, on $AdS$ there is symmetry breaking vacua with $\langle\phi\rangle=\pm\frac{1}{\a\sqrt{\l}}$, which have lower energy than the symmetry preserving vacuum.

This is expected to persist in more general cases. The effective potential for a scalar vev $\Phi$ associated to an operator of dimension $\Delta$ in $d$ dimensions, on flat space is
\be
V_{conf}=V_0\Phi^\frac{d}{\Delta}  \sp V_0>0\,,
\ee
from scaling arguments. In the presence of small constant negative curvature, there is a further coupling
\be
V_{conf,c}=V_0\Phi^\frac{d}{\Delta}-\frac{1}{6\a^2}\Phi^{\frac{d-2}{\Delta}}+{\cal O}(\a^{-4})\,,
\ee
which is also minimized by the non-trivial vev
\be
|\Phi|=\left(\frac{d-2}{d}\frac{1}{6\a^2 V_0}\right)^\frac{\Delta}{2}\,.
\ee
In general, the effective potential is complicated, and in our holographic case, we find an infinite number of vev vacua.

\subsection{Comparing the free energy of the two-boundary solutions\label{83}}

We now consider solutions that are stretching between two UV boundaries, $B_+$ and $B_-$.
Each QFT on a UV boundary is defined by two parameters: The negative constant curvature of the space-time on which this QFT is living, $R^{UV}$, and $\f_-$ which is the source of the scalar operator $\mathcal{O}$. Each theory on its own has therefore one dimensionless parameter, $\mathcal{R}$ defined by
\be \label{AUVcurve}
\mathcal{R} =  R^{UV} |\f_-|^{-2/\Delta_-}\,.
\ee
For two-boundary QFTs (which we call QFT$_+$ and QFT$_-$), we have four independent dimensionful parameters,
\be\label{indeps}
R^{UV}_+\,, \; R^{UV}_-\,, \; \f_-^{+}\,,  \; \f_-^{-}\,.
\ee
From there,  we can make three dimensionless parameters.
Two dimensionless curvatures\footnote{In our example, the dimension $\Delta_+$ is the same at both boundaries, but in general cases, it will be different.}, $\mathcal{R}_+$ and  $\mathcal{R}_-$
\be \label{dimRif}
\mathcal{R}_+=R^{UV}_+ (\f_-^{+})^{-2/\D_-}\sp \mathcal{R}_-=R^{UV}_- (\f_-^{-})^{-2/\D_-}\,,
\ee
and the ratio of the two relevant coupling constants, which can be found from the asymptotic sources of the $\f$ field
\be\label{Eratio}
\mathcal{U}\equiv \Big(\frac{\f_-^{+}}{\f_-^{-}}\Big)^{1/\Delta_-}\,.
\ee
\begin{figure}[!hb]
\centering
\begin{subfigure}{0.49\textwidth}
\includegraphics[width=1\textwidth]{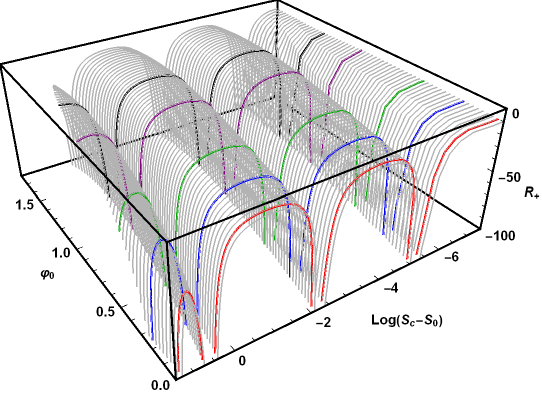}
\caption{\footnotesize{}}\label{RUT}
\end{subfigure}
\begin{subfigure}{0.49\textwidth}
\includegraphics[width=1\textwidth]{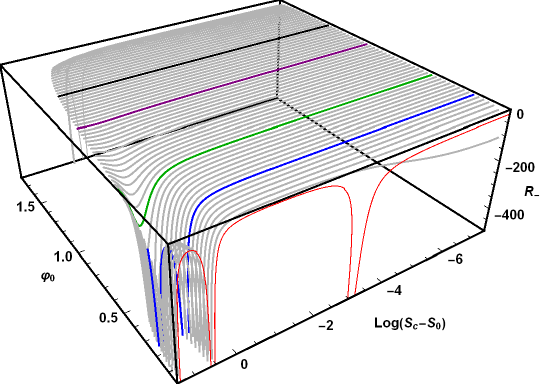}
\caption{\footnotesize{}}\label{RDT}
\end{subfigure}
\caption{\footnotesize{(a), (b): $\mathcal{R}_+$ and $\mathcal{R}_-$ inside the green region in map \ref{map}. The red, blue, green, purple, and brown curves belong to the specific values of $\f_0=0.03, 0.24, 0.51, 0.99, 1.32$. Notice that neither $\mathcal{R}_+$ nor $\mathcal{R}_-$ reach zero but have a negative maximum.} }
\end{figure}
\begin{figure}[!ht]
\centering
\includegraphics[width=0.52\textwidth]{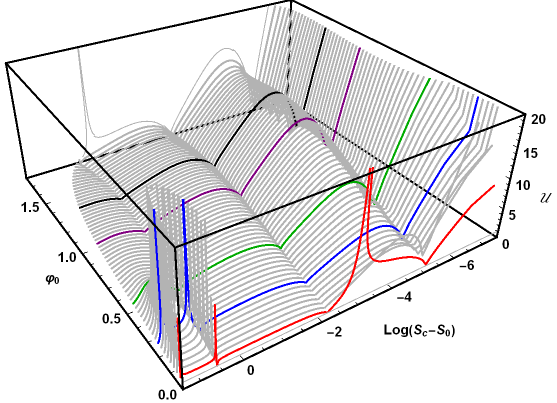}
\caption{\footnotesize{ The dimensionless ratio of couplings $\mathcal{U}=\Big|\frac{\f_-^{u}}{\f_-^{d}}\Big|^{1/\Delta_-}$ inside the green region in map \ref{map}.}}\label{RFFT}
\end{figure}

\begin{figure}[!t]
\centering
\includegraphics[width=0.52\textwidth]{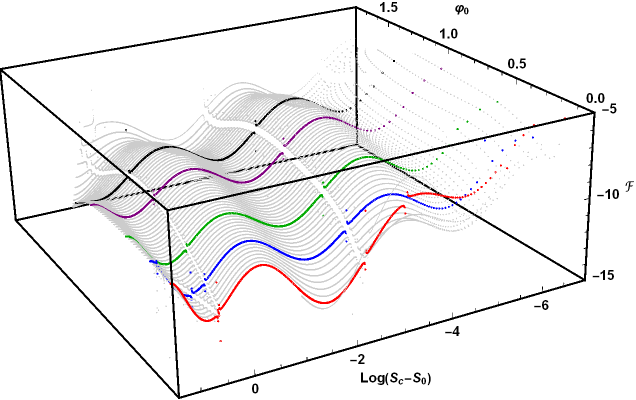}
\caption{\footnotesize{ Free energy for UV-UV solutions in the green region of the map \ref{map}. As $S_0\rightarrow S_c$ i.e. as the number of A-bounces is increasing the free energy becomes larger.} }\label{freeUVUV}
\end{figure}

We now extract all the above UV information for solutions inside the green region of the map \ref{map}. The results are given in figures \ref{RUT}--\ref{freeUVUV}. To draw the figures, we first choose a point ($\f_0, S_0$) in the up-right quarter of the map \ref{map}\footnote{We use the two symmetries of the equations in (\ref{sym1}) and (\ref{sym2}) to reduce the green region to a single quarter.}. For a fixed value of $\f_0$, there is a critical value of $S_0$ which we call $S_c(\f_0)$.
For $S_0>S_c$, the solution runs off to $\f=\pm\infty$ and has therefore a single boundary.
Therefore, for a given $\f_0$, two-boundary solutions exist only if $S_0\in[-S_c(\f_0),S_c(\f_0)]$.
Moreover, as $-S_0$ gives the solution running in the opposite direction, ($u\to -u$),  it is enough to restrict $0\leq S_0\leq S_c(\f_0)$.

We use $\log(S_c-S_0)$ and $\f_0$ to parametrize the solutions. Figure \ref{RUT} and \ref{RDT} show the values of $\mathcal{R}_+$ and $\mathcal{R}_-$. Figure \ref{RFFT} shows the dimensionless ratio of the couplings $\mathcal{U}$ in \eqref{Eratio} and figure \ref{freeUVUV} is the renormalized free energy in equation \eqref{frw6}.
The behavior of each figure is explained by the following
facts:

\begin{itemize}
\item A vev-driven solution occurs when $\mathcal{R}_+ \rightarrow -\infty$ or $\mathcal{R}_- \rightarrow -\infty$ because of the relation \eqref{dimRif}. Everywhere in figures \ref{RUT} or \ref{RDT},  the dimensionless curvatures asymptote to $-\infty$,  when we have a vev-driven solution.

\item  When $\log(S_c-S_0)$ takes negative values,  we have more and more vev-driven solutions.
This can be observed for example in figure \ref{RUT}.

\item For a holographic theory with two QFT boundaries, we can compute the dimensionless ratio of the two couplings (\ref{Eratio}), see figure \ref{RFFT}. In this figure, when $|\f_-^+|\rightarrow 0$ or $|\f_-^-|\rightarrow +\infty$ (equivalently in figure \ref{RUT} $\mathcal{R}_+\rightarrow -\infty$ or in figure \ref{RDT} $\mathcal{R}_-\rightarrow 0$) then $\mathcal{U}\rightarrow 0$.

\item Using the results of the previous section, i.e. equation \eqref{frw6}, we can plot the free energy of all solutions with two boundaries. As discussed earlier, for a vev-driven solution, ($\mathcal{R}_+\rightarrow -\infty$ or $\mathcal{R}_-\rightarrow -\infty$), the free energy tends to a constant value. The small jumps in figure \ref{freeUVUV} show this fact.

    We should note that in figures \ref{RUT} and \ref{RDT} $\mathcal{R}_+$ or $\mathcal{R}_-$ never reach zero, but have a maximum negative value, therefore the free energy is finite everywhere in our plot.

\item Solutions with ($\mathcal{R}_+\rightarrow -\infty$ or $\mathcal{R}_-\rightarrow -\infty$) are holographic RG interfaces. On one side of the interface, there is the (UV) CFT without a relevant coupling. On that side, the scalar flow is generated by the vev only. The vev is triggered by the presence of the interface.

\end{itemize}

\begin{figure}[!ht]
\centering
\begin{subfigure}{0.49\textwidth}
\includegraphics[width=1\textwidth]{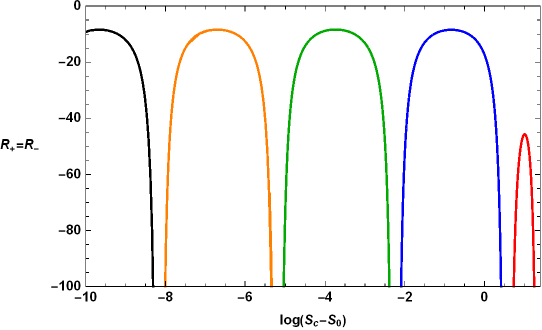}
\caption{}\label{UVD1}
\end{subfigure}
\centering
\begin{subfigure}{0.46\textwidth}
\includegraphics[width=1\textwidth]{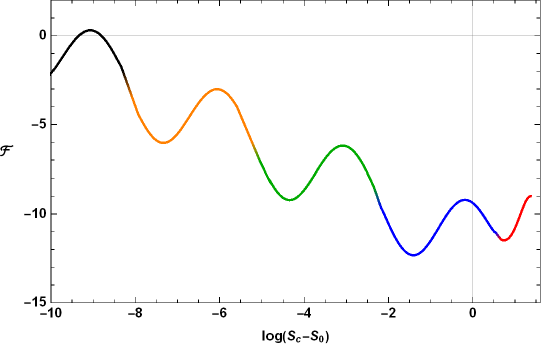}
\caption{}\label{UVD2}
\end{subfigure}
\centering
\begin{subfigure}{0.49\textwidth}
\includegraphics[width=1\textwidth]{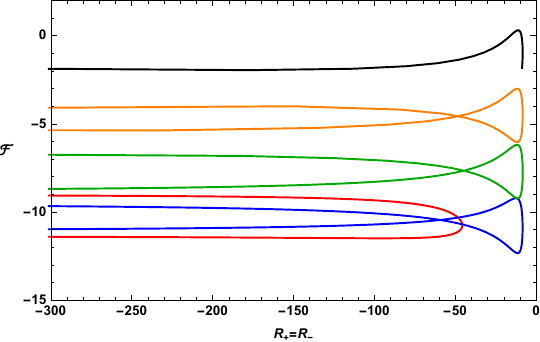}
\caption{}\label{UVD3}
\end{subfigure}
\caption{\footnotesize{These plots  show features of  symmetric solutions, corresponding to $\f_0=0$ in map \ref{map}. These all have $\mathcal{R}_+=\mathcal{R}_-$. (a): Shows the values of curvature as we change the IR parameter $S_0$, here $S_c=3.94$. (b) Renormalized free energy. (c): Free energy in terms of $\mathcal{R}_+$. Different colors show different branches in figure (a).}}
\end{figure}
\begin{figure}[!ht]
\centering
\includegraphics[width=0.5\textwidth]{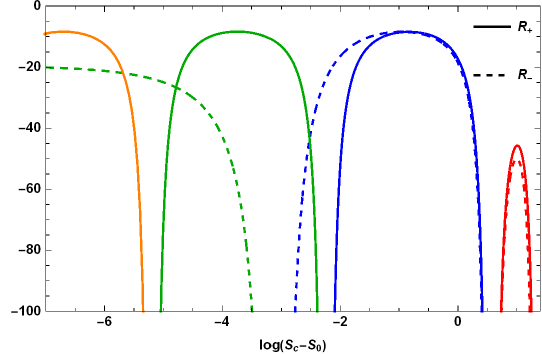}
\caption{\footnotesize{Curvature as a function of the IR parameter for (slightly) asymmetric solutions. For $\f_0=0.03$ (here $S_c=3.91$), the solid curves $\mathcal{R}_+$ and the dashed curves $\mathcal{R}_-$ have overlaps in some discrete points.}}\label{UVD4}
\end{figure}

To find which solution is dominant from the free energy point of view, we should compare solutions with the same curvatures on each boundary\footnote{There is also a third dimensionless, source, $\mathcal{U}$, defined in (\ref{Eratio}). This has to be matched also. However, as shown in \cite{Ghodsi:2022umc} there is a more general class of solutions where $\f_-, R^{UV}$ are rescaled at one of the boundaries, without changing the free energy. This leaves $\mathcal{R}_+$, $\mathcal{R}_-$ invariant, but rescales $\mathcal{U}$ at will.}.
 We will analyze the special case of solutions that have $\mathcal{R}_+=\mathcal{R}_-$.  We shall comment at the end on the general case where $\mathcal{R}_+\not=\mathcal{R}_-$. There are two classes of solutions with $\mathcal{R}_+=\mathcal{R}_-$. The first are those that sit on the vertical ($\varphi_0=0$) axis on the map \ref{map} and are symmetric under the interchange of the two boundaries. Indeed, because of the $\varphi\to -\varphi$  symmetry of the scalar potential, the corresponding geometries are symmetric with respect to the exchange of the two boundaries.
Figure \ref{UVD1} shows the value of the curvatures for these solutions. Now, the vertical asymptotes correspond to vev-vev solutions.

The second class involves other families of geometries with $\mathcal{R}_+=\mathcal{R}_-$: some of them are left-right symmetric, namely the one which sits on the horizontal axis of figure \ref{map}; some of them are not symmetric, as we shall see in an example below.

For the solutions displayed  \ref{UVD1},  as one moves toward the boundary of the green region in map \ref{map} (solutions with many numbers of A-bounces) the free energy increases slightly, see figure \ref{UVD2}. This is shown also in figure \ref{UVD3} when we draw the free energy in terms of the dimensionless curvatures.

Moving away from $\f_0=0$ in the map \ref{map}, for example for $\f_0=0.03$, as we change $S_0$ and read the values of $\mathcal{R}_+$ and $\mathcal{R}_-$, we find discrete points where accidentally $\mathcal{R}_+=\mathcal{R}_-$, see figure \ref{UVD4}.

To find a solution with minimum free energy, we should find all solutions with $\mathcal{R}_+=\mathcal{R}_-$ in the green region of the map \ref{map}.
\begin{figure}[!ht]
\centering
\begin{subfigure}{0.49\textwidth}
\includegraphics[width=1\textwidth]{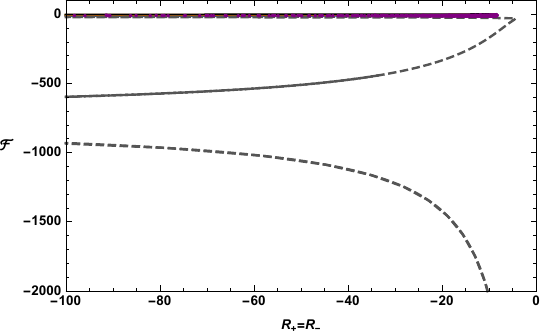}
\caption{}\label{EQR3}
\end{subfigure}
\begin{subfigure}{0.475\textwidth}
\includegraphics[width=1\textwidth]{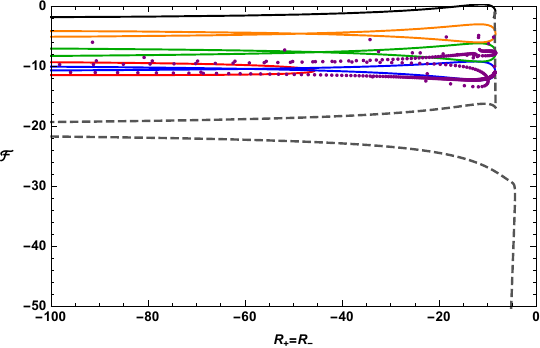}
\caption{}\label{EQR1}
\end{subfigure}
\caption{\footnotesize{(a) The plot of free energy as a function  of $\mathcal{R}_+=\mathcal{R}_-$  for two boundary solutions. The barely visible colored lines correspond to the connected two-boundary solutions.
The grey dashed lines correspond to the free energies of the disconnected two-boundary solutions.
(b) A blow-up of the top region of figure (a). The solid curves correspond to the symmetric ($S_0=0$, $\f_0>0$) connected two-boundary solutions. The purple dots correspond to the asymmetric connected two-boundary solutions. Both of these types of solutions belong to the green region of the map \ref{map} with $\mathcal{R}_+=\mathcal{R}_-$.
The dashed grey curves that appear here near the colored lines correspond to disconnected two-boundary solutions but are not visible in figure (a) as they are very close to the colored lines.
 Note that near $\mathcal{R}_+=\mathcal{R}_-\rightarrow 0$ only disconnected solutions exist and the dashed curve finally approaches $\mathcal{F}\rightarrow -\infty$.}}\label{EQR2}
\end{figure}
We have found these solutions in figure \ref{EQR2}. The solid curves are the parameters of the symmetric solutions. The purple dots are the other solutions with $\f_0>0$.

Some observations on these results are in order.

\begin{itemize}

\item  Two boundary solutions do not exist for sufficiently small $\mathcal{R}_+=\mathcal{R}_-$. This is clearly visible in figures \ref{UVD1} for symmetric solutions and \ref{UVD4} for all solutions.
    This is to be expected: if such solutions existed for arbitrarily small  $\mathcal{R}_{+}=\mathcal{R}_-$, then by continuity they would exist also for $\mathcal{R}_{+}=\mathcal{R}_-=0$. But we know independently, that in that case there are no two-boundary solutions.

\item The free energy of these solutions, is shown in figure \ref{EQR1}. The situation is rather complex. From about $\mathcal{R}_{+}=\mathcal{R}_-<-50$ the red solid branch of symmetric solutions dominates. But for smaller curvatures, it is the violet non-symmetric solutions that dominate. However, they do not form a continuous line and therefore the phase diagram seems chaotic. Below, we shall see that this is not relevant, after all, for the dominant saddle points.

\end{itemize}

\subsubsection{Disconnected two-boundary solutions}

In the case of two-boundary solutions, there is a further possibility, namely disconnected solutions with two boundaries.
In our case, such solutions exist because we have one-boundary solutions.
The question here is whether such saddle points are comparable to the two-boundary solutions.

In particular, the issue is whether the solutions have matching sources on all boundaries.
We do impose the matching on sources on the $AdS_{d+1}$ $B_{\pm}$ boundaries, but the issue is what happens at $B_3$. As we argued in section \ref{bc},
this issue is determined by the slice ansatz. When the slice is an $ AdS_d$  space, then the slice ansatz is equivalent to imposing an $O(1,d)$ symmetry in the theory. It is the symmetry that replaces the boundary conditions on the $B_3$ boundary and in this case, it is interpreted as the conformal symmetry of the interface.
This definitely matches them with the disconnected solutions and therefore a comparison of their free energies is imposed.\footnote{In cases where the negative constant curvature manifold is compact, the same conclusion holds. In such cases, there is no $B_3$ boundary and one must consider also the disconnected solutions.
In the case, where the negative constant-curvature manifold has infinite volume and a boundary, but does not have the maximal symmetry of $AdS_d$, then the issue of symmetry is not clear. If the manifold is also conformally flat, one may make a similar argument as before, but in the general case, we do not know how to address this issue.}

Therefore, we should also consider solutions with two disconnected boundaries. These are solutions constructed out of a pair of two UV-Reg solutions with the same UV data.
For a given value of $\mathcal{R}_{\pm}$, there is a non-zero number of one-boundary saddle points with the same source (but different vevs). If, however, we choose the one with the lowest free energy, both for $+$ and $-$, this is the one with the lowest combined free energy in the disconnected sector. For fixed $\mathcal{R}_+,\mathcal{R}_-$ this is the one we shall compare with the dominant one of the two-boundary connected solutions.

In particular, when  $\mathcal{R}_+=\mathcal{R}_-$, the disconnected solution has a free energy that is twice the dominant one-boundary solution with the same curvature, in  \eqref{fr19}.

We have considered the free energy of these solutions in figure \ref{EQR2} as a dashed curve (for simplicity we have considered just the dominant part of this free energy, i.e. the green curve in figure \ref{FRZ}).
\begin{figure}[!ht]
\centering
\begin{subfigure}{0.47\textwidth}
\includegraphics[width=1\textwidth]{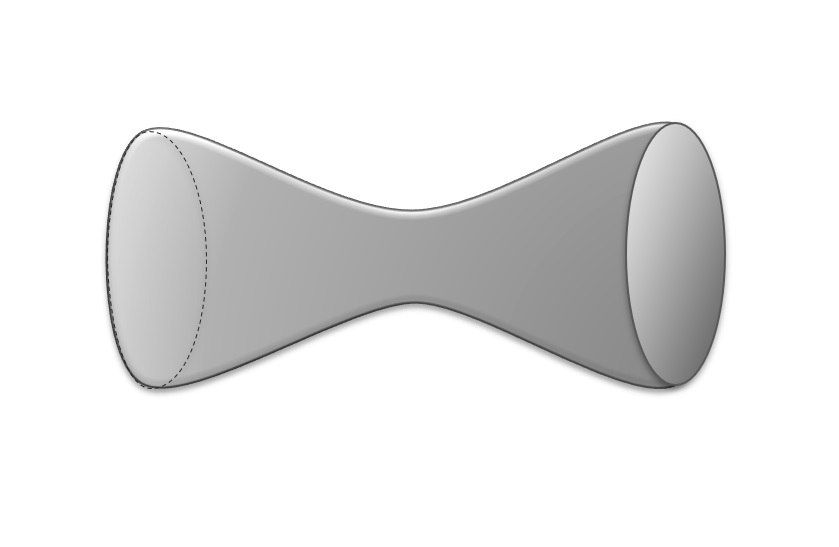}
\caption{}\label{worm1}
\end{subfigure}
\begin{subfigure}{0.47\textwidth}
\includegraphics[width=1\textwidth]{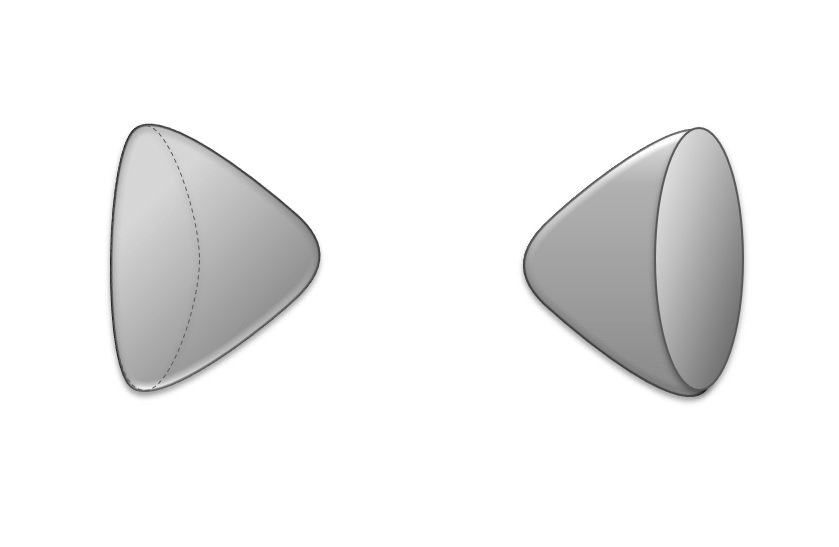}
\caption{}\label{worm2a}
\end{subfigure}
\caption{\footnotesize{A schematic representation of the two-boundary and disconnected two-boundary solutions.}}
\end{figure}

By comparing the free energy of all these solutions, we observe that the dominant solution (one with the minimum free energy) always is the solution with the dominant two disconnected boundaries.

When  $\mathcal{R}_+=\mathcal{R}_-\to 0$ (see the right-hand side of figure \ref{FRT}) only disconnected solutions exist.

This fact has interesting implications for the two-boundary solutions as descriptions of conformal interfaces between confining holographic theories.
Such theories for given values of their boundary data are described by several competing saddle points. Many of them correspond to connected two-boundary solutions with free energies ${\cal F}^{conn}_i(\mathcal{R}_+,\mathcal{R}_-)$. Many others are given by the factorized saddle points obtained from two of the many single-boundary solutions, with free energy

\be
{\cal F}_{disc}^{ij}(\mathcal{R}_+,\mathcal{R}_-)={\cal F}_{one-bound}^i(\mathcal{R}_+)+{\cal F}_{one-bound}^j(\mathcal{R}_-)\,.
\ee
We find that in all cases
\be
{\rm min}\{{\cal F}_{disc}^{ij}(\mathcal{R}_+,\mathcal{R}_-)\}~~<~~{\rm min}\{{\cal F}^{conn}_i(\mathcal{R}_+,\mathcal{R}_-)\}\,.
\ee

 A cross-correlator in an interface theory, corresponds to one operator inserted on one side of the interface and the other on the other side. In the theories above, such a correlator has no contribution from the disconnected solution, but has non-zero contributions from the two-boundary connected solutions. If $Z$ is the partition function of the theory and $Z_{OO}$ is the unnormalized, connected cross-correlator then schematically
\be
Z=\sum_{ij}e^{-{\cal F}^{ij}_{disc}}+\sum_{i}e^{-{\cal F}_{conn}^i}\sp Z_{OO}=\sum_{i}\langle OO\rangle_i~e^{-{\cal F}_{conn}^i}\,,
\ee
and the normalized cross-correlator is
\be
\langle OO\rangle_c =\frac{Z_{OO}}{Z}\simeq \sum_{i}\langle OO\rangle_i~e^{-({\cal F}_{conn}^i-{\cal F}_{disc,min})}+\cdots\,,
 \ee
 where the dots indicate exponentially suppressed contributions compared to the leading terms.

This indicates that the connected correlators vanish exponentially at large $N$ as $e^{-cN^2}$ where $c$ and order one constant.
 We do not understand the dual quantum field theory origin of such suppression, although it has the tell-tale dependence associated with NS$_5$ branes in string theory.

 Although we have made our analysis with two-boundary solutions with $\mathcal{R}_+=\mathcal{R}_-$, similar remarks hold when $\mathcal{R}_+\not=\mathcal{R}_-$.

\subsubsection{Two-boundary solutions that are vev on one or both boundaries}

There are many solutions where $\mR_{+}$, or $\mR_-$ diverge towards $-\infty$. Such solutions have a similar interpretation as those discussed in section \ref{cftvev}.
On the side of the interface where the source vanishes, one has the CFT. On the other side, one has the QFT with a non-trivial coupling constant. Such interfaces are known as RG interfaces, \cite{RG1}-\cite{RG4},\cite{C1}. The structure of their free energies, as well as the dominant solutions, follow verbatim the discussion in the previous section, with a few differences. There is no one-sided vev solution where also the vev vanishes. Therefore, for connected RG interface solutions, there is always symmetry breaking of the conformal symmetry on the CFT side.
Also, when constructing the disconnected two-boundary solutions, there is no single boundary zero vev CFT solution.

 Another interesting subset is double vev connected solutions that can be obtained when both $\mR_{\pm}\to -\infty$. In such solutions, both sides of the interface are vev solutions. Therefore, we have an interface separating two CFT sides\footnote{In our example here this is the same CFT. In more general situations, with a potential having many maxima, like the one in \cite{Ghodsi:2022umc} solutions exist where the FTs on the two sides can be distinct.}.  They can be obtained from the subset of solutions with $\mR_{+}=\mR_-\to -\infty$ and may be visualized in figure \ref{UVD3} as the points at which the various colored lines intersect with the vertical axis on the left.

 In this case, there exists also a connected two-boundary solution with a constant scalar, discussed in appendix \ref{fcft}, which has zero vevs on both sides. As shown in the same appendix, the zero vev solution is subdominant compared to the leading non-zero vev solution as can be seen in figure \ref{CFT1}.
 Here also, the disconnected solution is again dominant compared with the dominant connected solution.

\subsection{Comparing the free energy of the no-boundary solutions}
We can compare the free energy density of (no-boundary) Reg-Reg solutions using the relation we found in \eqref{frw9}. As already discussed in section \ref{noboso},  the Reg-Reg solutions are classified by the value of the $S_\infty^{(1)}$ parameter at $\f\rightarrow +\infty$.

\begin{figure}[!ht]
\centering
\includegraphics[width=0.5\textwidth]{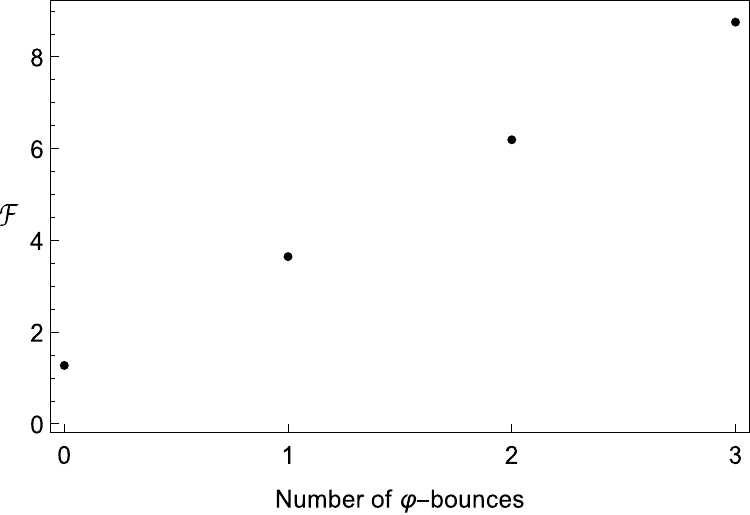}
\caption{\footnotesize{The free energy density of the Reg-Reg solutions in figures \ref{regreg}--\ref{regreg3}.}}\label{regreg4}
\end{figure}

For special discrete values of this parameter,  we find Reg-Reg solutions with different numbers of $\f$-bounces, as shown in figures \ref{regreg}--\ref{regreg3}. The free energy for these four solutions in figures \ref{regreg}--\ref{regreg3} are plotted in figure \ref{regreg4}.
We observe that the free energy is an increasing function of the number of $\f$-bounces.
Therefore the dominant solution is the one with no $\f$ bounces, that interpolates between $\f=+\infty$ and $\f=-\infty$.
 This solution is shown in figure  \ref{regreg}.

\section{Free energy of wormhole solutions\label{worm2}}

When the geometry of the slices is compact, there is no side boundary, $B_3$ and we have wormhole-like solutions. In this case, the values of the free energy are different as the renormalization procedure changes. Removing the contribution of the GHY term of the  side boundary from the on-shell action in appendix \ref{osa}, we find  that the  action (\ref{fr1}),  for one-boundary solutions and with appropriate regularization is:
\be \label{WH1}
S_{on-shell}=\frac{2M_P^{d-1}}{d} V_{\hepsilon}(\a) \Big(- R^{(\zeta)}Ue^{(d-2)A}+d(d-1)e^{dA}\dot{A}\Big)\Big|_{u_0}^{u_+}\,.
\ee
Compare this expression with (\ref{fr4}), which was obtained including the side boundary $B_3$: the difference is a factor $d$ in the coefficient of the first term, while the second term is unchanged. This can be traced back to the contribution of the GHY term on the $B_3$ boundary in the non-compact case.

To obtain a finite renormalized free energy, we shall  need to consider the following counter-terms in $d=4$
\be\label{WH2}
S_{ct}= -M_P^3\int d^4 x \sqrt{\gamma}\Big(
\frac{6}{\ell}+\frac{\D}{2\ell}\f^2+\frac{\ell}{2} R^{(\gamma)}+\frac{\ell^3}{48}(R^{(\gamma)})^2 \log\omega\epsilon
\Big)^{u_+=-\ell\log\epsilon}\,.
\ee
From the above results we obtain the following relations for renormalized free-energy densities:
\begin{itemize}
\item One-boundary solution
\be \label{WH3}
\mathcal{F}=\frac{F_{ren}}{M_P^3\ell^3 \bar V_{\hepsilon}(1)}=
\frac{3}{2}
 \left(1-96 \frac{\mcr}{\mathcal{R}^2}+ 48 \frac{\mbr}{ \mathcal{R}}\right)\,.
\ee
Figures \ref{cofe1} and \ref{cofe2} show the behavior of the free energy density as a function of the free parameter of the UV-Reg solutions. The curves are very similar to the non-compact case in figures \ref{freeUVR}--\ref{ZfreeG}.
\begin{figure}[!ht]
\centering
\begin{subfigure}{0.49\textwidth}
\includegraphics[width=1\textwidth]{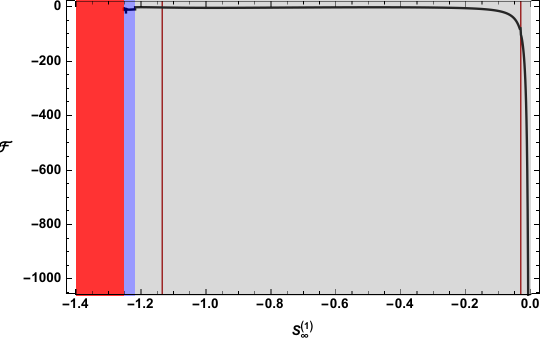}
\caption{\footnotesize{}}\label{cofe1}
\end{subfigure}
\begin{subfigure}{0.49\textwidth}
\includegraphics[width=1\textwidth]{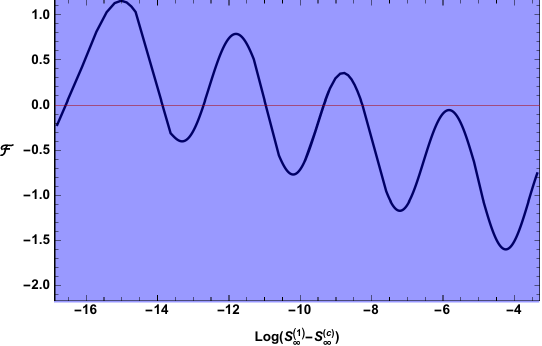}
\caption{\footnotesize{}}\label{cofe2}
\end{subfigure}
\caption{\footnotesize{(a): Free energy density for UV-Reg solutions with compact slice geometry. (b): The zoom of the blue region.}}
\end{figure}
\item Two-boundary solution
\begin{gather} \label{WH4}
\mathcal{F}=\frac{1}{M_P^3\ell^3 \bar V_{\hepsilon}(1)}\big({F}^{+}_{ren}-{F}^{-}_{ren}\big)=
3-144\Big(\frac{C^+(\mR_+,\mR_-)}{(\mathcal{R}_+)^2}+\frac{C^-(\mR_+,\mR_-)}{(\mathcal{R}_-)^2}\Big)\nn \\
+72\Big(\frac{\mathcal{B}_+(\mR_+,\mR_-)}{ \mathcal{R}_+}-\frac{\mathcal{B}_-(\mR_+,\mR_-)}{ \mathcal{R}_-}\Big)\,.
\end{gather}
Here, the only difference with the free energy density of non-compact cases is the coefficient of the last term. Figure \ref{freeUVUVnsb} shows a similar behavior as figure \ref{freeUVUV}.
\begin{figure}[!ht]
\centering
\includegraphics[width=0.52\textwidth]{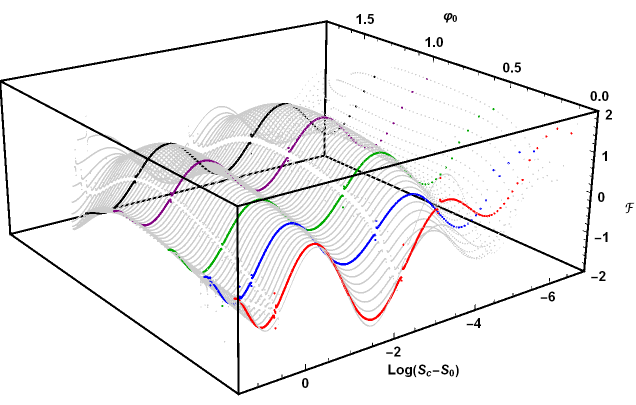}
\caption{\footnotesize{Free energy for wormhole solution with two UV boundaries.} }\label{freeUVUVnsb}
\end{figure}

\item No-boundary solution

Here, we should do the same steps as section \ref{FEnobo}, however, we must drop the side boundary term. The result is
\be
\mathcal{F}=\frac{F}{M_P^{d-1} \bar{V}_{\hepsilon}(1)}=-\frac{1}{d}\a^d R^{(\zeta)}\int_{u_1}^{u_2} du e^{(d-2)A}\,,\label{WH5}
\ee
so the only difference with the compact slices is in the overall coefficient of \eqref{WH5}.
\end{itemize}
We conclude this section by noting that the qualitative hierarchy of solutions with compact negative curvature slices is the same as with infinite volume slices.

\section*{Acknowledgements}
\addcontentsline{toc}{section}{Acknowledgements}

We would like to thank Roberto Emparan, Carlos Hoyos, Takaki Ishii,  Romuald Janik, Matti Jarvinen, Javier Mas, Vasilis Niarchos, Angel Paredes, Achilleas Porfyriadis, Edwan Preau, Alfonso Ramallo, Matthew Roberts, Jorge Russo, Christopher Rosen, Javier Subils.

This work is partially supported by the European MSCA grant HORIZON-MSCA-2022-PF-01-01 ``BlackHoleChaos" No.101105116 and by the H.F.R.I call ``Basic research Financing (Horizontal support of all Sciences)" under the National Recovery and Resilience Plan ``Greece 2.0" funded by the European Union -- NextGenerationEU (H.F.R.I. Project Number: 15384.) and by the In2p3 grant ``Extreme Dynamics".

\appendix

\begin{appendix}
\renewcommand{\theequation}{\thesection.\arabic{equation}}
\addcontentsline{toc}{section}{Appendices}
\section*{APPENDIX}

\section{Near-boundary asymptotics and the bulk-boundary dictionary\label{UVF}}

In this appendix, we present the near-boundary asymptotics of solutions we use in this paper.

In the present paper, we find regular solutions in the confining theory that contain either one or two UV fixed points ($AdS$-like boundary)\footnote{There are also Reg-Reg regular solutions but these solutions do not have any holographic dual QFT.}. We can obtain the physical parameters of the dual QFTs, which are living on these boundaries, by finding the expansions of the scale factor and scalar field near the boundaries.

We are interested in theories with a potential $V(\f)$ which has a maximum at $\varphi=0$. Upon the expansion near this point, the potential behaves as
\be\label{Vm2}
V =-\frac{d(d-1)}{\ell^2}-\frac12 m^2 \f^2+\mathcal{O}(\f^4)\sp m^2=\frac{(d-\Delta)\Delta}{\ell^2}\,,
\ee
where $\ell$ and $\Delta$ are two constant parameters.
Near the maximum of the potential, the solution of equations of motion \eqref{eq:EOM7} and \eqref{eq:EOM8} for $W$ and $S$ has two branches \cite{C}. For the minus branch as $\f\rightarrow 0^+$ the solutions have the following expansions
\begin{gather}\label{WLU}
W_-=\frac{2(d-1)}{\ell}+\frac{\D_-}{2\ell}\f^2+\frac{\mathcal{R}}{d\ell}|\f|^{\frac{2}{\D_-}}+\frac{\mcr}{\ell}|\f|^{\frac{d}{\D_-}}+\cdots\,,\\
S_-=\frac{\D_-}{\ell}|\f|+\frac{d\mcr}{\D_-\ell}|\f|^{\frac{d}{\D_-}-1}+\cdots\,,\label{SLU}
\end{gather}
where dots stand for higher power expansion terms and $\mathcal{R}$ and $\mcr$ are two constants of integration and we have defined
\be\label{ACfact}
\D_\pm=\frac{d}{2}\pm\sqrt{\frac{d^2}{4} - m^2\ell^2}\,.
\ee
Since $0<m^2<d^2/4\ell^2$ therefore $0< \D_- < d/2$ and $d/2< \D_+ < d$.
Moreover, there is a plus branch which is described by the following expansions
\begin{gather}\label{Wplus}
W_+=\frac{2(d-1)}{\ell}+\frac{\D_+}{2\ell}\f^2+\frac{\mathcal{R}}{d\ell}|\f|^{\frac{2}{\D_+}}+\cdots\,,\\
S_+=\frac{\D_+}{\ell}|\f|+\cdots\,.\label{Splus}
\end{gather}
The plus branch as shown in \cite{Pap}, is the upper envelope of the family of minus branch solutions parameterized by $\mcr$.

Having the \eqref{WLU} and \eqref{SLU} expansions, we can solve $\f(u)$ and $A(u)$ from \eqref{eq:defSc}--\eqref{eq:defWc} to obtain the scalar field and scale factor of the minus branch
 as $u\rightarrow +\infty$
\begin{gather}
 \f(u) =\f_- \ell^{\Delta_-}e^{-\Delta_-u / \ell} + \frac{ d\mcr \, |\f_-|^{\Delta_+ / \Delta_-}}{\Delta_-(d-2 \Delta_-)} \, \ell^{\Delta_+} e^{-\Delta_+ u /\ell} + \ldots \, ,\label{phiLU} \\
A(u) = {A}_- +\frac{u}{\ell} - \frac{\f_-^2 \, \ell^{2 \Delta_-}}{8(d-1)} e^{-2\Delta_- u / \ell}  -\frac{\mathcal{R}|\f_-|^{2/\Delta_-} \, \ell^2}{4d(d-1)} e^{-2u/\ell} \nonumber\\
  \hphantom{=} \ - \frac{\Delta_+ \mcr |\f_-|^{d/\Delta_-} \, \ell^d}{d(d-1)(d-2 \Delta_-)}e^{-du/\ell} +\ldots \,,\label{ALU}
\end{gather}
where $\f_-$ and $A_-$ are other constants of integration. Here $\f_-$ is the source and $\mcr$ is related to the vacuum expectation value of operator $O$ by
\be \label{vevo}
\langle O\rangle =\frac{ d\mcr}{\Delta_-}|\f_-|^{\Delta_+ / \Delta_-}\,.
\ee
Also, $\mathcal{R}$ is the ``dimensionless'' curvature parameter,  is related to the physical  curvature $R^{UV}$ seen by the dual field theory by:
\be \label{ifr13-app}
R^{UV}=\mathcal{R} |\f_- |^{\frac{2}{\D_-}}\,.
\ee

\section{Expansion around A-bounces\label{bounce}}

In addition to the UV fixed points discussed in the previous section,  solutions may have A-bounces and we can sort them according to the number of A-bounces.

We show that in confining models, there are two groups of solutions, one with at least one A-bounce and the other without any A-bounce.
An A-bounce is defined as a point where $W(\f)=0$ so the scale factor reaches a minimum (or a maximum) but since $S(\f)\neq0$ the flow does not stop at this point. As $\f\rightarrow \f_0^+$ the scale factor decreases and then after the A-bounce it increases.
The expansions of $W, S$ and $T$ are given by
\be
\label{AturnW}
W =\Big(\frac{(d-1) S_0}{d} + \frac{2V_0 }{dS_0}\Big)(\f-\f_0)+\Big(\frac{d+1}{2d S_0} - \frac{V_0 }{dS_0^3}\Big)V_1 (\f-\f_0)^2+\mathcal{O}((\f-\f_0)^3)\,,
\ee
\be\
S = S_0+\frac{V_1}{S_0}(\f-\f_0)+\Big( \frac{S_0}{4} + \frac{V_0}{ 2 ( d-1) S_0} - \frac{V_1^2}{ 2 S_0^3} + \frac{V_2}{ S_0}\Big)(\f-\f_0)^2
+  \mathcal{O}((\f-\f_0)^3)\,,  \label{AturnS}
\ee
\be
T = \big(V_0-\frac{S_0^2}{2}\big)\Big[1+\frac{(d-1) S_0^2 + 2 V_0}{2 (d-1) d S_0^2}(\f-\f_0)^2 +\mathcal{O}((\f-\f_0)^3)\Big] \,,
\label{AturnT}
\ee
where we have considered the expansion of the potential as
\be \label{vbas}
V(\f)=V_0+V_1 (\f-\f_0)+V_2(\f-\f_0)^2+\mathcal{O}((\f-\f_0)^3)\,.
\ee

\section{The  asymptotics as $\f\to\infty$}\label{avper}

In this appendix, we shall discuss the asymptotic solutions to equations (\ref{eq:EOM7}) and (\ref{eq:EOM8}), or equivalently (\ref{eq:EOM9}), when the scalar runs off to infinity.

We assume that as $\f\rightarrow +\infty$, the potential behaves exponentially as
\be \label{alfpot}
V\simeq -V_\infty e^{2a\f}+\cdots\,,
\ee
where $V_\infty$ is a positive constant, $a$ is any real number and the ellipsis implies subleading terms.
We shall solve asymptotically equation \eqref{eq:EOM9} for $S$, that we reproduce here,
\be
 dSS''-\frac{d}{2}S^2  - (S')^2=\frac{d}{d-1}V -{(d+2)}S'\frac{V'}{S} +dV''+\frac{{V'}^2}{S^2}\,.
\label{as1}
\ee
There are two independent classes of solutions for the above equation as $\f\to+\infty$.  We analyze these solutions in the next two subsections.

\subsection{The type 0 solution\label{a0}}

This is a leading and complete solution\footnote{This is the generic solution that contains all arbitrary integration constants. It is a valid solution in most cases as it is discussed in the text.}, for which the potential can be treated as a perturbation.
As we show below, this full solution can be written in the form
\be
S=\bar{S}_0+\cdots\,,
\label{aIR1a}
\ee
where $\bar{S}_0$ is the solution to the equation without a potential (i.e. $V=0$), and the ellipsis stands for sub-leading contributions.\footnote{Strictly speaking there are terms in the ellipsis that are leading compared with some parts in $S_0$. However, as we show the full perturbative expansion is well defined.}
The leading order solution $S_0$ satisfies
\be
d\bar{S}_0\bar{S}_0''-\frac{d}{2}\bar{S}_0^2  - (\bar{S}_0')^2=0\,.
\label{as2}
\ee
We parametrize the leading part inside $\bar{S}_0$ as
\be \bar{S}_0=c_1 e^{b_1\f}+\cdots\,,
 \label{aIR1d}
 \ee
and upon inserting it into \eqref{s2} we find
\be \label{aGbo}
b_1=a_G= \sqrt{\frac{d}{2(d-1)}}\,.
\ee
This implies that in order to consider the potential as a perturbation in \eqref{s1}, the value of $a$ in \eqref{lfpot} should be
\be \label{aalgb}
\frac{V(\f)}{S^2(\f)}\Big|_{\f\rightarrow +\infty}<1 \Rightarrow 2a<2b_1 \Rightarrow a<a_G\,,
\ee
in other words, the value of $a$ should be below the Gubser bound $a_G$.
We conclude, that this solution exists for all $a<a_G$, and this is what we assume henceforth.

 This type of solution was found many times in the past\footnote{It was recognized as a singular (and therefore holographically unacceptable) solution in \cite{iQCD}.}.
Note that the leading solution has already an arbitrary constant $c_1$. We expect also another arbitrary constant and it will appear at sub-leading orders. This can be seen from the exact solution of \eqref{s2}
\be\label{aExs0}
\bar{S}_0(\f)= C_1 \Big[\cosh\big(\sqrt{\frac{d-1}{2d}} \left(\f- C_2 \right)\big)\Big]^{\frac{d}{d-1}}\,,
\ee
where $C_1$ and $C_2$ are two constants of integration. We can keep the leading terms multiplying each constant  of integration as follows
\be \label{aapps0}
\bar{S}_0(\f)=
c_1 e^{\sqrt{\frac d{2(d-1)}}\varphi}+c_2 e^{-\frac{(d-2)\varphi}{\sqrt{2d(d-1)}}}+\cdots,
\ee
where $c_{1,2}$ are functions of $C_1$ and $C_2$ in (\ref{aExs0}) and the ellipsis denotes sub-leading exponential terms. This is the form of $\bar{S}_0$ we use from now to set up our perturbation theory.
To find the sub-leading terms of the solution, we assume a perturbation expansion for large $\f$.
\be
S=\bar{S}_0+\bar{S}_1+\cdots \sp \bar{S}_1\ll \bar{S}_0\,,
\label{as11}\ee
and from  equation (\ref{as1}) we obtain the linearized equation for $S_1$,
\be
\bar{S}_1''-\frac{2}{d}\frac{\bar{S}_0'}{\bar{S}_0}\bar{S}_1'+\frac{(\bar{S}_0''-\bar{S}_0)}{\bar{S}_0}\bar{S}_1-\Big(  \frac{1}{d-1}\frac{V}{\bar{S}_0} -\frac{(d+2)}{d}\frac{\bar{S}_0'}{\bar{S}_0}\frac{V'}{\bar{S}_0} +\frac{V''}{\bar{S}_0}+\frac{1}{d \bar{S}_0}\frac{{V'}^2}{\bar{S}_0^2}\Big)\simeq 0\,.
\label{as12}
\ee
To obtain the above equation we have neglected the non-linear terms in $\bar{S}_1$ because we assumed that they are negligible. This will be checked a posteriori once we find our solutions.

The solution to this equation contains a homogeneous piece proportional to the leading solution
as setting $\bar{S}_1\sim \bar{S}_0$ and neglecting the inhomogeneous piece leads back to (\ref{as2}).
It also contains a correction coming from the potential-dependent terms
\be\label{as13}
\bar{S}_1=c'_1 e^{\sqrt{\frac d{2(d-1)}}\varphi}+c'_2 e^{-\frac{(d-2)\varphi}{\sqrt{2d(d-1)}}}-\frac{a\big(4 a (d-1)+ \sqrt{2d(d-1)} -d\big)}{2c_1(2 a^2 (d-1) -d)} V_\infty e^{(2a-\sqrt{\frac{d}
{2(d-1)}})\f}\,.
\ee
Since the solutions of the homogeneous part are the same as $\bar{S}_0(\f)$ in \eqref{aapps0} we absorb the first and the second term in \eqref{as13} into the $\bar{S}_0(\f)$, and set without loss of generality  $c'_1=c'_2=0$ and write our solution to this order as
\be\label{as13a}
S=c_1 e^{\sqrt{\frac d{2(d-1)}}\varphi}+c_2 e^{-\frac{(d-2)\varphi}{\sqrt{2d(d-1)}}}-\frac{a\big(4 a (d-1)+ \sqrt{2d(d-1)} -d\big)}{2c_1(2 a^2 (d-1) -d)} V_\infty e^{(2a-\sqrt{\frac{d}
{2(d-1)}})\f}+\cdots\,.
\ee

\subsection{Type I and type II solutions\label{aI}}

There are special solutions whose leading exponential behavior is dictated by the leading behavior of the potential at large $\f$ in (\ref{alfpot}).
We parametrize now the leading behavior of $S, W$ and $T$ functions as
\be\label{as0fun}
S_0=S_\infty^{(0)} e^{b_2\f}\sp
W_0=W_\infty^{(0)} e^{b_2\f}\sp
T_0=T_\infty^{(0)} e^{2b_2\f}\,,
\ee
and we obtain two possibilities by inserting $S_0$ in  \eqref{as0fun} into \eqref{as1}

\begin{itemize}

\item The type I solution:
\be \label{as1GV}
S_\infty^{(0)}=\pm\sqrt{\frac{2V_\infty}{d-1}} \sp
b_2=a \,,
\ee
with
\be\label{as1WT}
W_{\infty}^{(0)}= \pm a\sqrt{8(d-1)V_\infty}\sp
T_{\infty}^{(0)}=d(2a^2-\frac{1}{d-1})V_\infty\,.
\ee

\item The type II solution:
\be \label{as2GV}
S_\infty^{(0)}=\pm 2 a \sqrt{\frac{(d-1) V_\infty}{d-2 a^2 (d-1)}}\sp b_2=a \,,
\ee
with
\be\label{as2WT}
W_{\infty}^{(0)}= \pm 2 \sqrt{\frac{(d-1)V_\infty}{d-2 a^2 (d-1)}} \sp T_{\infty}^{(0)}=0\,.
\ee
\end{itemize}
It is important to state here, that both solutions above are exact solutions to equation \eqref{as1}, assuming that the potential is given by its leading term in \eqref{alfpot}.
Also, note that the type II solutions exist if $|a|<a_G$.

We now consider a perturbation around the solution \eqref{as0fun} as follows
\be\label{as3GV0}
S=S_\infty^{(0)} e^{a\f}+\delta S(\f)\,,
\ee
where upon insertion \eqref{as3GV0} into \eqref{as1} we obtain the following linearized second-order differential equation
\be
\label{as3GV1}
\delta S''-\frac{2a}{d}\Big(1+\frac{(d+2)V_\infty}{{S_\infty^{(0)}}^2}\Big) \delta S'
+\Big(a^2-1+\frac{2a^2 (d+2)V_\infty}{d{S_\infty^{(0)}}^2}+\frac{8a^2 V_\infty^2}{d{S_\infty^{(0)}}^4}\Big)\delta S\simeq 0\,.
\ee
In the equation above we have dropped the non-linear terms in $\delta S$, as they are sub-leading and we shall check this a posteriori.
Since  \eqref{as3GV1} is a second-order differential equation with constant coefficients, we shall find three different answers, depending on the value of $a$ and $d$.
We denote by $\l_{1,2}$ the two roots of the characteristic polynomial
\be
\label{as3GV1a}
\l^2-\frac{2a}{d}\Big(1+\frac{(d+2)V_\infty}{{S_\infty^{(0)}}^2}\Big) \l
+\Big(a^2-1+\frac{2a^2 (d+2)V_\infty}{d{S_\infty^{(0)}}^2}+\frac{8a^2 V_\infty^2}{d{S_\infty^{(0)}}^4}\Big)\,.
\ee
We obtain the following cases
\be\label{as3GV}
\delta S=
\begin{cases}
S_{\infty}^{(1)} e^{\l_1\f}+S_{\infty}^{(2)} e^{\l_2\f}\,,
& \l_1 \neq \l_2\in \mathbb{R}\,,
\\ \cr
e^{\l\f}\big(S_{\infty}^{(1)}\f +S_{\infty}^{(2)}\big)\,, &
\l_1=\l_2=\l \in \mathbb{R}\,, \\  \cr
e^{\l\f}\big(S_{\infty}^{(1)}\cos(\omega\f) +S_{\infty}^{(2)}\sin(\omega\f)\big)\,, &
\l_{1,2}=\l\pm i\omega,\quad \l,\omega \in \mathbb{R}\,, \\
\end{cases}
\ee
where $S_{\infty}^{(1)}$ and $S_{\infty}^{(2)}$ are free constants of integration.
Moreover, for both type I and II solutions, we find the following expansions for $W$
\be\label{as4GV}
W\!=\! W_{\infty}^{(0)} e^{a\f}+\!
\begin{cases}
W_{\infty}^{(1)} e^{\l_1\f}+W_{\infty}^{(2)} e^{\l_2\f}+\cdots\,,
& \l_1 \neq \l_2\in \mathbb{R}\,,
\\ \cr
e^{\l\f}\big(W_{\infty}^{(1)}\f +W_{\infty}^{(2)}\big)+\cdots\,, &
\l_1=\l_2=\l \in \mathbb{R}\,, \\  \cr
e^{\l\f}\big(W_{\infty}^{(1)}\cos(\omega\f) +W_{\infty}^{(2)}\sin(\omega\f)\big)+\cdots\,, &
\l_{1,2}=\l\pm i\omega, \, \l,\omega \!\in\! \mathbb{R}, \\
\end{cases}
\ee
and for $T$
\be\label{as5GV}
T\!=\! T_{\infty}^{(0)} e^{2a\f}+\!
\begin{cases}
T_{\infty}^{(1)} e^{(a+\l_1)\f}+T_{\infty}^{(2)} e^{(a+\l_2)\f}+\cdots\,,
& \l_1 \neq \l_2\in \mathbb{R}\,,
\\ \cr
e^{(a+\l)\f}\big(T_{\infty}^{(1)}\f +T_{\infty}^{(2)}\big)+\cdots\,, &
\l_1=\l_2=\l  \in \mathbb{R}\,, \\  \cr
e^{(a+\l)\f}\big(T_{\infty}^{(1)}\!\cos(\omega\f) +T_{\infty}^{(2)}\!\sin(\omega\f)\big)+\cdots\,, &
\l_{1,2}=\l\pm i\omega, \l,\omega \!\in\! \mathbb{R}. \\
\end{cases}
\ee
The various parameters appearing in the solutions  (\ref{as3GV})--(\ref{as5GV}), are as follows:

For all of the solutions to be consistent, the subleading solutions must be smaller than the leading ones.
This requires that $Re(\l_i)<a$ for a solution to be acceptable. Otherwise, it is not. Each acceptable solution gives an extra integration constant.

\subsubsection{Type I}
For the type I solution, with the values given in \eqref{as1GV}, we find the following values for $\l_1$ and $\l_2$
\begin{gather} \label{as6GV}
 \l_1= \frac{a (d+1)-\sqrt{a^2 (d-9) (d-1)+4}}{2}\,, \\
 \l_2= \frac{a (d+1)+\sqrt{a^2 (d-9) (d-1)+4}}{2}\,.\label{at1lam2}
\end{gather}
Defining
\be \label{aad}
a_d=\frac{2}{\sqrt{(9-d)(d-1)}}\,,
\ee
we have classified the allowed region where the sub-leading terms are consistent, ($\l_{1,2}<a$) in tables \ref{type1d8} and \ref{type1d9}.

Knowing the expansion of $S(\f)$ we have the coefficients of $W(\f)$ and $T(\f)$ in \eqref{as4GV} and \eqref{as5GV} as follows
\begin{gather}
W_\infty^{(1)}= -\frac{d-1}{d} \left(a (d-3)+\sqrt{a^2 (d-9) (d-1)+4}\right) S_\infty^{(1)}\,,\label{ayy1}\\
W_\infty^{(2)}=-\frac{d-1}{d} \left(a (d-3)-\sqrt{a^2 (d-9) (d-1)+4}\right) S_\infty^{(2)}\,,\label{ayy2}
\end{gather}
and
\begin{gather}
T_\infty^{(1)}= \mp
\sqrt{\frac{2V_\infty}{d-1}}\left(a (d-1) \big(a (d-3)+\sqrt{a^2 (d-9) (d-1)+4}\big)+1\right)
S_\infty^{(1)}\,,\label{ayy3} \\
T_\infty^{(2)}=
\mp
\sqrt{\frac{2V_\infty}{d-1}}\left(a (d-1) \big(a (d-3)-\sqrt{a^2 (d-9) (d-1)+4}\big)+1\right)
S_\infty^{(2)}\,.\label{ayy4}
\end{gather}
We should note that when both $\l_1$ and $\l_2$ are complex numbers, i.e.
\be \label{asad3}
|a|>a_d \sp 1<d\leq 8\,,
\ee
the solution for $\delta S$ is given by the third line of \eqref{as3GV}
\be  \label{asad4}
\delta S=e^{\frac{a(d+1)}{2}\f}\big(S_{\infty}^{(1)}\cos(\omega\f) +S_{\infty}^{(2)}\sin(\omega\f)\big)\sp
\omega=\sqrt{\frac{a^2}{a^2_d}-1}\,.
\ee
The above result shows that $\delta S$ is an acceptable sub-leading only if $a<-a_d$.

The boundaries of the cases in the tables above must be analyzed separately.
At two specific values  $a=\pm a_d$ we have $\l_1=\l_2$ from \eqref{as6GV} and \eqref{at1lam2}, so the solution for $\delta S$ is given by the second line of \eqref{as3GV}. These cases exist as far as $1<d\leq 8$. We obtain

\begin{itemize}
\item $a=a_d$

Here we find that
\be \label{asad1}
\delta S= e^{\frac{d+1}{2}a_d \f}\big(S_{\infty}^{(1)}\f +S_{\infty}^{(2)}\big)\,,
\ee
which shows that $\delta S$ is not a perturbation around the leading term and therefore the type I solution has no integration constants at $a=a_G$.

\item $a=-a_d$

Here we obtain
\be \label{asad2}
\delta S= e^{-\frac{d+1}{2}a_d \f}\big(S_{\infty}^{(1)}\f +S_{\infty}^{(2)}\big)\,,
\ee
which both terms are allowed as a perturbation around the leading term and therefore the solution has two integration constants.
\end{itemize}

We also have two other specific values $a=\pm a_C$.
\begin{itemize}
\item $a=a_C$

By solving equation \eqref{as3GV1} we find that
\be\label{aspac1}
\delta S= S_{\infty}^{(1)} e^{a_C\f}+S_{\infty}^{(2)} e^{d a_C\f}\,.
\ee
The first term is of the same order as the leading term whereas the second is larger and must be dropped.
The equality indicates the presence of resonance. Considering
\be \label{axac1}
S= \sqrt{\frac{2V_\infty}{d-1}} e^{a_C \f} f(\f)\,,
\ee
and inserting into the equation \eqref{as1} gives
\be \label{axac2}
f''-\frac{1}{d}\frac{{f'}^2}{f^2}-\sqrt{\frac{d-1}{2}}\big(1+
\frac{2}{d}(1-f^2)\big)\frac{f'}{f^2}-\frac{d-1}{2d}\frac{(f^2-1)^2}{f^3}=0\,,
\ee
which by solving $f(\f)$ as a series near $\f= +\infty$ we find the solution for $S(\f)$ as
\be
S=  \sqrt{\frac{2V_\infty}{d-1}} e^{a_C \f}\Big(1+
\sum_{m=0}^{\infty}\sum_{n=m+1}^{\infty}S_{m,n}\frac{\log^m(\f/\f_0)}{(\sqrt{2(d-1)}\f)^{n}}\Big
)\,.
\label{alogs1}
\ee
Some of the starting coefficients  are
\begin{gather}
S_{0,1}=\frac{d}{2}\sp S_{0,2}=0\sp S_{0,3}=\frac{1}{32 }d (d-4)(d-52)\,,
\\
S_{0,4}=-\frac{1}{128}d(d-4)(d-20)(d+76)\sp S_{1,2}={d}\,,
\\
 S_{1,3}=-{2d}\sp S_{1,4}=\frac{1}{16}d ( 3 d(d-56)+688)\,,
\\
S_{2,3}={2d}  \sp S_{2,4}={10 d}\sp S_{3,4}={4d}\,.
\end{gather}
Note that the coefficients of the leading logs are $S_{m,m+1}=2^{m-1}$.

Therefore, at $a=a_c$ there is just one constant of integration.
The first few terms of $S, W$ and $T$ are given by
\be \label{aspac1S}
S=   \sqrt{\frac{2V_\infty}{d-1}} e^{a_C \f}\Big(1+\frac{\sqrt{2}d}{4\sqrt{d-1}}\frac{1}{\f}+
\frac{d}{2 (d-1)}\frac{\log(\f/\f_0)}{\f^2}
+\cdots\Big)\,,
\ee
\be \label{aWac}
W= \sqrt{V_\infty} e^{a_C \f}\Big(2-\frac{d-2}{\sqrt{2(d-1)}}\frac{1}{\f}
+\frac{(d-4)}{4 \f^2}
-\frac{d-2}{d-1}\frac{\log(\f/\f_0)}{\f^2}+\cdots
\Big)\,,
\ee
\be \label{aTac}
T=-V_\infty e^{2a_C\f}\Big(
\frac{d}{\sqrt{2(d-1)}}\frac{1}{\f}-\frac{3d (d-4) }{8(d-1)}\frac{1}{\f^2}+\frac{d}{d-1}\frac{\log(\f/\f_0)}{\f^2}+\cdots
\Big)\,.
\ee
The leading term in \eqref{aTac} shows that the slice curvature of this solution is negative.
\item $a=-a_C$

Here we find that
\be
\delta S=
S_{\infty}^{(1)} e^{-d a_C\f}+S_{\infty}^{(2)} e^{-a_C\f}\,.\label{aspac2}
\ee
The first term is a consistent solution but we should drop the second term which is the same as the leading order. In this case, we find
\be \label{aspac2S}
S= \sqrt{\frac{2V_\infty}{d-1}}e^{-a_C \f}\big(1-\frac{\sqrt{2}d}{4\sqrt{d-1}}\frac{1}{\f}
+\frac{d}{2 (d-1)}\frac{\log(\f/\f_0)}{\f^2}+\cdots\big)+S_{\infty}^{(1)} e^{-d a_C\f}+\cdots.
\ee
So this a solution with two constants of integration, $\f_0$ and $S_{\infty}^{(1)}$. Moreover, we find that
\begin{gather} \label{aWacm}
W=- \sqrt{V_\infty} e^{-a_C \f}\Big(2+\frac{d-2}{\sqrt{2(d-1)}}\frac{1}{\f}
+\frac{(d-4)}{4 \f^2}
-\frac{d-2}{d-1}\frac{\log(\f/\f_0)}{\f^2}+\cdots
\Big)\nn \\
-e^{-d a_C\f}\Big(S_\infty^{(1)} \frac{\sqrt{2(d-1)}}{d} +\cdots\Big)\,,
\end{gather}
and
\begin{gather} \label{aTacm}
T=V_\infty e^{-2a_C\f}\Big(
\frac{d}{\sqrt{2(d-1)}}\frac{1}{\f}+\big(\frac{3d (d-4) }{8(d-1)}\big)\frac{1}{\f^2}-\frac{d}{d-1}\frac{\log(\f/\f_0)}{\f^2}+\cdots
\Big)\nn \\
+ \sqrt{V_\infty}e^{-(d+1)a_C\f}\Big(\frac{d}{2} S_{\infty}^{(1)}+\cdots\Big)\,.
\end{gather}
In this case, the leading term of $T$ shows that the slice curvature of the solution is positive.

\end{itemize}

\subsubsection{Type II}
In this case, as equation \eqref{as2GV} shows, to have a real coefficient for the leading term we should consider
\be \label{acontii}
-a_G<a<a_G\,.
\ee
We obtain the following values for $\l_1$ and $\l_2$ in the perturbed solution \eqref{as3GV}
\be  \label{as8GV}
\l_1 = \frac{1}{a (d-1)}-a\,,
\ee
\be
\l_2 = \frac{d}{2a(d-1)}\,.  \label{at2lam2}
\ee
Using the expansion of $S(\f)$ in \eqref{as3GV}, we find the coefficients of $W(\f)$ and $T(\f)$ in \eqref{as4GV} and \eqref{as5GV} as follows
\be \label{azz1}
W_\infty^{(1)}= -\frac{d-2}{a d} S_\infty^{(1)}\sp
W_\infty^{(2)}= -\frac{2a(d-1)}{d} S_\infty^{(2)}\,,
\ee
and
\be \label{azz2}
T_{\infty}^{(1)}=\mp\frac{ \left(2 a^2 (d-1)+d-2\right) }{a (d-1)}\sqrt{\frac{(d-1)V_\infty}{d-2 a^2 (d-1)}}S_{\infty}^{(1)} \sp
T_\infty^{(2)}=0\,.
\ee


\section{The regular solutions\label{secRS}}

In this appendix, we find the IR-asymptotics of the ``regular" solutions in section \ref{REGSOL}.  Regularity here means that the solution has a subleading behavior compared with the generic solution as $\f\to\infty$.

This is only possible if the potential respects the Gubser's bound at large values of $\f$. In other words, the value of $a$ in \eqref{ppp} or \eqref{dpoti} should be $a<a_G$.
As $\f\rightarrow +\infty$ the series solution for equations of motion \eqref{eq:EOM4}, \eqref{eq:EOM5} and \eqref{eq:EOM6} are given by (see also section \ref{vper})
\be\label{infexp1a}
W = W^{(0)}_\infty e^{ a \f}  \Big(1+4a^2 w_1 \f^2 e^{-2a\f}+2a w_2 \f e^{-2a\f}+\cdots\Big)
+ W^{(1)}_\infty e^{ a \l \f}+\cdots\,,
\ee
\be\label{infexp2a}
S = S^{(0)}_\infty e^{ a \f}\Big(1+4a^2 s_1 \f^2 e^{-2a\f}+2a s_2 \f e^{-2a\f}+\cdots\Big)+S^{(1)}_\infty e^{ a \l \f}+\cdots\,,
\ee
\be \label{infexp3a}
T = T^{(0)}_\infty e^{2 a \f}\Big(1+4a^2 t_1 \f^2 e^{-2a\f}+2a t_2 \f e^{-2a\f}+\cdots\Big)+T^{(1)}_\infty e^{ a (\l+1) \f}+\cdots\,.
\ee
We assume that the solutions are real  and
\be\label{shar1}
\l<1\,.
\ee
All the expansion coefficients $w_i, s_i$ and $t_i$, $i=1, 2, ...$ are functions of $S_\infty^{(0)}$ and $W_\infty^{(0)}$. For example, the first coefficients are as follows
\begin{gather} \label{bas1}
w_1=-s_1 = \frac{b (d-1)^2 d}{2 a^2 \ell^2 (-2 (d-1)^2 {S_\infty^{(0)}}^2 + 2 a (d-1) d S_\infty^{(0)} W_\infty^{(0)} + d {W_\infty^{(0)}}^2)}\,, \\
t_1 = \frac{b (d-1)^2 d^2 S_\infty^{(0)} (S_\infty^{(0)} - a  W_\infty^{(0)})}{2 a^2 \ell^2 T_\infty^{(0)} (-2 (d-1)^2 {S_\infty^{(0)}}^2 +
   2 a (d-1) d S_\infty^{(0)}  W_\infty^{(0)} + d { W_\infty^{(0)}}^2)}\,. \label{bas2}
\end{gather}
Moreover, we have
\be\label{bas3}
T_\infty^{(0)}= -\frac14 \Big(\frac{d(d-1)}{\ell^2} + 2 {S_\infty^{(0)}}^2 - \frac{d }{d-1}{W_\infty^{(0)}}^2\Big)\,.
\ee
For the above expansions, we have two types of solutions:

\begin{itemize}

\item {\bf{Type I (deconfined solution)}}

Solving equations of motion gives the leading coefficients of the expansions as
\begin{gather}\label{deco1}
S^{(0)}_\infty =\pm \frac{\sqrt{d}}{\sqrt{2}\ell}\sp
W^{(0)}_\infty =\pm \frac{(d-1)\sqrt{2d} a}{\ell}\,,
 \\
T^{(0)}_\infty = -\frac{(1 - 2 a^2 (d-1)) d^2}{4\ell^2}\,,
\end{gather}
and the next leading coefficients are
\begin{gather}
\label{deco2}
W_\infty^{(1)} =-\frac{d-1}{d}\Big(\sqrt{a^2 (d-9) (d-1)+4}+a (d-3)\Big) S_\infty^{(1)}\,, \\ \label{deco3}
T_\infty^{(1)} =\mp\frac{\sqrt{d}}{\sqrt{2}\ell}\Big(a (d-1) \big(\sqrt{a^2 (d-9) (d-1)+4}+a (d-3)\big)+1\Big) S_\infty^{(1)}\,,
\end{gather}
where $S_{\infty}^{(1)}$ is a free parameter and
\be
\l = \frac{a(d+1)-\sqrt{a^2 (d-9) (d-1)+4}}{2 a}\,. \label{deco4}
\ee
The reality of the solutions and conditions in \eqref{shar1} are implying that
\be \label{consol1}
a^2<\frac{1}{2(d-1)}\,.
\ee
\item {\bf{Type II (confined solution)}}

We have also a second solution with the following coefficients for the leading and the next leading terms
\begin{gather}\label{wsinf0}
S^{(0)}_\infty =\pm\frac{a (d-1)}{\ell \sqrt{2a^2 \left(\frac{1}{d}-1\right)+1}}\sp W^{(0)}_\infty = \pm \frac{ (d-1)}{\ell \sqrt{2a^2 \left(\frac{1}{d}-1\right)+1}}\sp T^{(0)}_\infty =0\,,
\end{gather}
and
\begin{gather}
W^{(1)}_\infty = -\frac{(d-2)}{d a}S^{(1)}_\infty\sp T^{(1)}_\infty =
\mp\frac{\left(2 a^2 (d-1)+d-2\right)}{2 \big( \ell a \sqrt{2a^2 \left(\frac{1}{d}-1\right)+1}\big)}S^{(1)}_\infty\,.\label{wsinf1}
\end{gather}
However, for this solution we find
\be
\l = -\frac{a^2 (d-1)-1}{a^2 (d-1)}\,.
\ee
We should note that to find the next to leading terms, we have used the following condition
\be \label{mconsol2}
2 a^2 (d-1)+d-2\geq 0\,,
\ee
otherwise, we do not find a solution.
Moreover, the reality of the solutions gives
\be\label{rela1}
2 a^2 (d-1)<d\,.
\ee
The above conditions and \eqref{shar1} leads us to a restriction on the value of $a$
\be \label{consol2}
\frac{1}{2(d-1)}<a^2<\frac{d}{2(d-1)}\,.
\ee
We should note that because of \eqref{mconsol2}, to have negative curvature slices, for $S_{\infty}^{(1)}<0$ we should choose the lower signs in \eqref{wsinf0} and \eqref{wsinf1} and vice versa.

\end{itemize}

\section{Solutions with many A-loops}\label{loops}

The numerical solutions with many oscillations in the scale factor have two general properties:
\begin{itemize}
\item At the oscillation region, the scale factor $A(u)$ has small amplitude oscillations around a fixed value.

\item The oscillations of $\f$ are in a region in which the potential \eqref{ppp} can be approximated by ($0\leq \f \lesssim 2$)
    \end{itemize}
\begin{gather}\label{ml1}
V(\f)\approx -V_0-m^2\f^2-\l\f^4+\mathcal{O}\left(\varphi^6\right)\,,\\
V_0=
-\frac{(d-1) d}{\ell^2}\sp
m^2=\frac{(d-1) d \left(a^2+b\right)}{\ell^2} \sp
\l=-\frac{a^4 d(d-1)}{3 \ell^2}\,.
\label{ml2}
\end{gather}
We analyze this behavior here by first starting from the first-order equations
\be
\label{e1} S^2 - SW' + \frac{2}{d} T =0 \, ,
\ee
\be
\label{e22} \frac{d}{2(d-1)} W^2 -S^2 -2 T +2V =0 \, , \ee
\be
\label{e2} SS' - \frac{d}{2(d-1)} SW - V' = 0 \,.
\ee
We assume that $W$ is much smaller than the other functions in the region of $\f$ that we are interested and therefore we rewrite the first-order equations above as
\be
SS'-V'=\frac{d}{2(d-1)}SW\,,
\ee
\be\label{Ez3}
SW'-S^2+\frac{S^2-2V}{d}=\frac{W^2}{2(d-1)}\,,
\ee
\be
2T-2V+S^2=\frac{dW^2}{2(d-1)}\,,
\ee
where we collected the large terms on the left-hand side and the small terms on the right-hand side.
Setting to leading order the right-hand side to be negligible, and expanding the solution as
\be
S=S_0+S_1+S_2+\cdots\,,
\ee
and similarly, for the other functions, we calculate the leading order solution

\begin{figure}[!b]
\centering
\begin{subfigure}{0.49\textwidth}
\includegraphics[width=1\textwidth]{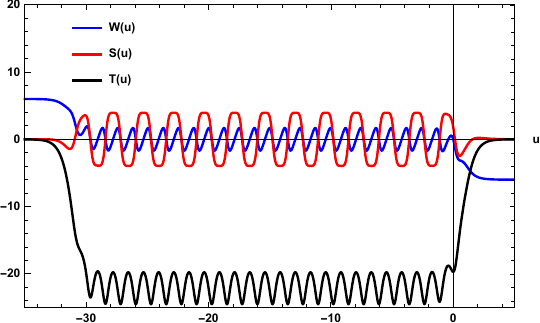}
\caption{}
\end{subfigure}\\
\begin{subfigure}{0.49\textwidth}
\includegraphics[width=1\textwidth]{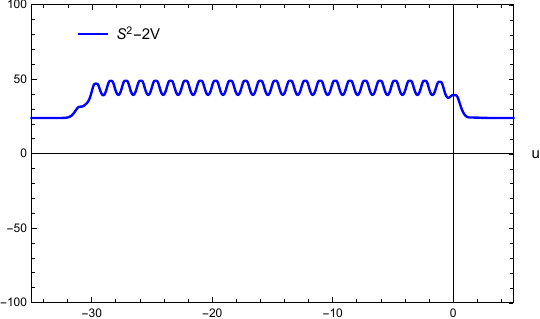}
\caption{}
\end{subfigure}
\begin{subfigure}{0.49\textwidth}
\includegraphics[width=1\textwidth]{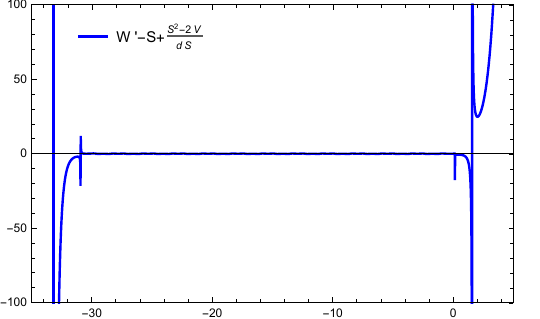}
\caption{}
\end{subfigure}
\caption{\footnotesize{(a) $W, S$ and $T$ as a function of $u$. (b) Shows the function $S^2-2V$. It should give the approximate value of $c$ in \eqref{Ez1}. (b) $W'-S+\frac{S^2-2V}{d S}$ as a function of $u$. It is zero, consistent with equation \eqref{Ez2}, or ignoring the right-hand side of \eqref{Ez3}.}}
\label{AA}
\end{figure}
\be\label{Ez1}
S_0^2=2V+c\,,
\ee
where $S_0$ is the leading approximation to $S$ and $c$ is a constant of integration.
Then,
\be\label{Ez2}
W'-S+\frac{S^2-2V}{d S}=\frac{W^2}{2(d-1)S}\approx 0 \quad \Rightarrow\quad W_0'=\frac{2V+\frac{d+1}{d}c}{\sqrt{2V+c}}\,,
\ee
which be integrated to give
\be
W_0(\f)=W(\f_0)+\int_{\f_0}^{\f}dv\frac{2V(v)+\frac{d+1}{d}c}{\sqrt{2V(v)+c}}\,,
\ee
and the slice curvature
\be
T_0=V-S_0^2=-c\,.
\ee
We can rewrite $S_0^2$ as
\be
S_0^2=2V(\f)-2V(\f_0)+S(\f_0)^2\sp c=S(\f_0)^2-2V(\f_0)\,,
\label{e13}\ee
for any $\f_0$ in the domain of validity of the solution.

We can solve for the subleading solutions to find
\be
S_1(\f)=\frac{d}{2(d-1)S_0(\f)}\int_{\f_0}^{\f}dv~S_0(v)W_0(v)\,,
\ee
\be
T_1=- \frac{d}{2(d-1)}\int_{\f_0}^{\f}dv~S_0(v)W_0(v)+\frac{d~W_0^2(\f)}{4(d-1)}\,.
\ee
To understand how good this approximation is for these solutions we present the plots in figure \ref{AA}.
In \ref{AA}(a), we portray the three full functions $S, W, T$ as functions of the $u$ variable for the solution over the whole range.
In \ref{AA}(b), we observe that $S^2-2V$ is almost constant up to small oscillations, in the regime of validity of the approximation.
In \ref{AA}(c) we plot $W'-S+\frac{S^2-2V}{d S}$ which according to equations should be equal to $\frac{W^2}{2(d-1)S}$ it is clear that $W^2$ can be approximated with zero in the range of validity of the approximation.

\begin{figure}[!ht]
\centering
\begin{subfigure}{0.49\textwidth}
\includegraphics[width=1\textwidth]{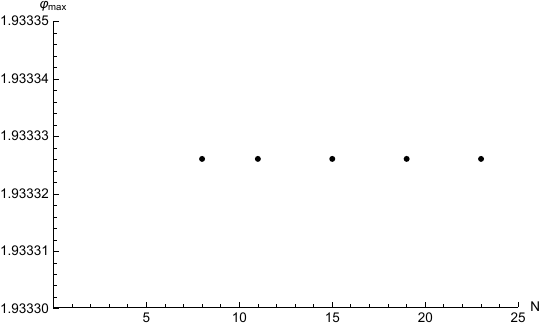}
\caption{}
\end{subfigure}\\
\begin{subfigure}{0.49\textwidth}
\includegraphics[width=1\textwidth]{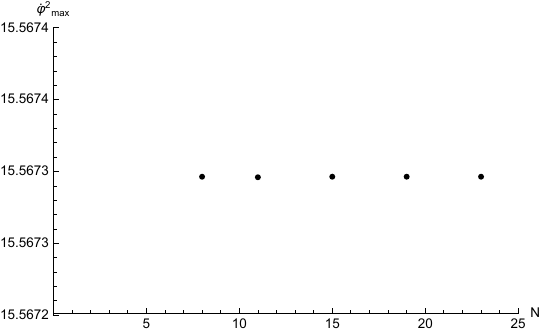}
\caption{}
\end{subfigure}
\begin{subfigure}{0.49\textwidth}
\includegraphics[width=1\textwidth]{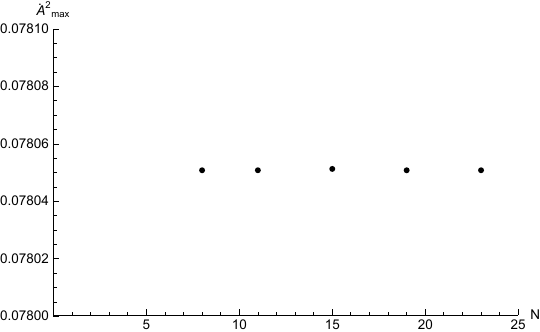}
\caption{}
\end{subfigure}
\caption{\footnotesize{(a), (b) and (c) are graphs for $\f_{max}, {\dot\f}^2_{max}$ and $\dot{A}^2_{max}$ for $N=8, 11, 15, 19, 23$ number of oscillations. Notice that the changes are very small in all graphs.}}\label{nos123}
\end{figure}

As the number of oscillations of the solution varies some of the properties do not change.
In particular, the maximum of $\dot\f^2$ oscillations, the maximum of $\dot A^2\sim W^2$ oscillations, as well as the maximum value of $\f$, $\f_{\rm max}$, do not change.
This can be seen in the plots of figure \ref{nos123} where $N$ in the horizontal axis counts the total number of oscillations of the exponent $A$ of the scale factor.

We continue by now studying the equation in the $u$ coordinate,
\begin{gather}
\label{OS1} 2(d-1) \ddot{A} + \dot{\f}^2 + \frac{2}{d} e^{-2A} R^{(\zeta)} =0 \,,
\\
\label{OS2} d(d-1) \dot{A}^2 - \frac{1}{2} \dot{\f}^2 + V - e^{-2A} R^{(\zeta)} =0 \,,
\\
\label{OS3} \ddot{\f} +d \dot{A} \dot{\f} - V' = 0 \,.
\end{gather}
Ignoring the $\dot{\f}\dot{A}$ in \eqref{OS3}, the equation of motion for $\f$ becomes
\be \label{ml3}
\ddot{\f}-V'\approx 0\rightarrow \dot{\f}^2\approx 2\int V' d\f+c_\f=2V(\f)-2V(\f_{\rm max}) \,,
\ee
which matches what we found in the first-order formalism. $\f=\f_{\rm max}$ is the maximal value that $\f$ can reach during the oscillation.
 Note that this agrees with (\ref{e13}) as at $\f_0\to \f_{\rm max}$ we have $S(\f_{\rm max})=0$.

\begin{figure}[!b]
\begin{center}
\includegraphics[width = 8cm]{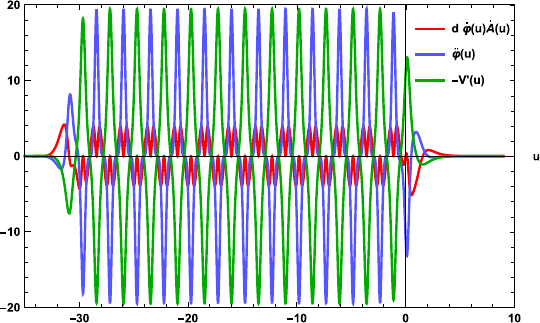}
\caption{\footnotesize{Various terms in equation \eqref{OS3}} for a solution with many loops, plotted as a function of the coordinate $u$. The amplitude of the oscillations of $\dot{\f}\dot{A}$ is much smaller than the other terms. The horizontal axis is the $u$ coordinate.}\label{3terms}
\end{center}
\end{figure}

We can check the above arguments by looking at the numerical results for a solution with many loops in figure \ref{3terms} and for which we have compared the three terms in equation \eqref{OS3} and found that $\dot{\f}\dot{A}$ is negligible compared to the other two terms.

\begin{figure}[!t]
\begin{center}
\includegraphics[width = 8cm]{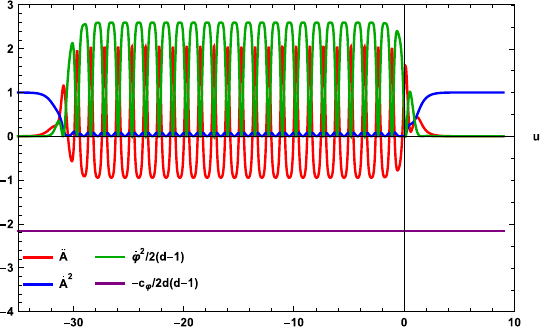}
\caption{\footnotesize{Various terms in equation \eqref{OS4}} for a solution with many loops, plotted as a function of the coordinate $u$. The amplitude of the oscillations of $\dot{A}^2$ is much smaller than the other terms. The horizontal axis is the $u$ coordinate.}\label{fterms}
\end{center}
\end{figure}

In the region where \eqref{ml3} is valid, we can combine equations \eqref{OS1} and \eqref{OS2} to make a curvature independent equation as follows
\be \label{OS4}
\ddot{A}+\dot{A}^2+\frac{1}{2(d-1)}\dot{\f}^2-\frac{c_\f}{2d(d-1)}\approx 0\,.
\ee
This equation shows that if $\f(u)$ has an oscillation with frequency $\omega$ then $A(u)$ should have $2\omega$ frequency. The comparison of the different terms in \eqref{OS4} is shown in figure \ref{fterms}. We observe that $\dot{A}^2$ is much smaller than the other terms in \eqref{OS4}, in agreement with a similar statement about $W$ in the first-order formalism.

Solving \eqref{ml3} for potential \eqref{ml1}, we obtain
\begin{gather} \label{ml4}
u-u_0 \approx\int_{\f_{\rm max}}^{\f} \frac{d \f}{\sqrt{2\l(\f_{\rm max}^2-\f^2)(m^2+\l\f_{\rm max}^2+\l\f^2)}}\nn \\
\approx -\frac{1}{\sqrt{4\l\f_{\rm max}^2+2m^2}}F\big(\arccos\frac{\f(u)}{\f_{\rm max}
}\,\big|\,\frac{\f_{\rm max}}{\sqrt{2\f_{\rm max}^2+\frac{m^2}{\l}}}\big)\,,
\end{gather}
where
\be
F\big(\arccos\frac{\f(u)}{\f_{\rm max}}\,\big|\,\frac{\f_{\rm max}}{\sqrt{2\f_{\rm max}^2+\frac{m^2}{\l}}}\big)=\int_0^{\arccos\frac{\f(u)}{\f_{\rm max}}}\Big[1-(\frac{\f_{\rm max}}{\sqrt{2\f_{\rm max}^2+\frac{m^2}{\l}}})\sin^2\theta\Big]^{-\frac12}d\theta\,,
\ee
is the Elliptic function. By inverting \eqref{ml4}
\be \label{ml5}
\f(u)\approx
\f_{\rm max} \,\text{cn}\Big(\sqrt{4\l\f_{\rm max}^2+2m^2}(u-u_0) ,\frac{\f_{\rm max}}{\sqrt{2\f_{\rm max}^2+\frac{m^2}{\l}}}\Big)\,,
\ee
and $\text{cn}(\a u,k)$ is the Jacobi elliptic function. The period of oscillations of \eqref{ml5} is given by
\be \label{ml6}
T=\frac{4}{\sqrt{4\l\f_{\rm max}^2+2m^2}} F\big(\frac{\pi}{2}\,|\,\frac{\f_{\rm max}}{\sqrt{2\f_{\rm max}^2+\frac{m^2}{\l}}}\big)\,.
\ee
\begin{figure}[!t]
\begin{center}
\includegraphics[width = 8cm]{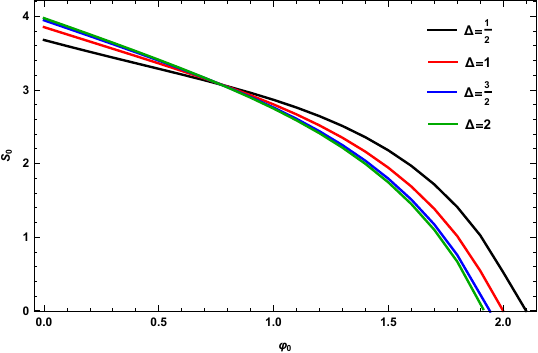}
\caption{\footnotesize{The boundary between UV-UV and UV-Reg solutions for four various values of $\D_-$. The value at which $S_0=0$ on each curve shows the value of $\f_{\rm max}$ in \eqref{ml5}.}}\label{bonds}
\end{center}
\end{figure}
\begin{figure}[!t]
\begin{center}
\includegraphics[width = 8cm]{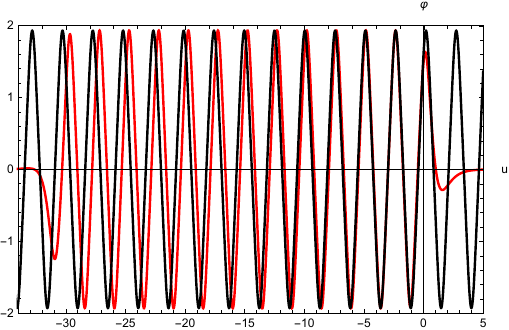}
\caption{\footnotesize{The black curve is the analytic solution of $\f(u)$ in \eqref{ml5}. The red curve is a UV-UV solution with a finite number of oscillations. The frequency and amplitude of oscillations are approximately fitted in the first few oscillations. The horizontal axis is the $u$ coordinate.}}\label{osc}
\end{center}
\end{figure}
In our numerical solutions $\f_{\rm max}\approx 2.0$ therefore the frequency of the oscillations would be $\omega=2\pi/T\approx 2.5$. For various choices of $\D$, we have shown the boundary of the green-blue region in figure \ref{bonds}. The value at which $S_0=0$ on each curve shows the value of $\f_{\rm max}$ in the above computations.

We present the comparison of the approximate analytic relation \eqref{ml5} with a real UV-UV solution with a finite number of oscillations in figure \ref{osc}.


\section{Dimensional reduction of solutions\label{reduction}}

In this appendix, we discuss in more detail the correspondence between solutions of the uplifted theory and the lower-dimensional theory.

As we already mentioned, the potential in the confining theory is equal to the uplifted one as $\f\rightarrow \pm\infty$ or when the $S^n$ shrinks to zero size. The shrinking is happening either regularly or singularly. For each case, below we find the relation between the parameters of the two theories. A review of solutions and asymptotic expansions of the uplifted theory is given in appendix \ref{rev}, but for more details see \cite{Ghodsi:2023pej}.

\subsection{Reduction near a regular shrinking sphere}\label{ssr}

Consider a regular end-point in uplifted theory at $\tu=\tu_0$ which corresponds to a regular shrinking sphere. This end-point was discussed in section 3 of  \cite{Ghodsi:2023pej}.

 The expansion of the sphere's scale factor, $A_2$,  is given in \eqref{rega2}. For simplicity, we rewrite it as
\be\label{ud2}
e^{2A_2(\tu)}=\tilde{s}_2 (\tu-\tu_0)^2+ \tilde{s}_4(\tu-\tu_0)^4+\cdots.
\ee
By inserting \eqref{ud2} into \eqref{ud1} and integration we find the relation between the holographic coordinate in uplifted theory, $\tu$, and the holographic coordinate of the confining theory, $u$, as
\be\label{ud3}
{{u}}-{{u}_0}=
\frac{(d-1) \tilde{s}_2^{\frac{n}{2 (d-1)}} (\tu-\tu_0)^{\frac{n}{d-1}+1}}{d+n-1}
+\frac{n \tilde{s}_4  \tilde{s}_2^{\frac{n}{2 (d-1)}-1} (\tu-\tu_0)^{\frac{n}{d-1}+3}}{2 (3 d+n-3)}+\cdots,
\ee
or inversely
\begin{gather}\label{ud4}
\tu-\tu_0 =
\tilde{s}_2^{-\frac{n}{2 (d+n-1)}}\big(\frac{d+n-1}{d-1}\big)^{\frac{d-1}{d+n-1}}({{u}}-{{u}_0})^{\frac{d-1}{d+n-1}}\nn \\
-\frac{n \tilde{s}_4 \tilde{s}_2^{\frac{3 (d-1)}{2 (d+n-1)}-\frac{5}{2}}}{3d+n-3}\big(\frac{d+n-1}{d-1}\big)^{\frac{3(d-1)}{d+n-1}}({{u}}-{{u}_0})^{\frac{3(d-1)}{d+n-1}}+\cdots.
\end{gather}
Now by using the relation \eqref{zz2a} for scalar field $\f$, and equations \eqref{ud2} and \eqref{ud4} we read the functionality of the scalar field near $u=u_0$ in the confining theory
\begin{gather}\label{ud5}
{\f}(u) =-\sqrt{\frac{2n(d+n-1)}{d-1}}A_2(u) \nn\\
=-\sqrt{\frac{n(d-1)}{2(d+n-1)}}\log\big(\frac{\tilde{s}_2(d+n-1)^2 ({{u}}-{{u}_0})^2}{(d-1)^2}\big)\nn \\
-\frac{3 \tilde{s}_4 \sqrt{n} (d-1)^{\frac{n-3 d+3}{2 (d+n-1)}} (d+n-1)^{\frac{5 d+n-5}{2 (d+n-1)}} \tilde{s}_2^{-\frac{d+2 n-1}{d+n-1}} ({{u}}-{{u}_0})^{\frac{2 (d-1)}{d+n-1}}}{\sqrt{2} (3 d+n-3)}+\cdots,
\end{gather}
or inversely
\be\label{ud6}
{{u}}-{{u}_0}=
\frac{(d-1) e^{-\sqrt{\frac{d+n-1}{2n(d-1)}}{\f}}}{\sqrt{\tilde{s}_2} (d+n-1)}
-\frac{3 \tilde{s}_4 (d-1)}{2 \tilde{s}_2^{\frac52} (3 d+n-3)}
e^{-\frac{3d+n-3}{\sqrt{2n(d-1)(d+n-1)}}{\f}}+\cdots.
\ee
Using the definition \eqref{eq:defSc}, we can read $S({\f})$ by getting a $u$ derivative of \eqref{ud5} and then by substitution of \eqref{ud6}
\be\label{ud7}
S({\f})=
-\sqrt{\frac{n (d+n-1)}{d-1}}\Big(\sqrt{2\tilde{s}_2}e^{\sqrt{\frac{d+n-1}{2n(d-1)}}{\f}}
+\frac{3\tilde{s}_4}{\sqrt{2\tilde{s}_2^3}}e^{\frac{(n-d+1)}{\sqrt{2(d-1) n (d+n-1)}}{\f}}+\cdots\Big).
\ee
We can rewrite \eqref{ud7} as
\be\label{ud8}
S(\f) = S^{(0)}_\infty e^{ a \f}+S^{(1)}_\infty e^{ a \l \f}+\cdots,
\ee
where after inserting the coefficients $\tilde{s}_2$ and $\tilde{s}_4$ from expansion \eqref{rega2} and using the relation \eqref{abd} for $a$ and the relation \eqref{barr} for curvature $R_2$ of the sphere, we obtain
\begin{gather}
\l = \frac{a^2 (d-1)-1}{a^2 (d-1)}\sp
a =\sqrt{\frac{d+n-1}{ 2n(d-1)}}\,,
\nn \\
S^{(0)}_\infty =-\frac{a (d-1)}{{\ell} \sqrt{2a^2 \left(\frac{1}{d}-1\right)+1}}\,, \nn \\
S^{(1)}_\infty =
\frac{2 a {\ell} \Big(\tell^2 R_1 +
   2(d-1) \big(a^2 (d-1)-1\big)  \big(a_0 (2 a^2 d(d-1)-1) +
2 a^2 \tell^2 R_1\big)\Big)}{a_0 \tell^2 \big(2 a^2 (d-1) + d-2\big) \big(1- 2 a^2 (d-1)\big)^2 \big(2a^2 \left(\frac{1}{d}-1\right)+1\big)^{-\frac12}}\,,
\label{ud9}
\end{gather}
where $a_0=e^{2A_1(\tu_0)}$.
So in this way, we find a relation between the free parameter $a_0$ in the uplifted theory and $S^{(1)}_\infty$ in the confining theory.

We can also read $W({\f})$ directly from relation \eqref{w}, which gives the following expansion
\be \label{ud11}
W(\f) = W^{(0)}_\infty e^{ a {\f}} + W^{(1)}_\infty e^{ a \l {\f}}+\cdots,
\ee
where
\be \label{ud12}
W^{(0)}_\infty = - \frac{ (d-1)}{{\ell} \sqrt{2a^2 \left(\frac{1}{d}-1\right)+1}} \sp
W^{(1)}_\infty = -\frac{(d-2)}{d a}S^{(1)}_\infty\,.
\ee
Moreover, using the relation \eqref{t} we read
\be \label{ud14}
T(\f) = T^{(0)}_\infty e^{2 a {\f}} + T^{(1)}_\infty e^{ a (\l+1) {\f}}+\cdots,
\ee
with
\be \label{ud15}
T^{(0)}_\infty =0 \sp
T^{(1)}_\infty =
\frac{2 a^2 (d-1)+d-2}{2{\ell} a \sqrt{2a^2 \left(\frac{1}{d}-1\right)+1}}S^{(1)}_\infty\,.
\ee
We should note that since
\be \label{ud16}
2 a^2 (d-1)+d-2>0\,,
\ee
to have a negative curvature solution i.e. $T<0$,  we should consider
\be \label{ud17}
S_{\infty}^{(1)}<0\,.
\ee
Looking at \eqref{ud9} we observe that
\be \label{ud18}
\frac{dS_\infty^{(1)}}{d a_0}=-\frac{2 a{\ell} R_1}{a_0^2 \left(2 a^2 (d-1)+d-2\right)}\sqrt{\frac{d-2 a^2 (d-1)}{d}}
 >0\,,
\ee
which means that $S_{\infty}^{(1)}$ is monotonically decreasing function of $a_0$.

As we discussed in appendix \ref{rev}, in uplifted theory when the sphere shrinks to zero size regularly, we find three types of solutions. The parameter $a_0=e^{2A_1(\tu_0)}$ classifies the solutions to the regular to boundary or (R, B)-type, the product space solution, and the regular to singular or (R, A)-type. According to this analysis, we conclude that:
\begin{itemize}
\item The (R, B)-type:

This regular solution (see fig \ref{shrink1}) in the uplifted theory  is labeled by  $+\infty>a_0\geq a_0^c$ where
\be \label{a0cr}
a_0^c=-\frac{\tell^2 R_1}{d (d+n)}\,.
\ee
In confining theory by \eqref{ud9} this corresponds to region
\be \label{dcons1}
S_{\infty}^{(1)max}>S_{\infty}^{(1)}> S_{\infty}^{(1)c}\,,
\ee
where
\be\label{sinfc}
S^{(1)c}_{\infty}=
-\frac{ {\ell}(d+n)}{(d-1) \tell^2} \sqrt{\frac{2(n-1) (d+n-1)}{d}} \,,
\ee
and
\begin{gather}\label{smaxinf}
S^{(1)max}_\infty
=\begin{cases}
(1-\frac{d}{n+1})S_{\infty}^{(1)c}\,, & d<n+1\,,\vspace{5mm} \\
0\,, & d\geq n+1\,,
\end{cases}
\end{gather}
where we have assumed that $S_\infty^{(1)}<0$.

\item The (R, A)-type:

This singular solution (see fig \ref{shrink2}) in the uplifted theory is labeled by  $a_0^c>a_0>0$. In confining theory this corresponds to
\be \label{dcons2}
0>S_{\infty}^{(1)c}>S_{\infty}^{(1)}>-\infty\,.
\ee

\item $AdS_d\times AdS_{n+1}$

At $a_0=a_0^c$ we have the product space solution (fig \ref{shrink0}) therefore  $S^{(1)}_\infty$ is given by \eqref{sinfc} or
\be \label{sprod1}
S^{(1)}_\infty=S^{(1)c}_\infty=\frac{2 a  \left(1-2 a^2 (d-1) d\right){\ell} }{\left(1-2 a^2 (d-1)\right)^2\tell^2}\sqrt{2a^2 (\frac{1}{d}-1)+1}\,.
\ee

\item Global $AdS_{d+n+1}$:

As we discussed in appendix \ref{rev} the global $AdS_{d+n+1}$ is an exact solution of equations of motion. This solution is a specific point among the (R, B)-type solutions. The metric is
\be \label{gmetr1}
ds^2= d\tu^2 + a_0 \cosh^2\frac{\tu-\tu_0}{\tell} ds^2_{AdS_d} + s_0 \sinh^2\frac{\tu-\tu_0}{\tell} d\Omega_n^2\,,
\ee
where
\be  \label{gmetr2}
a_0=-\frac{\tell^2 R_1}{d(d-1)}\sp
s_0=\frac{\tell^2 R_2}{n(n-1)}\,.
\ee
This solution corresponds to
\be \label{sglob1}
S^{(1)glob}_\infty = -\frac{2 a (d-1){\ell}}{ \left(1-2 a^2 (d-1) \right)^2\tell^2}\sqrt{2 a^2 (\frac{1}{d}-1)+1}=\frac{n}{d+n}S_{\infty}^{(1)c}\,.
\ee
\subsection{Reduction near a singular shrinking sphere}
Analysis of the end-points in uplifted theory shows that there are two types of solutions when $S^n$ shrinks to a zero size in a singular way: (S, B)-type (fig \ref{AB1}) and (S, A)-type (fig \ref{TypeS4}).  For these solutions at $\tu=\tu_0$, the scale factors behave as
\begin{gather}\label{sA1}
A_1(\tu) = \lambda_1 \log(\tu-\tu_0)+\frac12 \log a_0+\cdots\,,\\
A_2(\tu) = \lambda_2 \log(\tu-\tu_0)+\frac12 \log s_0+\cdots\,, \label{sA2}
\end{gather}
where $a_0$ and $s_0$ are two free constants and $\l_1$ and $\l_2$ are given by
\be
\l_1=\frac{d-\sqrt{dn(d+n-1)}}{d(d+n)}\sp
\l_2=\frac{n+\sqrt{dn(d+n-1)}}{n(d+n)}\,.
\ee
 Doing the same steps as the previous section from the uplifted theory and to the leading term of expansions we find
\be \label{sA3}
{u}-{u}_0= \frac{(d-1) s_0^{\frac{n}{2(d-1)}} (\tu-\tu_0)^{\frac{n\lambda_2 }{d-1}+1}}{d+n\lambda_2 -1}+\cdots\,,
\ee
and the scalar field as a function of $u$ is
\be \label{sA4}
{\f}(u)=
{\frac{-\sqrt{2}(d-1) \lambda_2}{d+\lambda_2 n-1}} \big(\sqrt{\frac{n^2}{d-1}+n}\big) \log \Big(\frac{({u}-{u}_0)s_0^{\frac{1}{2\l_2}} (d+\lambda_2 n-1)}{d-1}\Big)+\cdots\,.
\ee
Finally we can read $S({\f})$ as
\be \label{sssh}
S({\f})=-
\frac{\sqrt{2} \sqrt{\frac{n^2}{d-1}+n} \big(\sqrt{d n (d+n-1)}+n\big)s_0^{\frac{n (d+n)}{2 (\sqrt{d n (d+n-1)}+n)}} }{n (d+n)}e^{\sqrt{\frac{d}{2(d-1)}} {\varphi}}+\cdots\,.
\ee
Comparing the leading term of \eqref{sssh} with the leading term \eqref{ud7} shows a different behavior.
In section \ref{vper}, we observed a similar functionality for $S(\f)$. This shows that singular solutions from the uplifted solutions are related to singular solutions in the confining theory upon the reduction on $S^n$ as $\f\rightarrow \pm \infty$. Inversely we can say that the singularity of a solution in a confining theory can be described by a singular shrinking of a sphere in the uplifted theory.

\end{itemize}
\subsection{Reduction near the $AdS$-like boundary}
In uplifted theory, we have solutions with an $AdS$-like boundary. The leading terms of the scale factors near this boundary at $\tilde{u}\rightarrow +\infty$ are
\be  \label{Adbd1}
A_1(\tu)=\bar{A}_1+\frac{\tu}{\tell}+\cdots\sp
A_2(\tu)=\bar{A}_2+\frac{\tu}{\tell}+\cdots\,.
\ee
Using \eqref{chvab} we can read the holographic coordinate $u$ in the confining theory as
\be \label{Adbd2}
u=\frac{(d-1) \tell }{n}e^{\frac{n }{(d-1) \tell}(\bar{A}_2 \tell+\tu)}+\cdots\,.
\ee
We can read also the scalar field from \eqref{zz2a} (we choose the negative sign here), and then we obtain the following relation
\be \label{Adbd3}
u=\frac{(d-1) \tell }{n}e^{- \sqrt{\frac{n}{2(d-1) (d+n-1)}}\varphi}+\cdots\,.
\ee
Therefore as $\tu\rightarrow +\infty$
\be \label{Adbd4}
u\rightarrow +\infty \sp \f\rightarrow -\infty\,.
\ee
The function $A(u)$ in confining metric \eqref{zz1a} becomes
\be \label{Adbd5}
A=A_1+\frac{n}{d-1}A_2=\bar{A}_1-\bar{A}_2+\frac{(d+n-1) }{n}\log \frac{n u}{(d-1) \tell}+\cdots\,,
\ee
which is not similar to a scale factor near an $AdS$-like boundary. This is expected, since as $\f\rightarrow -\infty$ the potential in confining theory i.e. \eqref{zz3} behaves as
\be \label{Adbd7}
V(\f, \chi)\rightarrow 0\,.
\ee
Moreover,
\be \label{Adbd6}
S(\f)=\frac{d\f}{du}=-\frac{1}{\tell }\sqrt{\frac{2n(d+n-1)}{d-1}}e^{\sqrt{\frac{n}{2(d-1) (d+n-1)}}\varphi}+\cdots\,.
\ee
We should note that for the potential which we are considering in \eqref{vt1} we have $-\infty<V\leq -\frac{d(d-1)}{\ell^2}$, therefore we should not expect to see the behaviors in \eqref{Adbd5} or \eqref{Adbd6} as $u\rightarrow +\infty$. The only possibility is that the potential is a perturbation in the confining theory at large $\f$. However, in section \ref{vper} we already observed that in such a case as $\f\rightarrow -\infty$ we have
$S(\f)\sim e^{-\sqrt{\frac{d}{2(d-1)}}\f}$
which is not compatible with \eqref{Adbd6}.

\subsection{Deconfined holographic theories on $AdS$ \label{dcth}}

As shown in \cite{GK}, we can construct a deconfining theory on $AdS$ by reducing the $d+n+1$-dimensional theory on a torus $T^n$. Then the geometry of the uplifted theory is described by an $AdS_{d+n+1}$ space which is sliced by an $AdS_d$ and a torus $T^n$
\be \label{Tn1}
ds^2 = d\tu^2 + e^{2A_1(\tu)} ds^2_{AdS_d} + e^{2A_2(\tu)} ds^2_{T^n}\,.
\ee
Since the torus is flat we should insert $R_2=0$ in the potential \eqref{zz3}  therefore, now the sub-leading term is dominant. Once again we can consider the following potential for scalar field $\f$ in $d+1$-dimensional theory
\be \label{dvt1}
{V}(\f)=
-\frac{d(d-1)}{{{\ell}}^2}\left(b\f^2+\cosh^2(a\f)\right)\sp
b=\frac{\Delta(d-\Delta)}{2 d (d-1)}-a^2\,.
\ee
To match the potential of the uplifted and confining theories at large $\tilde{\f}$ we should have
\be
2a =\sqrt{\frac{2n}{(d-1)(d+n-1)}}\sp
\frac{(d+n)(d+n-1)}{\tell^2}=\frac{d(d-1)}{4{{\ell}}^2}
\,.\label{dabd}
\ee
Here, the value of $a$ for all values of $n,d>1$ lies in the red region of figure \ref{CDCM} which means that the theory in $d+1$ dimensions is deconfining.

Near the boundary at $\tu\rightarrow -\infty$, we can find solutions in which the scale factor of the torus $T^n$ is shrinking while the scale factor of $AdS_d$ tends to a constant value. The expansions of the scale factors then are obtained by
\be \label{Tn2}
A_1(\tu) = \frac{1}{2} \log\Big[-\frac{\tell^2 R_1}{d(d+n)}+a_1 e^{\frac{(\sqrt{8+n}-\sqrt{n})\sqrt{d+n}}{2}\frac{\tu}{\tell}}+\cdots\Big]\,,
\ee
\be
A_2(\tu) = \sqrt{\frac{d+n}{n}}\frac{\tu}{\tell}+\frac{a_1 d^2  (d+n) }{ \sqrt{n} \left(3 \sqrt{n}+\sqrt{n+8}\right) \tell^2 R_1}e^{\frac{\left(\sqrt{n+8}-\sqrt{n}\right) \sqrt{d+n}}{2}\frac{\tu}{\tell}}+\cdots\,.
\label{Tn3}
\ee
We now do the same steps as the confining theory to find the related solutions to the confining theory. The final result for $S(\f)$ is given by the following series expansion
\be \label{Tn4}
S = S^{(0)}_\infty e^{ a \f}+S^{(1)}_\infty e^{ a \l \f}+\cdots\,,
\ee
where
\be \label{Tn5}
S^{(0)}_\infty = -\frac{\sqrt{d}}{\sqrt{2}{\ell}}\sp
S^{(1)}_\infty =\frac{a_1 d^{\frac72} (\l -1)^2 (d-\l )}{8 \sqrt{2} {\ell}^3 R_1 (2 d-\l -1) \left(d (\l -3)-\l^2+\l +2\right)}\,,
\ee
and
\be \label{Tn6}
\l = \frac{a(d+1)-\sqrt{a^2 (d-9) (d-1)+4}}{2 a}\,.
\ee

\section{Conifolds in Einstein-dilaton theory - A review\label{rev}}

In this appendix, we review very briefly the results of \cite{Ghodsi:2023pej}.
Consider the Einstein gravity with a cosmological constant in a $d+n+1$ dimensional bulk space-time parametrized by coordinates $\tilde{x}^a\equiv (\tu, \tilde{x}^\mu)$ where $\tu$ is the holographic coordinate
\be
S= M_P^{d+n-1} \int d\tilde{u}\, d^{d+n}\tilde{x} \sqrt{-\tilde g} \Big(
\tilde R + \frac{(d+n)(d+n-1)}{\tell^2}\Big)\,,
\label{nA2}
\ee
where $M_P$ is the $d+n+1$ dimensional Plank mass and $\tilde{\ell}$ is a length scale.
We consider the metric as a domain wall solution and slices that are a product of Einstein manifolds i.e.
\be\label{j1}
ds^2=d\tu^2+e^{2A_1(\tu)}\zeta^{1}_{\a\b} d\tilde{x}^{\a} d\tilde{x}^{\b} + e^{2A_2(\tu)}\zeta^{2}_{\m\n} d\tilde{x}^{\m} d\tilde{x}^{\n}\,.
\ee
Here $\zeta^1$ and $\zeta^2$ are the $AdS_d$ and $S^n$ metrics respectively. The non-trivial components of Einstein's equation are
\be
 \label{eq1}
 \big( d \dot{A_1}+n\dot A_2\big)^2 - d \dot{A}_1^2-n\dot{A}_2^2 - e^{-2A_1}
R_1- e^{-2A_2}R_2    = \frac{(d+n)(d+n-1)}{\tell^2}\,,
\ee
\be
 \label{eq2}
 (d+n-1) \big(d \ddot{A_1}+n\ddot{A_2}\big) +dn (\dot{A_1} - \dot{A_2})^2 + e^{-2A_1} R_1 +e^{-2A_2} R_2 = 0\,,
 \ee
 \be
 \label{eq3}
 \ddot{A_1} + \dot{A_1} ( d\dot{A_1}+n\dot{A_2}) - \frac{1}{d} e^{-2A_1} R_1 = \ddot{A_2} + \dot{A_2} ( d\dot{A_1}+n\dot{A_2})- \frac{1}{n} e^{-2A_2} R_2\,,
\ee
where $R_1$ and $R_2$ are the scalar curvatures of $AdS_d$ and $S^n$.

\subsection{Exact solutions\label{ESS}}
Before the analysis of the possible solutions, we should emphasize two exact solutions  of the equations of motion \eqref{eq1}--\eqref{eq3}:
\begin{itemize}

\item {\bf{$AdS_d \times AdS_{n+1}$ (product space) solution}}

In this case, the scale factors of $AdS_d$ and $S^n$ are given by
\be \label{exsol2}
e^{2A_1(\tu)}=-\frac{\tell^2 R_1}{d (d+n)}\sp
e^{2A_2(\tu)}=\frac{\tell^2 R_2}{(n-1)(d+n)}\sinh^2\Big(\sqrt{\frac{d+n}{n}}\frac{\tu-\tu_0}{\tell}\Big)\,,
\ee
which means that the metric describes a product space $AdS_d \times AdS_{n+1}$. This is a regular solution.

\item {\bf{Global $AdS_{d+n+1}$ solution}}

The next exact solution is the global $AdS$ solution
\be \label{glob0}
ds^2= d\tu^2 + e^{2\bar{A}_1} \cosh^2\frac{\tu-\tu_0}{\tell} ds^2_{AdS_d} + e^{2\bar{A}_2} \sinh^2\frac{\tu-\tu_0}{\tell} d\Omega_n^2\,,
\ee
where equations of motion fix the coefficients to
\be \label{glob1}
e^{2\bar{A}_1}=-\frac{\tell^2 R_1}{d(d-1)}\sp
e^{2\bar{A}_2}=\frac{\tell^2 R_2}{n(n-1)}\,.
\ee

\end{itemize}
\subsection{Regular and singular solutions}
The analysis of solutions in \cite{Ghodsi:2023pej} led us to four classes of end-points for solutions
\begin{itemize}

\item {\bf{B:}} An $AdS$-like boundary\footnote{In our solutions we have only one boundary, which we consider to be located at $\tu=+\infty$.} where both sphere and $AdS$ sizes diverge.

\item {\bf{R:}} A regular end-point where the sphere shrinks to a zero size and the $AdS$ scale factor goes to a constant value.

\item {\bf{A:}} This is a singular end-point where the sphere size diverges while the $AdS$ size vanishes.

\item {\bf{S:}}  This is another singular end-point where the sphere size vanishes while the $AdS$ size diverges.

\end{itemize}

According to the above possible end-points, we find the following types of solutions. Each solution is characterized by its end-points:

\begin{itemize}
\item (R, B)--type: This is a regular class of solutions.

\item (R, A)--type, (S, B)--type, (A, B)--type, (A, A)--type and (S, A)--type: These are all singular solutions.
\end{itemize}
 In the following, we  review different types of solutions:

\begin{itemize}

\item {\bf{(R, B)--type:}} This is a solution that starts from a regular end-point at $\tu=\tu_0$ and asymptotes to an $AdS$ boundary at $\tu\rightarrow +\infty$. At the end-point, the scale factor of the sphere is zero but the $AdS$ space has a finite size.
The expansion of scale factors near a regular solution at $\tilde{u}=\tilde{u}_0$ is given by
\begin{gather}\label{rega1}
e^{2A_1(\tu)} = a_0 + \frac{a_0 d (d + n) + \tell^2 R_1}{d \tell^2 (1 + n)} (\tu-\tu_0)^2 \nn \\
 -\frac{(a_0 d (d + n) +  \tell^2 R_1) (a_0 d (d - n-4) (d + n) + (d-3) \tell^2 R_1)}{ 3 a_0 d^2 \tell^4 (1 + n)^2 (3 + n)} (\tu-\tu_0)^4 \nn \\
 + \mathcal{O}(\tu-\tu_0)^6\,, \\
e^{2A_2(\tu)} = \frac{R_2}{n(n-1)} (\tu-\tu_0)^2
+\frac{(a_0 (d - d^2 + n + n^2) - \tell^2 R_1) R_2}{3 a_0 \tell^2 n^2 (n^2-1)} (\tu-\tu_0)^4\nn \\
+ \mathcal{O}(\tu-\tu_0)^6\,.\label{rega2}
\end{gather}
The quantity $a_0$ is a non-zero positive (but otherwise arbitrary) constant.
The regular (R, B)-type solution exists as far as $a_0>a_0^c$ where
\be \label{acprod}
a^c_0\equiv -\frac{\tell^2 R_1}{d (d+n)}\,.
\ee
An example of this type is sketched in figure \ref{shrink1}.
\begin{figure}[!t]
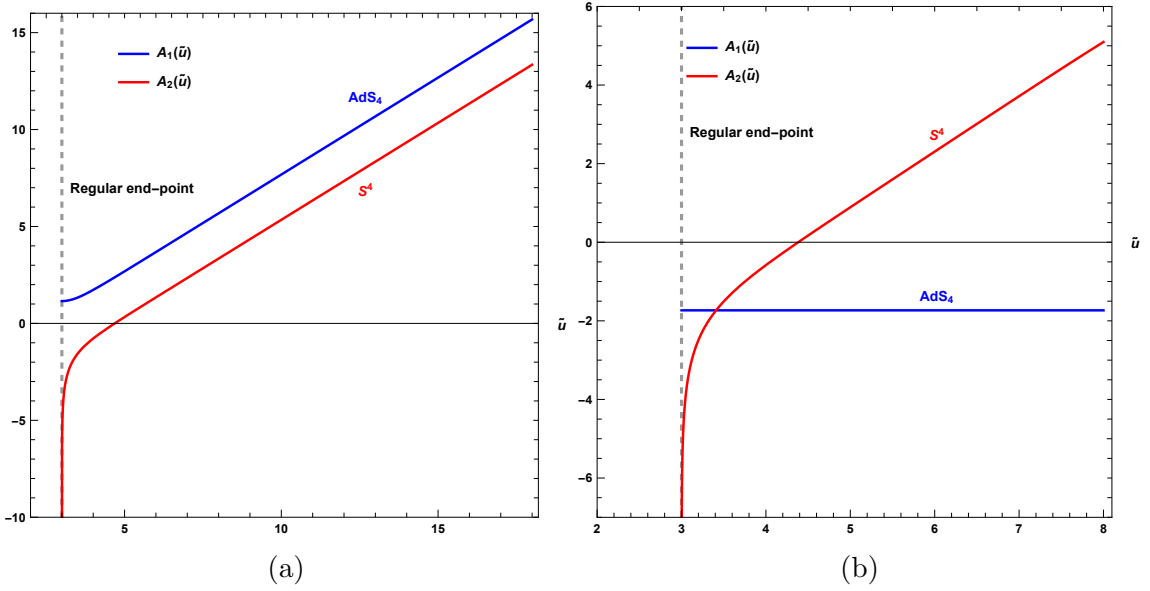

\centering
\begin{subfigure}{0.49\textwidth}
\includegraphics[width=1\textwidth]{figures/shrink1}
\caption{}\label{shrink1}
\end{subfigure}
\begin{subfigure}{0.49\textwidth}
\includegraphics[width=1\textwidth]{figures/shrink0}
\caption{}\label{shrink0}
\end{subfigure}
\caption{\footnotesize{(a): (R, B)--type: The scale factors of $AdS$ and $S$ (blue and red curves), start at a regular end-point (dashed line). At this point, the sphere scale factor shrinks to a zero size but $AdS$ has a finite non-zero size. Both scale factors reach the $AdS$ boundary. (b): $AdS_d\times AdS_{n+1}$ solution \eqref{exsol2}.}}
\end{figure}
We should note that the product space solution $AdS_d\times AdS_{n+1}$ is a single solution corresponding to choosing $a_0=a_0^c$ from \eqref{acprod}.
Figure \ref{shrink0} shows this solution.

\item {\bf{(R, A)--type:}}  If we choose  $a_0<a_0^c$,  then although we start from a regular end-point at $\tu=\tu_0$, the $AdS$ scale factor decreases until it reaches zero at a finite $\tu>\tu_0$. This is a singular end-point. An example of this singular solution is given in figure \ref{shrink2}.

\begin{figure}[!t]
\begin{center}
\begin{subfigure}{0.49\textwidth}
\includegraphics[width=1\textwidth]{figures/shrink2}
\caption{}\label{shrink2}
\end{subfigure}
\begin{subfigure}{0.47\textwidth}
\includegraphics[width=1\textwidth]{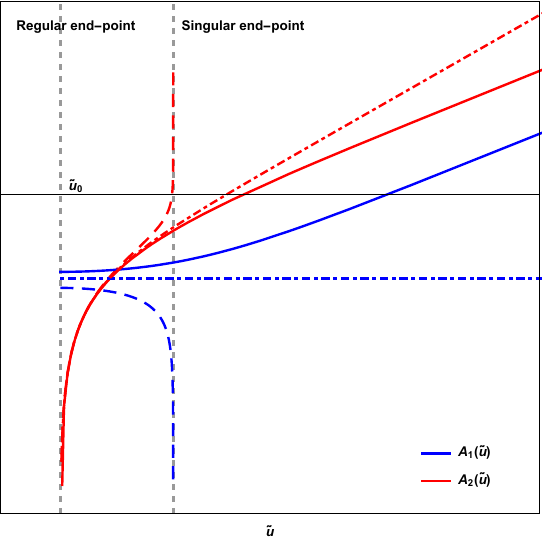}
\caption{}\label{UVIR1}
\end{subfigure}
\end{center}
\caption{\footnotesize{(a): (R, A)--type solution is a singular solution that starts at a regular end-point (left dashed line) and reaches a singular end-point (right dashed line). (b): Transition between solutions as we decrease the initial value of $a_0$. The solid curves show an example of (R, B)--type. By decreasing $a_0$ and at a specific point $a_0=a_0^c$ we have the $AdS_{d}\times AdS_{n+1}$ solution (dot-dashed curves). Below that point, all solutions are the (R, A)--type.}}
\end{figure}
\end{itemize}

At an arbitrary regular end-point $\tu_0$, as we decrease the scale factor of $AdS_d$ i.e. $a_0=e^{2A_1(\tu_0)}$, we observe the transition between the above solutions. This is sketched in figure \ref{UVIR1}.

We also have two other singular solutions which have a shrinking sphere:
\begin{figure}[!t]
\begin{center}
\begin{subfigure}{0.49\textwidth}
\includegraphics[width=1\textwidth]{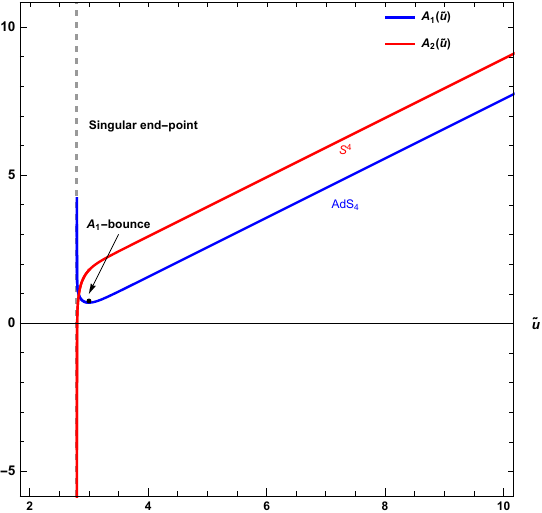}
\caption{}\label{AB1}
\end{subfigure}
\begin{subfigure}{0.49\textwidth}
\centering
\includegraphics[width=1\textwidth]{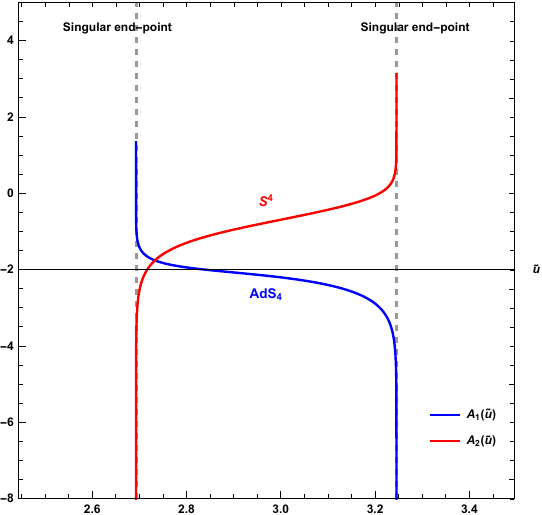}
\caption{\footnotesize{}}\label{TypeS4}
\end{subfigure}
\end{center}
\caption{(a): \footnotesize{(S, B)--type: Left to the $A_1$-bounce there is a singular IR end-point. Both scale factors reach the UV boundary. (b): An example of (S, A)--type, a monotonic solution with two singular end-points.}}
\end{figure}
\begin{figure}[!t]
\begin{center}
\begin{subfigure}{0.49\textwidth}
\includegraphics[width=1\textwidth]{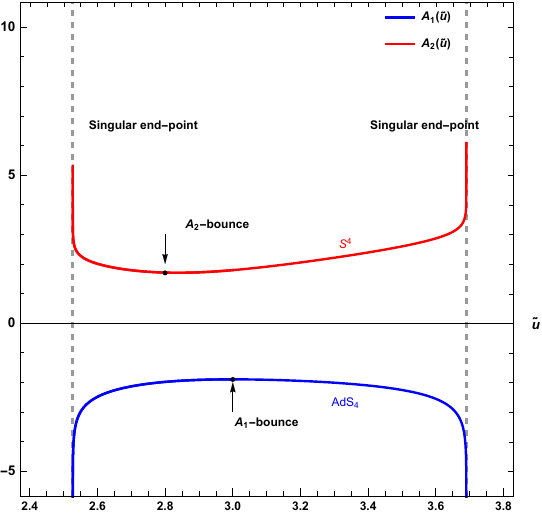}
\caption{}\label{AB2}
\end{subfigure}
\begin{subfigure}{0.49\textwidth}
\includegraphics[width=1\textwidth]{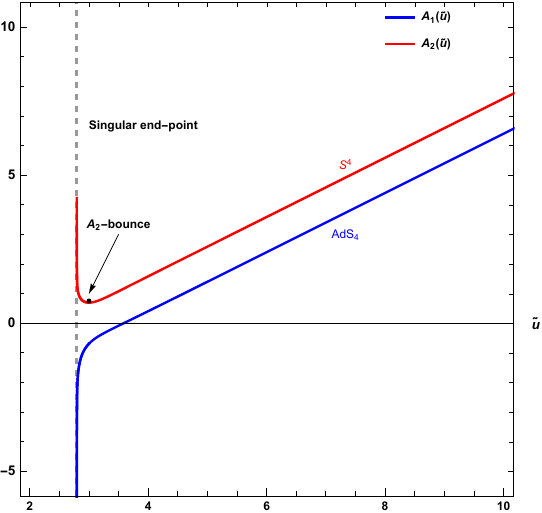}
\caption{}\label{AB3}
\end{subfigure}
\end{center}
\caption{(a): \footnotesize{(a): (A, A)--type: An example of solutions with two singular end-points.  (b): (A, B)--type: Left to the $A_2$-bounce there is a singular end-point where the $AdS$ scale factor is zero but the sphere scale factor diverges. Both scale factors reach the $AdS$ boundary.}}
\end{figure}

{\bf{(S, B)--type:}} Figure \ref{AB1} shows a solution with an $A_1$-bounce. On the left-hand side of the bounce, the solution has a singular end-point i.e. the sphere shrinks but the scale factor of $AdS$ diverges. On the right-hand side, there is an $AdS$ boundary at $\tu\rightarrow +\infty$.

{\bf{(S, A)--type:}} These are solutions with two singular end-points. At the left end-point, the $AdS_d$ scale factor diverges but $S^n$ shrinks. However at the right end-point, the $AdS_d$ shrinks and $S^n$ diverges, see figure \ref{TypeS4}.

The are also two other singular solutions which $AdS$ is shrinking now:

{\bf{(A, A)--type:}} This is a solution with one $A_1$-bounce and one $A_2$-bounce, see figure \ref{AB2}. On both sides of these bounces the scale factor of the sphere is diverging but for $AdS$ space it shrinks to zero and so on both sides, we have singular end-points.

{\bf{(A, B)--type:}} Figure \ref{AB3} shows an example of solutions with just one $A_2$-bounce.
On the left-hand side of the sphere bounce, the solution has a singular end-point where the scale factor of the sphere diverges but the $AdS$ scale shrinks to zero. On the right-hand side, there is an $AdS$ boundary as $\tu\rightarrow +\infty$.

\section{On-shell action}\label{osa}
The action on the gravity side contains three parts
\be\label{H1}
S=S_1+S_2+S_3\,,
\ee
\be
S_1\equiv M_P^{d-1}\int du d^{d}x\sqrt{ g}\left[ R-{\frac12}(\pa{\f})^2- V({\f})\right]\,,
\label{H1a}\ee
\be
S_2\equiv 2M_P^{d-1} \int_{{B}_+\cup{B}_-} d^d x \sqrt{\gamma} K\,,
\label{H1b}\ee
\be
S_3\equiv 2M_P^{d-1} \int_{B_3} d^d \tilde{x} \sqrt{\tilde{\gamma}} \tilde{K}\,.
\label{H1c}\ee
$S_1$ is the bulk action. $S_2$ and $S_3$ contain the GHY boundary terms. Here $B_+$ and $B_-$ are the asymptotically $AdS_{d+1}$ boundaries and we may have two, one, or zero such boundaries.
Their metric $\gamma$ is the induced metric, proportional to the metric $\zeta$ of the slice, defined in (\ref{eq:metric}), (\ref{eq:Rzeta}).
They appear for $u\to \pm\infty$, and we shall regularize them by putting them at $u_{\pm}$. In the end, we take $u_{\pm}\to\pm\infty$.

$S_3$ is the GHY term for the side boundary $B_3$, which is the union of the boundaries of the slices.
It is an asymptotically $AdS_{d}$ boundary times the $u$ coordinate.
If the slices are compact (and therefore have no boundary), then there is no $B_3$ boundary.
The renormalization in that case gives a different result for the on-shell action and we present this in section \ref{worm2}.

Without loss of generality, we consider this metric $\zeta_{\m\n}$ to be $AdS_d$ in Poincar\'e coordinates.
The results below, do not depend on this choice, as long as the slice metric has an asymptotically $AdS_d$ boundary.
We obtain
\be \label{H2}
ds^2=du^2+e^{2A(u)}\frac{\a^2}{\xi^2}\big(d\xi^2+\eta_{ab}dx^adx^b\big) \,,
\ee
where $\eta_{ab}$, $a,b=1,2,\cdots,d-1$ is the $(d-1)$-dimensional Minkowski metric.
Therefore, the slice boundary $B_3$ is located at $\xi=0$, and the regulated $B_3$ boundary at $\xi=\hepsilon$.

The induced metric, normal vector, and extrinsic curvature of each boundary are as follows:
\begin{itemize}
\item (regularized) $B_+$ and $B_-$ boundaries: $x=\{\xi,x^a\}$
\be \label{H3}
ds^2_{\gamma^{\pm}}=\gamma^{\pm}_{\mu\nu}dx^{\mu} dx^{\nu}=
e^{2A(u_{\pm})}\zeta_{\m\n}dx^{\m}dx^{\nu}= e^{2A(u_{\pm})}\frac{\a^2}{\xi^2}\big(d\xi^2+\eta_{ab}dx^adx^b\big)
\,,
\ee
\be  \label{H4}
 n^{\pm}_M=\pm\delta_M^u \sp K^{\pm}_{\mu\nu}=\mp\dot{A}(u_{\pm})\zeta_{\mu\nu} \sp K^{\pm}=\mp d\dot{A}(u_{\pm})\,.
\ee
where $M,N$ are bulk indices and $\mu,\nu$ are boundary indices.

\item (regularized) $B_3$ boundary: $\tilde{x}=\{u,x^a\}$
\be \label{H5}
ds^2_{\tilde{\g}}=du^2+e^{2A(u)}\frac{\a^2}{\hepsilon^2}\eta_{ab} dx^adx^b \sp \tilde{n}_M=\frac{\a e^A}{\hepsilon}\delta_M^\xi \,,
\ee
\be \label{H6}
\tilde{K}_{uu}=0 \sp \tilde{K}_{ab}=-\frac{\a e^{A(u)}}{\hepsilon^2}\eta_{ab}\sp \tilde{K}=-\frac{1}{\a}(d-1)e^{-A(u)} .
\ee

\end{itemize}

$\bullet$ {\bf{On-shell action for solutions with one-boundary}}\vspace{0.3cm}

To compute the on-shell action we notice that the equation of motion from the variation of the metric gives
\be \label{H7}
R=\frac12 (\p\f)^2+\frac{d+1}{d-1}V(\f)\,.
\ee
Using the above relation together with \eqref{eq:EOM1} and \eqref{eq:EOM2}, the regularized on-shell action for the sum of $S_1$ and $S_2$ in \eqref{H1} simplifies to (here we have considered a solution with one boundary at $u=u_+$ and ending at $u=u_0$)
\be \label{H8}
S_1+S_2=\frac{2M_P^{d-1}}{d}\int_{\hepsilon}^{+\infty} d\xi \int d^{d-1} x\sqrt{\zeta}\Big(
R^{(\zeta)}\int_{u_0}^{u_+} du\,  e^{(d-2)A}+d(d-1)e^{dA}\dot{A}\Big{|}^{u_+}_{u_0}\Big)\,,
\ee
where $\zeta=det \zeta_{\mu\nu}$ and $R^{(\z)}$ is given by \eqref{eq:Rzeta}.
Moreover, for the side boundary $B_3$ we find
\be \label{H9}
S_{3}=-2(d-1)M_P^{d-1}\int d^{d-1} x\int_{u_0}^{u_+}du\frac{\a^{d-2}}{\hepsilon^{d-1}}e^{(d-2)A} \,.
\ee
We introduce a scalar function $U(u)$ such that
\be \label{H10}
(d-2)\dot{A}U+\dot{U}=-1 \Rightarrow e^{(d-2)A}=-\frac{d}{du}\Big(U e^{(d-2)A}\Big)\,.
\ee
In this way, we can write
\begin{gather} \label{H11}
S_1+S_2
= \frac{2M_P^{d-1}}{d}\int_\epsilon^{+\infty} d\xi \int d^{d-1} x \sqrt{\zeta}\Big(-R^{(\zeta)}Ue^{(d-2)A}\Big|_{u_0}^{u_+}+d(d-1)e^{dA}\dot{A}\Big|^{u_+}_{u_0}\Big)\nn \\
= \frac{2M_P^{d-1}}{d(d-1)}\frac{\a^d}{\epsilon^{d-1}} \int d^{d-1} x \Big(-R^{(\zeta)}Ue^{(d-2)A}+d(d-1)e^{dA}\dot{A}\Big)\Big|^{u_+}_{u_0}\,,
\end{gather}
and
\begin{gather} \label{H12}
S_3=2(d-1)M_P^{d-1}\frac{\a^{d-2}}{\hepsilon^{d-1}}\int d^{d-1} xUe^{(d-2)A}\Big|_{u_0}^{+\infty} \nn \\
= \frac{2}{d}M_P^{d-1}\frac{\a^{d}}{\hepsilon^{d-1}}\int d^{d-1} x
\Big(-R^{(\zeta)}U e^{(d-2)A}\Big)\Big|_{u_0}^{+\infty} \,.
\end{gather}
Therefore, the total on-shell action is
\begin{gather}
S_{on-shell} =S_1+S_2+S_3 \nn \\
=\frac{2M_P^{d-1}}{d(d-1)}\frac{\a^d}{\hepsilon^{d-1}} \int d^{d-1} x \Big(-d R^{(\zeta)}Ue^{(d-2)A}+d(d-1)e^{dA}\dot{A}\Big)\Big|^{u_+}_{u_0}\nn \\
=\frac{2M_P^{d-1}}{d} V_{\hepsilon} (\a) \Big(-d R^{(\zeta)}Ue^{(d-2)A}+d(d-1)e^{dA}\dot{A}\Big)\Big|^{u_+}_{u_0}\,,\label{H13}
\end{gather}
where in the last line we have made explicit the  regularized volume of the slices
\be \label{H14}
V_{\hepsilon}(\alpha)=\int_{\hepsilon}^{+\infty} d\xi \int d^{d-1}x \sqrt{\zeta}=
\int d^{d-1}x \frac{\a^d}{(d-1)\hepsilon^{d-1}}\equiv \bar V_{\hepsilon}(1) \a^{d}\,.
\ee
so that our formulae are valid for any slice manifold with constant negative curvature and an asymptotically $AdS_{d}$ boundary.

\vspace{0.3cm}
$\bullet$ {\bf{On-shell action for solutions with two boundaries}}\vspace{0.3cm}

To obtain the on-shell action in this case, we must do the same steps as the one-boundary solution. The only change is to replace the end-point with the second boundary i.e. $u_0\rightarrow u_-$ in the equations above. Then the final results would be

\be \label{H15}
S_{on-shell} =\frac{2M_P^{d-1}}{d} V_{\hepsilon}(\a) \Big(-d R^{(\zeta)}Ue^{(d-2)A}+d(d-1)e^{dA}\dot{A}\Big)\Big|^{u_+}_{u_-}\,.
\ee

$\bullet$ {\bf{On-shell action for solutions with compact slice geometries}}\vspace{0.3cm}

When slices are compact, there is no side boundary, and therefore the $S_3$ part is absent in the computation of the on-shell action. For example for wormhole solutions stretched between two asymptotic boundaries we find the on-shell action as
\be \label{H16}
S_{on-shell} =\frac{2M_P^{d-1}}{d} V_{\hepsilon}(\a) \Big(- R^{(\zeta)}Ue^{(d-2)A}+d(d-1)e^{dA}\dot{A}\Big)\Big|^{u_+}_{u_-}\,,
\ee
where the only difference with \eqref{H15} is in the coefficient of the first term.

\section{Free energy of the CFT\label{fcft}}
In this appendix, we compute the free energy corresponding to the CFT solution with constant negative curvature slices (without any scalar vev on both $B_{\pm}$).

We consider the case that the scalar field is fixed at the maximum of the potential at $\f=0$ with   $V(0)=-\frac{d(d-1)}{\ell^2}$.
The general CFT solution with $\f=0$ with $V'(0)=0$ to the equations (\ref{eq:EOM1})-(\ref{eq:EOM3}) is
\be\label{I1}
e^{A(u)}=e^\frac{u-u_0}{\ell}+\frac{|R^{(\zeta)}|\ell^2}{4d(d-1)}e^{-\frac{u-u_0}{\ell}}\geq \frac{\ell\sqrt{|R^{(\zeta)}|}}{\sqrt{d(d-1)}}\,,
\ee
with $u_0$ an arbitrary constant.
The solution has two boundaries at $u\to \pm{\infty}$.

\begin{figure}[!t]
\begin{center}
\begin{subfigure}{0.49\textwidth}
\includegraphics[width=1\textwidth]{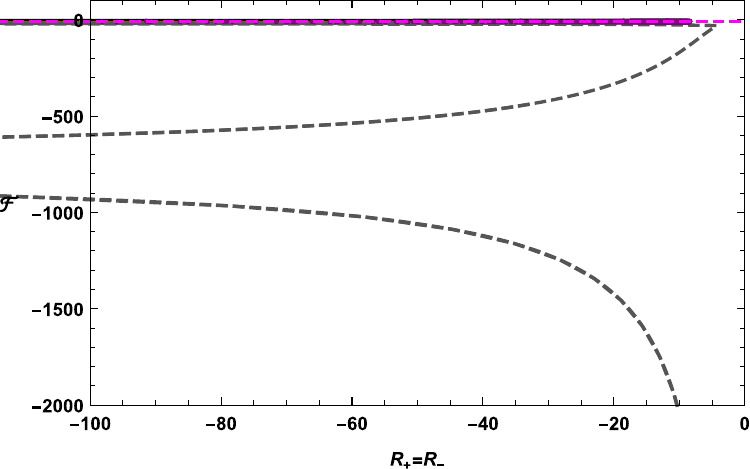}
\caption{}\label{CFT1a}
\end{subfigure}
\begin{subfigure}{0.48\textwidth}
\includegraphics[width=1\textwidth]{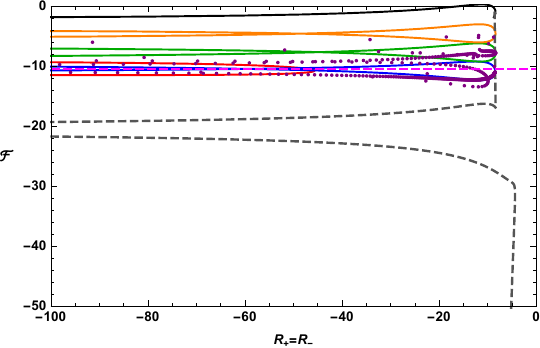}
\caption{}\label{CFT1}
\end{subfigure}
\end{center}
\caption{\footnotesize{The free energy of the solution with $\mathcal{R}_+=\mathcal{R}_-$. The free energy of solutions with two vevs corresponds to the values where $\mathcal{R}_+=\mathcal{R}_-\rightarrow -\infty$ (here large negative values). The magenta dashed line indicates the free energy of the CFT solution. Figure (b) is the zoomed region of figure (a) near this dashed line.}}
\end{figure}

Since we know the exact solution in \eqref{I1}, it would be simpler to compute the free energy directly (without the introduction of the scalar function $U(u)$). For solutions with two boundaries  at $u=u_{\pm}\rightarrow \pm\infty$, and with non-compact $B_3$ boundary, the regulated free energy is
\be \label{I2}
F_{reg}=-\frac{2M_P^{d-1}}{d}\int_{\hepsilon}^{+\infty} \!\! d\xi \int d^{d-1} x\sqrt{\zeta}\Big(
d R^{(\zeta)}\int_{u_-}^{u_+} du\,  e^{(d-2)A}+d(d-1)e^{dA}\dot{A}\Big{|}^{u_+}_{u_-}\Big)\,,
\ee
whereupon inserting \eqref{I1} for $d=4$ we obtain
\begin{gather}
F_{reg}=M_P^3\int_{\hepsilon}^{+\infty} \!\! d\xi \int d^{3} x\sqrt{\zeta}\Big(-\frac{6}{\ell}e^{\frac{4(u-u_0)}{\ell}}-\frac34 \ell R^{(\z)}e^{\frac{2(u-u_0)}{\ell}}+\frac{1}{12}\ell^2 (R^{(\z)})^2 u \nn \\
+\frac{\ell^5 (R^{(\z)})^3}{3072}e^{\frac{-2(u-u_0)}{\ell}}+\frac{\ell^7 (R^{(\z)})^4}{884736}e^{\frac{-4(u-u_0)}{\ell}}\Big)_{u_-}^{u_+}\,.\label{I3}
\end{gather}
To cancel the divergences we should include the following counter-terms on both boundaries at $u=u_{\pm}$
\be\label{I4}
F^{\pm}_{ct}=\pm M_P^3\int d^4 x \sqrt{\gamma^{\pm}}\Big(
\frac{6}{\ell}+\frac{5\ell}{4} R^{(\gamma^\pm)}+\frac{\ell^3}{12}(R^{(\gamma^\pm)})^2 \log\omega\epsilon
\Big)\Big|_{u_{\pm}}\,,
\ee
where $\gamma^{\pm}$ is the induced metric on $u_\pm$.
Adding \eqref{I3} and \eqref{I4} one finds the renormalized free energy as
\be \label{I5}
F_{ren}=\lim_{u_{\pm}\to \pm\infty}(F_{reg}+F_{ct}^+ + F_{ct}^-)=M_P^3\int_{\hepsilon}^{+\infty} \!\! d\xi \int d^{3} x\sqrt{\zeta}\Big(\frac{1}{96} \ell^3 (R^{(\z)})^2 (-7 + 16 \log\omega)\Big)\,.
\ee
In the scheme of $\omega=1$ and by using \eqref{H14} and \eqref{eq:Rzeta} we obtain
\be  \label{I6}
F_{ren}=-\frac{21}{2}M_P^3 \ell^3 {\bar{V}}_{\hepsilon}(1)\,.
\ee
To compare the free energy of the CFT solution with those solutions which are vevs on both sides,  we have created figure \ref{CFT1}. In this figure, the free energy for solutions with two vevs corresponds to the values where $\mathcal{R}_+=\mathcal{R}_-\rightarrow -\infty$. The magenta dashed line indicates the free energy of the CFT solution without vevs.

We conclude from figure \ref{CFT1} that out of all competing connected two-boundary geometries, the lowest free energy is the one belonging to the bottom part of the red branch. Therefore the first vev solution is the dominant saddle point. When we allow disconnected solutions (dashed grey curves), figure
\ref{CFT1}  indicates that the overall dominant solution is the leading disconnected solution with nontrivial vevs.

To conclude, in the two-sided case, the maximally symmetric solution with trivial scalar is not the dominant vev-vev solution: rather, a non-zero vev is dynamically generated on both sides and the dominant saddle is the product of two disconnected one-boundary solutions.

\begin{figure}[!t]
\begin{center}
\begin{subfigure}{0.49\textwidth}
\includegraphics[width=1\textwidth]{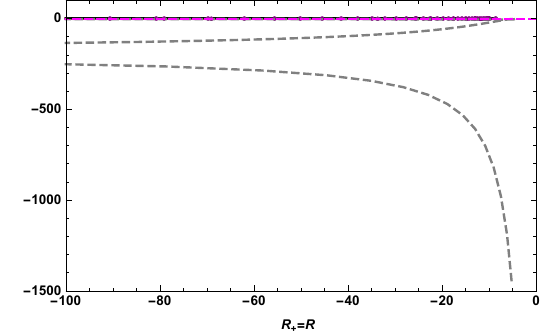}
\caption{}\label{CFT2a}
\end{subfigure}
\begin{subfigure}{0.48\textwidth}
\includegraphics[width=1\textwidth]{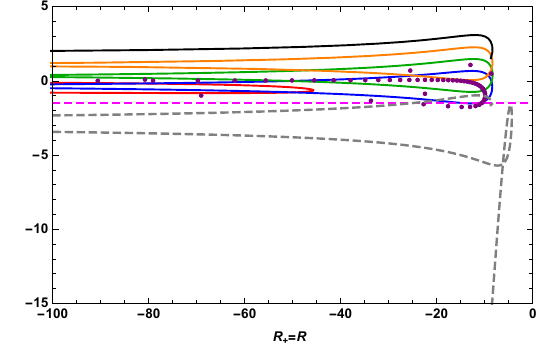}
\caption{}\label{CFT2}
\end{subfigure}
\end{center}
\caption{\footnotesize{The free energy of solutions with two vevs corresponds to the values where $\mathcal{R}_+=\mathcal{R}_-\rightarrow -\infty$ (here large negative values). The magenta dashed lines indicate the free energy of the CFT solution.}}\end{figure}

In the case of compact slice geometries the regulated free energy is different and it becomes
\be \label{I7}
F_{reg}=-\frac{2M_P^{d-1}}{d}\int_{\hepsilon}^{+\infty} \!\! d\xi \int d^{d-1} x\sqrt{\zeta}\Big(
 R^{(\zeta)}\int_{u_-}^{u_+} du\,  e^{(d-2)A}+d(d-1)e^{dA}\dot{A}\Big{|}^{u_+}_{u_-}\Big)\,.
\ee
The counter-terms are
\be\label{I8}
F^{\pm}_{ct}=\pm M_P^3\int d^4 x \sqrt{\gamma^{\pm}}\Big(
\frac{6}{\ell}+\frac{1\ell}{2} R^{(\gamma^\pm)}+\frac{\ell^3}{48}(R^{(\gamma^\pm)})^2 \log\omega\epsilon
\Big)^{u_{\pm}}\,.
\ee
Then the renormalized free energy would be
\be \label{I9}
F_{ren}=M_P^3\int_{\hepsilon}^{+\infty} \!\! d\xi \int d^{3} x\sqrt{\zeta}\Big(\frac{1}{96} \ell^3 (R^{(\z)})^2 (-1 + 4 \log\omega)\Big)\,.
\ee
In the scheme of $\omega=1$
\be \label{I10}
F_{ren}=-\frac{3}{2}M_P^3 \ell^3 {\bar{V}}_{\hepsilon}(1)\,.
\ee
To compare the free energy of the CFT solution (dashed magenta line) with those solutions with two vevs, we have marked on figure \ref{CFT2} with a dashed line the CFT solution free energy.
It is clear that here, the CFT two-boundary solution dominates all other connected solutions with non-trivial vevs. However, the disconnected solution with non-trivial vevs is still the dominant one overall.

\end{appendix}



\begin{thebibliography}{100}




\bibitem{Witten:1998zw}
E.~Witten,
``Anti-de Sitter space, thermal phase transition, and confinement in gauge theories,''
\hrj{10.4310/ATMP.1998.v2.n3.a3}{Adv. Theor. Math. Phys. \textbf{2} (1998), 505-532};
\hre{hep-th}{9803131}.

\bibitem{KK}
  E.~Kiritsis and C.~Kounnas,
  {\em ``Infrared regularization of superstring theory and the one loop calculation of coupling constants,''}
 \hrj{10.1016/0550-3213(95)00156-M}{Nucl.\ Phys.\ B {\bf 442} (1995) 472};
  \hre{hep-th}{9501020}.\\
{\em ``Curved four-dimensional space-times as infrared regulator in superstring theories,''}
\hrj{10.1016/0920-5632(95)00441-B}{Nucl.\ Phys.\ Proc.\ Suppl.\  {\bf 41} (1995) 331};
  \hre{hep-th}{9410212}.


\bibitem{CW}
C.~G.~Callan, Jr. and F.~Wilczek,
{\em ``Infrared behavior at negative curvature,''}
\hrj{10.1016/0550-3213(90)90451-I}{Nucl. Phys. B \textbf{340} (1990), 366-386}.

\bibitem{A2}
O.~Aharony, M.~Berkooz, D.~Tong and S.~Yankielowicz,
{\em ``Confinement in Anti-de Sitter Space,''}
\hrj{10.1007/JHEP02(2013)076}{JHEP \textbf{02} (2013), 076};
\hri{1210.5195}{[hep-th]}.

\bibitem{Maldacena:2004rf}
J.~M.~Maldacena and L.~Maoz,
{\em ``Wormholes in AdS,''}
\hrj{10.1088/1126-6708/2004/02/053}{JHEP \textbf{02} (2004), 053};
\hre{hep-th}{0401024}.

\bibitem{Buchel:2002wf}
A.~Buchel,
{\em ``Gauge / gravity correspondence in accelerating universe,''}
\hrj{10.1103/PhysRevD.65.125015}{Phys. Rev. D \textbf{65} (2002), 125015};
\hre{hep-th}{0203041}.


\bibitem{Marolf:2010tg}
D.~Marolf, M.~Rangamani and M.~Van Raamsdonk,
{\em ``Holographic models of de Sitter QFTs,''}
\hrj{10.1088/0264-9381/28/10/105015}{Class. Quant. Grav. \textbf{28} (2011), 105015};
\hri{1007.3996}{[hep-th]}.

\bibitem{Blackman:2011in}
J.~Blackman, M.~B.~McDermott and M.~Van Raamsdonk,
{\em ``Acceleration-Induced Deconfinement Transitions in de Sitter Spacetime,''}
\hrj{10.1007/JHEP08(2011)064}{JHEP \textbf{08} (2011), 064};
\hri{1105.0440}{[hep-th]}.

\bibitem{C}
J.~K.~Ghosh, E.~Kiritsis, F.~Nitti and L.~T.~Witkowski,
{\em``Holographic RG flows on curved manifolds and quantum phase transitions,''}
\hrj{10.1007/JHEP05(2018)034}{JHEP \textbf{05} (2018), 034};
\hri{1711.08462}{[hep-th]}.




\bibitem{F}
J.~K.~Ghosh, E.~Kiritsis, F.~Nitti and L.~T.~Witkowski,
{\em ``Holographic RG flows on curved manifolds and the $F$-theorem,''}
\hrj{10.1007/JHEP02(2019)055}{JHEP \textbf{02} (2019), 055};
\hri{1810.12318}{[hep-th]}

\bibitem{Ghosh:2020qsx}
J.~K.~Ghosh, E.~Kiritsis, F.~Nitti and L.~T.~Witkowski,
{\em ``Back-reaction in massless de Sitter QFTs: holography, gravitational DBI action and f(R) gravity,''}
\hrj{10.1088/1475-7516/2020/07/040}{JCAP \textbf{07} (2020), 040};
\hri{2003.09435}{[hep-th]}.

\bibitem{s2s2}
E.~Kiritsis, F.~Nitti and E.~Pr\'eau,
{\em ``Holographic QFTs on $S^{2}\times S^{2}$, spontaneous symmetry breaking and Efimov saddle points,''}
\hrj{10.1007/JHEP08(2020)138}{JHEP \textbf{08} (2020), 138};
\hri{2005.09054}{ [hep-th]}.


\bibitem{s3}
E.~Kiritsis and C.~Litos,
{\em ``Holographic RG flows on Squashed S$^{3}$,''}
\hrj{10.1007/JHEP12(2022)161}{JHEP \textbf{12} (2022), 161};
\hri{2209.14342}{ [hep-th]}.


\bibitem{Ghosh:2021lua}
J.~K.~Ghosh, E.~Kiritsis, F.~Nitti and L.~T.~Witkowski,
{\em ``Revisiting Coleman-de Luccia transitions in the AdS regime using holography,''}
\hrj{10.1007/JHEP09(2021)065}{JHEP \textbf{09} (2021), 065};
\hri{2102.11881}{[hep-th]}.

\bibitem{Ghodsi:2022umc}
A.~Ghodsi, J.~K.~Ghosh, E.~Kiritsis, F.~Nitti and V.~Nourry,
{\em ``Holographic QFTs on AdS$_d$, wormholes and holographic interfaces,''}
\hrj{10.1007/JHEP01(2023)121}{JHEP \textbf{01} (2023), 121};
\hri{2209.12094}{ [hep-th]}.


\bibitem{Ghodsi:2023pej}
A.~Ghodsi, E.~Kiritsis and F.~Nitti,
{\em ``Holographic CFTs on AdS$_{d}\times{}$ S$^{n}$ and conformal defects,''}
\hrj{10.1007/JHEP10(2023)188}{JHEP \textbf{10} (2023), 188};
\hri{2309.04880}{[hep-th]}.

\bibitem{Donos:2017sba}
A.~Donos, J.~P.~Gauntlett, C.~Rosen and O.~Sosa-Rodriguez,
{\em ``Boomerang RG flows with intermediate conformal invariance,''}
\hrj{10.1007/JHEP04(2018)017}{JHEP \textbf{04} (2018), 017};
\hri{1712.08017}{[hep-th]}.


\bibitem{multirg}
  E.~Kiritsis, F.~Nitti and L.~S.~Pimenta,
  {\em ``Exotic RG Flows from Holography,''}
\hrj{10.1002/prop.201600120}{  Fortsch. Phys. \textbf{65} (2017) no.2, 1600120};
\hri{1611.05493}{[hep-th]};\\
 F.~Nitti, L.~Silva Pimenta and D.~A.~Steer,
{\em ``On multi-field flows in gravity and holography,''}
\hrj{10.1007/JHEP07(2018)022}{JHEP \textbf{07} (2018), 022};
\hri{1711.10969 }{[hep-th]}.





  \bibitem{iQCD}
 U.~Gursoy    and E.~Kiritsis,
  {\em ``Exploring improved holographic theories for QCD: Part I,''}
  \hrj{10.1088/1126-6708/2008/02/032}{JHEP {\bf 0802} (2008) 032};
\hri{0707.1324}{[hep-th]};\\
  U.~Gursoy, E.~Kiritsis and F.~Nitti,
  {\em ``Exploring improved holographic theories for QCD: Part II,''}
  \hrj{10.1088/1126-6708/2008/02/019}{JHEP {\bf 0802} (2008) 019};
  \hri{0707.1349}{[hep-th]};\\
  U.~Gursoy, E.~Kiritsis, L.~Mazzanti, G.~Michalogiorgakis and F.~Nitti,
  {\em ``Improved Holographic QCD,''}
  \hrj{10.1007/978-3-642-04864-7}{Lect.\ Notes Phys.\  {\bf 828} (2011) 79};
\hri{1006.5461}{[hep-th]}.

\bibitem{thermo}
U.~Gursoy, E.~Kiritsis, L.~Mazzanti and F.~Nitti,
{\em ``Deconfinement and Gluon Plasma Dynamics in Improved Holographic QCD,''}
\hrj{10.1103/PhysRevLett.101.181601}{Phys. Rev. Lett. \textbf{101} (2008), 181601};
\hri{0804.0899}{ [hep-th]};\\
U.~Gursoy, E.~Kiritsis, L.~Mazzanti and F.~Nitti,
  {\em ``Holography and Thermodynamics of 5D Dilaton-gravity,''}
  \hrj{10.1088/1126-6708/2009/05/033}{JHEP {\bf 0905}, 033 (2009)};
\hri{0812.0792}{[hep-th]}.


\bibitem{KS}
Y.~Kinar, E.~Schreiber and J.~Sonnenschein,
{\em ``Q anti-Q potential from strings in curved space-time: Classical results,''}
\hrj{10.1016/S0550-3213(99)00652-5}{Nucl. Phys. B \textbf{566} (2000), 103-125};
\hre{hep-th}{9811192}.



\bibitem{GK}
B.~Gouteraux and E.~Kiritsis,
{\em ``Generalized Holographic Quantum Criticality at Finite Density,}
\hrj{10.1007/JHEP12(2011)036}{JHEP \textbf{12}, 036 (2011)};
\hri{1107.2116}{[hep-th]}.

\bibitem{Gouteraux:2011qh}
B.~Gouteraux, J.~Smolic, M.~Smolic, K.~Skenderis and M.~Taylor,
{\em ``Holography for Einstein-Maxwell-dilaton theories from generalized dimensional reduction,''}
\hrj{10.1007/JHEP01(2012)089}{JHEP \textbf{01}, 089 (2012)}
\hri{arXiv:1110.2320}{[hep-th]}.

\bibitem{Gubser}
S.~S.~Gubser,
  {\em ``Curvature singularities: The good, the bad, and the naked,''}
\hrj{10.4310/ATMP.2000.v4.n3.a6}{ Adv.\ Theor.\ Math.\ Phys.\  {\bf 4} (2000) 679-745};
  \hre{hep-th}{0002160}.

\bibitem{BP}
P.~Betzios, E.~Kiritsis and O.~Papadoulaki,
{\em ``Euclidean Wormholes and Holography,''}
\hrj{10.1007/JHEP06(2019)042}{JHEP \textbf{06} (2019), 042}.
\hri{1903.05658}{ [hep-th]]}.
{\em ``Interacting systems and wormholes,''}
\hrj{10.1007/JHEP02(2022)126}{JHEP \textbf{02} (2022), 126};
\hri{2110.14655Z}{ [hep-th]}.



\bibitem{Son}
D.~B.~Kaplan, J.~W.~Lee, D.~T.~Son and M.~A.~Stephanov,
{\em ``Conformality Lost,''}
\hrj{10.1103/PhysRevD.80.125005}{Phys. Rev. D \textbf{80} (2009), 125005};
\hri{0905.4752}{ [hep-th]}.


\bibitem{E1}
K.~Jensen, A.~Karch, D.~T.~Son and E.~G.~Thompson,
{\em ``Holographic Berezinskii-Kosterlitz-Thouless Transitions,''}
\hrj{10.1103/PhysRevLett.105.041601}{Phys. Rev. Lett. \textbf{105} (2010), 041601};
\hri{1002.3159}{ [hep-th]}.


\bibitem{E2}
N.~Iqbal, H.~Liu and M.~Mezei,
{\em ``Quantum phase transitions in semilocal quantum liquids,''}
\hrj{10.1103/PhysRevD.91.025024}{Phys. Rev. D \textbf{91} (2015) no.2, 025024};
\hri{1108.0425}{ [hep-th]}.

\bibitem{JK}
M.~Jarvinen and E.~Kiritsis,
{\em ``Holographic Models for QCD in the Veneziano Limit,''}
\hrj{10.1007/JHEP03(2012)002}{JHEP \textbf{03} (2012), 002};
\hri{1112.1261}{ [hep-ph]};\\
  M.~Jarvinen,
{\em ``Massive holographic QCD in the Veneziano limit,''}
\hrj{10.1007/JHEP07(2015)033}{JHEP \textbf{07} (2015), 033};
\hri{1501.07272}{ [hep-ph]}.


\bibitem{Bak}
D.~Bak, M.~Gutperle and S.~Hirano,
{\em ``A Dilatonic deformation of AdS(5) and its field theory dual,''}
\hrj{10.1088/1126-6708/2003/05/072}{JHEP \textbf{05} (2003), 072};
\hre{hep-th}{0304129}.

\bibitem{CF}
A.~B.~Clark, D.~Z.~Freedman, A.~Karch and M.~Schnabl,
{\em ``Dual of the Janus solution: An interface conformal field theory,''}
\hrj{10.1103/PhysRevD.71.066003}{Phys. Rev. D \textbf{71} (2005), 066003};
\hre{hep-th}{0407073}.

\bibitem{DH2}
E.~D'Hoker, J.~Estes and M.~Gutperle,
{\em``Ten-dimensional supersymmetric Janus solutions,''}
\hrj{10.1016/j.nuclphysb.2006.08.017}{Nucl. Phys. B \textbf{757} (2006), 79-116};
\hre{0603012}{hep-th}.

\bibitem{GW1}
D.~Gaiotto and E.~Witten,
{\em ``Janus Configurations, Chern-Simons Couplings, And The theta-Angle in N=4 Super Yang-Mills Theory,''}
\hrj{10.1007/JHEP06(2010)097}{JHEP \textbf{06} (2010), 097};
\hri{0804.2907 }{[hep-th]}.

\bibitem{GW2}
D.~Gaiotto and E.~Witten,
{\em ``S-Duality of Boundary Conditions In N=4 Super Yang-Mills Theory,''}
\hrj{10.4310/ATMP.2009.v13.n3.a5}{Adv. Theor. Math. Phys. \textbf{13} (2009) no.3, 721-896};
\hri{0807.3720 }{[hep-th]}.


\bibitem{DH}
E.~D'Hoker, J.~Estes and M.~Gutperle,
{\em ``Exact half-BPS Type IIB interface solutions. I. Local solution and supersymmetric Janus,''}
\hrj{10.1088/1126-6708/2007/06/021}{JHEP \textbf{06} (2007), 021};
\hri{0705.0022}{ [hep-th]}.

\bibitem{JO}
K.~Jensen and A.~O'Bannon,
{\em ``Holography, Entanglement Entropy, and Conformal Field Theories with Boundaries or Defects,''}
\hrj{10.1103/PhysRevD.88.106006}{Phys. Rev. D \textbf{88} (2013) no.10, 106006};
\hri{1309.4523}{[hep-th]}.

\bibitem{C1}
I.~Arav, K.~C.~M.~Cheung, J.~P.~Gauntlett, M.~M.~Roberts and C.~Rosen,
{\em ``Spatially modulated and supersymmetric mass deformations of $ \mathcal{N} $ = 4 SYM,''}
\hrj{10.1007/JHEP11(2020)156}{JHEP \textbf{11} (2020), 156};
\hri{2007.15095}{ [hep-th]}.

\bibitem{C2}
I.~Arav, K.~C.~M.~Cheung, J.~P.~Gauntlett, M.~M.~Roberts and C.~Rosen,
{\em ``A new family of $AdS_4$ S-folds in type IIB string theory,''}
\hrj{10.1007/JHEP05(2021)222}{JHEP \textbf{05} (2021), 222};
\hri{2101.07264}{ [hep-th]}.

\bibitem{KR}
A.~Karch and L.~Randall,
{\em ``Locally localized gravity,''}
\hrj{10.1088/1126-6708/2001/05/008}{JHEP \textbf{05} (2001), 008};
\hre{hep-th}{0011156}.


\bibitem{BCFT1}
T.~Takayanagi,
{\em ``Holographic Dual of BCFT,''}
\hrj{10.1103/PhysRevLett.107.101602}{Phys. Rev. Lett. \textbf{107} (2011), 101602};
\hri{1105.5165}{ [hep-th]}.

\bibitem{BCFT2}
M.~Fujita, T.~Takayanagi and E.~Tonni,
{\em ``Aspects of AdS/BCFT,''}
\hrj{10.1007/JHEP11(2011)043}{JHEP \textbf{11} (2011), 043};
\hri{1108.5152}{[hep-th]}.

\bibitem{Gut}
M.~Gutperle and J.~Samani,
{\em ``Holographic RG-flows and Boundary CFTs,''}
\hrj{doi:10.1103/PhysRevD.86.106007}{Phys. Rev. D \textbf{86} (2012), 106007};
\hri{1207.7325}{[hep-th]}.









\bibitem{limit} S. D. Glazek and K. G. Wilson {\em Renormalization of overlapping transverse
divergences in a model light-front Hamiltonian}, Phys. Rev. D 47, 4657 (1993);\\
 P.~F.~Bedaque, H.~W.~Hammer and U.~van Kolck,
{\em ``Effective theory of the triton,''}
\hrj{10.1016/S0375-9474(00)00205-0}{Nucl. Phys. A \textbf{676} (2000), 357-370};
\hre{nucl-th}{9906032};\\
A.~LeClair, J.~M.~Roman and G.~Sierra,
{\em ``Log periodic behavior of finite size effects in field theories with RG limit cycles,''}
\hrj{10.1016/j.nuclphysb.2004.08.033}{Nucl. Phys. B \textbf{700} (2004), 407-435};
\hre{hep-th}{0312141}.








\bibitem{RG1}
I.~Brunner and D.~Roggenkamp,
{\em ``Defects and bulk perturbations of boundary Landau-Ginzburg orbifolds,''}
\hrj{10.1088/1126-6708/2008/04/001}{JHEP \textbf{04} (2008), 001}
\hri{0712.0188}{ [hep-th]}.

\bibitem{RG2}
D.~Gaiotto,
{\em ``Domain Walls for Two-Dimensional Renormalization Group Flows,''}
\hrj{10.1007/JHEP12(2012)103}{JHEP \textbf{12} (2012), 103}
\hri{1201.0767}{ [hep-th]}.

\bibitem{RG3}
A.~Konechny and C.~Schmidt-Colinet,
{\em ``Entropy of conformal perturbation defects,''}
\hrj{10.1088/1751-8113/47/48/485401}{J. Phys. A \textbf{47} (2014) no.48, 485401}
\hri{1407.6444} {[hep-th]}.


\bibitem{RG4}
A.~Konechny,
{\em ``RG boundaries and interfaces in Ising field theory,''}
\hrj{10.1088/1751-8121/aa60f6}{J. Phys. A \textbf{50} (2017) no.14, 145403}
\hri{1610.07489}{ [hep-th]};
{\em ``Properties of RG interfaces for 2D boundary flows,''}
\hrj{10.1007/JHEP05(2021)178}{JHEP \textbf{05} (2021), 178}
\hri{2012.12361}{ [hep-th]}.




\bibitem{Pap}
I.~Papadimitriou,
{\em ``Multi-Trace Deformations in AdS/CFT: Exploring the Vacuum Structure of the Deformed CFT,''}
\hrj{10.1088/1126-6708/2007/05/075}{JHEP \textbf{05} (2007), 075}
\hre{hep-th}{0703152}.










\end{thebibliography}
\end{document}